\documentclass[12pt,a4paper,reqno]{book}

\usepackage[headheight=20mm,top=25mm,inner=20mm,outer=20mm,centering]{geometry}%

\usepackage[nouppercase]{scrpage2}

\usepackage[english]{babel}

\usepackage{amsmath}
\usepackage{amssymb}
\usepackage{amsfonts}
\usepackage{mathrsfs}

\usepackage{palatino}

\usepackage{graphicx}

\usepackage{bbm}

\usepackage{hyperref}

\usepackage{mathtools}

\usepackage{slashed}
\usepackage{braket}

\usepackage{dsfont}

\usepackage{accents}

\usepackage{booktabs}

\usepackage{tikz} 
\usetikzlibrary{calc}
\usetikzlibrary{arrows}
\usetikzlibrary{trees}
\usetikzlibrary{decorations.pathmorphing}
\usetikzlibrary{decorations.markings}

\usepackage{comment}

\usepackage{xspace}
\hypersetup{
    colorlinks,
    linkcolor={red!50!black},
    citecolor={green!50!black},
    urlcolor={blue!50!black}
}

\usepackage{subfig}

\usepackage[font=small]{caption}

\makeatletter
 \let\old@startsection=\@startsection
 \let\oldl@section=\l@section
 \renewcommand{\@startsection}[6]{\old@startsection{#1}{#2}{#3}{#4}{#5}{#6\mathversion{bold}}}
 \renewcommand{\l@section}[2]{\oldl@section{\mathversion{bold}#1}{#2}}
\makeatother

\DeclareMathOperator{\Str}{Str}
\DeclareMathOperator{\tr}{tr}

\newcommand{\CFT}{\textup{CFT}}

\newcommand{\Smat}{\mathcal{S}}

\newcommand{\comm}[2]{[#1,#2]}
\newcommand{\acomm}[2]{\{#1,#2\}}

\newcommand{\alg}[1]{\mathfrak{#1}}

\newcommand{\su}{\alg{su}}
\newcommand{\so}{\alg{so}}
\newcommand{\sls}{\alg{sl}}

\newcommand{\psu}{\alg{psu}}

\newcommand{\ag}{\alg{g}}

\newcommand{\gen}[1]{\mathbf{#1}}

\renewcommand{\leq}{\leqslant}

\newcommand{\alert}[1]{{\color[rgb]{1,0,0} #1}}

\newcommand{\adsfive}{AdS$_5\times$S$^5$}

\newcommand{\bos}{\alg{b}}
\newcommand{\fer}{\alg{f}}

\newcommand{\lagr}{\mathscr{L}}

\usepackage[us,24hr]{datetime}

\def\Xint#1{\mathchoice
  {\XXint\displaystyle\textstyle{#1}}%
  {\XXint\textstyle\scriptstyle{#1}}%
  {\XXint\scriptstyle\scriptscriptstyle{#1}}%
  {\XXint\scriptscriptstyle\scriptscriptstyle{#1}}%
  \!\int}
\def\XXint#1#2#3{{\setbox0=\hbox{$#1{#2#3}{\int}$}
    \vcenter{\hbox{$#2#3$}}\kern-.5\wd0}}
\def\pint{\;\Xint-}

\newcommand{\ce}{\text{c.e.}}

\newcommand{\de}{\textup{d}}

\newcommand{\suA}{\bullet}

\newcommand{\algSL}{\alg{sl}}

\newcommand{\algSU}{\alg{su}}

\newcommand{\genQ}{\gen{Q}}

\newcommand{\1}{\mathbf{1}}

\newcommand{\ie}{\textit{i.e.}\xspace}
\newcommand{\eg}{\textit{e.g.}\xspace}

\newcommand{\cchi}{\underline{\chi}{}}
\newcommand{\cbchi}{\underline{\widetilde{\chi}}{}}

\newcommand{\sL}{\mbox{\tiny L}}
\newcommand{\sR}{\mbox{\tiny R}}
\newcommand{\sI}{\mbox{\tiny I}}
\newcommand{\sJ}{\mbox{\tiny J}}

\newcommand{\smallL}{\sL}
\newcommand{\smallR}{\sR}

\newcommand{\vacuum}{\mathbf{0}}

\newcommand{\st}{\text{st}}

\newcommand{\ratiosigma}{\sigma^-}

\newcommand{\BES}{\text{BES}}

\newcommand{\AFS}{\text{AFS}}
\newcommand{\HL}{\text{HL}}

\newcommand{\curvearrowurl}{\curvearrowleft}
\newcommand{\curvearrowdlr}{\rotatebox[origin=c]{180}{$\curvearrowleft$}}
\newcommand{\inturl}{\;{\int \negthickspace \negthickspace \negthickspace\negthickspace \negthinspace \curvearrowurl}\mbox{ }\,} 
\newcommand{\intdlr}{\;{\int \negthickspace \negthickspace \negthickspace \negthinspace \curvearrowdlr}\mbox{ }\,} 
\newcommand{\ointc}{\;{\int \negthickspace \negthickspace \negthickspace \negthinspace \circlearrowleft}\mbox{ }\,}

\newcommand{\CDD}{\scriptscriptstyle\text{CDD}}

\newcommand{\Fup}{F^{\rotatebox[origin=c]{360}{$\scriptstyle\curvearrowleft$}}}
\newcommand{\Fdw}{F^{\rotatebox[origin=c]{180}{$\scriptstyle\curvearrowleft$}}}

\tikzset{
particle/.style={draw=blue,line width=1, postaction={decorate},
    decoration={markings,mark=at position .5 with {\arrow[draw=blue]{>}}}},
antiparticle/.style={draw=blue,line width=1, postaction={decorate},
    decoration={markings,mark=at position .5 with {\arrow[draw=blue]{<}}}},
particlecross/.style={draw=red,line width=1,dashed, postaction={decorate},
    decoration={markings,mark=at position .5 with {\arrow[draw=red]{>}}}},
antiparticlecross/.style={draw=red,line width=1,dashed, postaction={decorate},
    decoration={markings,mark=at position .5 with {\arrow[draw=red]{<}}}},
nopart/.style={draw=none}
 }

\tikzstyle dynkin node=[very thick,shape=circle,draw,inner sep=0pt,minimum size=5mm]
\tikzstyle dynkin line=[very thick]
\tikzstyle inverse line=[gray,line width=1.46pt,line cap=round, dash pattern=on 0pt off 2\pgflinewidth]
\tikzstyle red phase=[thick,red,decoration={snake,amplitude=0.1mm,segment length=1.6mm},decorate]
\tikzstyle blue phase=[thick,blue,decoration={snake,amplitude=0.1mm,segment length=0.9mm},decorate]
\tikzstyle green phase=[thick,darkgreen,decoration={snake,amplitude=0.1mm,segment length=0.9mm},decorate]
\tikzstyle brown phase=[thick,orange,decoration={snake,amplitude=0.1mm,segment length=0.9mm},decorate]

\tikzstyle arrow=[thick,rounded corners=18pt,-latex]
\tikzstyle box=[draw,rounded corners,outer sep=4pt]
\tikzstyle B node=[outer sep=0pt]
\tikzstyle Q node=[inner sep=1pt,outer sep=0pt]

\pgfdeclarelayer{background}
\pgfdeclarelayer{foreground}
\pgfsetlayers{background,main,foreground}

\newcommand{\adsthree}{AdS$_3\times$S$^3\times$T$^4$}

\newcommand{\perm}{\mathbf{\pi}}

\usepackage{color}

\definecolor{grey}{rgb}{0.4,0.4,0.5}
\definecolor{darkgreen}{rgb}{0,0.5,0}
\definecolor{darkred}{rgb}{0.6,0.0,0}
\definecolor{lightbrown}{rgb}{1,0.9,0.8}
\definecolor{brown}{rgb}{0.6,0.3,0.3}
\definecolor{darkblue}{rgb}{0,0,0.8}
\definecolor{darkmagenta}{rgb}{0.5,0,0.5}

\def\cT{{\mathcal T}}

\def\de{\delta }

\def\Xint#1{\mathchoice
  {\XXint\displaystyle\textstyle{#1}}%
  {\XXint\textstyle\scriptstyle{#1}}%
  {\XXint\scriptstyle\scriptscriptstyle{#1}}%
  {\XXint\scriptscriptstyle\scriptscriptstyle{#1}}%
  \!\int}
\def\XXint#1#2#3{{\setbox0=\hbox{$#1{#2#3}{\int}$}
    \vcenter{\hbox{$#2#3$}}\kern-.5\wd0}}
\def\pint{\;\Xint-}

\def\be{\begin{equation}}
\def\ee{\end{equation}}

\newcommand{\bea}{\begin{eqnarray}}
\newcommand{\eea}{\end{eqnarray}}
\newcommand{\bei}{\begin{itemize}}
\newcommand{\eei}{\end{itemize}}
\newcommand{\bee}{\begin{enumerate}}
\newcommand{\eee}{\end{enumerate}}

\newcommand{\ads}{${\rm  AdS}_5\times {\rm S}^5\ $}

\def\ov{\over}

\def\la{\label}

\def\a {\alpha}
\def\b {\beta}
\def\s {\sigma}
\def\g {\gamma}
\def\om {\omega}

\def\p{\phi}
\def\vp{\varphi}
\def\vk{\varkappa}

\def\e{\eta}
\def\t{\theta}
\def\z{\zeta}
\def\eps{\epsilon}
\def\r {\rho}
\def\G{\Gamma}

\def\pa {\partial}

\newcommand{\op}{\mathcal{O}}
\newcommand{\optilde}{\widetilde{\mathcal{O}}}
\newcommand{\opinv}{\op^{\text{inv}}}
\newcommand{\gb}{\alg{g_b}}
\newcommand{\gf}{\alg{g_f}}

\newcommand{\Ab}{A^\alg{b}}
\newcommand{\Af}{A^\alg{f}}

\newcommand{\ul}[1]{\underline{#1}}

\newcommand{\ferm}[4]{#1_{#2 \, \ul{#3}}^{\ \, \ul{#4}}}
\def\bg{\boldsymbol{\gamma}}

\def\L{\mathscr L}
\def\mI{\mathbbm{1}}

\def\qdp{\upsilon}
\def\dpnu{\nu}

\newcommand{\etaadsfive}{(AdS$_5\times$S$^5)_\eta$}

\newcommand{\T}{\Theta}

\newcommand{\genQind}[3]{\gen{Q}^{#1 \, \ul{#2}}_{\ \, \ul{#3}}}

\newcommand{\as}{\alg{s}}

\begin{document}

\frontmatter

\pagestyle{empty}
\thispagestyle{empty}

\begin{flushright}\footnotesize\ttfamily
Imperial-TP-RB-2016-03
\\
\end{flushright}
\vspace{5em}

\begin{center}
\textbf{\Large\mathversion{bold} Integrable strings for AdS/CFT}

\vspace{2em}

\textrm{\large Riccardo Borsato} 

\vspace{2em}

{\itshape
The Blackett Laboratory, Imperial College,\\  SW7 2AZ, London,
U.K.\\[0.2cm]
}

\vspace{1em}

\texttt{R.Borsato@imperial.ac.uk}


\end{center}

\vspace{6em}
{\small 

\begin{center}
\textbf{Abstract}
\end{center}

\vspace{10pt}

\noindent
In this PhD thesis we review some aspects of integrable models related to string backgrounds or their deformations.
In the first part we develop methods to obtain exact results in the AdS$_3/$CFT$_2$ correspondence. We consider the \adsthree \ background with pure Ramond-Ramond flux and we find the all-loop worldsheet S-matrix by exploiting the symmetries of the model in light-cone gauge. 
As we naturally include the massless modes on the worldsheet, we derive the full set of Bethe-Yang equations.
In the massive sector we give also a spin-chain description and we write down solutions compatible with crossing for the scalar factors which are not constrained by simmetries.
In the second part of the thesis we consider the so-called ``$\eta$-deformation'' of the superstring on \adsfive.
We first discuss the effects of the deformation at the level of the bosonic $\sigma$-model, and we match the tree-level worldsheet scattering processes to the expansion of the $q$-deformed S-matrix. To identify the missing Ramond-Ramond fields we then compute the action quadratic in fermions, and we provide an alternative derivation by looking at the kappa-symmetry variations. The resulting background fields do not solve the equations of motion of type IIB supergravity and we comment on this.

}

\mainmatter

\tableofcontents
\cleardoublepage
\pagestyle{scrheadings}


${\phantom{.}{}}$

\begin{flushright}

\vspace{4.5cm}

\textbf{\textit{ To Anna.
\\
To my family.}}

\vspace{3.5cm}
{\footnotesize
Marco Polo descrive un ponte, pietra per pietra. 
\\
``Ma qual \`e la pietra che sostiene il ponte?'' chiede
Kublai Kan.
\\
``Il ponte non \`e sostenuto da questa o quella pietra,'' risponde Marco, 
\\``ma dalla linea dell'arco che esse formano.''
\\
Kublai Kan rimane silenzioso, riflettendo. Poi soggiunge: 
\\
``Perch\'e mi parli delle pietre? \`E solo dell'arco che
m'importa.''
\\
Polo risponde: ``Senza pietre non c'\`e arco.''

\vspace{24pt}

\textit{Le citt\`a invisibili, Italo Calvino}

\vspace{96pt}

Marco Polo describes a bridge, stone by stone.
\\
``But which is the stone that supports the bridge?''
Kublai Khan asks.
\\
``The bridge is not supported by one stone or another,''
Marco answers, 
\\``but by the line of the arch that they
form.''
\\
Kublai Khan remains silent, reflecting. Then he adds:
\\
``Why do you speak to me of the stones? It is only the arch
that matters to me.''
\\
Polo answers: ``Without stones there is no arch.''

\vspace{24pt}

\textit{Invisible Cities, Italo Calvino}
}

\end{flushright}

\chapter{Introduction}\label{ch:intro}

Together with great achievements, general relativity and quantum field theory come with unresolved problems.
Among others, these include quantisation of gravity and a feasible description of strongly-coupled gauge theories.
Recent developments have proved that string theory is a useful framework to investigate open issues in theoretical Physics.
One sign of this success may be seen in the discovery of dualities, through which we are able to relate seemingly different concepts.
In the most celebrated ``holographic duality'', gravity and gauge theories turn out to be two sides of the same coin.
A prominent role in the study of this correspondence has been played by Integrability.
The term ``Integrability'' is very broad, and actually collects many different concepts---from classical integrable models to factorisation of scattering in two-dimensional quantum field theories, et cetera.
For the moment we point out that methods borrowed from Integrability allow one to obtain exact results that go beyond the usual perturbative analysis, thus giving stringent tests for holography.
Interestingly, Integrability provides a new language to describe both the string and the gauge theory forming the holographic pair. 
In this introductory chapter we review some of these achievements and explain how this thesis fits in this context.


\bigskip

Maldacena's proposal~\cite{Maldacena:1997re} currently known as the AdS/CFT correspondence is a concrete version of the holographic principle anticipated by 't~Hooft in 1974~\cite{'tHooft:1973jz}.
AdS/CFT conjectures the equivalence between a gravity theory living in an (asymptotically) Anti-de Sitter (AdS) spacetime in $d+1$ dimensions, and a gauge theory---or more precisely a conformal field theory (CFT)---in flat $d$-dimensional Minkowski spacetime.
Often one refers to AdS as the \emph{bulk} and interprets the gauge theory as living on the \emph{boundary} of this spacetime.

The best understood example of this conjecture is the pair AdS$_5$/CFT$_4$.
On the one side we have string theory on the ten-dimensional background {\adsfive}, the product of a five-dimensional Anti-de Sitter and a five-dimensional sphere. 
On the other side we find $\mathcal{N}=4$ super Yang-Mills (SYM), the maximally supersymmetric gauge theory in four dimensions.
Although the equivalence is believed to hold precisely at any point in the parameter space, it becomes more testable in the \emph{planar} or \emph{large-$N$} limit.

For the gauge theory, $N$ is the number of colors of the gauge group $SU(N)$, and as pointed out already by 't~Hooft~\cite{'tHooft:1973jz} sending $N\to \infty$ is a way to simplify the problem while keeping some of its non-trivial features. 
In some sense it is an approximation along a direction different from the usual perturbation theory, the latter being an expansion in the number of loops. 
At large $N$ an indefinite number of loops remains, but at leading order only \emph{planar} graphs survive.
These are the Feynman graphs that can be drawn on genus-zero surfaces like the sphere, as opposed to the ones that can be drawn only on surfaces with handles.
To be more precise,  when $N\to \infty$ we have to send the Yang-Mills coupling $g_{\text{YM}}$ to zero in such a way that the effective coupling $\lambda=g_{\text{YM}}^2 N$ remains finite. 
After this limit is taken, one may still implement the usual perturbation theory by performing an expansion at small values of $\lambda$---the point $\lambda=0$ corresponding to the free theory.

On the string side, the planar limit corresponds to considering free, \ie non interacting, strings. 
More precisely, for $N$ large the string coupling constant $g_s$ is related to the 't~Hooft coupling $\lambda$ as $g_s=\lambda/4 \pi N$ and it tends to zero, while the tension $g=\sqrt{\lambda}/2 \pi$ remains finite.
These relations show one of the most exciting features of the AdS/CFT correspondence, namely that it is a weak/strong duality.
In fact, the regime in which the string theory is more tractable is not for small values of $\lambda$, but rather for $\lambda\gg 1$. 
This is when the tension is large and the string moves like a rigid object.
Therefore, if we decide to make our life easier by considering the ``simple'' regime for the string, this actually corresponds to the---usually unaccessible---gauge theory at strong coupling, and vice versa!

\bigskip

The AdS/CFT correspondence is made quantitatively more precise by saying how to match observables on the two sides of the duality.
In particular, conformal dimensions of operators correspond to the energies of the dual string configurations~\cite{Witten:1998qj,Gubser:1998bc}.
Integrability for gauge and string theory has the power of computing \emph{exactly} the dependence of these observables on the effective coupling $\lambda$.
We are then able to go beyond a perturbative expansion at weak or at strong coupling, and we can actually interpolate the spectrum between the two sides for any finite value of $\lambda$.
The achievement is not just computational, but also conceptual. In fact, with these methods we find a unified description of both the gauge and the string theory in a single \emph{quantum integrable model} in $1+1$ dimensions.
From the point of view of the gauge theory, the interpretation is that of a spin-chain with long-range interactions; the different flavors of the field content of $\mathcal{N}=4$ SYM correspond in fact to the directions of the spins. 
For the string, the quantum integrable model is the one arising on the worldsheet after gauge-fixing, where the excitations now correspond to the bosonic and fermionic coordinates that parameterise the spacetime in which the superstring is living.
Integrability allows one to compute the all-loop S-matrix governing the scattering of the excitations, on the spin-chain or on the worldsheet.
The remarkable fact is that both sides of the AdS/CFT correspondence lead to the same result.

\paragraph{Gauge theory and Integrability}
The first hint about the presence of Integrability in the large-$N$ limit appeared on the gauge theory side\footnote{Integrability in the context of four-dimensional gauge theories appeared already in~\cite{Lipatov:1993yb,Faddeev:1994zg}, where it was shown that it manifests itself in some specific regimes of Quantum Chromodynamics (QCD).}. 
In their seminal paper~\cite{Minahan:2002ve}, Minahan and Zarembo showed that the problem of finding the spectrum of the gauge theory can be rephrased in terms of an \emph{integrable spin-chain}.
Let us say a few words about this.

Because of conformal symmetry, interesting observables to consider are the conformal dimensions of gauge-invariant operators.
These are formed by taking traces of products of fields, where the trace is needed to ensure gauge invariance\footnote{In the large-$N$ limit it is enough to consider single-trace operators, as at leading order the conformal dimension of multi-trace operators is additive.}.
The one-dimensional object that we get by taking this product already suggests how a spin-chain comes into the game.
The various fields here play the role of the spins pointing in different directions in the space of flavors. 
Because of the cyclicity of the trace, we can anticipate that what we consider are \emph{periodic} spin-chains.
The analogy becomes more precise when one computes loop corrections to the dilatation operator.
For simplicity let us focus on an ``$\su(2)$ sector'' with just two scalar fields of $\mathcal{N}=4$ SYM, that we will interpret as ``spin up'' and ``spin down''. One finds that at one loop the operator measuring the anomalous dimension mixes operators that differ by permutations of fields sitting at neighbouring sites.
Its expression actually matches that of the Hamiltonian of the Heisenberg spin-chain, solved by Bethe with a method that is commonly called the \emph{Bethe Ansatz}~\cite{Bethe:1931hc}.
It is then really nice to discover that we can compute anomalous dimensions by using the same diagonalisation techniques.

The story is not restricted to this $\su(2)$ sector, as the complete dilatation operator at one loop still has the form of an integrable Hamiltonian~\cite{Beisert:2003jj}.
Also higher loops can be accounted for~\cite{Beisert:2003tq,Beisert:2003jj}, and one finds that at higher orders interactions become more and more non-local, meaning that not only nearest-neighbour sites are coupled, but also next-to-nearest neighbour, et cetera.
We refer to~\cite{Rej:2010ju} for a review.

\bigskip

In the Heisenberg spin-chain that arises at one-loop, an S-matrix can be defined to describe the scattering of the magnons.
It turns out that the key object on which one should focus to obtain all-loop results is not the Hamiltonian---which becomes more and more complicated at higher orders---but rather the S-matrix.
In fact, this can be fixed even at finite values of the effective coupling by imposing compatibility with the symmetries~\cite{Beisert:2005tm}, which for $\mathcal{N}=4$ SYM are given by two copies of a central extension of the Lie superalgebra $\su(2|2)$.
The possibility of bootstrapping the S-matrix is a consequence, on the one hand, of the presence of these powerful symmetries, and on the other hand of the knowledge of the exact dependence of the central charges on the momenta of the excitations and on the effective coupling~\cite{Beisert:2005tm}.
It is clear that this bootstrapping method relies on the assumption that Integrability extends to all loop orders; 
let us review here some important points and refer to Chapter~\ref{ch:S-matrix-T4} for a more detailed discussion.

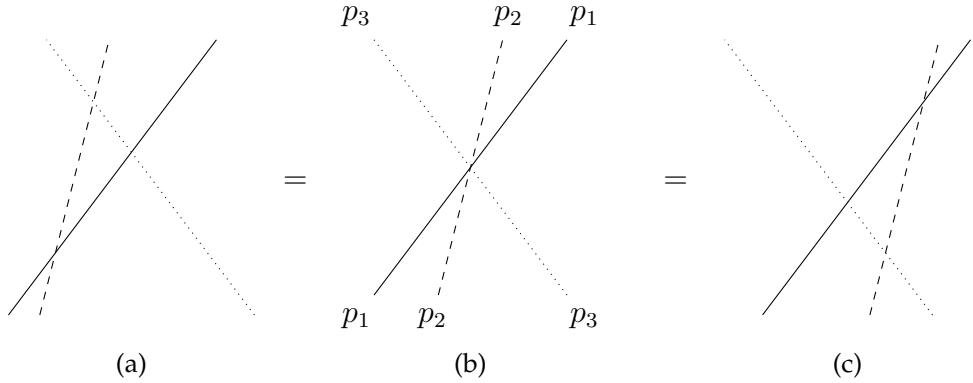
\begin{figure}[t]
  \centering
\hspace{-0.75cm}
 \subfloat[\label{fig:YB-left-Intro}]{
  \begin{tikzpicture}[%
    box/.style={outer sep=1pt},
    Q node/.style={inner sep=1pt,outer sep=0pt},
    arrow/.style={-latex}
    ]%
    \node [box] (p1in) at ($(-1.5cm,-2cm)+(0.5cm,0cm)$) {};
    \node [box] (p2in) at (-0.5cm,-2cm) {};
    \node [box] (p3in) at ($(+1.5cm,-2cm)+(1cm,0cm)$) {};

    \node [box] (p1out) at ($(+1.5cm,2cm)+(0.5cm,0cm)$) {};
    \node [box] (p2out) at (+0.5cm,2cm) {};
    \node [box] (p3out) at ($(-1.5cm,2cm)+(1cm,0cm)$) {};

	\draw (p1in) -- (p1out);
	\draw [dashed] (p2in) -- (p2out);
	\draw [dotted] (p3in) -- (p3out);

  \end{tikzpicture}
}
\raisebox{2cm}{$=$}
\hspace{0cm}
 \subfloat[\label{fig:YB-central-Intro}]{
  \begin{tikzpicture}[%
    box/.style={outer sep=1pt},
    Q node/.style={inner sep=1pt,outer sep=0pt},
    arrow/.style={-latex}
    ]%
    \node [box] (p1in) at (-1.5cm,-2cm) {$p_1$};
    \node [box] (p2in) at (-0.5cm,-2cm) {$p_2$};
    \node [box] (p3in) at (+1.5cm,-2cm) {$p_3$};

    \node [box] (p1out) at (+1.5cm,2cm) {$p_1$};
    \node [box] (p2out) at (+0.5cm,2cm) {$p_2$};
    \node [box] (p3out) at (-1.5cm,2cm) {$p_3$};

	\draw (p1in) -- (p1out);
	\draw [dashed] (p2in) -- (p2out);
	\draw [dotted] (p3in) -- (p3out);

  \end{tikzpicture}
}
\hspace{0.5cm}
\raisebox{2cm}{$=$}
 \subfloat[\label{fig:YB-right-Intro}]{
  \begin{tikzpicture}[%
    box/.style={outer sep=1pt},
    Q node/.style={inner sep=1pt,outer sep=0pt},
    arrow/.style={-latex}
    ]%
    \node [box] (p1in) at ($(-1.5cm,-2cm)+(0cm,0cm)$) {};
    \node [box] (p2in) at ($(-0.5cm,-2cm)+(0.5cm,0cm)$) {};
    \node [box] (p3in) at ($(+1.5cm,-2cm)+(-0.5cm,0cm)$) {};

    \node [box] (p1out) at ($(+1.5cm,2cm)+(0cm,0cm)$) {};
    \node [box] (p2out) at ($(+0.5cm,2cm)+(0.5cm,0cm)$) {};
    \node [box] (p3out) at ($(-1.5cm,2cm)+(-0.5cm,0cm)$) {};

	\draw (p1in) -- (p1out);
	\draw [dashed] (p2in) -- (p2out);
	\draw [dotted] (p3in) -- (p3out);
 
  \end{tikzpicture}
}
 \caption{The vertical axis corresponds to propagation in time, while the horizontal axis parameterises space. A process like the one in the center involving three particles may be factorised as a sequence of two-body scatterings in two possible ways, as in the left or right figure. Consistency between the two factorisations yields the Yang-Baxter equation.}
  \label{fig:Yang-Baxter-Intro}
\end{figure}

The S-matrix that is considered dictates just $2\to 2$ scattering.
This is enough for integrable models, since the number of particles is conserved and generic $N\to N$ scatterings can be derived just from the knowledge of the two-body S-matrix. 
This is a crucial property of integrable theories that goes under the name of \emph{factorisation} of scattering, and we refer to~\cite{Dorey:1996gd} for a nice review. 
The idea is that thanks to the large amount of symmetry generators, one is allowed to move the wave packets of the excitations independently, to disentangle interactions in such a way that only two particles are involved every time an interaction takes place. 
If this is possible, then any generic process can be reinterpreted as a sequence of two-body interactions.
In Figure~\ref{fig:Yang-Baxter-Intro} we show how this works in the example of the scattering of three particles.
Notice that in this case we have two different possibilities to achieve factorisation.
It is clear that for consistency it should not matter which choice of factorisation we pick.
This imposes a constraint on the two-body S-matrix $\mathcal{S}$ that goes under the name of \emph{Yang-Baxter equation}.
It is found by equating the left and right processes in Figure~\ref{fig:Yang-Baxter-Intro}
$$
\mathcal{S}_{23}\cdot\mathcal{S}_{13}\cdot\mathcal{S}_{12}
=
\mathcal{S}_{12}\cdot\mathcal{S}_{13}\cdot\mathcal{S}_{23}\,,
$$
and one can check that this is enough to ensure consistency also for factorisation of scattering with more than three particles.
 
Symmetries actually allow one to fix the S-matrix only up to an overall scalar function, which in this context is called the ``dressing factor''~\cite{Arutyunov:2004vx}.
Without this factor ``just'' ratios of scattering elements would be known exactly in the effective coupling $\lambda$. 
However, the dressing factor for $\mathcal{N}=4$ SYM is actually known~\cite{Beisert:2006ez}, and it can be found by solving the equation obtained by imposing \emph{crossing invariance}~\cite{Janik:2006dc,Arutyunov:2006iu}, which relates physical processes to the ones in which one particle is analytically continued to an unphysical channel.

As we will explain in more detail later, in this thesis we will show how it is possible to apply similar methods to a specific instance of the AdS$_3$/CFT$_2$ correspondence, allowing us to find an all-loop S-matrix for that case. 
This opens the possibility of implementing the same program that proved to be successful for AdS$_5$/CFT$_4$, and solve another dual pair exactly in the planar limit, suggesting that the presence of Integrability might be more general than expected.
Rather than the ones used for $\mathcal{N}=4$ SYM, the methods we exploit are actually borrowed from the description of strings on {\adsfive}.
Let us briefly review the situation there.

\paragraph{String theory and Integrability}
In parallel to the findings for the gauge theory, developments were achieved also on the string theory side of the AdS/CFT correspondence.
The string is described as a non-linear $\sigma$-model on the background {\adsfive}, and thanks to the realisation in terms of the supercoset $\text{PSU}(2,2|4)/(\text{SO}(4,1)\times \text{SO}(5))$ it is possible to write down its action to all orders in the fields~\cite{Metsaev:1998it}.

Integrability starts appearing at the classical level.
In fact, the equations of motion for the superstring on the background {\adsfive} admit a formulation in terms of a Lax connection $L_\a(z,\tau,\sigma)$~\cite{Bena:2003wd}, which depends on the worldsheet coordinates $(\tau,\sigma)$ and a \emph{spectral parameter} that we denote by $z$. It is a way to encode the dynamics of the model, as the flatness condition 
$$
\pa_\tau L_\sigma -\pa_\sigma  L_\tau - [L_\tau,L_\sigma]=0\,,
$$
provides the equations of motion for the string.
The Lax connection is of primary importance, since expanding the trace of its path-ordered exponential---that goes under the name of transfer matrix---around any value of the spectral parameter generates the complete tower of conserved charges of the system.
These charges can be used to construct solutions of the equations of motion.
Classical integrability is inherited also by reduced models, obtained by confining the motion of the string to specific dynamics, and one of the finite-dimensional integrable models that may be recovered is \eg the Neumann model~\cite{Arutyunov:2003uj,Arutyunov:2003za}.

\bigskip

The action for the string has a local invariance, which generates unphysical modes that should be removed by fixing a gauge.
It turns out that, to make contact with the description of the gauge theory, the proper gauge choice is a combination of the light-cone gauge for the bosonic coordinates and a specific ``kappa-gauge'' for the fermions. We will review this procedure in Chapter~\ref{ch:strings-light-cone-gauge}.
The Hamiltonian of the light-cone gauge-fixed two-dimensional model is highly non-linear, and a standard way to study it is by implementing the usual expansion in powers of fields.
The quadratic Hamiltonian turns out to be the free Hamiltonian for eight massive bosons and eight massive fermions of unit mass.
From the quartic contribution one can extract the tree-level two-body scattering processes~\cite{Klose:2006zd} that satisfy the classical Yang-Baxter equation---the semiclassical limit of the equation that we encountered before.
Loop contributions may be taken into account to construct the S-matrix perturbatively~\cite{Klose:2007rz}, confirming again consistency with factorised scattering.
Let us stress again that now the perturbative expansion is performed for large values of $\lambda$, as opposed to the small-$\lambda$ expansion used for the perturbation theory in $\mathcal{N}=4$ SYM.

The perturbative results suggest that the approach used on the gauge-theory side to construct an all-loop S-matrix may be considered also for the string.
In this context the scattering will involve excitations on the worldsheet, and rather than a spin-chain we now encounter a field theory in $1+1$ dimensions, to be quantised to all orders.
For the string there exists actually a derivation of the exact eigenvalues of the central charges~\cite{Arutyunov:2006ak} that are crucial in the all-loop construction.
Exploiting the symmetries, one finds an S-matrix that is essentially the same as the one derived from the point of view of the gauge theory~\cite{Arutyunov:2006yd}---the two objects being related by a change of the two-particle basis.
This S-matrix is supposed to describe to all-loops the scattering of the excitations on the worldsheet.
The results rely on the assumption that Integrability extends from the classical to the quantum level
; however, the strongest indication of its validity is that the all-loop S-matrix matches with the perturbative results of both the string and the gauge theory.


\bigskip

The program of finding the S-matrix is so important because its knowledge allows one to construct the \emph{Bethe-Yang equations} by imposing periodicity of the wave-function.
These are the equations that one should solve to compute the spectrum of the theory~\cite{Beisert:2005fw}.
Let us mention that the Bethe-Yang equations derived from the all-loop S-matrix actually describe the spectrum only in the so-called \emph{asymptotic limit}.
In fact, as anticipated, from the point of view of the gauge theory higher-loop corrections introduce long-range interactions, and these eventually lead to virtual particles travelling all around the spin-chain.
These wrapping interactions give contributions that are exponentially suppressed in the length of the spin-chain, and become important for precision computations when this is finite.
The same issue has a counterpart on the string side.
In fact, in order to define asymptotic states on the worldsheet and an S-matrix, one has to consider the limit of large length of the light-cone gauge-fixed string.
In both cases, therefore, to compute the exact spectrum one has to account for finite-size corrections~\cite{Ambjorn:2005wa}.

A way to incorporate the wrapping corrections is to use the trick of the mirror model, first introduced by Zamolodchikov in the context of relativistic integrable systems~\cite{Zamolodchikov:1989cf}.
Rather than considering a model with finite length, one chooses to reinterpret the problem as the one of finding the spectrum for another model, with infinite size but at finite temperature.
The treatment can be done with the method of the Thermodynamic Bethe Ansatz (TBA)~\cite{Yang:1968rm}, which can be applied to the case of the ground state of {\adsfive}~\cite{Arutyunov:2007tc,Arutyunov:2009ur,Bombardelli:2009ns} as well as the excited states~\cite{Gromov:2009tv,Gromov:2009zb,Arutyunov:2009ax}, allowing one to obtain numerical results for the spectrum with arbitrary precision.
Thanks to the inclusion of the finite-size effects~\cite{Arutyunov:2010gb,Balog:2010xa}, it was possible to match with perturbative results at \emph{five} loops in the gauge theory~\cite{Eden:2012fe}.
A more recent and refined version of the TBA is the Quantum Spectral Curve~\cite{Gromov:2013pga}, through which it is possible to efficiently obtain \emph{analytic} results for anomalous dimensions of operators up to \emph{ten} loops~\cite{Marboe:2014gma}.

\paragraph{AdS$_3$/CFT$_2$}

The striking presence of Integrability for AdS$_5/$CFT$_4$ at large-$N$ and the great success achieved raise the natural question of whether it is possible to apply the same methods also to other instances of the AdS/CFT correspondence.
We may wonder whether also lower-dimensional and less supersymmetric models still have the chance of being solvable.
The answer was shown to be positive for the ABJM theory~\cite{Aharony:2008ug}, see~\cite{Klose:2010ki} for a review.
The case that is of interest for this thesis is rather AdS$_3$/CFT$_2$.

AdS$_3$ gravity was actually the first \emph{ante litteram} example of the holographic duality.
In 1986 Brown and Henneaux showed that its asymptotic symmetry algebra---the gauge transformations that leave the field configurations invariant at the boundary---coincides with the Virasoro algebra, that is the symmetry of two-dimensional CFTs~\cite{Brown:1986nw}.

On the one hand, gravity in three dimensions is remarkably simpler than the one we experience in our world, and it can be seen as an easier set-up to investigate some questions.
An example of this is that it does not contain a propagating graviton.
On the other hand, this does not make it a trivial theory at all.
As shown by Ba\~{n}ados, Teitelboim and Zanelli, gravity in three dimensions with a negative cosmological constant admits black hole solutions ~\cite{Banados:1992wn}.
These are locally  isomorphic to empty AdS$_3$, differing from it because of global identifications~\cite{Banados:1992gq}.
Also these black holes follow the famous Bekenstein-Hawking area-law for the entropy~\cite{Bekenstein:1972tm,Bekenstein:1973ur}, making AdS$_3$ a nice playground to further understand the nature of these objects.
For black holes whose near-horizon geometry is (locally) AdS$_3$, it was actually possible to derive the area-law  by performing a micro-state counting in the dual CFT$_2$~\cite{Strominger:1997eq}.
This computation generalises the one for black holes arising in string theory, as considered in~\cite{Strominger:1996sh,Callan:1996dv,Maldacena:1996ky}. 
Let us mention that AdS$_3$/CFT$_2$ appears also in this context because of a particular D-brane construction, the D1-D5 system.
In Chapters~\ref{ch:symm-repr-T4},~\ref{ch:S-matrix-T4} and~\ref{ch:massive-sector-T4} we will actually study strings propagating on the background that arises as the near-horizon limit of D1-D5.
Let us briefly review some facts about backgrounds that are relevant for AdS$_3$/CFT$_2$.

The backgrounds that we want to consider here preserve a total of 16 supercharges, and are AdS$_3\times$S$^3\times$S$^3\times$S$^1$ and AdS$_3\times$S$^3\times$T$^4$.
The former is actually a family of backgrounds, as the amount of supersymmetry is preserved if the radii of AdS and the two three-spheres S$^3_{(1)}$ and S$^3_{(2)}$ satisfy the constraint
$$
{R^{-2}_{\text{AdS}}}={R^{-2}_{{(1)}}}+{R^{-2}_{{(2)}}}\,.
$$
We then find a family parameterised by a continuous parameter $\alpha=R^{2}_{\text{AdS}}/R^{2}_{{(1)}}$, where $0<\a<1$.
The algebra of isometries is given by $\alg{d}(2,1;\a)_{\sL}\oplus \alg{d}(2,1;\a)_{\sR}$, where the labels for the two copies of the exceptional Lie superalgebra~\cite{Frappat:1996pb} refer to the Left and Right movers of the dual CFT$_2$.
The background {\adsthree} may be understood as a contraction of the previous case, when we blow up the radius of one of the S$^3$'s and then compactify these directions together with the remaining S$^1$ to get a four-dimensional torus. At the level of the algebra this is achieved by a proper $\a\to0$ limit, or alternatively $\a\to1$.
In this case the algebra of isometries is $\psu(1,1|2)_{\sL}\oplus \psu(1,1|2)_{\sR}$.

The above backgrounds provide a rich structure since they can be supported by a mixture of Ramond-Ramond (RR) and Neveu-Schwarz--Neveu-Schwarz (NSNS) fluxes, where a parameter permits to interpolate between the pure RR and the pure NSNS backgrounds.
The latter case was solved by using methods of representations of chiral algebras~\cite{Giveon:1998ns,Elitzur:1998mm,Maldacena:2000hw,Maldacena:2000kv}.
On the contrary the case of pure RR cannot be addressed with these techniques~\cite{Berkovits:1999im} and we will argue that the right language to study it in the planar limit is indeed Integrability.

One of the main challenges of AdS$_3/$CFT$_2$ is that the gauge theories dual to the above backgrounds are not as well understood as the example of $\mathcal{N}=4$ SYM in four dimensions.
Maldacena argued that the dual CFT$_2$ should be found at the infra-red fixed point of the Higgs branch of the dual gauge theory~\cite{Maldacena:1997re}.
In the case of {\adsthree} the finite-dimensional algebra mentioned before is completed to \emph{small} $\mathcal{N}=(4,4)$ superconformal symmetry~\cite{Seiberg:1999xz}, while for AdS$_3\times$S$^3\times$S$^3\times$S$^1$ one finds \emph{large} $\mathcal{N}=(4,4)$~\cite{Boonstra:1998yu}.
Constructions of long-range spin-chains for the dual CFT$_2$'s are unfortunately lacking.
A first proposal of a weakly coupled spin-chain description of the CFT$_2$ dual to {\adsthree} appeared in~\cite{Pakman:2009mi}, for a different and more recent description see~\cite{Sax:2014mea}.
In this thesis we show that addressing the problem on the string theory side of the correspondence allows us to derive the desired all-loop S-matrix.

One of the new features common to backgrounds relevant for the AdS$_3$/CFT$_2$ correspondence is the presence of \emph{massless} worldsheet excitations, corresponding to flat directions\footnote{For AdS$_3\times$S$^3\times$S$^3\times$S$^1$ directions corresponding to massless modes are the circle $S^1$ and a linear combination of the two equators of the S$^3$'s. For {\adsthree} the flat directions correspond to the torus.}. For some time they have been elusive in the Integrability description, but we will show that they can be naturally included in it.
The massive sectors of AdS$_3\times$S$^3\times$S$^3\times$S$^1$ and {\adsthree} may be described respectively by the cosets
$$
\frac{\text{D}(2,1;\a)_{\sL}\times \text{D}(2,1;\a)_{\sR}}{\text{SO}(1,2)\times \text{SO}(3) \times \text{SO}(3)}\,,
\qquad\qquad
\frac{\text{PSU}(1,1|2)_{\sL}\times \text{PSU}(1,1|2)_{\sR}}{\text{SO}(1,2)\times \text{SO}(3) }\,,
$$
and following the method of Ref.~\cite{Metsaev:1998it} one can construct the action for the non-linear $\sigma$-models on AdS$_3\times$S$^3\times$S$^3$ and AdS$_3\times$S$^3$~\cite{Park:1998un,Pesando:1998wm,Rahmfeld:1998zn}.
The missing flat directions can then be re-inserted by hand, and agreement with the Green-Schwarz action can be shown in a certain kappa-gauge for fermions~\cite{Babichenko:2009dk}.

Classical Integrability for the pure RR backgrounds\footnote{It is interesting that classical Integrability was extended also to the case in which a $B$-field is present in the background~\cite{Cagnazzo:2012se}.} was demonstrated in~\cite{Babichenko:2009dk}.
In fact, these cosets enjoy a $\mathbb{Z}_4$ symmetry that allows one to borrow the construction for the Lax representation of the {\adsfive} background~\cite{Bena:2003wd}, see also~\cite{Adam:2007ws,Zarembo:2010sg,Zarembo:2010yz}.

In this thesis we will consider the case of strings on the pure RR {\adsthree} background, and by assuming that Integrability extends from the classical to the quantum level we will derive an all-loop S-matrix, in the spirit of what was done for strings on {\adsfive}.
For the case of mixed flux see~\cite{Lloyd:2014bsa}, and for results on AdS$_3\times$S$^3\times$S$^3\times$S$^1$ see~\cite{Borsato:2012ud,Borsato:2012ss,Borsato:2015mma}.

\paragraph{Deforming {\adsfive}}

Together with the study of lower-dimensional AdS models, one may wonder whether it is possible to deform the superstring on {\adsfive} and its dual $\mathcal{N}=4$ SYM, to relax some of the symmetries while preserving the integrable structure.
This would teach us what the conditions are under which Integrability is still present, and it would allow us to study cases that are less special than the maximally supersymmetric theory in four dimensions.

Examples of deformations of the $\sigma$-model that preserve its classical Integrability are orbifolds of either AdS$_5$ or S$^5$~\cite{Beisert:2005he,Solovyov:2007pw}, where  the fields living on the worldsheet are identified through the action of a discrete subgroup of the bosonic isometries.
Another class of deformations is generated by the so-called ``TsT-transformations'', that can be implemented any time the background possesses at least two commuting isometries.
Let us call $\phi_1$ and $\phi_2$ the two directions on which these isometries act as shifts. The TsT-transformation is a sequence of a T-duality, a shift, and a T-duality on $\phi_1$ and $\phi_2$.
The first T-duality transformation acts on $\phi_1$, producing the dual coordinate $\tilde{\phi}_1$; the shift is implemented on $\phi_2$ as $\phi_2\to \phi_2 + \gamma \tilde{\phi}_1$; to conclude, one performs another T-duality along $\tilde{\phi}_1$~\cite{Alday:2005ww}.
Multi-parameter deformations are made possible by the various choices of pairs of U$(1)$-isometries used to implement the TsT-transformation, and in general they can break all supersymmetries~\cite{Frolov:2005dj}. A restriction to a one-parameter real deformation of the sphere reproduces the Lunin-Maldacena background~\cite{Lunin:2005jy}, which preserves $\mathcal{N}=1$ supersymmetry.
The effects of these classes of deformations on the gauge theory and on the quantum integrable model have also been studied, and we refer to~\cite{vanTongeren:2013gva} for a review on this.

A different approach consists of deforming the symmetry algebra by a continuous parameter.
The case we want to discuss is generally referred to as $q$-deformation, where $q$ is indeed the deformation parameter.
This deformation replaces a Lie algebra $\alg{f}$ by its quantum group version $U_q(\alg{f})$, which we will just denote by $\alg{f}_q$.
To show how this works in a simple example\footnote{For higher-rank algebras, the deformed commutation relations in the Serre-Chevalley basis must be supplemented by the $q$-deformed Serre relations.}, let us consider the case of the $\alg{sl}(2)$ algebra where we denote by $\gen{S}_3$ the Cartan element and by $\gen{S}_\pm$ the positive and negative roots, \ie the ladder operators. The $\alg{sl}_q(2)$ relations are given by
$$
[\gen{S}_3,\gen{S}_\pm] = \pm 2 \gen{S}_\pm\,,
\qquad\quad
[\gen{S}_+,\gen{S}_-]=\frac{q^{\gen{S}_3}-q^{-\gen{S}_3}}{q-q^{-1}}\,,
$$
meaning that the deformation modifies the right-hand-side of the commutation relation of the two ladder operators.
Sending the deformation parameter $q \to 1$ we recover the undeformed algebra.
The $q$-deformation is not just a beautiful mathematical construction, it is also physically motivated.
The most famous realisation of it is the XXZ spin-chain~\cite{Faddeev:1996iy}. In fact, allowing for anisotropy (\ie  a different coupling related to $q$ for the spins in the $z$-direction) one obtains a $q$-deformation, in the sense presented above, of the XXX spin-chain.

The interest for this type of deformation in the context of AdS/CFT first sparkled when Beisert and Koroteev studied the $q$-deformation of the R-matrix of the Hubbard model~\cite{Beisert:2008tw}, see also~\cite{Beisert:2010kk,Beisert:2011wq}.
After solving the crossing equation for the dressing factor, it was possible to define an all-loop S-matrix for the $q$-deformation of the integrable model describing the dual pair of AdS$_5$/CFT$_4$~\cite{Hoare:2011wr}.
The case considered was that of $q$ being a root of unity, and it was shown that the ``vertex to IRF'' transformation can be used to restore unitarity of the corresponding S-matrix~\cite{Hoare:2013ysa}.
Much progress has been made, and thanks to the TBA construction of~\cite{Arutyunov:2012zt,Arutyunov:2012ai} it is even possible to compute the spectrum at finite size.
We want to stress that all this work was pursued just by using the description of the deformed quantum integrable model, bypassing the meaning of this deformation for both the gauge and the string theory.

The gap was filled on the string side by Delduc, Magro and Vicedo, who proposed a method to deform the action for the superstring on {\adsfive}~\cite{Delduc:2013qra}.  This realises a $q$-deformation of the symmetry algebra of the classical charges~\cite{Delduc:2014kha}, where now the deformation parameter is \emph{real}.
It is a generalisation of deformations valid for bosonic cosets~\cite{Delduc:2013fga} and it is of the type of the Yang-Baxter $\sigma$-model of Klim\v{c}\'ik~\cite{Klimcik:2002zj,Klimcik:2008eq}.
It is sometimes referred to as ``$\eta$-deformation'', where $\eta$ is a deformation parameter that is related to $q$. 
The limit $\eta\to0$ gives back the undeformed model.
The remarkable fact is that by construction the deformation procedure preserves the classical Integrability of the original model.
In this thesis we will study this deformation when it is applied to strings on {\adsfive}, and we will compare it to the S-matrix of Beisert and Koroteev.

Let us mention that recently a new method was studied, going under the name of ``$\lambda$-deformation''.
It was first introduced by gauging a combination of a principal chiral model and a Wess-Zumino-Witten model~\cite{Sfetsos:2013wia}, and it was then extended to strings on symmetric spaces~\cite{Hollowood:2014rla} and on {\adsfive}~\cite{Hollowood:2014qma}. 
There is evidence that it realises the $q$-deformation in the case of $q$ being root of unity~\cite{Hollowood:2015dpa},
and it was shown to be related to the $\eta$-deformation~\cite{Vicedo:2015pna,Hoare:2015gda} by the Poisson-Lie T-duality of~\cite{Klimcik:1995ux,Klimcik:1995dy}.

To conclude this paragraph let us point out that it is still unclear how to construct the duals of these $\sigma$-models, in other words how to $q$-deform $\mathcal{N}=4$ SYM.
The result is expected to be a non-commutative gauge theory, and it would be extremely interesting to build it explicitly.

\paragraph{About this thesis}
This thesis contains some of the author's contributions to the research on Integrability applied to AdS/CFT.
Part of this work has been devoted to the understanding of lower dimensional examples, and we will present in particular the derivation of an all-loop S-matrix for the case of {\adsthree}.
A different direction was motivated by questions on the $\eta$-deformation of strings on {\adsfive}.

We start in \textbf{Chapter~\ref{ch:strings-light-cone-gauge}} with a review on basic notions that will be useful for the remaining chapters of the thesis.
We begin with a discussion on bosonic strings and on how to fix light-cone gauge. We follow~\cite{Arutyunov:2009ga}, but we include also the possibility in which a background $B$-field is present.
The main consequences of the light-cone gauge-fixing are explained.
We then extend the discussion to include fermions. We present the generic action for type IIB superstring at quadratic order in fermions, and explain how to fix a proper kappa-gauge.
After presenting the \emph{decompactification limit} of the worldsheet---necessary to define asymptotic states---we discuss the large-tension expansion, equivalent to the usual expansion in powers of fields on the worldsheet.
We review also perturbative quantisation and the corresponding scattering theory, whose all-loop generalisation will be a major topic in the following.

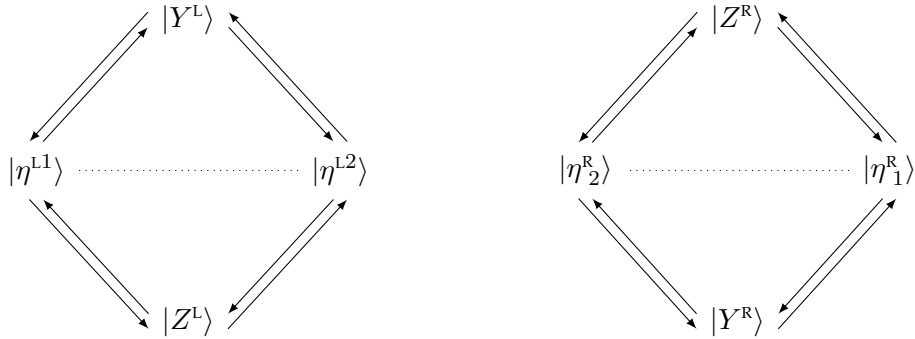
\begin{figure}[t]
  \centering
  \begin{tikzpicture}[%
    box/.style={outer sep=1pt},
    Q node/.style={inner sep=1pt,outer sep=0pt},
    arrow/.style={-latex}
    ]%

    \node [box] (PhiM) at ( 0  , 2cm) {\small $\ket{Y^{\sL}}$};
    \node [box] (PsiP) at (-2cm, 0cm) {\small $\ket{\eta^{\sL 1}}$};
    \node [box] (PsiM) at (+2cm, 0cm) {\small $\ket{\eta^{\sL 2}}$};
    \node [box] (PhiP) at ( 0  ,-2cm) {\small $\ket{Z^{\sL}}$};

    \newcommand{\horshift}{0.09cm,0cm}
    \newcommand{\vershift}{0cm,0.10cm}
 
    \draw [arrow] ($(PhiM.west) +(\vershift)$) -- ($(PsiP.north)-(\horshift)$) node [pos=0.5,anchor=south east,Q node] {};
    \draw [arrow] ($(PsiP.north)+(\horshift)$) -- ($(PhiM.west) -(\vershift)$) node [pos=0.5,anchor=north west,Q node] {};

    \draw [arrow] ($(PsiM.south)-(\horshift)$) -- ($(PhiP.east) +(\vershift)$) node [pos=0.5,anchor=south east,Q node] {};
    \draw [arrow] ($(PhiP.east) -(\vershift)$) -- ($(PsiM.south)+(\horshift)$) node [pos=0.5,anchor=north west,Q node] {};

    \draw [arrow] ($(PhiM.east) -(\vershift)$) -- ($(PsiM.north)-(\horshift)$) node [pos=0.5,anchor=north east,Q node] {};
    \draw [arrow] ($(PsiM.north)+(\horshift)$) -- ($(PhiM.east) +(\vershift)$) node [pos=0.5,anchor=south west,Q node] {};

    \draw [arrow] ($(PsiP.south)-(\horshift)$) -- ($(PhiP.west) -(\vershift)$) node [pos=0.5,anchor=north east,Q node] {};
    \draw [arrow] ($(PhiP.west) +(\vershift)$) -- ($(PsiP.south)+(\horshift)$) node [pos=0.5,anchor=south west,Q node] {};
       
    \draw [dotted] (PsiM) -- (PsiP) node [pos=0.65,anchor=north west,Q node] {};

  \end{tikzpicture}
\hspace{2cm}
  \begin{tikzpicture}[%
    box/.style={outer sep=1pt},
    Q node/.style={inner sep=1pt,outer sep=0pt},
    arrow/.style={-latex}
    ]%

    \node [box] (PhiM) at ( 0  , 2cm) {\small $\ket{Z^{\sR}}$};
    \node [box] (PsiP) at (-2cm, 0cm) {\small $\ket{\eta^{\sR}_{\  2}}$};
    \node [box] (PsiM) at (+2cm, 0cm) {\small $\ket{\eta^{\sR}_{\  1}}$};
    \node [box] (PhiP) at ( 0  ,-2cm) {\small $\ket{Y^{\sR}}$};

    \newcommand{\horshift}{0.09cm,0cm}
    \newcommand{\vershift}{0cm,0.10cm}
 
    \draw [arrow] ($(PhiM.west) +(\vershift)$) -- ($(PsiP.north)-(\horshift)$) node [pos=0.5,anchor=south east,Q node] {};
    \draw [arrow] ($(PsiP.north)+(\horshift)$) -- ($(PhiM.west) -(\vershift)$) node [pos=0.5,anchor=north west,Q node] {};

    \draw [arrow] ($(PsiM.south)-(\horshift)$) -- ($(PhiP.east) +(\vershift)$) node [pos=0.5,anchor=south east,Q node] {};
    \draw [arrow] ($(PhiP.east) -(\vershift)$) -- ($(PsiM.south)+(\horshift)$) node [pos=0.5,anchor=north west,Q node] {};

    \draw [arrow] ($(PhiM.east) -(\vershift)$) -- ($(PsiM.north)-(\horshift)$) node [pos=0.5,anchor=north east,Q node] {};
    \draw [arrow] ($(PsiM.north)+(\horshift)$) -- ($(PhiM.east) +(\vershift)$) node [pos=0.5,anchor=south west,Q node] {};

    \draw [arrow] ($(PsiP.south)-(\horshift)$) -- ($(PhiP.west) -(\vershift)$) node [pos=0.5,anchor=north east,Q node] {};
    \draw [arrow] ($(PhiP.west) +(\vershift)$) -- ($(PsiP.south)+(\horshift)$) node [pos=0.5,anchor=south west,Q node] {};
       
    \draw [dotted] (PsiM) -- (PsiP) node [pos=0.65,anchor=north west,Q node] {};
  \end{tikzpicture}
  \caption{The Left and Right massive modules. Excitations $Z^{\sL,\sR}$ correspond to transverse directions in AdS$_3$, while $Y^{\sL,\sR}$ in S$^3$. Fermions are denoted by $\eta$. The arrows correspond to supercharges, while the dotted lines correspond to the action of an $\su(2)$.}
  \label{fig:massive-Intro}
\end{figure}

\begin{figure}[t]
  \centering
  \begin{tikzpicture}[%
    box/.style={outer sep=1pt},
    Q node/.style={inner sep=1pt,outer sep=0pt},
    arrow/.style={-latex}
    ]%
\newcommand{\xshiftmasslessone}{-4cm}
\begin{scope}[xshift=\xshiftmasslessone]
    \node [box] (PhiM) at ( 0  , 2cm) {\small $\ket{\chi^{1}}$};
    \node [box] (PsiP) at (-2cm, 0cm) {\small $\ket{T^{11}}$};
    \node [box] (PsiM) at (+2cm, 0cm) {\small $\ket{T^{21}}$};
    \node [box] (PhiP) at ( 0  ,-2cm) {\small $\ket{\widetilde{\chi}^{1}}$};

    \newcommand{\horshift}{0.09cm,0cm}
    \newcommand{\vershift}{0cm,0.10cm}
 
    \draw [arrow] ($(PhiM.west) +(\vershift)$) -- ($(PsiP.north)-(\horshift)$) node [pos=0.5,anchor=south east,Q node] {};
    \draw [arrow] ($(PsiP.north)+(\horshift)$) -- ($(PhiM.west) -(\vershift)$) node [pos=0.5,anchor=north west,Q node] {};

    \draw [arrow] ($(PsiM.south)-(\horshift)$) -- ($(PhiP.east) +(\vershift)$) node [pos=0.5,anchor=south east,Q node] {};
    \draw [arrow] ($(PhiP.east) -(\vershift)$) -- ($(PsiM.south)+(\horshift)$) node [pos=0.5,anchor=north west,Q node] {};

    \draw [arrow] ($(PhiM.east) -(\vershift)$) -- ($(PsiM.north)-(\horshift)$) node [pos=0.5,anchor=north east,Q node] {};
    \draw [arrow] ($(PsiM.north)+(\horshift)$) -- ($(PhiM.east) +(\vershift)$) node [pos=0.5,anchor=south west,Q node] {};

    \draw [arrow] ($(PsiP.south)-(\horshift)$) -- ($(PhiP.west) -(\vershift)$) node [pos=0.5,anchor=north east,Q node] {};
    \draw [arrow] ($(PhiP.west) +(\vershift)$) -- ($(PsiP.south)+(\horshift)$) node [pos=0.5,anchor=south west,Q node] {};

    \draw [dotted] (PsiM) -- (PsiP) node [pos=0.65,anchor=north west,Q node] {};
\end{scope}
%
%
%
\newcommand{\xshiftmasslesstwo}{1cm}
\newcommand{\yshiftmasslesstwo}{1cm}
\begin{scope}[xshift=\xshiftmasslesstwo,yshift=\yshiftmasslesstwo]

    \node [box] (PhiM) at ( 0  , 2cm) {\small $\ket{\chi^{2}}$};
    \node [box] (PsiP) at (-2cm, 0cm) {\small $\ket{T^{12}}$};
    \node [box] (PsiM) at (+2cm, 0cm) {\small $\ket{T^{22}}$};
    \node [box] (PhiP) at ( 0  ,-2cm) {\small $\ket{\widetilde{\chi}^{2}}$};

    \newcommand{\horshift}{0.09cm,0cm}
    \newcommand{\vershift}{0cm,0.10cm}
 
    \draw [arrow] ($(PhiM.west) +(\vershift)$) -- ($(PsiP.north)-(\horshift)$) node [pos=0.5,anchor=south east,Q node] {};
    \draw [arrow] ($(PsiP.north)+(\horshift)$) -- ($(PhiM.west) -(\vershift)$) node [pos=0.5,anchor=north west,Q node] {};

    \draw [arrow] ($(PsiM.south)-(\horshift)$) -- ($(PhiP.east) +(\vershift)$) node [pos=0.5,anchor=south east,Q node] {};
    \draw [arrow] ($(PhiP.east) -(\vershift)$) -- ($(PsiM.south)+(\horshift)$) node [pos=0.5,anchor=north west,Q node] {};

    \draw [arrow] ($(PhiM.east) -(\vershift)$) -- ($(PsiM.north)-(\horshift)$) node [pos=0.5,anchor=north east,Q node] {};
    \draw [arrow] ($(PsiM.north)+(\horshift)$) -- ($(PhiM.east) +(\vershift)$) node [pos=0.5,anchor=south west,Q node] {};

    \draw [arrow] ($(PsiP.south)-(\horshift)$) -- ($(PhiP.west) -(\vershift)$) node [pos=0.5,anchor=north east,Q node] {};
    \draw [arrow] ($(PhiP.west) +(\vershift)$) -- ($(PsiP.south)+(\horshift)$) node [pos=0.5,anchor=south west,Q node] {};

    \draw [dotted] (PsiM) -- (PsiP) node [pos=0.65,anchor=south west,Q node] {};
    
    \draw [dashed] ($(PhiM.west)+({-0.1cm,0.1cm})$) -- ($(PhiM.east)-({\xshiftmasslesstwo,\yshiftmasslesstwo})+({\xshiftmasslessone,0cm})+({-0.1cm,0.2cm})$);
    \draw [dashed] (PsiM) -- ($(PsiM.east)-({\xshiftmasslesstwo,\yshiftmasslesstwo})+({\xshiftmasslessone,0cm})$);
    \draw [dashed] (PsiP) -- ($(PsiP.east)-({\xshiftmasslesstwo,\yshiftmasslesstwo})+({\xshiftmasslessone,0cm})+({-0.1cm,0.1cm})$);
    \draw [dashed] ($(PhiP.south west)+({0cm,0.2cm})$) -- ($(PhiP.east)-({\xshiftmasslesstwo,\yshiftmasslesstwo})+({\xshiftmasslessone,0cm})+({0cm,-0.1cm})$) node [pos=0.45,anchor=north west,Q node] {};
\end{scope}
%
  \end{tikzpicture}
  \caption{The massless module. $T^{\dot{a}a}$ are excitations on T$^4$, and the fermions are denoted by $\chi^a$ and $\tilde{\chi}^a$. Dotted and dashed lines correspond to the actions of two $\su(2)$ algebras.}
  \label{fig:massless-Intro}
\end{figure}
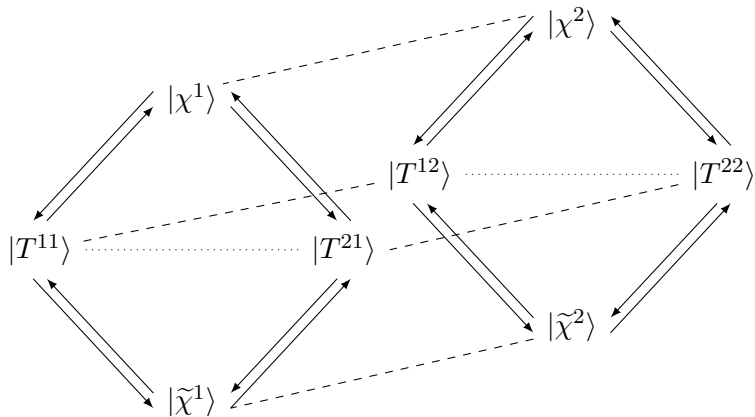

\textbf{Chapter~\ref{ch:symm-repr-T4}} is the first one specifically devoted to AdS$_3$/CFT$_2$. 
We consider the background {\adsthree} and we study the centrally-extended symmetry algebra $\mathcal{A}$ of the charges commuting with the light-cone Hamiltonian.
We derive the exact momentum-dependence of the central charges, and then we study the representation of $\mathcal{A}$ under which the excitations are organised, in a limit in which the dispersion relation is relativistic.
The analysis shows that this representation is \emph{reducible}, a feature of {\adsthree} that was not there for the known case of {\adsfive}. 
We find a total of three irreducible representations, labelled by the eigenvalue of an angular momentum in AdS$_3\times$S$^3$. 
Figure~\ref{fig:massive-Intro} shows the two \emph{massive} representations, where this eigenvalue takes value $+1$ and $-1$ on Left and Right excitations respectively. 
Here Left and Right refer to the two copies of $\psu(1,1|2)$, that are isometries of the background.
The algebra $\mathcal{A}$ was first identified in~\cite{Borsato:2013qpa} from the point of view of the spin-chain with symmetry $\psu(1,1|2)_{\sL}\oplus \psu(1,1|2)_{\sR}$. The excitations on this spin-chain correspond to the massive worldsheet excitations of {\adsthree}.
In this chapter we take instead the point of view of the string theory description and we follow~\cite{Borsato:2014exa,Borsato:2014hja}, where \emph{massless} excitations---see Figure~\ref{fig:massless-Intro}---were finally included. 
Using the knowledge of the central charges and arguments of representation theory, we generalise the representations to all loops in the large-$N$ limit.
We further study these representations and introduce the notion of Left-Right symmetry.

In \textbf{Chapter~\ref{ch:S-matrix-T4}} we  impose compatibility with symmetries and bootstrap the all-loop S-matrix for the worldsheet excitations, as done in~\cite{Borsato:2014exa,Borsato:2014hja}. 
Remarkably, the S-matrix satisfies the Yang-Baxter equation, confirming compatibility with the assumption of factorisation of scattering.
The S-matrix is actually fixed completely up to some dressing factors that cannot be found from symmetries. Taking into account the constraints of unitarity and of Left-Right symmetry, we find a total of four unspecified functions.
Further constraints are imposed on them by the crossing equations, that we derive.
We then explain how to impose the periodicity condition on the wave-function to derive the Bethe-Yang equations.
We guide the reader through the diagonalisation procedure, introducing the various complications in different steps, until the nesting procedure is used.
We conclude by presenting the complete\footnote{This result has appeared in~\cite{Borsato:2016kbm}.} set of Bethe-Yang equations for {\adsthree}.

We restrict our attention to the massive sector\footnote{The massive sector of {\adsthree} has been discussed in detail also in the thesis of A. Sfondrini~\cite{Sfondrini:2014via}, to which we refer for an alternative presentation.} in \textbf{Chapter~\ref{ch:massive-sector-T4}}.
First we show that the previous results are closely related to a spin-chain description, following~\cite{Borsato:2013qpa}. 
This spin-chain needs to be dynamical---the interactions change its length---in order to correctly account for the central extension of the algebra. 
An all-loop S-matrix can be determined, which is related to the worldsheet S-matrix by a similarity transformation.
We also present solutions to the crossing equations for the dressing factors of the massive sector, and we provide some checks for their validity.
These solutions and the corresponding discussion were first presented in~\cite{Borsato:2013hoa}.
By taking a proper thermodynamical limit in the regime of large string tension, we also recover the so-called ``finite-gap equations'' from the Bethe-Yang equations, repeating the calculation in~\cite{Borsato:2013qpa}.
We conclude by referring to the independent perturbative calculations that confirm our all-loop results.

\begin{figure}[t]
  \centering
\includegraphics[width=0.4\textwidth]{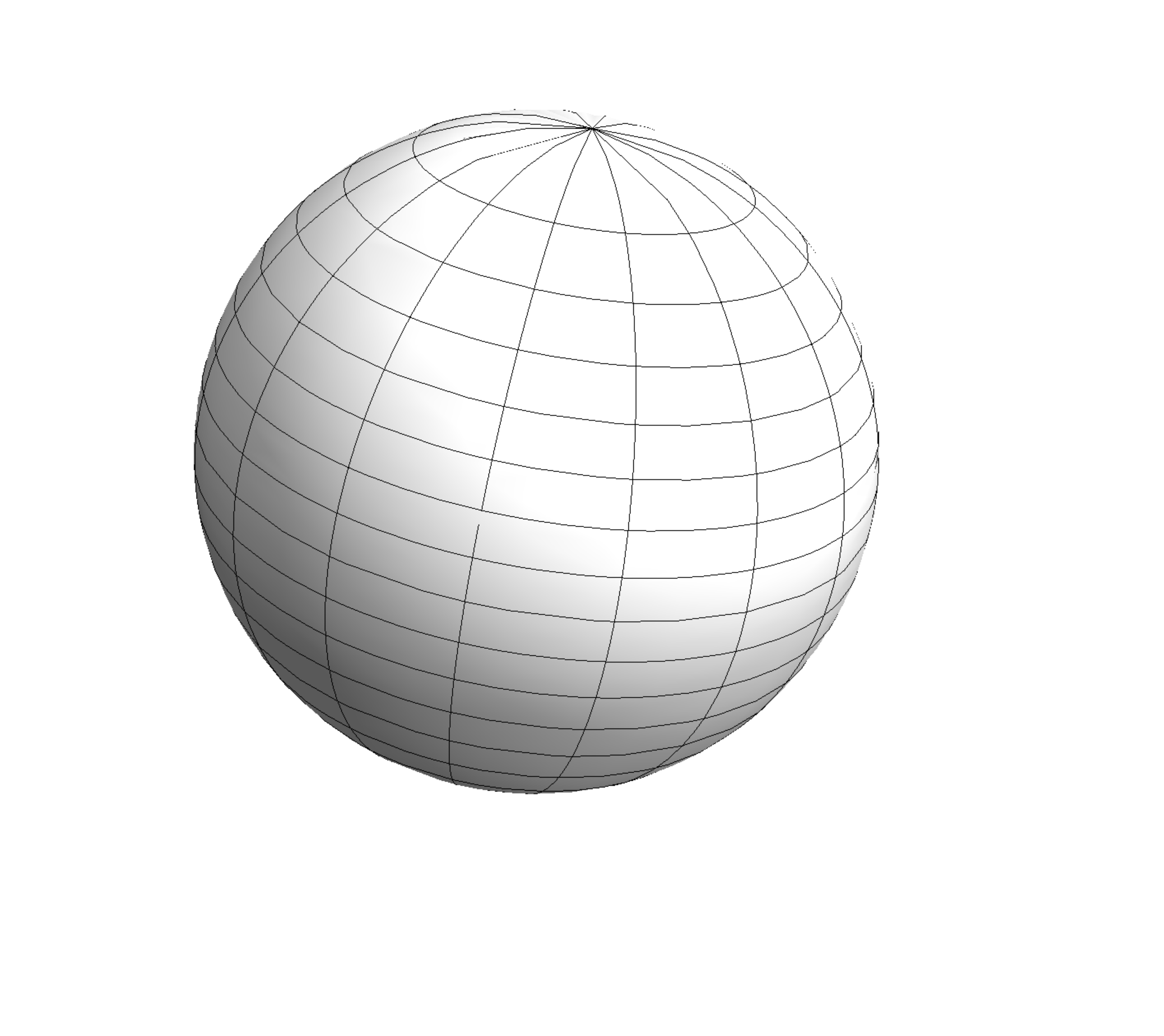}
%
%
\includegraphics[width=0.4\textwidth]{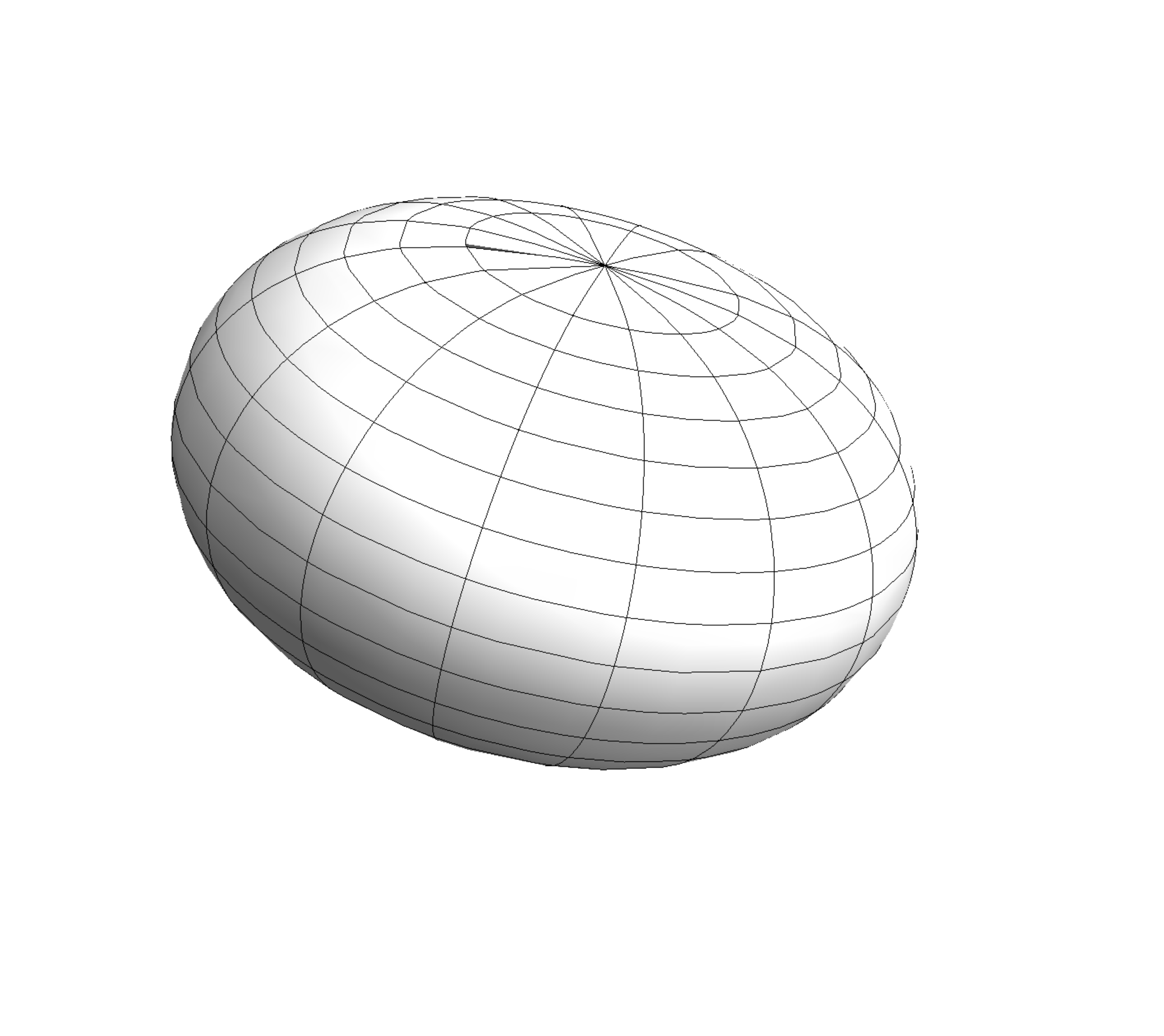}
  \caption{When we apply the $\eta$-deformation to a two-dimensional sphere---left figure---we find that its effect is a squashing---right figure.}
  \label{fig:eta-def-sphere-Intro}
\end{figure}

In \textbf{Chapter~\ref{ch:qAdS5Bos}} we begin the investigation of the $\eta$-deformation of the string on {\adsfive}.
Here we restrict to the bosonic model.
After a brief introduction to the undeformed model and to the deformation procedure, we derive the results first obtained in~\cite{Arutyunov:2013ega}.
We find that the background metric is deformed and a $B$-field is generated.
A representation of the squashing-effect of the deformation in the case of a two-dimensional sphere may be seen in Figure~\ref{fig:eta-def-sphere-Intro}.
The bosonic action is studied perturbatively by computing the tree-level S-matrix for the scattering of bosonic worldsheet excitations.
The result allows us to succesfully match with the large-tension limit of the all-loop S-matrix found by fixing the $\psu_q(2|2)_{\ce}$ symmetry.
In particular, we can relate the two deformation parameters $\eta$ and $q$ on the two sides.
We conclude the chapter with some concluding remarks.

In \textbf{Chapter~\ref{ch:qAdS5Fer}} we want to address the question of whether the deformed metric and $B$-field can be completed to a full type IIB supergravity background. 
With this motivation we compute the action of the deformed coset at quadratic order in fermions, as done in~\cite{Arutyunov:2015qva}.
We cure the---only apparent---mismatch with the standard form of the Green-Schwarz action by implementing proper field redefinitions on the bosonic and fermionic coordinates.
From the action we extract the couplings to the odd-rank tensors that should correspond to the Ramond-Ramond fields multiplied by the exponential of the dilaton.
We also compute the kappa-symmetry variations of the bosonic and fermionic coordinates, and of the worldsheet metric at leading order. From this computation we confirm the results obtained from the Lagrangian, and we show that they are \emph{not} compatible with the equations of motion of type IIB supergravity.
We conclude the chapter with a discussion of these findings.

\chapter{Strings in light-cone gauge}\label{ch:strings-light-cone-gauge}

\medskip

This chapter serves as an introduction and a review of notions that are needed to derive the results in the rest of the thesis.
In fact, we will use the same methods for strings on both the {\adsthree} background and on the $\eta$-deformed {\adsfive}. We find then useful to present here a slightly more general discussion valid for both cases.

We explain how to fix \emph{uniform light-cone gauge} for bosonic and fermionic degrees of freedom in the action of a freely propagating string.
The need of fixing a gauge that removes some unphysical bosonic degrees of freedom comes from reparameterisation invariance on the worldsheet. At the same time, another local symmetry called ``kappa-symmetry''---now parameterised by Grassmann quantities---suggests that half of the fermions should be gauged away.
Clearly, different gauge fixings are possible, all being equivalent in the sense that the physical observables that we compute will not depend on any particular choice.
However, it is obvious that some of them may be more convenient than others. 
The type of gauge-fixing used for backgrounds relevant for the AdS/CFT correspondence appears to provide models that are solvable by non-perturbative methods.

This gauge is a generalisation of what was first introduced in flat space in~\cite{goddard1973quantum}. 
In fact the procedure is quite general and the only necessary requirement to impose it is the presence of two commuting isometries---in our case these are shifts of time and an angle.
Although other choices are possible---one might choose the angle being in Anti-de Sitter~\cite{Metsaev:2000yu}---the most convenient one for AdS/CFT is to combine into the light-cone coordinates the time of AdS and an angle of the compact space. The procedure we present here was used to gauge-fix the $\sigma$-model describing the string on {\adsfive}~\cite{Arutyunov:2004yx,Arutyunov:2005hd,Arutyunov:2006gs} and corresponds to the one used to study spinning strings~\cite{Kruczenski:2004kw,Kruczenski:2004cn}.

We start with the gauge-fixing procedure for bosons in Section~\ref{sec:Bos-string-lcg} and then extend it to fermions in Section~\ref{sec:fermions-type-IIB}.
In Section~\ref{sec:decomp-limit} we explain how to define an S-matrix that governs scattering of worldsheet excitations---in the limit of long strings---and we provide a discussion on perturbation theory.
We refer to~\cite{Arutyunov:2009ga} for a more detailed review on these topics.

\section{Bosonic strings}\label{sec:Bos-string-lcg}
Restricting to the bosonic model, we can already capture the essential features of the gauge-fixing procedure. 
The string moves on a target manifold parameterised by ten coordinates $X^M, \ M=0,\ldots,9$.
Two of them---for defineteness $X^0\equiv t$ that is time and $X^5 \equiv \phi$ that for us will be an angle of a compact manifold---correspond to the abelian isometries of the full action that we will exploit to fix light-cone gauge. 
Invariance under shifts of two such coordinates results in a dependence of the action on just their derivatives.

A rank$-2$ symmetric tensor $G_{MN}$ defines a metric on the target space, that we assume to be written in ``block form''
\begin{equation}
\begin{aligned}
{\rm d}s^2&= G_{MN} {\rm d}X^M{\rm d}X^N\\
&= G_{tt} {\rm d}t^2+G_{\phi\phi} {\rm d}\phi^2+G_{\mu\nu} {\rm d}X^\mu{\rm d}X^\nu\,,
\end{aligned}
\end{equation}
where $X^\mu$ are the eight transversal coordinates and $G_{tt}<0$.
In general one might have also a rank$-2$ anti-symmetric tensor $B_{MN}$. We include this possibility, as it will be needed in Chapter~\ref{ch:qAdS5Bos} and Chapter~\ref{ch:qAdS5Fer}.
The action for the bosonic string then takes the form of the Polyakov action
\begin{equation}\label{eq:bos-str-action}
\begin{aligned}
S^{\bos}&= \int_{-{\frac{L}{2}}}^{\frac{L}{2}} \, {\rm d}\sigma {\rm d} \tau\  \lagr^{\bos}\,,
\\
\lagr^{\bos}&=-\frac{g}{2}   \left( \, \gamma^{\alpha\beta} \partial_\alpha X^M \partial_\beta X^N G_{MN} -\epsilon^{\alpha\beta} \partial_\alpha X^M \partial_\beta X^N B_{MN} \right)\,.
\end{aligned}
\end{equation}
Here $\tau$ and $\sigma$ are respectively the timelike and spatial coordinates parameterising the worldsheet, for which we use Greek indices $\a,\b$. For closed strings $\sigma\in[-{\frac{L}{2}},{\frac{L}{2}}]$ parameterises a circle of length $L$ and periodic boundary conditions for the fields are used.
The symmetric tensor $\gamma^{\a\b}=h^{\a\b}\sqrt{-h}$ is the Weyl-invariant combination\footnote{With abuse of language we will always refer to it as just the worldsheet metric.} of the world-sheet metric $h_{\a\b}$, and for us the component $\gamma^{\tau\tau}<0$.
For the anti-symmetric tensor $\epsilon^{\alpha\beta}$ we use the convention $\epsilon^{\tau\sigma}=1$.
The whole action is multiplied by $g$, that plays the role of the string tension.

We use first-order formalism and introduce conjugate momenta
\begin{equation}
p_M = \frac{\delta S^{\bos}}{\delta \dot{X}^M} = - g \gamma^{\tau\beta} \partial_\beta X^N G_{MN} + g X^{'N} B_{MN}\,,
\end{equation}
where we are using the shorthand notation $\dot{X}\equiv\pa_{\tau}X(\tau,\sigma),\ X'\equiv\pa_{\sigma}X(\tau,\sigma)$.
Using $\det\gamma^{\a\b}=-1$ we can rewrite the action as
\begin{equation}\label{eq:bos-act-I-ord}
S^{\bos}=  \int_{-{\frac{L}{2}}}^{\frac{L}{2}} \, {\rm d}\sigma {\rm d} \tau \left( p_M \dot{X}^M + \frac{\gamma^{\tau\sigma}}{\gamma^{\tau\tau}} C_1 + \frac{1}{2g \gamma^{\tau\tau}} C_2  \right),
\end{equation}
where $C_1, C_2$ are the Virasoro constraints. They explicitly read as
\begin{equation}
\begin{aligned}
C_1 &= p_M X'^{M}, \\ 
C_2 &= G^{MN} p_M p_N+ g^2 X'^{M} X'^{N} G_{MN} - 2 g\, p_M X'^{Q} G^{MN} B_{NQ} + g^2 X'^{P} X'^{Q} B_{MP} B_{NQ} G^{MN} . 
\end{aligned}
\end{equation}
The components of $\g^{\a\b}$ are Lagrange multipliers, implying that we should solve the equations $C_1=0$ and $C_2=0$ in a certain gauge.
It is convenient to introduce light-cone coordinates $x^+$ and $x^-$ as linear combinations of $t,\phi$~\cite{Arutyunov:2006gs}
\begin{equation}\label{eq:lc-coord}
x^+= (1-a)\, t +a\,  \phi, \qquad\quad x^-=\phi-t.
\end{equation}
To be more general, we make the combination defining $x^+$ dependent on a generic parameter $a$. The above combinations have been chosen in such a way that the conjugate momentum\footnote{We use a different convention from~\cite{Arutyunov:2009ga} for what we call $p_+$ and $p_-$.} of $x^+$ is the sum of the conjugate momenta of $t$ and $\phi$
\begin{equation}
p_+=\frac{\delta S}{\delta \dot{x}^+}=p_t+p_\phi\,,\qquad
p_-=\frac{\delta S}{\delta \dot{x}^-}=-a\, p_t+(1-a)\,p_\phi\,.
\end{equation}
In these coordinates the two Virasoro constraints are rewritten as
\begin{equation}\label{eq:Vira-constr-bos}
\begin{aligned}
C_1 =& p_+ x'^{+}+p_- x'^{-}+p_\mu X'^{\mu}\,,\\
C_2 =& G^{++} p_+^2 +2 G^{+-} p_+ p_- + G^{--} p_-^2 \\
& + g^2 G_{--} (x'^-)^2 + 2 g^2 G_{+-} x'^+ x'^- + g^2 G_{++} (x'^+)^2 + \mathcal{H}^{\bos}_x\, ,
\end{aligned}
\end{equation}
where we have assumed that the $B$-field vanishes along light-cone directions---as this is valid for the examples that we will consider---and
\begin{equation}
\begin{aligned}
\nonumber
G^{++} &= a^2 G_{\phi\phi}^{-1} + (a-1)^2 G_{tt}^{-1}, \quad &G^{+-}& = a G_{\phi\phi}^{-1} + (a-1) G_{tt}^{-1}, \quad G^{--} = G_{\phi\phi}^{-1} + G_{tt}^{-1}, \\
G_{--} &= (a-1)^2 G_{\phi\phi} + a^2 G_{tt}, \quad &G_{+-} &= -(a-1) G_{\phi\phi} - a G_{tt}, \quad G_{++} = G_{\phi\phi} + G_{tt}\,.
\end{aligned}
\end{equation}
In $C_2$ we have collected all expressions that depend only on the transverse coordinates $X^\mu$ into the object
\begin{equation}
\mathcal{H}^{\bos}_x=G^{\mu\nu} p_\mu p_\nu+ g^2 X'^{\mu} X'^{\nu} G_{\mu\nu} - 2 g p_\mu X'^{\rho} G^{\mu\nu} B_{\nu\rho} + g^2 X'^{\lambda} X'^{\rho} B_{\mu\lambda} B_{\nu\rho} G^{\mu\nu} .
\end{equation}
The \emph{uniform light-cone gauge} is achieved by fixing
\begin{equation}\label{eq:unif-lcg}
x^+= \tau+a \,m \, \sigma, \qquad\quad p_-=1,
\end{equation}
where we allow the coordinate $\phi$ to wind $m$ times around the circle $\phi({\frac{L}{2}})-\phi(-{\frac{L}{2}})=2\pi \, m$.
The name ``uniform'' comes from the fact that we choose $p_-$ to be independent of $\sigma$, and this choice makes this light-cone momentum uniformly distributed along the string.
Thanks to this gauge, the term $p_M\dot{X}^M=p_+\dot{x}^++p_-\dot{x}^-+p_\mu\dot{X}^\mu$ in the action~\eqref{eq:bos-act-I-ord} is simplified, and we are led to identify the light-cone momentum $p_+$ with the Hamiltonian (density) of the gauge-fixed model\footnote{We have dropped the total derivative term $\dot{x}^-$.}
\begin{equation}
S^{\bos}_{\text{g.f.}}=  \int_{-{\frac{L}{2}}}^{\frac{L}{2}} \, {\rm d}\sigma {\rm d} \tau \, \left( p_\mu\dot{X}^\mu -\mathcal{H}^{\bos} \right)\,,
\qquad
\mathcal{H}^{\bos}=-p_+(X^\mu,p_\mu)\,,
\end{equation}
once the Virasoro constraints are satisfied.
In this gauge the first Virasoro constraint $C_1=0$ may be used to solve for $x'^-$ as
\begin{equation}
x'^-= - p_\mu X'^{\mu}-a\, m\, p_+.
\end{equation}
Notice that only the \emph{derivative} of this light-cone coordinate can be written as a local expression of the transverse fields.
Since we are describing closed strings, we should actually impose the following periodicity condition
\begin{equation}
2\pi \, m = x^-(L/2)-x^-(-L/2) = \int_{-{\frac{L}{2}}}^{\frac{L}{2}} \, {\rm d}\sigma\, x'^-= -\int_{-{\frac{L}{2}}}^{\frac{L}{2}} \, {\rm d}\sigma  \, p_\mu X'^\mu+a\, m\, \int_{-{\frac{L}{2}}}^{\frac{L}{2}} \, {\rm d}\sigma\, \mathcal{H}^{\bos}\,,
\end{equation}
that we call \emph{level-matching} condition.
We recognise that the above is a constraint involving the woldsheet momentum
\begin{equation}
p_{\text{ws}}= -\int_{-{\frac{L}{2}}}^{\frac{L}{2}} \, {\rm d}\sigma  \, p_\mu X'^\mu\,,
\end{equation}
which is the charge associated to shifts of the worldsheet coordinate $\sigma$, under which the action is invariant.
From now on we will just consider the case of zero winding $m=0$, as it yields a well-defined large-tension limit, see Section~\ref{sec:decomp-limit}.
In this case the level-matching condition imposes that the worldsheet momentum must vanish for physical configurations
\begin{equation}
p_{\text{ws}}=0, \qquad \quad (\text{when } m=0)\,.
\end{equation}
In Chapter~\ref{ch:symm-repr-T4} we actually use a method where we first need to relax the level-matching condition, meaning that we allow for the configuations to have non-vanishing worldsheet momentum. The above condition is then imposed only at the end, as a constraint on the states of the Hilbert space.

\medskip

Solving the second Virasoro constraint $C_2=0$, we find explicitly the light-cone Hamiltonian (density). The solution to this quadratic equation that yields a positive Hamiltonian is
\begin{equation}\label{eq:lc-Hamilt-bos}
\mathcal{H}^{\bos}=-p_+=\frac{G^{+-}+\sqrt{ (G^{+-})^2- G^{++} \left(G^{--}+g^2 G_{--}  (x'^-)^2 +\mathcal{H}^{\bos}_x\right)} }{ G^{++}}\,.
\end{equation}
To relate the Hamiltonian on the worldsheet to the \emph{spacetime} energy of the string, let us note that---because of the invariance of the action under shifts of $t$ and $\phi$---we can define two conserved quantities
\begin{equation}
E=-\int_{-{\frac{L}{2}}}^{\frac{L}{2}} {\rm d}\sigma \, p_t\,,
\qquad\quad
J=\int_{-{\frac{L}{2}}}^{\frac{L}{2}} {\rm d}\sigma \, p_\phi\,.
\end{equation}
The first of them is the spacetime energy, while the second measures the angular momentum in the direction of $\phi$.
After going to light-cone coordinates, these are combined into
\begin{equation}\label{eq:total-light-cone-mom-P+P-}
P_+= \int_{-{\frac{L}{2}}}^{\frac{L}{2}} {\rm d} \sigma \, p_+ = J-E\,,
\qquad
P_-= \int_{-{\frac{L}{2}}}^{\frac{L}{2}} {\rm d} \sigma \, p_- = (1-a)\,J+a\,E\,.
\end{equation}
On the one hand, we immediately discover the relation between the light-cone Hamiltonian and the spacetime charges $E$ and $J$. On the other hand, using~\eqref{eq:unif-lcg} we find how these fix the length $L$ of the string
\begin{equation}\label{eq:Ham-En-L-P-}
\int_{-{\frac{L}{2}}}^{\frac{L}{2}} {\rm d} \sigma \, \mathcal{H}^{\bos} = E-J\,,
\qquad
L = P_-=(1-a)\,J+a\,E\,.
\end{equation}
The first of these equations justifies the choice of the gauge. From the point of view of the AdS/CFT correspondence it is indeed desirable to compute the spacetime energy $E$, that is then related by a simple formula to the Hamiltonian on the worldsheet.
The second of the above equations shows that the Hamiltonian secretely depends on $P_-$ as well, although just through the integration limits.
The length of the string is a gauge-dependent quantity, as it is confirmed by the explicit $a$-dependence.

After this discussion on the gauge-fixing procedure for the bosonic model, let us now include also the fermionic degrees of freedom.

\section{Fermions and type IIB}\label{sec:fermions-type-IIB}
When symmetries allow for a supercoset description, the action of the string may be computed perturbatively in powers of fermions as it was done for the case of {\adsfive} in~\cite{Metsaev:1998it} following the ideas of~\cite{Henneaux:1984mh}.
For our discussions we will need only the contribution to the action at \emph{quadratic} order in fermions.
In order to be more general and account also for cases in which a coset description is not valid, we review the Green-Schwarz action for the superstring~\cite{Green:1983wt}.

We work in type IIB, where we have two sets of 32-components Majorana-Weyl fermions $\Theta_I$ labelled by $I=1,2$. In most expressions we write only these labels and we omit the spinor indices, on which the ten-dimensional Gamma-matrices are acting. We get a total of 32 real degrees of freedom after imposing the chirality and the Majorana conditions
\begin{equation}
\G_{11}\Theta_I=\Theta_I\,,
\qquad\qquad
\bar{\Theta}_I=\Theta^t_I \mathcal{C}\,.
\end{equation}
In the above equations, $\G_{11}$ is constructed by multiplying all the $32\times32$ rank-1 Gamma-matrices, and $\mathcal{C}$ is the charge conjugation matrix.
The barred version of the fermions is defined in the standard way $\bar{\Theta}_I\equiv\Theta^\dagger_I \G^0$.

The Green-Schwarz action of type II superstring~\cite{Grisaru1985116} may be found order by order in fermions, and its explicit form is known to fourth order in $\Theta$~\cite{Wulff:2013kga}. For us it will be enough to stop at second order~\cite{Grisaru1985116,Cvetic:1999zs} 
\be\label{eq:IIB-action-theta2}
\begin{aligned}
S^{\fer^2}&= \int_{-{\frac{L}{2}}}^{\frac{L}{2}} \, {\rm d}\sigma {\rm d} \tau\
\lagr^{\fer^2}\,,
\\
\lagr^{\fer^2}&=-\frac{g}{2} 
\ i\, \bar{\Theta}_I \left( \g^{\a\b} \delta^{IJ} +\epsilon^{\a\b} \sigma_3^{IJ} \right) {e}^m_\a \G_m \, {D}^{JK}_\b \Theta_K\,.
\end{aligned}
\ee
In type IIB the operator ${D}^{IJ}_\a$ acting on $\Theta$ has the following expression 
\be
\begin{aligned}
{D}^{IJ}_\a  = &
\delta^{IJ} \left( \pa_\a  -\frac{1}{4} {\omega}^{mn}_\a \G_{mn}  \right)
+\frac{1}{8} \sigma_3^{IJ} {e}^m_\a {H}_{mnp} \G^{np}
\\
&-\frac{1}{8} e^{\varphi} \left( \epsilon^{IJ} \G^p {F}^{(1)}_p + \frac{1}{3!}\sigma_1^{IJ} \G^{pqr} {F}^{(3)}_{pqr} + \frac{1}{2\cdot5!}\epsilon^{IJ} \G^{pqrst} {F}^{(5)}_{pqrst}   \right) {e}^m_\a \G_m.
\end{aligned}
\ee
In the equations above, $e^m_\a=\pa_{\a}X^Me^m_M$ is the pullback of the vielbein on the worldsheet, and it is related to the spacetime metric as
\be
G_{MN}=e^m_Me^n_N\eta_{mn}\,.
\ee
The spin connection $\omega^{mn}_{\a}=\pa_{\a}X^M\omega_{M}^{mn}$ satisfies the equation
\be\label{eq:spi-conn-vielb}
{\omega}_M^{mn}=
- {e}^{N \, [m} \left( \pa_M {e}^{n]}_N - \pa_N {e}^{n]}_M + {e}^{n] \, P} {e}_M^p \pa_P {e}_{Np} \right),
\ee
where the factor $1/2$ is included in the anti-symmetrisation of the indices $m,n$.
Also the field-strength of the $B$-field appears in the fermionic action
\be
H_{MNP}=3\pa_{[M}B_{NP]}=\pa_M B_{NP}+\pa_N B_{PM}+\pa_P B_{MN}\,.
\ee
The quantities denoted by $F^{(n)}$ are the Ramond-Ramond field-strengths and $\varphi$ is the dilaton.
The whole set of fields satisfies the supergravity equations of motion~\cite{Bergshoeff:1985su}.
We refer to Appendix~\ref{app:IIBsugra} where we collect these equations.

\medskip

The Green-Schwarz action presented above enjoys a local fermionic symmetry called ``kappa-symmetry''~\cite{Green:1983wt,Grisaru1985116}. This is a generalisation of the symmetry found for superparticles~\cite{Siegel1983397} and it allows one to gauge away half of the fermions, thus recovering the correct number of physical degrees of freedom.
At lowest order, the kappa-variation is implemented on the bosonic and fermionic coordinates as
\be\label{eq:kappa-var-32}
\begin{aligned}
 \delta_{\kappa}X^M &= - \frac{i}{2} \ \bar{\T}_I \G^M  \delta_{\kappa} \T_I + \mathcal{O}(\T^3)\,,
\qquad &&\G^M={e}^{Mm}  \G_m\,,
\\
\delta_{\kappa} \T_I &= -\frac{1}{4} (\delta^{IJ} \gamma^{\a\b} - \sigma_3^{IJ} \epsilon^{\a\b})  \G_\b  {K}_{\a J}+ \mathcal{O}(\T^2)\,,
\qquad &&\G_\b={e}_{\b}^m  \G_m\,,
\end{aligned}
\ee
where we have introduced local fermionic parameters ${K}_{\a I}$ with chirality opposite to the one of the fermions $\G_{11}{K}_{\a I}=-{K}_{\a I}$.
Together with the kappa-variation of the worldsheet metric
\be
\begin{aligned}
\delta_\kappa \g^{\a\b}&=
 2i\ \Pi^{IJ\, \a\a'}\Pi^{JK\, \b\b'}
\ \bar{{K}}_{I\a'}{D}^{KL}_{\b'}\Theta_{L}+ \mathcal{O}(\T^3),
\\
\Pi^{IJ\, \a\a'}&\equiv\frac{\delta^{IJ}\g^{\a\a'}+\sigma_3^{IJ}\epsilon^{\a\a'}}{2}\,,
\end{aligned}
\ee
one finds invariance of the action under kappa-symmetry $\delta_{\kappa}(S^\bos+S^{\fer^2})=0$ at first order in $\Theta$.

Let us use this freedom to gauge away half of the fermions.
We consider the Gamma-matrices $\G_0$ and $\G_5$---corresponding to the coordinates $t$ and $\phi$ used in Section~\ref{sec:Bos-string-lcg} to fix the gauge for bosonic strings---and we define the combinations\footnote{Another definition that seems natural from the point of view of a generic $a$-gauge is $\G^+=(1-a)\G^0+a\G^5, \ \G^-=-\G^0+\G^5$.}
\be\label{eq:defin-Gamma-pm}
\G^{\pm}=\frac{1}{2} (\G^5\pm\G^0)\,.
\ee
Kappa-symmetry is fixed by imposing~\cite{Metsaev:2000yu}
\be\label{eq:kappa-lcg}
\G^+\T_I=0\quad \implies\quad \bar{\T}_I\G^+=0\,.
\ee
This gauge simplifies considerably the form of the Lagrangian. To start, in this gauge all terms containing an even number of Gamma-matrices in the light-cone directions vanish, as it is seen by using the identity
\be
\G^+\G^-+\G^-\G^+=\mathbf{1}_{32}\,.
\ee
Moreover, the motivation for choosing this gauge is that at leading order in the usual perturbative expansion in powers of fields it gives a non-vanishing and standard kinetic term for fermions, see Section~\ref{sec:decomp-limit}.


\medskip

Let us first show how to generalise the procedure of Section~\ref{sec:Bos-string-lcg} by including the fermionic contributions.
We first define an effective metric $\hat{G}_{MN}$ and an effective $B$-field $\hat{B}_{MN}$ containing all the couplings to the fermions that do not involve derivatives on them
\be
\begin{aligned}
\hat{G}_{MN}=&{G}_{MN} + i   \,\bar{\Theta}_I \, \, {e}^m_{(M} \G_m \Bigg[ -\frac{1}{4}\delta^{IJ} {\omega}^{pq}_{N)} \G_{pq}    
+ \frac{1}{8} \sigma_3^{IJ}{e}^n_{N)} H_{npq} \G^{pq}    \\
& -\frac{1}{8} e^\varphi \left(\eps^{IJ} \G^p F^{(1)}_p +\frac{1}{3!}\sigma_1^{IJ} \G^{pqr} F^{(3)}_{pqr} +\frac{1}{2\cdot5!} \eps^{IJ}\G^{pqrst} F^{(5)}_{pqrst}  \right) {e}^n_{N)} \G_n      \Bigg]\Theta_J\, ,
\\
\hat{B}_{MN}=&{B}_{MN} - i   \,\sigma_3^{IK}\bar{\Theta}_I \, \, {e}^m_{[M} \G_m \Bigg[ -\frac{1}{4}\delta^{KJ} {\omega}^{pq}_{N]} \G_{pq}    
+ \frac{1}{8} \sigma_3^{KJ}{e}^n_{N]} H_{npq} \G^{pq}    \\
& -\frac{1}{8} e^\varphi \left(\eps^{KJ} \G^p F^{(1)}_p +\frac{1}{3!}\sigma_1^{KJ} \G^{pqr} F^{(3)}_{pqr} +\frac{1}{2\cdot5!} \eps^{KJ}\G^{pqrst} F^{(5)}_{pqrst}  \right) {e}^n_{N]} \G_n      \Bigg]\Theta_J\, .
\end{aligned}
\ee
This allows us to rewrite the sum of the bosonic and fermionic Lagrangians as
\be
\begin{aligned}
\lagr^{\bos}+\lagr^{\fer^2}=-\frac{g}{2}   \Bigg( &\, \gamma^{\alpha\beta} \partial_\alpha X^M \partial_\beta X^N \hat{G}_{MN} -\epsilon^{\alpha\beta} \partial_\alpha X^M \partial_\beta X^N \hat{B}_{MN} \\
&+i\, \bar{\Theta}_I \left( \g^{\a\b} \delta^{IJ} +\epsilon^{\a\b} \sigma_3^{IJ} \right) {e}^m_\a \G_m \, {\pa}_\b \Theta_J
\Bigg)\,.
\end{aligned}
\ee
The momenta $p_M$ conjugate to the bosonic coordinates $X^M$ receive fermionic corrections, that using the above rewriting are
\be
\begin{aligned}
p_M =&- g \gamma^{\tau\beta} \partial_\beta X^N \hat{G}_{MN} + g X^{'N} \hat{B}_{MN}
\\
&-g\frac{i}{2}\,  \bar{\Theta}_I \left(  \g^{\tau\b}  \delta^{IJ} \G_M \, {\pa}_\b \Theta_J
+\sigma_3^{IJ}   \G_M \, \Theta_J'\right)\,.
\end{aligned}
\ee
After inverting the above relation for $\dot{X}^M$ we find that the Lagrangian is
\be\label{eq:lagr-I-order-form-fer}
\begin{aligned}
\lagr^{\bos}+\lagr^{\fer^2}=&p_M\dot{X}^M +\frac{i}{2}p_M\bar{\Theta}_I\G^M\dot{\T}_I
\\
&+\frac{i}{2}g\, \sigma_3^{IJ} \, X'^M\bar{\T}_I\G_M\dot{\T}_J +\frac{i}{2}g\,  B_{MN}X'^M\bar{\T}_I \G^N\dot{\T}_I
\\& + \frac{\gamma^{\tau\sigma}}{\gamma^{\tau\tau}} C_1 + \frac{1}{2g \gamma^{\tau\tau}} C_2\,.
\end{aligned}
\ee
At second order in fermions, the two Virasoro constraints read as
\begin{equation}\label{eq:Vira-bos-and-fer}
\begin{aligned}
C_1 =& p_M X'^{M}+\frac{i}{2}p_M\bar{\T}_I\G^M\T_I'+\frac{i}{2}g\, \sigma_3^{IJ}\, X'^M \bar{\T}_I\G_M\T_J'+\frac{i}{2}g\,  B_{MN}X'^M\bar{\T}_I \G^N\T_I', \\ 
C_2 =&  \hat{G}^{MN} p_M p_N+ g^2 X'^{M} X'^{N} \hat{G}_{MN} - 2 g\, p_M X'^{Q} \hat{G}^{MN} \hat{B}_{NQ} + g^2 X'^{P} X'^{Q} \hat{B}_{MP} \hat{B}_{NQ} \hat{G}^{MN}\\
&+ig^2\, X'^M \bar{\T}_I\G_M\T_I'+ig\, \sigma_3^{IJ} p_M\bar{\T}_I\G^M \T_J'-ig^2 \, {B}_{MP}X'^P \sigma_3^{IJ} \bar{\T}_I \G^M \T_J'. 
\end{aligned}
\end{equation}
At this point we introduce bosonic light-cone coordinates as in~\eqref{eq:lc-coord} and fix the gauge as in~\eqref{eq:unif-lcg}. 
Together with the gauge fixing for the fermions~\eqref{eq:kappa-lcg}, we then find the gauge-fixed Lagrangian at order $\Theta^2$.
Now $x'^-$ and $p_+$ must be determined by solving the Virasoro constraints $C_1=0,\ C_2=0$ that include the fermionic contributions as in~\eqref{eq:Vira-bos-and-fer}.
The gauge-fixed Lagrangian\footnote{We have assumed as in the previous section that the $B$-field vanishes along light-cone coordinates.}
\be\label{eq:gauge-fix-lagr-I-order-form-fer}
\begin{aligned}
\left(\lagr^{\bos}+\lagr^{\fer^2}\right)_{\text{g.f.}}=&p_\mu\dot{X}^\mu
+\frac{i}{2}\bar{\Theta}_I\left[\delta^{IJ}\left(p_+\G^{\check{+}}+p_-\G^{\check{-}}\right)+g\, \sigma_3^{IJ} \, X'^-\G_{\check{-}}\right]\dot{\T}_I
\\
&+p_+
\,,
\end{aligned}
\ee
shows that the Hamiltonian for the gauge-fixed model remains to be related to the momentum conjugate to $x^+$, namely $\mathcal{H}=-p_+(X^\mu,p_\mu,\T_I)$.
In the kinetic term for fermions of the gauge-fixed Lagrangian, Gamma-matrices with transverse indices disappear thanks to the kappa-gauge~\eqref{eq:kappa-lcg}.
We defined Gamma-matrices with checks on the indices to distinguish them from the ones introduced in~\eqref{eq:defin-Gamma-pm}, as now we consider linear combinations of Gamma-matrices with \emph{curved} indices $\G_M=e_M^m\G_m$
\be
\begin{aligned}
&\G^{\check{+}}=\phantom{-}a\G^\phi+(1-a)\G^t\,,
\qquad
&&\G^{\check{-}}=\G^\phi-\G^t\,,
\\
&\G_{\check{-}}=-a\G_t+(1-a)\G_\phi\,,
\qquad
&&\G_{\check{+}}=\G_t+\G_\phi\,.
\end{aligned}
\ee
The kinetic term for the fermions defines a Poisson structure that in general is not canonical. One may choose to keep this or rather implement field redefinitions to recast the kinetic term in the standard form.

The description simplifies when we use the usual perturbative expansion in powers of fields.
We explain how to implement it in the next section, after presenting the decompactification limit.

\section{Decompactification limit and quantisation}\label{sec:decomp-limit}
After fixing light-cone gauge for bosons and fermions as in the previous sections, we obtain a Hamiltonian on a cilinder, defining the time evolution of a closed strings.
In general this model is complicated because of the non-linear nature of the interactions. 

The first step that we take is the so-called \emph{decompactification limit}. It essentially consists in taking the length of the string to be very large $L\gg 1$.
The model originally defined on the cylinder becomes then a problem on the two-dimensional plane.
When this limit is taken, one should replace the periodic boundary conditions for the fields with the ones decaying at infinity.
The strategy is then to solve the model in the $L\to \infty$ limit, and to take into account the finite-length corrections in a later step.
\begin{figure}[t]
  \centering
\includegraphics[width=0.3\textwidth]{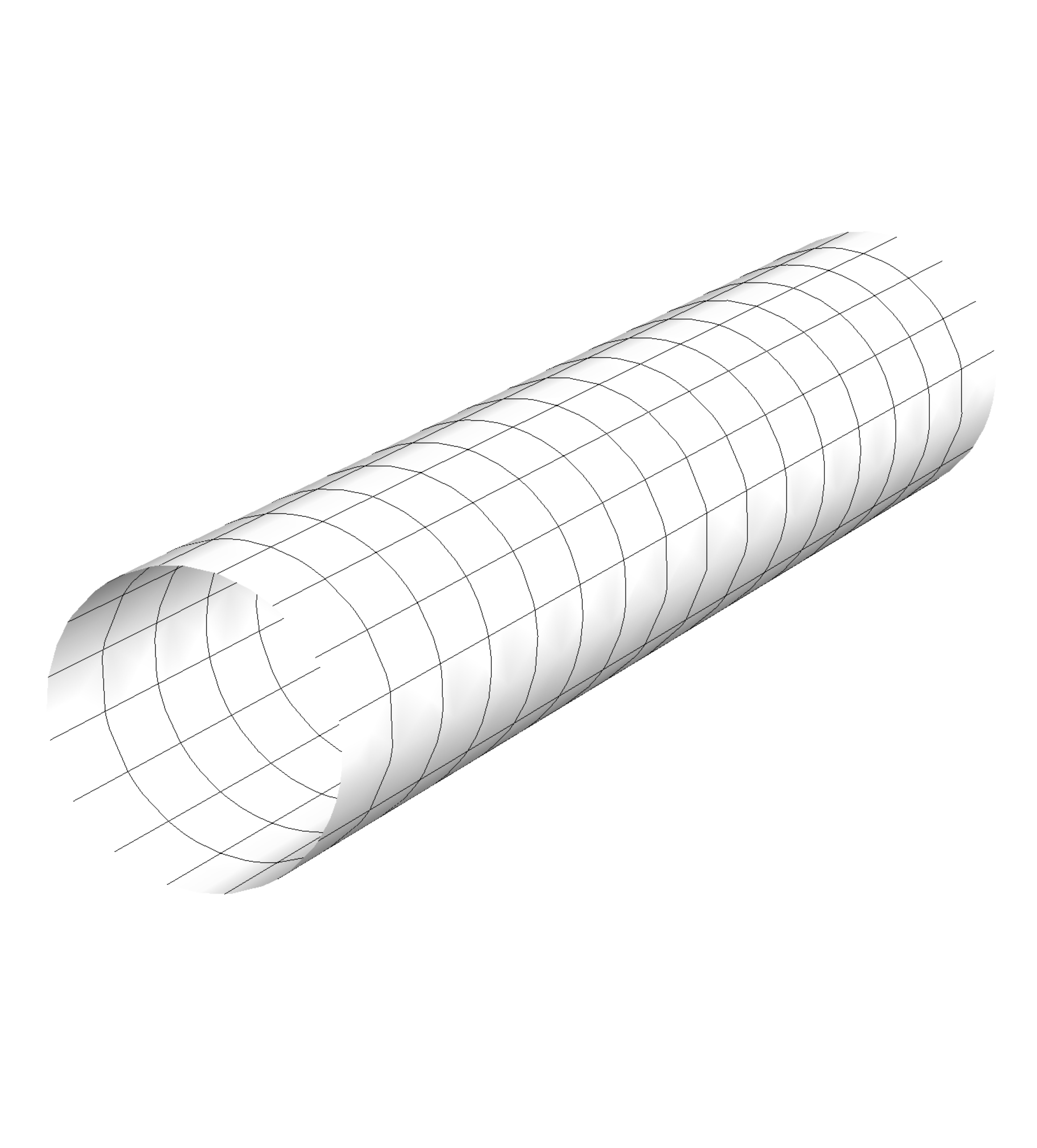}
  \begin{tikzpicture}[%
    box/.style={outer sep=1pt},
    Q node/.style={inner sep=1pt,outer sep=0pt},
    arrow/.style={-latex}
    ]%
	\node [box] at ( 0  , -3cm) {\raisebox{2cm}{$\xrightarrow{P_-\to \infty}$}};
 \end{tikzpicture}
\includegraphics[width=0.4\textwidth]{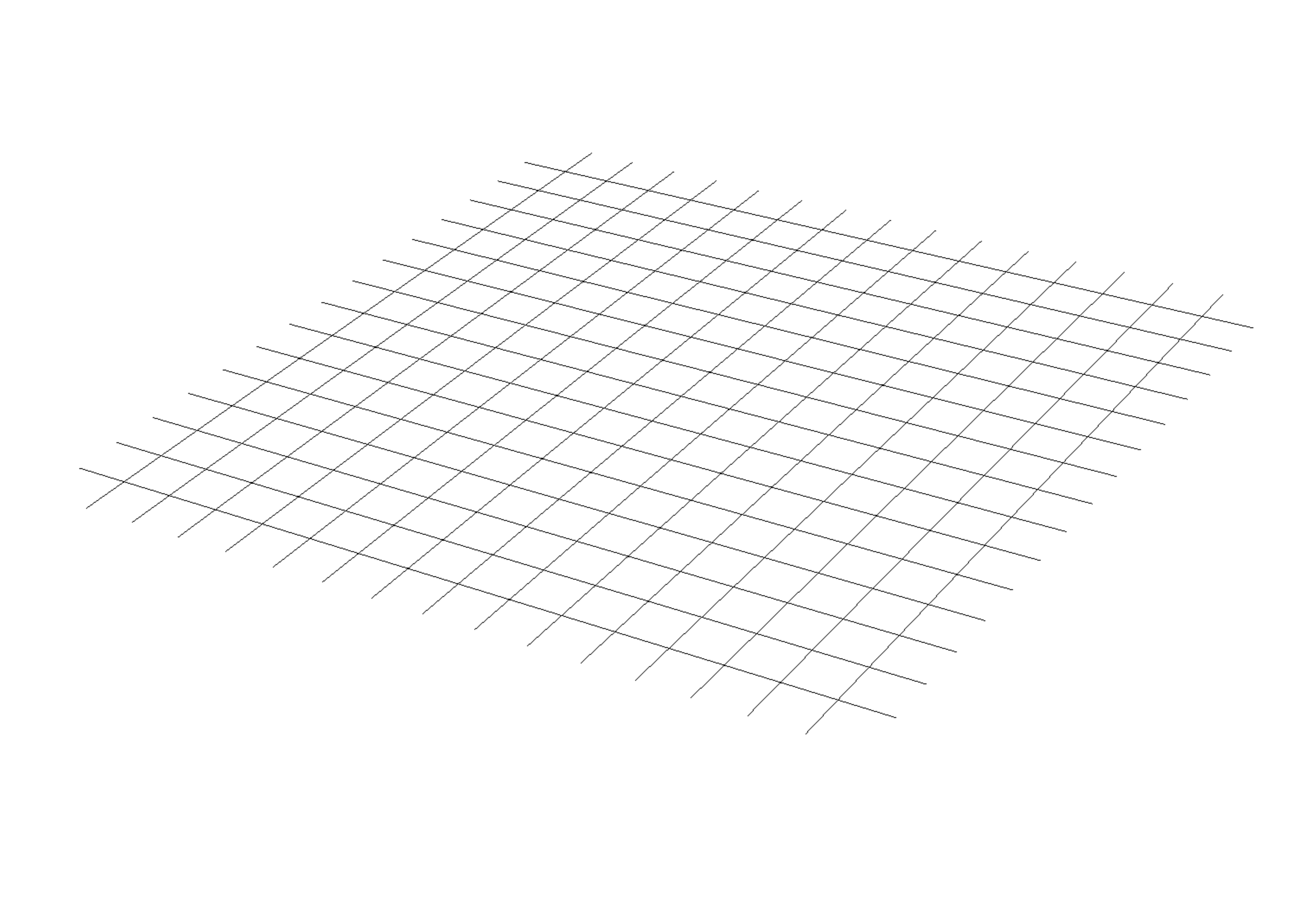}
  \caption{In the decompactification limit we take the total light-cone momentum $P_-$ to be very large. This is equivalent to taking the limit of infinite length of the string. The original model defined on a cylinder lives now on a plane.}
  \label{fig:dec-limit}
\end{figure}
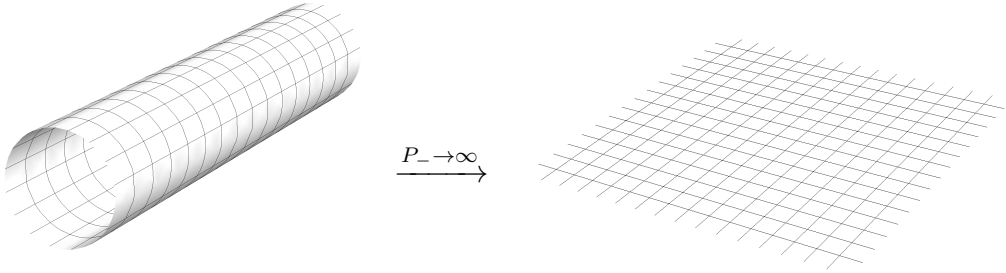

The technical reason why the decompactification limit can be taken is that in uniform light-cone gauge the length of the string $L=P_-$ is equal to the total momentum conjugate to $x^-$, see~\eqref{eq:Ham-En-L-P-}. The momentum $P_-$ enters the light-cone Hamiltonian only through the integration limits for the worldsheet coordinate $\sigma$, therefore sending $P_-\to \infty$ has really just the effect of decompactifying the cylinder.

The light-cone Hamiltonian is expressed in terms of the target-space charges as $E-J$. This means that in order to get configurations with finite worldsheet energy, we should take both $E$ and $J$ to be large, in such a way that their difference is finite.

\subsection{Large tension expansion}\label{sec:large-tens-exp}
In~\cite{Berenstein:2002jq} Berenstein, Maldacena and Nastase (BMN) showed that there exists a limit of AdS$_n\times$S$^n$ spaces that reproduces plane-wave geometries. For the $\sigma$-model on {\adsfive} this matches the plane-wave background of~\cite{Blau:2001ne,Blau:2002dy}.
In the light-cone gauge-fixed theory, this limit is equivalent to the usual expansion in powers of fields truncated at leading order.
In this section we want to look at the ``near-BMN limit'', where we take this expansion beyond the leading order.

In fact, in our case we can look at it as a large-tension expansion.
To implement it one has to rescale the worldsheet coordinate $\sigma\to g \sigma$ and then the bosonic and fermionic fields as
\be\label{eq:field-expansion}
X^\mu\to \frac{1}{\sqrt{g}} X^\mu\,,
\qquad
p_\mu\to \frac{1}{\sqrt{g}} p_\mu\,,
\qquad
\T_I\to \frac{1}{\sqrt{g}} \T_I\,.
\ee
In the action, inverse powers of the string tension $g$ organise the contributions at different powers in the fields
\begin{equation}
\begin{aligned}
S_{\text{g.f.}}= \int_{-\infty}^\infty {\rm d}\tau {\rm d} \sigma \, &\left( \lagr_2+\frac{1}{g}\lagr_4+\frac{1}{g^2}\lagr_6+ \cdots \right).
\end{aligned}
\end{equation}
Here $\lagr_n$ is the contribution to the Lagrangian containing $n$ powers of the physical fields.
Using~\eqref{eq:gauge-fix-lagr-I-order-form-fer}, at lowest order we find simply
\begin{equation}\label{eq:quad-lagr-str}
\begin{aligned}
\lagr_2= p_\mu\dot{X}^\mu +i\, \bar{\Theta}_I\G_0\dot{\T}_I-\mathcal{H}_2.
\end{aligned}
\end{equation}
The first two terms define a canonical Poisson structure for bosons and fermions\footnote{The kinetic term for fermions comes from the term $\frac{i}{2}p_M\bar{\Theta}_I\G^M\dot{\T}_I=\frac{i}{2}p_+\bar{\T}_I[(1-a)\G^t+a\G^\phi]\dot{\T}+\frac{i}{2}p_-\bar{\T}_I[-\G^t+\G^\phi]\dot{\T}$ in~\eqref{eq:lagr-I-order-form-fer}, where Gamma matrices along transverse directions do not contribute thanks to the gauge-fixing for fermions~\eqref{eq:kappa-lcg}. At leading order we have to consider just the contribution of $p_-=1$, and we assume that $e_t^0\sim1, e_\phi^5\sim1$ expanding in transverse bosons.}.

The form of the quadratic Hamiltonian depends on the specific theory considered. For the case of {\adsfive} (this will be true also for its $\eta$-deformations), $\mathcal{H}_2$ is the Hamiltonian for eight free massive bosons and eight free massive fermions, see Section~\ref{sec:quartic-action-lcg-etaads}.
For the case of {\adsthree} we find instead a collection of four bosons and four fermions that are massive, plus four bosons and four fermions that are massless, see Section~\ref{sec:quadr-charges-T4}. The massless fields are a consequence of the presence of the four-dimensional torus in the background. 

The higher order contributions to the Lagrangian define the interactions of the fields, which are organised in inverse powers of the string tension $g$.

We notice that under this rescaling of the physical fields, the quantity $x'^-$ that solves the constraint $C_1$ in~\eqref{eq:Vira-bos-and-fer} has the form
\be\label{eq:xminus-rescaled-g}
x'^-=-\frac{1}{g} \left( p_\mu X'^{\mu}+i\, \bar{\T}_I\G_0\T_I' \right)+\mathcal{O}(1/g^2)\,,
\ee
and the leading contribution is at order $1/g$.
Let us now discuss quantisation of the model.

\medskip

\subsection{Perturbative quantisation}
Here we address the quantisation of the two-dimensional quantum field theory that we find on the worldsheet after the gauge-fixing and the decompactification limit.
Assuming that a canonical Poisson structure for both bosons and fermions of the classical theory was achieved, in the quantised theory we can write equal-time commutation and anti-commutation relations
\be
[X^\mu(\sigma,\tau),p_\nu(\sigma',\tau)]=i\, \delta^\mu_\nu \, \delta(\sigma-\sigma')\,,
\qquad
\{ \T^{\ul{a}}(\sigma,\tau),\T^\dagger_{\ul{b}}(\sigma',\tau) \} = \delta^{\ul{a}}_{\ul{b}}\, \delta(\sigma-\sigma')\,.
\ee
Here $\ul{a},\ul{b}$ are indices that span all the eight complex fermionic degrees of freedom, remaining after gauge fixing.
One may introduce oscillators for the bosonic fields
\be
\begin{aligned}
X^\mu(\sigma,\tau) &= \frac{1}{\sqrt{2\pi}} \int {\rm d} p\, \frac{1}{\sqrt{2\, \omega(p)}} \left( e^{ip\sigma} a^\mu(p,\tau) + e^{-ip\sigma} a^{\mu\dagger}(p,\tau) \right)\,,
\\
p_\mu(\sigma,\tau) &= \frac{1}{\sqrt{2\pi}} \int {\rm d} p \ \frac{i}{2}\, \sqrt{2\, \omega(p)} \left( e^{-ip\sigma} a^\dagger_\mu(p,\tau) - e^{ip\sigma} a_{\mu}(p,\tau) \right)\,,
\end{aligned}
\ee
in such a way that the creation and annihilation operators satisfy canonical commutation relations
\be
[a^\mu(p,\tau),a^\dagger_\nu(p',\tau)]=\delta^\mu_\nu \, \delta(p-p')\,.
\ee
The explicit form of the frequency $\omega(p)$ is dictated by the quadratic Hamiltonian $\mathcal{H}_2$.
Similarly, for fermions we may write
\be
\T^{\ul{a}}(\sigma,\tau)= \frac{e^{i\, \phi_{\ul{a}}}}{\sqrt{2\pi}} \int \frac{{\rm d} p}{\sqrt{\omega(p)}}\, \left( e^{ip\sigma} f(p)\, a^{\ul{a}}(p,\tau) + e^{-ip\sigma} g(p)\, a^{{\ul{a}}\dagger}(p,\tau) \right)\,,
\ee
where we have the freedom of choosing a phase $\phi_{\ul{a}}$, and we have introduced the wave-function parameters $f(p),\ g(p)$.
The creation and annihilation operators satisfy canonical anti-commutation relations 
\be
\{a^{\ul{a}}(p,\tau),a^\dagger_{\ul{b}}(p',\tau)\}=\delta^{\ul{a}}_{\ul{b}} \, \delta(p-p')\,,
\ee
if these functions satisfy\footnote{Typically one also sets $f^2(p)-g^2(p)=m$, so that  $\T^\dagger_I \T_I$ in $\mathcal{H}_2$ generates a mass term written for the oscillators $a,a^\dagger$ multiplied by the mass $m$.}
\be 
f^2(p)+g^2(p)=\omega(p)\,,
\qquad
\frac{f(-p)g(-p)}{\omega(-p)}=-\frac{f(p)g(p)}{\omega(p)}\,.
\ee
For simplicity, let us collect all bosonic and fermionic oscillators together and label them by $k=(\mu,\ul{a})$. The time evolution for these operators is dictated by
\be\label{eq:inter-op-time-ev}
\dot{a}^{k}(p,\tau)=i\, [\gen{H}(a^\dagger,a),a^k(p,\tau)]\,,
\ee
and similarly for creation operators. Here $\gen{H}(a^\dagger,a)$ is the full Hamiltonian written in terms of the oscillators.
Because of the complicated nature of the interactions, one prefers to formulate the problem in terms of \emph{scattering}.

We do not try to describe interactions at any time $\tau$, but rather we focus on the in- and out-operators that evolve freely and coincide with the ones of the interacting theory at $\tau=-\infty$ and $\tau=+\infty$
\be\label{eq:bound-cond-in-out}
a|_{\tau=-\infty} = a_{\text{in}}|_{\tau=-\infty}\,,
\qquad
a|_{\tau=+\infty} = a_{\text{out}}|_{\tau=+\infty}\,.
\ee 
They create in- and out-states
\be
\begin{aligned}
\ket{p_1,p_2,\ldots,p_n}^{\text{in}}_{k_1,k_2,\ldots,k_n}&=a^\dagger_{\text{in},k_1}(p_1)\cdots a^\dagger_{\text{in},k_n}(p_n)\ket{\vacuum}\,,
\\
\ket{p_1,p_2,\ldots,p_n}^{\text{out}}_{k_1,k_2,\ldots,k_n}&=a^\dagger_{\text{out},k_1}(p_1)\cdots a^\dagger_{\text{out},k_n}(p_n)\ket{\vacuum}\,,
\end{aligned}
\ee
from the vacuum $\ket{\vacuum}$, which is killed by annihilation operators.
These operators are particularly simple because by definition interactions are switched off
\be\label{eq:free-op-time-ev}
\begin{aligned}
\dot{a}^{k}_{\text{in}}(p,\tau)&=i\, [\gen{H}_{2}(a^\dagger_{\text{in}},a_{\text{in}}),a^k_{\text{in}}(p,\tau)]\,,
\\
\dot{a}^{k}_{\text{out}}(p,\tau)&=i\, [\gen{H}_{2}(a^\dagger_{\text{out}},a_{\text{out}}),a^k_{\text{out}}(p,\tau)]\,,
\end{aligned}
\ee
meaning that their time evolution is dictated just by the quadratic Hamiltonian $\gen{H}_2$.

Since we want all pairs of creation and annihilation operators $(a^\dagger_{\text{in}},a_{\text{in}})$,$(a^\dagger_{\text{out}},a_{\text{out}})$ and $(a^\dagger,a)$ to satisfy canonical commutation relations, they all must be related by unitarity operators.
In particular, in- and out-operators are related to the interacting operators by
\be
\begin{aligned}
a(p,\tau)&= \mathbb{U}_{\text{in}}^\dagger(\tau) \cdot a_{\text{in}}(p,\tau) \cdot \mathbb{U}_{\text{in}}(\tau)\,,
\\
a(p,\tau)&= \mathbb{U}_{\text{out}}(\tau) \cdot a_{\text{out}}(p,\tau) \cdot \mathbb{U}_{\text{out}}^\dagger(\tau)\,,
\end{aligned}
\ee
where we require $\mathbb{U}_{\text{in}}(\tau=-\infty)={1},\ \mathbb{U}_{\text{out}}(\tau=+\infty)={1}$ to respect the boundary conditions~\eqref{eq:bound-cond-in-out}.
The unitary operator that we call $\mathbb{S}$ is actually the most interesting of them, as it relates in- and out-operators
\be
a_{\text{in}}(p,\tau)=\mathbb{S}\cdot a_{\text{out}}(p,\tau) \cdot \mathbb{S}^\dagger\,,
\qquad
\mathbb{S}\ket{\vacuum}=\ket{\vacuum}\,.
\ee
From this definition we have that the map between in- and out-states is given by the S-matrix
\be
\ket{p_1,p_2,\ldots,p_n}^{\text{in}}_{k_1,k_2,\ldots,k_n}=\mathbb{S}\ket{p_1,p_2,\ldots,p_n}^{\text{out}}_{k_1,k_2,\ldots,k_n},
\ee
and consistency of the above relations implies
\be
\mathbb{S}=\mathbb{U}_{\text{in}}(\tau)\cdot \mathbb{U}_{\text{out}}(\tau).
\ee
Let us mention that the time dependence on the right hand side is only apparent; in fact, the in- and out-operators are free and evolve with the same time dependence, that cancels. This means that we may evaluate the expression at any preferred value of $\tau$.
The three unitary operators are determined by imposing that the time evolutions~\eqref{eq:inter-op-time-ev} and~\eqref{eq:free-op-time-ev} are respected. For $\mathbb{U}_{\text{in}},\mathbb{U}_{\text{out}}$ one can check that
\be
\begin{aligned}
\mathbb{U}_{\text{in}}(\tau) &= \mathcal{T} \text{exp} \left( -i\int_{-\infty}^\tau {\rm d}\tau'\ \gen{V}\left(a_{\text{in}}^\dagger(\tau'),a_{\text{in}}(\tau')\right) \right)\,,
\\
\mathbb{U}_{\text{out}}(\tau) &= \mathcal{T} \text{exp} \left( -i\int_{\tau}^{+\infty} {\rm d}\tau'\ \gen{V}\left(a_{\text{out}}^\dagger(\tau'),a_{\text{out}}(\tau')\right) \right)\,,
\end{aligned}
\ee
solves the desired equations,
where we have introduced the potential $\gen{V}=\gen{H}-\gen{H}_2$, and $\mathcal{T}\text{exp}$ is the time-ordered exponential.
For evaluating the S-matrix we can use the fact that boundary conditions simplify the formulae and write two equivalent results
\be\label{eq:S-mat-Texp}
\begin{aligned}
\mathbb{S}&=\mathbb{U}_{\text{in}}(+\infty)=\mathcal{T} \text{exp} \left( -i\int_{-\infty}^{\infty} {\rm d}\tau'\ \gen{V}\left(a_{\text{in}}^\dagger(\tau'),a_{\text{in}}(\tau')\right) \right)\,,
\\
&=\mathbb{U}_{\text{out}}(-\infty)=\mathcal{T} \text{exp} \left( -i\int_{-\infty}^{\infty} {\rm d}\tau'\ \gen{V}\left(a_{\text{out}}^\dagger(\tau'),a_{\text{out}}(\tau')\right) \right)\,.
\end{aligned}
\ee
We conclude pointing out that perturbation theory is a useful tool to compute the scattering processes. 
We get approximate and simpler results if we define the T-matrix as
\be\label{eq:def-Tmat}
\mathbb{S}=1+\frac{i}{g} \mathbb{T}\,,
\ee
and we take the large-tension expansion of~\eqref{eq:S-mat-Texp}. At leading order we obtain
\be\label{eq:pert-Tmat}
\mathbb{T} = -g\int_{-\infty}^\infty{\rm d} \tau'\ \gen{V}(\tau')+\ldots \,,
\ee
where $\gen{V}=1/g\, \gen{H}_4+\mathcal{O}(1/g^{2})$.
We then recover the known fact that the quartic Hamiltonian provides the $2\to 2$ tree-level scattering elements.
Subleading contributions in inverse powers of $g$ for the matrix $\mathbb{T}$ will give the quantum corrections for the $2\to 2 $ scattering processes.
In the next chapter we show how in some cases non-perturbative methods may be used to account for quantum corrections to all orders.

\chapter{Symmetries of \adsthree}\label{ch:symm-repr-T4}

In this chapter we study the 1+1-dimensional model that emerges as a description for strings on the {\adsthree} background, after fixing light-cone gauge on the worldsheet and taking the decompactification limit.
The symmetry algebra of the original model---the isometries of this background are given by $\psu(1,1|2)_{\sL}\oplus\psu(1,1|2)_{\sR}$---is broken to a smaller algebra under the gauge-fixing procedure explained in Chapter~\ref{ch:strings-light-cone-gauge}. The generators that commute with the light-cone Hamiltonian close into the superalgebra that we call $\mathcal{A}$. The explicit commutation relations are presented in the next section.
Here it is enough to say that the vector space underlying $\mathcal{A}$ can be decomposed as
$$
\psu(1|1)^4 \oplus \alg{u}(1)^2 \oplus \so(4) \oplus \alg{u}(1)^2\,.
$$
The four copies of $\psu(1|1)$ provide  a total of eight real supercharges. As it can be seen in~\eqref{eq:cealgebra}, their anti-commutators yield the $\alg{u}(1)$ central charges corresponding to the light-cone Hamiltonian $\gen{H}$ and an angular momentum $\gen{M}$ in AdS$_3\times$S$^3$. 
The $\so(4)$ symmetry is present only in the decompactification limit, where we have to consider the zero-winding sector for the torus and use vanishing boundary conditions for the worldsheet fields. This $\so(4)$ may be decomposed into $\su(2)_\bullet \oplus \su(2)_\circ$, to show more conveniently that the supercharges transform in doublets of $\su(2)_\bullet$, and are not charged under $\su(2)_\circ$.

Let us introduce some terminology and say that \emph{on-shell}---when we consider states for which the total worldsheet momentum vanishes---these are the only generators appearing.
Going \emph{off-shell} we relax the condition on the momentum and we get $\mathcal{A}$, a central extension of the on-shell algebra.
The two new $\alg{u}(1)$ generators $\gen{C},\overline{\gen{C}}$ measure the momentum of the state and play a major role in the whole construction. 

For the reader's convenience, we start by presenting the commutation relations defining the algebra $\mathcal{A}$, and we explain how to rewrite its generators in terms of the elements of a smaller superalgebra, namely $\su(1|1)^2_{\ce}$.
The explicit form of the charges at quadratic order in the fields permits, on the one hand, to check the closure under the correct commutation relations. On the other hand, it suffices to derive the exact momentum dependence of the eigenvalues of the central charges $\gen{C},\overline{\gen{C}}$ .
We also study the representations of $\mathcal{A}$ under which the worldsheet excitations are organised, first in the near-BMN limit and then to all-loops. We also rewrite them as bi-fundamental representations of $\su(1|1)^2_{\ce}$, and we show that we can define a discrete ``\mbox{Left-Right} symmetry'' that will be crucial for constructing the S-matrix in the next chapter. We conclude with an explicit parameterisation of the action of the charges, as a function of the momenta of the worldsheet excitations.
We collect in an appendix the calculations of the gauge-fixed action needed to obtain these results.

\section{Off-shell symmetry algebra of \adsthree}\label{sec:SymmetryAlgebraT4}
To accustom the reader to the symmetry algebra $\mathcal{A}$ derived in~\cite{Borsato:2014exa}, we start by introducing the notation for the bosonic and fermionic charges, and we present the (anti)commutation relations that they satisfy.
To begin we have the anti-commutators
\begin{equation}
\label{eq:cealgebra}
\begin{aligned}
&\{\gen{Q}_{\smallL}^{\ \dot{a}},\overline{\gen{Q}}{}_{\smallL \dot{b}}\} =\frac{1}{2}\delta^{\dot{a}}_{\ \dot{b}}\,(\gen{H}+\gen{M}),
&\qquad &\{\gen{Q}_{\smallL}^{\ \dot{a}},{\gen{Q}}{}_{\smallR \dot{b}}\} =\delta^{\dot{a}}_{\ \dot{b}}\,\gen{C},
\\
&\{\gen{Q}_{\smallR  \dot{a}},\overline{\gen{Q}}{}_{\smallR}^{\ \dot{b}}\} =\frac{1}{2}\delta^{\ \dot{b}}_{\dot{a}}\,(\gen{H}-\gen{M}),
&\qquad &\{\overline{\gen{Q}}{}_{\smallL \dot{a}},\overline{\gen{Q}}{}_{\smallR}^{\ \dot{b}}\} =\delta^{\ \dot{b}}_{\dot{a}}\,\overline{\gen{C}}.
\end{aligned}
\end{equation}
Here $\gen{H},\gen{M},\gen{C},\overline{\gen{C}}$ are central elements of the algebra. The charge $\gen{H}$ corresponds to the Hamiltonian, and $\gen{M}$ to a combination of angular momenta in AdS$_3\times$S$^3$. 
The charges $\gen{C},\overline{\gen{C}}$ are related by complex conjugation and they appear only after relaxing the level matching condition, see Chapter~\ref{ch:strings-light-cone-gauge}. 
If we set $\gen{C}=\overline{\gen{C}}=0$ we remove the central extension, and the two copies Left (L) and Right (R) of the algebra decouple.

The supercharges are denoted by $\genQ$ and the bar means complex conjugation. The labels L or R are inherited from the superisometry algebra $\alg{psu}(1,1|2)_{\sL}\oplus\alg{psu}(1,1|2)_{\sR} $, where they refer to the chirality in the dual~$\CFT_2 $. 
The supercharges transform under the fundamental and anti-fundamental representations of $\su(2)_{\bullet}$, whose indices are denoted by $\dot{a}=1,2$ 
\begin{equation}
\comm{{\gen{J}_{\bullet {\dot{a}}}}^{\dot{b}}}{\gen{Q}_{\dot{c}}} = \delta^{\dot{b}}_{\ {\dot{c}}} \gen{Q}_{\dot{a}} - \frac{1}{2} \delta^{\ {\dot{b}}}_{\dot{a}} \gen{Q}_{\dot{c}},
\qquad
\comm{{\gen{J}_{\bullet {\dot{a}}}}^{\dot{b}}}{\gen{Q}^{\dot{c}}} = -\delta^{\ {\dot{c}}}_{\dot{a}} \gen{Q}^{\dot{b}} + \frac{1}{2} \delta^{\ {\dot{b}}}_{\dot{a}} \gen{Q}^{\dot{c}}. 
\end{equation}
Here ${\gen{J}_{\bullet {\dot{a}}}}^{\dot{b}}$ denotes the generators of $\su(2)_{\bullet}$. Together with the generators ${\gen{J}_{\circ a}}^b$ 
of $\su(2)_{\circ}$---under which the supercharges are not charged---they span the algebra $\so(4) = \su(2)_{\bullet} \oplus \su(2)_{\circ}$
\begin{equation}
\comm{{\gen{J}_{\bullet \dot{a}}}^{\dot{b}}
}{
{\gen{J}_{\bullet \dot{c}}}^{\dot{d}}}
 = 
\delta^{\dot{b}}_{\ \dot{c}}\, {\gen{J}_{\bullet \dot{a}}}^{\dot{d}}
-
\delta^{\dot{d}}_{\ \dot{a}}\, {\gen{J}_{\bullet \dot{c}}}^{\dot{b}},
\qquad
\comm{{\gen{J}_{\circ a}}^b}{{\gen{J}_{\circ c}}^d} = \delta^b_{\ c}\, {\gen{J}_{\circ a}}^d
-
\delta^d_{\ a}\, {\gen{J}_{\circ c}}^b.
\end{equation}
The whole set of (anti-)commutation relations defines the algebra $\mathcal{A}$, that we continue to study in more detail in the rest of the chapter.

\subsection{The symmetry algebra as a tensor product}\label{sec:AlgebraTensorProductT4}
Focusing on the subalgebra $\psu(1|1)^4_\ce\subset \mathcal{A}$, it is convenient to rewrite its generators---namely the supercharges and the central charges---in terms of generators of a smaller algebra,\footnote{This possibility has a counterpart in the case of \adsfive, where the generators that commute with the light-cone Hamiltonian close into two copies of $\su(2|2)_{\ce}$~\cite{Arutyunov:2006ak}. The S-matrix may be then written as a tensor product of two $\su(2|2)_{\ce}$-invariant S-matrices. In Section~\ref{sec:smat-tensor-prod} we will show how in the case of {\adsthree} we may rewrite an S-matrix compatible with $\psu(1|1)^4_\ce\subset \mathcal{A}$ as a tensor product of two $\su(1|1)^2_\ce$-invariant S-matrices.} that we call $\su(1|1)^2_\ce$.
Let us start from $\su(1|1)^2=\alg{su}(1|1)_{\sL}\oplus\alg{su}(1|1)_{\sR}$, defined as the sum of two copies of $\alg{su}(1|1)$ labelled by L and R
\begin{equation}\label{eq:comm-rel-su112}
\acomm{\mathbb{Q}_{\smallL}}{\overline{\mathbb{Q}}{}_{\smallL}} = \mathbb{H}_{\smallL},
\qquad
\acomm{\mathbb{Q}_{\smallR}}{\overline{\mathbb{Q}}{}_{\smallR}} = \mathbb{H}_{\smallR}.
\end{equation}
A central extension of this is the algebra $\su(1|1)^2_\ce$ that we want to consider. The two new central elements $\mathbb{C},\overline{\mathbb{C}}$ appear on the right hand side of the following anti-commutators mixing L and R~\cite{Borsato:2012ud}
\begin{equation}\label{eq:comm-rel-su112-ce}
\acomm{\mathbb{Q}_{\smallL}}{\mathbb{Q}_{\smallR}} = \mathbb{C}\,,
\qquad
\acomm{\overline{\mathbb{Q}}{}_{\smallL}}{\overline{\mathbb{Q}}{}_{\smallR}} = \overline{\mathbb{C}}\,.
\end{equation}
It is now easy to see that the supercharges of $\psu(1|1)^4$ appearing in the previous subsection may be constructed via the elements of $\su(1|1)^2_\ce$. 
Intuitively, we identify the $\su(2)_\bullet$ index ``$1$'' with the first space in a tensor product, and the index ``$2$'' with the second space\footnote{It is important to make this identification when Left supercharges have an upper $\su(2)_\bullet$ index, while for Right supercharges the index is lower. In fact, for this rewriting to work, if Left supercharges transform in the anti-fundamental representation of $\su(2)_\bullet$, then Right supercharges have to transform in the fundamental---or viceversa. Hermitian conjugation swaps fundamental and anti-fundamental representations.} and we write
\begin{equation}\label{eq:supercharges-tensor-product}
  \begin{aligned}
    \gen{Q}_{\smallL}^{\ 1} = \mathbb{Q}_{\smallL} \otimes \1 ,\, \qquad
    \overline{\gen{Q}}{}_{\smallL 1} = \overline{\mathbb{Q}}{}_{\smallL} \otimes \1 ,\, \qquad
    \gen{Q}_{\smallL}^{\ 2} = \Sigma \otimes \mathbb{Q}_{\smallL} ,\, \qquad
    \overline{\gen{Q}}{}_{\smallL 2} = \Sigma \otimes \overline{\mathbb{Q}}{}_{\smallL} , \\
    \gen{Q}_{\smallR 1} = \mathbb{Q}_{\smallR} \otimes \1 , \qquad
    \overline{\gen{Q}}{}_{\smallR}^{\ 1} = \overline{\mathbb{Q}}{}_{\smallR} \otimes \1 , \qquad
    \gen{Q}_{\smallR 2} = \Sigma \otimes \mathbb{Q}_{\smallR} , \qquad
    \overline{\gen{Q}}{}_{\smallR}^{\ 2} = \Sigma \otimes \overline{\mathbb{Q}}{}_{\smallR} .
  \end{aligned}
\end{equation}
The matrix $\Sigma$ is defined as the diagonal matrix taking value $+1$ on bosons and $-1$ on fermions. In this way we can take into account the odd nature of the supercharges while using the ordinary tensor product $\otimes$.
Following the same rule, for the central elements we first define
\begin{equation}\label{eq:supercharges-tensor-product-C}
  \begin{aligned}
    &\gen{H}_{\smallL}^{\ 1} = \mathbb{H}_{\smallL} \otimes \1 , \qquad &&
    \gen{H}_{\smallL}^{\ 2} = \1 \otimes \mathbb{H}{\smallL} , \qquad &&
    \gen{C}^{1} = \mathbb{C} \otimes \1 ,
\qquad &&
    \gen{C}^{2} = \1 \otimes \mathbb{C} , \\
    &\gen{H}_{\smallR}^{\ 1} = \mathbb{H}_{\smallR} \otimes \1 , \qquad &&
    \gen{H}_{\smallR}^{\ 2} = \1 \otimes \mathbb{H}_{\smallR} , \qquad &&
    \overline{\gen{C}}{}^{1} = \overline{\mathbb{C}} \otimes \1 ,
\qquad &&
    \overline{\gen{C}}{}^{2} = \1 \otimes \overline{\mathbb{C}} .
  \end{aligned}
\end{equation}
To reproduce the property that these generators are not charged under the $\su(2)_\bullet$ algebra, we identify the charges in the two spaces as
\begin{equation}
\gen{H}_{\sL}\equiv\gen{H}_{\smallL}^{\ 1}=\gen{H}_{\smallL}^{\ 2},
\quad
\gen{H}_{\sR}\equiv\gen{H}_{\smallR}^{\ 1}=\gen{H}_{\smallR}^{\ 2},
\qquad
\gen{C}\equiv\gen{C}^1=\gen{C}^2,
\quad
\overline{\gen{C}}\equiv\overline{\gen{C}}{}^1=\overline{\gen{C}}{}^2.
\end{equation}
Another consequence of this requirement is that the above generators become proportional to the identity operator on irreducible representations.

Using these identifications and the anti-commutation relations~\eqref{eq:comm-rel-su112}-\eqref{eq:comm-rel-su112-ce}, one can check that the anti-commutation relations~\eqref{eq:cealgebra} of $\psu(1|1)^4_\ce$ are satisfied, where we have
\begin{equation}
\gen{H}=\gen{H}_{\smallL}+\gen{H}_{\smallR},\qquad
\gen{M}=\gen{H}_{\smallL}-\gen{H}_{\smallR}\,.
\end{equation}
The tensor product construction presented here will be particularly useful when studying the representations of the algebra $\mathcal{A}$, and we refer to Section~\ref{sec:BiFundamentalRepresentationsT4} for further details.

\subsection{Charges quadratic in the fields}\label{sec:quadr-charges-T4}
We present the expressions for the bosonic and fermionic conserved charges that enter the superalgebra $\mathcal{A}$, as derived from the worldsheet Lagrangian.
We refer to Appendix~\ref{app:gauge-fixed-action-T4} for notation, and for the calculations of the gauged-fixed action following the general explanation of Chapter~\ref{ch:strings-light-cone-gauge}.
We parameterise the transverse directions of AdS$_3$ with complex coordinates $Z,\bar{Z}$ and the transverse directions of S$^3$ with $Y,\bar{Y}$, such that $Z^\dagger=\bar{Z},Y^\dagger=\bar{Y}$. The directions on T$^4$ are denoted by $X^{\dot{a}a}$. The index $\dot{a}=1,2$ corresponds to $\su(2)_{\bullet}$, while $a=1,2$ to $\su(2)_{\circ}$. The reality condition on these bosons is $(X^{11})^\dagger=X^{22}, (X^{12})^\dagger=-X^{21}$. 

Half of the fermions are denoted with the letter $\eta$ and carry a label L or R. Being charged under $\su(2)_{\bullet}$ they carry also an index $\dot{a}=1,2$.
The other half of the fermions are denoted with the letter $\chi$ and are equipped with a label $+$ or $-$ and the $\su(2)_{\circ}$ index $a=1,2$. In both cases a bar means charge conjugation. 
Later we show that the former are massive, the latter are massless. 

At quadratic order in the transverse fields\footnote{In this section we use bold face notation also for the charges written in terms of the fields.}, the light-cone Hamiltonian $\gen{H}$ and the angular momentum $\gen{M}$ are 
\begin{equation}\label{eq:quadr-Hamilt-fields-T4}
\begin{aligned}
&\gen{H}=\int{\rm d}\sigma\Bigg( 
2 P_{\bar{Z}}P_Z +\frac{1}{2}\bar{Z}Z + \frac{1}{2}\bar{Z}'Z'+2 P_{\bar{Y}}P_Y + \frac{1}{2}\bar{Y}Y +\frac{1}{2}\bar{Y}'Y'
\\
&\qquad\qquad\qquad\qquad+\bar{\eta}_{\sL\dot{a}}\eta_{\sL}^{\ \dot{a}}+\bar{\eta}_{\sR}^{\ \dot{a}}\eta_{\sR\dot{a}}
+{\eta}_{\sL}^{\ \dot{a}} \eta_{\sR \dot{a}}'-{\bar{\eta}}_{\sR}^{\ \dot{a}} \bar{\eta}_{\sL \dot{a}}'
\\
&\qquad\qquad\qquad\qquad+  P_{\dot{a}a}P^{\dot{a}a} +\frac{1}{4} X_{\dot{a}a}'X^{\dot{a}a'}
+{\chi}_{+}^{\ a} \chi_{- a}'-{\bar{\chi}}_{-}^{\ a} \bar{\chi}_{+ a}'
 \Bigg)\,,\\
\end{aligned}
\end{equation}
\begin{equation}
\begin{aligned}
&\gen{M}= \int{\rm d}\sigma\Bigg(iP_{\bar{Z}}Z-iP_Z\bar{Z} +iP_{\bar{Y}}Y-iP_Y\bar{Y} +\bar{\eta}_{\sL\dot{a}}\eta_{\sL}^{\ \dot{a}}-\bar{\eta}_{\sR}^{\ \dot{a}}\eta_{\sR\dot{a}}
\Bigg)\,.
\end{aligned}
\end{equation}
The Hamiltonian shows that the fields $Z,\bar{Z},Y,\bar{Y}$ parameterising AdS and the sphere are massive, with mass equal to $1$ in our units. They are accompanied by fermions $\eta$ with the same value of the mass. The fields $X^{\dot{a}a}$ that parameterise the torus are massless, as well as the fermions denoted by $\chi$.
Taking the Poisson bracket of a given charge with the various fields we may discover its action on them.
In particular, when we do it for the angular momentum $\gen{M}$---a central element of the algebra---we discover that it takes eigenvalues $\pm 1$ for massive fields, and $0$ for massless fields.
We learn that the representation of $\mathcal{A}$ is \emph{reducible}.

The knowledge of the supercharges allows us to compute the central charges $\gen{C},\overline{\gen{C}}$ exactly in the string tension $g$.
In order to do that, one has to keep exact expressions involving the light-cone coordinate $x^-$, that carries the information about the worldsheet momentum as showed in Chapter~\ref{ch:strings-light-cone-gauge}. On the other hand, we might want to perform an expansion in transverse fields, and we actually decide to stop the expansion at quadratic order. 
This is preferable from the point of view of the presentation, and it is enough for our purposes\footnote{For expressions at quartic order---in particular at first order in fermions and third order in bosons--- we refer to~\cite{Borsato:2014hja}.}. 
In this hybrid expansion, the supercharges read as
\begin{equation}\label{eq:supercharges-quadratic-T4}
\begin{aligned}
&\gen{Q}_{\smallL}^{\ {\dot{a}}}=\frac{e^{-\frac{\pi}{4}i}}{2}\int{\rm d}\sigma \ e^{\frac{i}{2}\, x^-}\Bigg(
 2P_{Z}\eta^{\ {\dot{a}}}_{\smallL}-i Z'\bar{\eta}^{\ {\dot{a}}}_{\smallR}+  iZ\eta^{\ {\dot{a}}}_{\smallL}
-\epsilon^{{\dot{a}\dot{b}}}\,
\big(2i P_{\bar{Y}} \bar{\eta}_{\smallL{\dot{b}}}- {\bar{Y}'}\eta_{\smallR {\dot{b}}}+ \bar{Y}\bar{\eta}_{\smallL {\dot{b}}}\big)\\
&
\qquad\qquad\qquad\qquad\qquad
\qquad\qquad\qquad\qquad\qquad
-2\epsilon^{{\dot{a}\dot{b}}} P_{{\dot{b}a}}\chi_{+}^{\ {a}}
-i(X^{{\dot{a}a}})'\, \bar{\chi}_{-{a}} \Bigg),\\
&\gen{Q}_{\smallR {\dot{a}}}=\frac{e^{-\frac{\pi}{4}i}}{2}\int{\rm d}\sigma \ e^{\frac{i}{2}\, x^-}\Bigg(
2P_{\bar{Z}}\eta_{\smallR {\dot{a}}}-i\bar{Z}'\bar{\eta}_{\smallL {\dot{a}}}+i\bar{Z}\eta_{\smallR {\dot{a}}}
+\epsilon_{{\dot{a}\dot{b}}}\,\big(2i {P}_Y\bar{\eta}^{\ {\dot{b}}}_{\smallR}-{Y'}\eta^{\ {\dot{b}}}_{\smallL}+ Y\bar{\eta}^{\ {\dot{b}}}_{\smallR}\big)\\
&\qquad\qquad\qquad\qquad\qquad\qquad\qquad\qquad\qquad\qquad
+2P_{{\dot{a}a}}\chi_{-}^{\ {a}}
-i \epsilon_{{\dot{a}\dot{b}}}(X^{{\dot{b}a}})'\, \bar{\chi}_{+{a}}\Bigg),
\end{aligned}
\end{equation}
while their Hermitian conjugates are found directly by
\begin{equation}
\overline{\gen{Q}}{}_{\smallL {\dot{a}}}= (\gen{Q}_{\smallL}^{\ {\dot{a}}})^\dagger,
\qquad\qquad
\overline{\gen{Q}}{}_{\smallR}^{\ {\dot{a}}}= (\gen{Q}_{\smallR {\dot{a}}})^\dagger.
\end{equation}
Using the canonical (anti-)commutation relations for the fields as in Appendix~\ref{app:gauge-fixed-action-T4}, one finds that the above supercharges indeed close into the algebra $\mathcal{A}$ defined by~\eqref{eq:cealgebra}.

We now want to derive the form of the generators $\gen{C},\overline{\gen{C}}$ introduced by the central extension.
Their exact eigenvalues are found thanks to the hybrid expansion of the supercharges, where expressions in $x^-$ have been kept exactly. In fact it is this light-cone coordinate that carries information on the worldsheet momentum, as it can be seen from the Virasoro constraints in Chapter~\ref{ch:strings-light-cone-gauge}.
Computing, for example, $\{\gen{Q}_{\smallL}^{\ 1},{\gen{Q}}{}_{\smallR 1}\}$ one finds\footnote{When we compute the anti-commutator of a Left and a Right supercharge, we should keep only terms at order zero in the transverse fermions, as higher order terms mix with fermionic corrections to the supercharges that we have dropped. This approximation does not prevent to find the result, since $\gen{C}$ is a central element and the knowledge of its eigenvalue on bosonic fields is enough.}
\begin{equation}
\begin{aligned}
&\gen{C}=-\frac{i}{4}\int{\rm d}\sigma \ e^{i\, x^-}\Bigg[ -2i\left( P_Z \bar{Z}' + P_{\bar{Z}} Z'+ P_Y \bar{Y}' + P_{\bar{Y}} Y'+ P_{\dot{a}a} X^{\dot{a}a'}\right)
 \\
&
\qquad\qquad\qquad\qquad\qquad
 +\partial_\sigma(\bar{Z}Z+\bar{Y}Y+X_{\dot{a}a} X^{\dot{a}a}) +\ldots
\Bigg]\\
&\phantom{\gen{C}={}}=\frac{g}{2}\ \int{\rm d}\sigma \ e^{i\, x^-} (x'^- +\text{total derivative})\,,
\end{aligned}
\end{equation}
where we have used the relation~\eqref{eq:xminus-rescaled-g} that solves one of the Virasoro conditions, and we have dropped a total derivative term.
The combination that appears is particularly nice and can be integrated as
\begin{equation}
-\frac{ig}{2}\, \int_{-\infty}^{+\infty}{\rm d}\sigma \ \frac{{\rm d}}{{\rm d} \sigma}e^{i\, x^-} =-\frac{ig}{2}\, \left(e^{i\, x^-(+\infty)} - e^{i\, x^-(-\infty)}\right) 
= -\frac{ig}{2}\, e^{i\, x^-(-\infty)} (e^{i p_{\text{ws}}}-1).
\end{equation}
Here $g$ is the string tension. To be more general, from now on we write this result in terms of a new effective coupling $h(g)$, that may be identified with $g$ in the semiclassical regime $h\sim g$.
These central charges may be then written in terms of the charge $\gen{P}$ measuring the worldsheet momentum as
\begin{equation}
\label{eq:allloop-centralcharges}
\gen{C}=+\frac{ih}{2}(e^{+i\gen{P}}-1),
\qquad\qquad
\overline{\gen{C}}=-\frac{ih}{2}(e^{-i\gen{P}}-1)\,,
\end{equation}
where we have fixed a normalisation for $e^{i\, x^-(-\infty)}$.
This is the key result that will allow us to find the exact S-matrix in Chapter~\ref{ch:S-matrix-T4}.
It is worth stressing that with this computation we were able to fix the exact momentum dependence of these central charges, and one may take into account higher order corrections in power of fields to check that the dependence is not modified.
This derivation is classical, and it would be interesting to explicitly show that the result is solid under quantum corrections, at least at the leading orders in the near-BMN limit.

The eigenvalues of the central charges $\gen{C},\overline{\gen{C}}$ that we have found match with those computed in the case of {\adsfive}~\cite{Arutyunov:2006ak}.
In the context of AdS$_5$/CFT$_4$, these central charges appear also in the construction of the gauge theory side, with exactly the same eigenvalues~\cite{Beisert:2005tm}.
This fact was a strong suggestion that they are not modified by quantum corrections.
We will assume that also in the case of {\adsthree} quantum corrections do not spoil the result found with the classical computation presented above.

\section{Representations of the off-shell symmetry algebra}\label{sec:RepresentationsT4}
In this section we want to study the representations of the off-shell symmetry algebra $\mathcal{A}$ that are relevant for {\adsthree}.
For simplicity we start by considering the near-BMN limit introduced in Section~\ref{sec:large-tens-exp}, and then we explain how we can extend the results to all-loops.
We further study these representations and we provide a parameterisation in terms of the momentum of the excitation.
\subsection{Near-BMN limit}
We start by considering the near-BMN limit, where we truncate the charges at the quadratic order in the fields. For the supercharges this means also that we will ignore the factor $e^{i\, x_-}$ in~\eqref{eq:supercharges-quadratic-T4}.
Introducing creation and annihilation operators, we rewrite the charges in momentum space and discuss the representations under which the excitations transform.
For the explicit map between fields and oscillators we refer to Appendix~\ref{app:oscillators}.
The tables below summarise the notation for the annihilation operators that we use. Creation operators are denoted with a dagger.
We have bosonic ladder operators $a$ carrying a label $z$ or $y$, to distinguish excitations on AdS or the sphere respectively, and a label L or R. 
As anticipated in the previous section, they create massive excitations on the worldsheet. 
The labels L,R appear also for the ladder operators $d$ of massive fermions, that also carry a $\su(2)_{\bullet}$ index.
For bosons of T$^4$, creation and annihilation operators will carry two indices, as they are charged under $\su(2)_{\bullet}$ and $\su(2)_{\circ}$.
They are massless excitations, and together with them we find also massless fermions, whose ladder operators $d,\tilde{d}$ carry just an $\su(2)_{\circ}$ index.

\vspace{12pt}
\noindent
Bosons: \quad
\begin{tabular}{c|c|c}
AdS$_3$ & S$^3$ & T$^4$
\\
\hline
$a_{\sL z}$, $a_{\sR z}$ & $a_{\sL y}$, $a_{\sR y}$ & $a_{\dot{a}a}$
\end{tabular}
\qquad\quad
Fermions: \quad
\begin{tabular}{c|c}
massive & massless
\\
\hline
\raisebox{-3pt}{$d_{\sL,\dot{a}}$, $d_{\sR}{}^{\dot{a}}$} & \raisebox{-3pt}{$d_a$, $\tilde{d}_a$}
\end{tabular}

\vspace{12pt}
When acting on the vacuum, we create the eight massive states that we denote as
\begin{equation}\label{eq:massive-states-BMN}
\ket{Z^{\sL,\sR}}=a^\dagger_{\sL,\sR\, z}\ket{\vacuum},\quad
\ket{Y^{\sL,\sR}}=a^\dagger_{\sL,\sR\, y}\ket{\vacuum},\quad
\ket{\eta^{\sL \dot{a}}}=d^{\ \dot{a} \dagger}_{\sL}\ket{\vacuum},\quad
\ket{\eta^{\sR}_{\ \dot{a}}}=d^{\dagger}_{\sR \dot{a}}\ket{\vacuum},
\end{equation}
and the eight massless ones
\begin{equation}\label{eq:massless-states-BMN}
\ket{T^{\dot{a}a}}=a^{\dot{a} a\dagger}\ket{\vacuum},\qquad
\ket{\chi^{a}}=d^{a\,\dagger}\ket{\vacuum},\qquad
\ket{\widetilde{\chi}^{a}}=\tilde{d}^{a\,\dagger}\ket{\vacuum}.
\end{equation}
The notation that we introduce here for states in the near-BMN limit is the same one that we will use for the exact representations, starting in Section~\ref{sec:exact-repr-T4}.
In terms of ladder operators, the central charges are written as
\begin{equation}\label{eq-central-charges-osc-T4}
\begin{aligned}
&\gen{H}= \int {\rm d} p \ \Bigg[ \omega_p \left( a_{\sL z}^\dagger a_{\sL z} +a_{\sL y}^\dagger a_{\sL y} + a_{\sR z}^\dagger a_{\sR z} +a_{\sR y}^\dagger a_{\sR y}
+ d_{\sL}^{\ {\dot{a}}\,\dagger}d_{\sL {\dot{a}}}  +d_{\sR{\dot{a}}}^\dagger d_{\sR}^{\ {\dot{a}}} \right)
\\
&\qquad\qquad\qquad\qquad\qquad
+\tilde{\omega}_p \left( a^{\dot{a}a\dagger} a_{\dot{a}a} 
+ d^{a\,\dagger}d_{a} + \tilde{d}^{a\,\dagger}\tilde{d}_{a}  \right)
\Bigg],\\
&\gen{M}= \int {\rm d} p \ \Bigg[  a_{\sL z}^\dagger a_{\sL z} +a_{\sL y}^\dagger a_{\sL y}  + d_{\sL}^{\ {\dot{a}}\,\dagger}d_{\sL {\dot{a}}}
-\left( a_{\sR z}^\dagger a_{\sR z} +a_{\sR y}^\dagger a_{\sR y}  +d_{\sR{\dot{a}}}^\dagger d_{\sR}^{\ {\dot{a}}} \right)
\Bigg],\\
&\gen{C}= -\frac{1}{2}\int {\rm d} p \ p \ \Bigg[ a_{\sL z}^\dagger a_{\sL z} +a_{\sL y}^\dagger a_{\sL y} + a_{\sR z}^\dagger a_{\sR z} +a_{\sR y}^\dagger a_{\sR y}
+ d_{\sL}^{\ {\dot{a}}\,\dagger}d_{\sL {\dot{a}}}  +d_{\sR{\dot{a}}}^\dagger d_{\sR}^{\ {\dot{a}}} 
\\
&\qquad\qquad\qquad\qquad\qquad
+ a^{\dot{a}a\dagger} a_{\dot{a}a} + d^{a\,\dagger}d_{a} + \tilde{d}^{a\,\dagger}\tilde{d}_{a}  
\Bigg].
\end{aligned}
\end{equation}
As expected, in the near-BMN limit the frequencies for massive $(\omega_p=\sqrt{1+p^2})$ and massless excitations $(\tilde{\omega}_p=|p|)$ show a relativistic dispersion relation. We will see later how this is deformed by the $h$-dependence, see~\eqref{eq:all-loop-disp-rel-T4}.
In the near-BMN limit, the element introduced by the central extension is essentially just the charge measuring the worldsheet momentum $\gen{C} \sim - \frac{1}{2} \gen{P}$.

The quadratic supercharges of Eq.~\eqref{eq:supercharges-quadratic-T4} take the form
\begin{equation}\label{eq-supercharges-osc-T4}
\begin{aligned}
&\gen{Q}_{\smallL}^{\ {\dot{a}}}= \int {\rm d} p \ \Bigg[
f_p (d_{\sL}^{\ {\dot{a}}\,\dagger} a_{\sL y} + \epsilon^{{\dot{a}\dot{b}}}\, a_{\sL
z}^\dagger  d_{\sL {\dot{b}}})
+g_p (a_{\sR y}^\dagger  d_{\sR}^{\ {\dot{a}}} +\epsilon^{{\dot{a}\dot{b}}}\, d_{\sR
{\dot{b}}}^\dagger  a_{\sR z}) \\
&\qquad\qquad\qquad\qquad\qquad\qquad\qquad\qquad\qquad\qquad
+\tilde{f}_p \left( \epsilon^{{\dot{a}\dot{b}}}\, \tilde{d}^{{a}\,\dagger}a_{{\dot{b}a}}+
a^{{\dot{a}a}\,\dagger}d_{{a}}\right)\Bigg],\\
&\gen{Q}_{\smallR {\dot{a}}}=\int {\rm d} p \ \Bigg[
f_p (d_{\sR {\dot{a}}}^\dagger  a_{\sR y} -\epsilon_{{\dot{a}\dot{b}}}\, a_{\sR z}^\dagger
d_{\sR}^{\ {\dot{b}}})
+g_p (a_{\sL y}^\dagger  d_{\sL {\dot{a}}} -\epsilon_{{\dot{a}\dot{b}}}\,  d_{\sL}^{\
{\dot{b}}\,\dagger} a_{\sL z})\\
&\qquad\qquad\qquad\qquad\qquad\qquad\qquad\qquad\qquad\qquad
 +\tilde{g}_p \left( d^{{a}\,\dagger}a_{{\dot{a}a}}-
\epsilon_{{\dot{a}\dot{b}}}\, a^{{\dot{b}a}\,\dagger}\tilde{d}_{{a}}\right)\Bigg].
\end{aligned}
\end{equation}
Here we have introduced the functions $f_p,g_p$ and $\tilde{f}_p,\tilde{g}_p$ for massive and massless excitations respectively, defined by
\begin{equation}
\begin{aligned}
f_p&=\sqrt{\frac{1+\omega_p}{2}},\qquad\qquad
&&g_p=-\frac{p}{2f_p},
\\
\tilde{f}_p&=\sqrt{\frac{\tilde{\omega}_p}{2}},\qquad\qquad
&&\tilde{g}_p=-\frac{p}{2\tilde{f}_p}.
\end{aligned}
\end{equation}
The action of the bosonic and the fermionic charges on the states of Equation~\eqref{eq:massive-states-BMN} and~\eqref{eq:massless-states-BMN} define the representation under which the excitations of {\adsthree} are organised.

It is clear that this representation is \emph{reducible}. We find three irreducible representations, that may be labelled by the eigenvalue $m$ of the central charge $\gen{M}$: $m=+1$ for Left massive, $m=-1$ for Right massive, and $m=0$ for massless excitations.

\subsection{All-loop representations}\label{sec:exact-repr-T4}
The study of the charges at quadratic level allowed for the understanding of the representations at the near-BMN order.
We now want to go beyond this limit and write down representations that are supposed to be valid to all loops.
In particular we want to reproduce the full non-linear momentum dependence of the charge $\gen{C}$, as in Eq.~\eqref{eq:allloop-centralcharges}.
Doing so we discover that the dispersion relation is modified, and for generic $h$ it is not relativistic.
The results rely on the assumption that the eigenvalues of the central charges $\gen{C},\overline{\gen{C}}$ derived from the classical computation remain unmodified at the quantum level.
As already pointed out, the main motivation for believing this is the fact that the same central charges were found on both sides of the AdS$_5$/CFT$_4$ dual pair~\cite{Beisert:2005tm,Arutyunov:2006ak}.

The key point of the construction is that each of the irreducible representations found in the near-BMN limit---Left-massive, Right-massive and massless---is a \emph{short} representation of $\psu(1|1)^4_{\ce}$. Even beyond the near-BMN limit the dimensionality remains the same, and they remain to be short. 
A generic short representation satisfies the important constraint relating the central charges\footnote{This equation is a consequence of the fact that in a short representation a highest weight state---defined as being annihilated by the raising operators $\overline{\gen{Q}}_{\sL\dot{a}},\gen{Q}_{\sR}^{\ \dot{a}}$---is annihilated also by a particular combination of lowering operators $\frac{1}{2}(\gen{H}-\gen{M})\gen{Q}_{\sL}^{\ \dot{a}}-\gen{C}\overline{\gen{Q}}_{\sR\dot{a}}$. Then the vanishing of the anti-commutator $\{\overline{\gen{Q}}_{\sL\dot{a}},\frac{1}{2}(\gen{H}-\gen{M})\gen{Q}_{\sL}^{\ \dot{a}}-\gen{C}\overline{\gen{Q}}_{\sR\dot{a}}\}$ yields the desired constraint on the central elements.}
\begin{equation}\label{eq:short-cond}
\gen{H}^2 = \gen{M}^2 + 4 \gen{C}\overline{\gen{C}}\,,
\end{equation}
called \emph{shortening condition}.
It allows us to solve immediately for the eigenvalue $E_p$ of the Hamiltonian $\gen{H}$, in terms of the eigenvalues $m$ and $\frac{ih}{2}(e^{i\, p}-1)$ of $\gen{M}$ and $\gen{C}$, yielding
\begin{equation}\label{eq:all-loop-disp-rel-T4}
E_p = \sqrt{ m^2 + 4 h^2 \sin^2 \frac{p}{2}}\,.
\end{equation}
This result is particularly important because it states what is the energy of a fundamental worldsheet excitation at a generic value of $h$. For this reason it is often referred to as the \emph{all-loop} dispersion relation.
$\gen{M}$ measures an angular momentum, and its eigenvalue will continue to take the integer values $m=+1,-1,0$. In other words it does not depend on $h$ and $p$ even beyond the near-BMN limit.

\medskip

We now proceed with the discussion on the \emph{exact} representations by presenting the action of the supercharges on the states. 
The result is written in terms of the coefficients $a_p,\bar{a}_p,b_p,\bar{b}_p$, that will be fixed in Section~\ref{sec:RepresentationCoefficientsT4} by requiring that we reproduce the eigenvalues~\eqref{eq:allloop-centralcharges} and~\eqref{eq:all-loop-disp-rel-T4}  of the central charges, and that we match with the results in the near-BMN limit once we rescale the momentum $p\to p/h$ and send $h\to \infty$. 
We show the action of the supercharges separately for each of the irreducible modules.

\begin{figure}[t]
  \centering
  \begin{tikzpicture}[%
    box/.style={outer sep=1pt},
    Q node/.style={inner sep=1pt,outer sep=0pt},
    arrow/.style={-latex}
    ]%

    \node [box] (PhiM) at ( 0  , 2cm) {\small $\ket{Y^{\sL}}$};
    \node [box] (PsiP) at (-2cm, 0cm) {\small $\ket{\eta^{\sL 1}}$};
    \node [box] (PsiM) at (+2cm, 0cm) {\small $\ket{\eta^{\sL 2}}$};
    \node [box] (PhiP) at ( 0  ,-2cm) {\small $\ket{Z^{\sL}}$};

    \newcommand{\horshift}{0.09cm,0cm}
    \newcommand{\vershift}{0cm,0.10cm}
 
    \draw [arrow] ($(PhiM.west) +(\vershift)$) -- ($(PsiP.north)-(\horshift)$) node [pos=0.5,anchor=south east,Q node] {\scriptsize $\gen{Q}^{\ 1}_{\sL},\overline{\gen{Q}}{}^{\ 1}_{\sR}$};
    \draw [arrow] ($(PsiP.north)+(\horshift)$) -- ($(PhiM.west) -(\vershift)$) node [pos=0.5,anchor=north west,Q node] {};

    \draw [arrow] ($(PsiM.south)-(\horshift)$) -- ($(PhiP.east) +(\vershift)$) node [pos=0.5,anchor=south east,Q node] {};
    \draw [arrow] ($(PhiP.east) -(\vershift)$) -- ($(PsiM.south)+(\horshift)$) node [pos=0.5,anchor=north west,Q node] {\scriptsize $\overline{\gen{Q}}{}_{\sL 1},\gen{Q}_{\sR 1}$};

    \draw [arrow] ($(PhiM.east) -(\vershift)$) -- ($(PsiM.north)-(\horshift)$) node [pos=0.5,anchor=north east,Q node] {};
    \draw [arrow] ($(PsiM.north)+(\horshift)$) -- ($(PhiM.east) +(\vershift)$) node [pos=0.5,anchor=south west,Q node] {\scriptsize $\overline{\gen{Q}}{}_{\sL 2},{\gen{Q}}{}_{\sR 2}$};

    \draw [arrow] ($(PsiP.south)-(\horshift)$) -- ($(PhiP.west) -(\vershift)$) node [pos=0.5,anchor=north east,Q node] {\scriptsize $
    \gen{Q}^{\ 2}_{\sL},\overline{\gen{Q}}{}^{\ 2}_{\sR}$};
    \draw [arrow] ($(PhiP.west) +(\vershift)$) -- ($(PsiP.south)+(\horshift)$) node [pos=0.5,anchor=south west,Q node] {};
       
    \draw [dotted] (PsiM) -- (PsiP) node [pos=0.65,anchor=north west,Q node] {\scriptsize $\gen{J}^{\ \ \dot{b}}_{\suA\dot{a}}$};

  \end{tikzpicture}
\hspace{2cm}
  \begin{tikzpicture}[%
    box/.style={outer sep=1pt},
    Q node/.style={inner sep=1pt,outer sep=0pt},
    arrow/.style={-latex}
    ]%

    \node [box] (PhiM) at ( 0  , 2cm) {\small $\ket{Z^{\sR}}$};
    \node [box] (PsiP) at (-2cm, 0cm) {\small $\ket{\eta^{\sR}_{\  2}}$};
    \node [box] (PsiM) at (+2cm, 0cm) {\small $\ket{\eta^{\sR}_{\  1}}$};
    \node [box] (PhiP) at ( 0  ,-2cm) {\small $\ket{Y^{\sR}}$};

    \newcommand{\horshift}{0.09cm,0cm}
    \newcommand{\vershift}{0cm,0.10cm}
 
    \draw [arrow] ($(PhiM.west) +(\vershift)$) -- ($(PsiP.north)-(\horshift)$) node [pos=0.5,anchor=south east,Q node] {\scriptsize $\gen{Q}^{\ 1}_{\sL},\overline{\gen{Q}}{}^{\ 1}_{\sR}$};
    \draw [arrow] ($(PsiP.north)+(\horshift)$) -- ($(PhiM.west) -(\vershift)$) node [pos=0.5,anchor=north west,Q node] {};

    \draw [arrow] ($(PsiM.south)-(\horshift)$) -- ($(PhiP.east) +(\vershift)$) node [pos=0.5,anchor=south east,Q node] {};
    \draw [arrow] ($(PhiP.east) -(\vershift)$) -- ($(PsiM.south)+(\horshift)$) node [pos=0.5,anchor=north west,Q node] {\scriptsize $\overline{\gen{Q}}{}_{\sL 1},\gen{Q}_{\sR 1}$};

    \draw [arrow] ($(PhiM.east) -(\vershift)$) -- ($(PsiM.north)-(\horshift)$) node [pos=0.5,anchor=north east,Q node] {};
    \draw [arrow] ($(PsiM.north)+(\horshift)$) -- ($(PhiM.east) +(\vershift)$) node [pos=0.5,anchor=south west,Q node] {\scriptsize $\overline{\gen{Q}}{}_{\sL 2},{\gen{Q}}{}_{\sR 2}$};

    \draw [arrow] ($(PsiP.south)-(\horshift)$) -- ($(PhiP.west) -(\vershift)$) node [pos=0.5,anchor=north east,Q node] {\scriptsize $
    \gen{Q}^{\ 2}_{\sL},\overline{\gen{Q}}{}^{\ 2}_{\sR}$};
    \draw [arrow] ($(PhiP.west) +(\vershift)$) -- ($(PsiP.south)+(\horshift)$) node [pos=0.5,anchor=south west,Q node] {};
       
    \draw [dotted] (PsiM) -- (PsiP) node [pos=0.65,anchor=north west,Q node] {\scriptsize $\gen{J}^{\ \ \dot{b}}_{\suA\dot{a}}$};

  \end{tikzpicture}
  \caption{The Left and Right massive modules. The supercharges indicated explicitly correspond to the outer arrows only. The two massive fermions within each module are related by $\su(2)_\bullet$ ladder operators.}
  \label{fig:massive}
\end{figure}
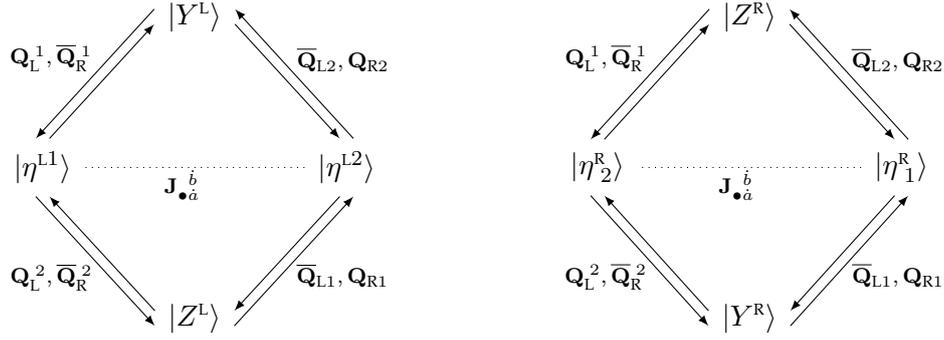

\paragraph{Massive representations}
The Left and Right modules are depicted in Figure~\ref{fig:massive}. Each of them has the shape of a square, where supercharges connect adjacent corners. The two corners hosting the fermions are related by $\su(2)_\bullet$ generators.
More explicitly, the action of the supercharges on the Left module is
\begin{equation}\label{eq:exact-repr-left-massive}
  \begin{aligned}
    \gen{Q}_{\sL}^{\ \dot{a}} \ket{Y_p^{\sL}} &= a_p \ket{\eta^{\sL \dot{a}}_p},
    \qquad
    &\gen{Q}_{\sL}^{\ \dot{a}} \ket{\eta^{\sL \dot{b}}_p} &= \epsilon^{\dot{a}\dot{b}} \, a_p \ket{Z_p^{\sL}}, \\
    \overline{\gen{Q}}{}_{\sL \dot{a}} \ket{Z_p^{\sL}} &=  - \epsilon_{\dot{a}\dot{b}}  \, \bar{a}_p \ket{\eta^{\sL \dot{b}}_p},
    \qquad
    &\overline{\gen{Q}}{}_{\sL \dot{a}} \ket{\eta^{\sL \dot{b}}_p}& =  \delta_{\dot{a}}^{\ \dot{b}}  \, \bar{a}_p \ket{Y_p^{\sL}}, \\[8pt]
    \gen{Q}_{\sR \dot{a}} \ket{Z^{\sL}_p} &= - \epsilon_{\dot{a}\dot{b}} \,  b_p \ket{\eta^{\sL \dot{b}}_p},
    \qquad
    &\gen{Q}_{\sR \dot{a}} \ket{\eta^{\sL \dot{b}}_p} &= \delta_{\dot{a}}^{\ \dot{b}} \, b_p \ket{Y^{\sL}_p},\\
    \overline{\gen{Q}}{}_{\sR}^{\ \dot{a}} \ket{Y^{\sL}_p} &= \bar{b}_p \ket{\eta^{\sL \dot{a}}_p},
    \qquad
    &\overline{\gen{Q}}{}_{\sR}^{\ \dot{a}} \ket{\eta^{\sL \dot{b}}_p} &= \epsilon^{\dot{a}\dot{b}} \,  \bar{b}_p \ket{Z^{\sL}_p}.
  \end{aligned}
\end{equation}
As it is clear also from the picture, if we define $\overline{\gen{Q}}{}_{\sL 1},\gen{Q}_{\sR 1}$ to be our raising operators, then the bosonic excitation $\ket{Y^{\sL}}$ of the sphere is the highest weight state of this module.
For the Right module the situation is different, as the highest weight state is the bosonic excitation $\ket{Z^{\sR}}$ of AdS.
The action of the supercharges in this case is\footnote{Although we have defined Right fermions with a lower $\su(2)_\bullet$ index in Eq.~\eqref{eq:massive-states-BMN}, here we prefer to raise it, to avoid collision with the label for the momentum of the excitation. We raise $\su(2)$ indices with the help of $\epsilon^{\dot{a}\dot{b}}$.}
\begin{equation}\label{eq:exact-repr-right-massive}
  \begin{aligned}
    &\gen{Q}_{\sL}^{\ \dot{a}} \ket{Z_p^{\sR}} =  b_p \ket{\eta^{\sR \dot{a}}_p},
    \qquad
    &\gen{Q}_{\sL}^{\ \dot{a}} \ket{\eta^{\sR \dot{b}}_p} &=- \epsilon^{\dot{a}\dot{b}} \,  b_p \ket{Y_p^{\sR}},\\
    &\overline{\gen{Q}}{}_{\sL \dot{a}} \ket{Y_p^{\sR}} = \epsilon_{\dot{a}\dot{b}} \,  \bar{b}_p \ket{\eta^{\sR \dot{b}}_p},
    \qquad
    &\overline{\gen{Q}}{}_{\sL \dot{a}} \ket{\eta^{\sR \dot{b}}_p} &= \delta_{\dot{a}}^{\ \dot{b}} \,  \bar{b}_p \ket{Z_p^{\sR}}, \\[8pt]
    &\gen{Q}_{\sR \dot{a}} \ket{Y_p^{\sR}} =  \epsilon_{\dot{a}\dot{b}} \,  a_p \ket{\eta^{\sR \dot{b}}_p},
    \qquad
    &\gen{Q}_{\sR \dot{a}} \ket{\eta^{\sR \dot{b}}_p} &= \delta_{\dot{a}}^{\ \dot{b}} \,  a_p \ket{Z_p^{\sR}}, \\
    &\overline{\gen{Q}}{}_{\sR}^{\ \dot{a}} \ket{Z_p^{\sR}} = \bar{a}_p \ket{\eta^{\sR \dot{a}}_p},
    \qquad
    &\overline{\gen{Q}}{}_{\sR}^{\ \dot{a}} \ket{\eta^{\sR \dot{b}}_p} &= - \epsilon^{\dot{a}\dot{b}}  \, \bar{a}_p \ket{Y_p^{\sR}}.
  \end{aligned}
\end{equation}
The above exact representations reproduce the ones found in the near-BMN limit after identifying $a_p\sim\bar{a}_p\sim f_p$ and $b_p\sim\bar{b}_p\sim g_p$.
When going on-shell one has to set also $b_p=\bar{b}_p=0$, with the result that only Left (Right) supercharges act non-trivially on Left (Right) states.

\begin{figure}[t]
  \centering
  \begin{tikzpicture}[%
    box/.style={outer sep=1pt},
    Q node/.style={inner sep=1pt,outer sep=0pt},
    arrow/.style={-latex}
    ]%
\newcommand{\xshiftmasslessone}{-4cm}
\begin{scope}[xshift=\xshiftmasslessone]
    \node [box] (PhiM) at ( 0  , 2cm) {\small $\ket{\chi^{1}}$};
    \node [box] (PsiP) at (-2cm, 0cm) {\small $\ket{T^{11}}$};
    \node [box] (PsiM) at (+2cm, 0cm) {\small $\ket{T^{21}}$};
    \node [box] (PhiP) at ( 0  ,-2cm) {\small $\ket{\widetilde{\chi}^{1}}$};

    \newcommand{\horshift}{0.09cm,0cm}
    \newcommand{\vershift}{0cm,0.10cm}
 
    \draw [arrow] ($(PhiM.west) +(\vershift)$) -- ($(PsiP.north)-(\horshift)$) node [pos=0.5,anchor=south east,Q node] {\scriptsize $\gen{Q}^{\ 1}_{\sL},\overline{\gen{Q}}{}_{\sR}^{\ 1}$};
    \draw [arrow] ($(PsiP.north)+(\horshift)$) -- ($(PhiM.west) -(\vershift)$) node [pos=0.5,anchor=north west,Q node] {};

    \draw [arrow] ($(PsiM.south)-(\horshift)$) -- ($(PhiP.east) +(\vershift)$) node [pos=0.5,anchor=south east,Q node] {};
    \draw [arrow] ($(PhiP.east) -(\vershift)$) -- ($(PsiM.south)+(\horshift)$) node [pos=0.5,anchor=north west,Q node] {
};

    \draw [arrow] ($(PhiM.east) -(\vershift)$) -- ($(PsiM.north)-(\horshift)$) node [pos=0.5,anchor=north east,Q node] {};
    \draw [arrow] ($(PsiM.north)+(\horshift)$) -- ($(PhiM.east) +(\vershift)$) node [pos=0.5,anchor=south west,Q node] {
};

    \draw [arrow] ($(PsiP.south)-(\horshift)$) -- ($(PhiP.west) -(\vershift)$) node [pos=0.5,anchor=north east,Q node] {\scriptsize $\gen{Q}^{\ 2}_{\sL},\overline{\gen{Q}}{}_{\sR}^{\ 2}$};
    \draw [arrow] ($(PhiP.west) +(\vershift)$) -- ($(PsiP.south)+(\horshift)$) node [pos=0.5,anchor=south west,Q node] {};

    \draw [dotted] (PsiM) -- (PsiP) node [pos=0.65,anchor=north west,Q node] {\scriptsize $\gen{J}^{\ \ \dot{b}}_{\suA\dot{a}}$};

\end{scope}
%
%
%
\newcommand{\xshiftmasslesstwo}{1cm}
\newcommand{\yshiftmasslesstwo}{1cm}
\begin{scope}[xshift=\xshiftmasslesstwo,yshift=\yshiftmasslesstwo]

    \node [box] (PhiM) at ( 0  , 2cm) {\small $\ket{\chi^{2}}$};
    \node [box] (PsiP) at (-2cm, 0cm) {\small $\ket{T^{12}}$};
    \node [box] (PsiM) at (+2cm, 0cm) {\small $\ket{T^{22}}$};
    \node [box] (PhiP) at ( 0  ,-2cm) {\small $\ket{\widetilde{\chi}^{2}}$};

    \newcommand{\horshift}{0.09cm,0cm}
    \newcommand{\vershift}{0cm,0.10cm}
 
    \draw [arrow] ($(PhiM.west) +(\vershift)$) -- ($(PsiP.north)-(\horshift)$) node [pos=0.5,anchor=south east,Q node] {
};
    \draw [arrow] ($(PsiP.north)+(\horshift)$) -- ($(PhiM.west) -(\vershift)$) node [pos=0.5,anchor=north west,Q node] {};

    \draw [arrow] ($(PsiM.south)-(\horshift)$) -- ($(PhiP.east) +(\vershift)$) node [pos=0.5,anchor=south east,Q node] {};
    \draw [arrow] ($(PhiP.east) -(\vershift)$) -- ($(PsiM.south)+(\horshift)$) node [pos=0.5,anchor=north west,Q node] {\scriptsize $\overline{\gen{Q}}{}_{\sL 1}, \gen{Q}_{\sR 1}$};

    \draw [arrow] ($(PhiM.east) -(\vershift)$) -- ($(PsiM.north)-(\horshift)$) node [pos=0.5,anchor=north east,Q node] {};
    \draw [arrow] ($(PsiM.north)+(\horshift)$) -- ($(PhiM.east) +(\vershift)$) node [pos=0.5,anchor=south west,Q node] {\scriptsize $\overline{\gen{Q}}{}_{\sL 2}, \gen{Q}_{\sR 2}$};

    \draw [arrow] ($(PsiP.south)-(\horshift)$) -- ($(PhiP.west) -(\vershift)$) node [pos=0.5,anchor=north east,Q node] {
};
    \draw [arrow] ($(PhiP.west) +(\vershift)$) -- ($(PsiP.south)+(\horshift)$) node [pos=0.5,anchor=south west,Q node] {};

    \draw [dotted] (PsiM) -- (PsiP) node [pos=0.65,anchor=south west,Q node] {
};
    
    \draw [dashed] ($(PhiM.west)+({-0.1cm,0.1cm})$) -- ($(PhiM.east)-({\xshiftmasslesstwo,\yshiftmasslesstwo})+({\xshiftmasslessone,0cm})+({-0.1cm,0.2cm})$);
    \draw [dashed] (PsiM) -- ($(PsiM.east)-({\xshiftmasslesstwo,\yshiftmasslesstwo})+({\xshiftmasslessone,0cm})$);
    \draw [dashed] (PsiP) -- ($(PsiP.east)-({\xshiftmasslesstwo,\yshiftmasslesstwo})+({\xshiftmasslessone,0cm})+({-0.1cm,0.1cm})$);
    \draw [dashed] ($(PhiP.south west)+({0cm,0.2cm})$) -- ($(PhiP.east)-({\xshiftmasslesstwo,\yshiftmasslesstwo})+({\xshiftmasslessone,0cm})+({0cm,-0.1cm})$) node [pos=0.45,anchor=north west,Q node] {\scriptsize $\gen{J}^{\ \ b}_{\circ a}$};

\end{scope}
%
  \end{tikzpicture}
  \caption{The massless module. It is composed of two short representations of $\psu(1|1)^4_{\ce}$ that are connected by the action of the $\su(2)_{\circ}$ generators.}
  \label{fig:massless}
\end{figure}
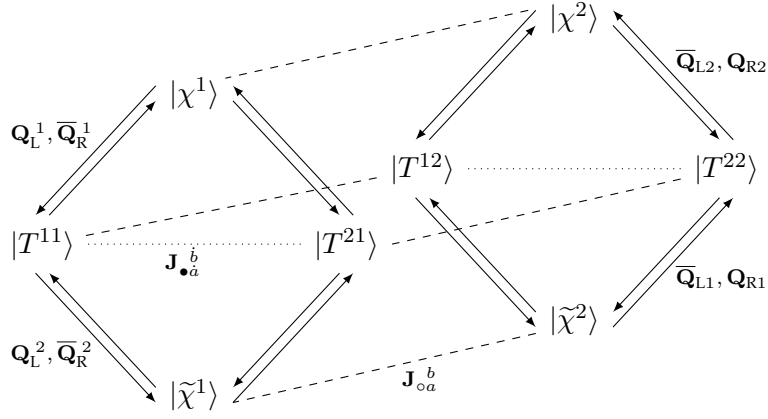

\paragraph{Massless representations}
Figure~\ref{fig:massless} shows the massless module, with the shape of a parallelepiped. It is obtained by gluing together two short $\psu(1|1)^4_{\ce}$ representations---with the shape of a square, like in the case of massive excitations---related by the action of $\su(2)_\circ$ generators.
The explicit action of the supercharges on the massless module is 
\begin{equation}\label{eq:exact-repr-massless}
  \begin{aligned}
    \gen{Q}_{\sL}^{\ \dot{a}} \ket{T^{\dot{b}a}_p}& = \epsilon^{\dot{a}\dot{b}} a_p \ket{\widetilde{\chi}^a_p},
    \qquad
    &\gen{Q}_{\sL}^{\ \dot{a}} \ket{\chi^{a}_p} \;&=  a_p \ket{T^{\dot{a}a}_p}, \\
    \overline{\gen{Q}}{}_{\sL \dot{a}} \ket{\widetilde{\chi}^{a}_p}\;& = -\epsilon_{\dot{a}\dot{b}} \bar{a}_p \ket{T^{\dot{b}a}_p},
    \qquad
    &\overline{\gen{Q}}{}_{\sL \dot{a}} \ket{T^{\dot{b}a}_p} &= \delta_{\dot{a}}^{\ \dot{b}} \bar{a}_p \ket{\chi^a_p}, \\[8pt]
    \gen{Q}_{\sR \dot{a}} \ket{T^{\dot{b}a}_p} &= \delta_{\dot{a}}^{\ \dot{b}} b_p \ket{\chi^a_p},
    \qquad
    &\gen{Q}_{\sR \dot{a}} \ket{\widetilde{\chi}^a_p} \;&= -\epsilon_{\dot{a}\dot{b}} b_p \ket{T^{\dot{b}a}_p}, \\
    \overline{\gen{Q}}{}_{\sR}^{\ \dot{a}} \ket{\chi^a_p}\;& = \bar{b}_p \ket{T^{\dot{a}a}_p},
    \qquad
    &\overline{\gen{Q}}{}_{\sR}^{\ \dot{a}} \ket{T^{\dot{b}a}_p} &= \epsilon^{\dot{a}\dot{b}} \bar{b}_p \ket{\widetilde{\chi}^a_p}.
  \end{aligned}
\end{equation}
Masslessness of the excitations is encoded in the fact that they are  annihilated by~$\gen{M}$, which results in a constraint on the representation coefficients\footnote{We stress that the coefficients $a_p,\bar{a}_p,b_p,\bar{b}_p$ appearing for the massless module are different from the ones for the massive modules. The dependence on the eigenvalue $m$ is not written explicitly not to burden the notation.}
\begin{equation}
\label{eq:masslessness}
|a_p|^2 = |b_p|^2.
\end{equation}
To match with the near-BMN limit one has to take $a_p\sim\bar{a}_p\sim \tilde{f}_p$ and $b_p\sim\bar{b}_p\sim \tilde{g}_p$.
Differently from the massive case, on-shell all supercharges annihilate massless excitations.

\subsection{Bi-fundamental representations}\label{sec:BiFundamentalRepresentationsT4}
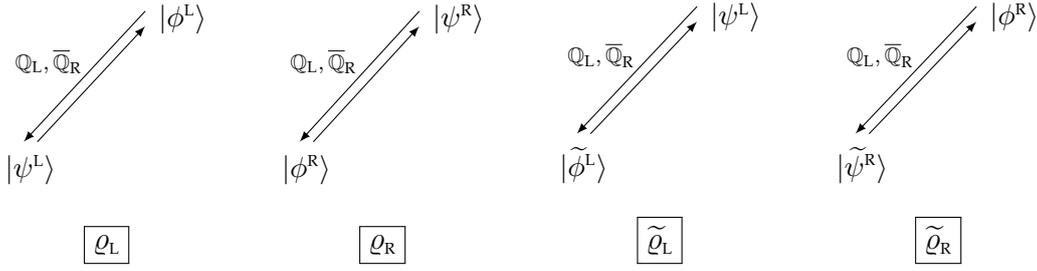
\begin{figure}[t]
  \centering
  \begin{tikzpicture}[%
    box/.style={outer sep=1pt},
    Q node/.style={inner sep=1pt,outer sep=0pt},
    arrow/.style={-latex}
    ]%

    \node [box] (PhiM) at ( 0  , 2cm) {\small $\ket{\phi^{\sL}}$};
    \node [box] (PsiP) at (-2cm, 0cm) {\small $\ket{\psi^{\sL}}$};

    \newcommand{\horshift}{0.09cm,0cm}
    \newcommand{\vershift}{0cm,0.10cm}
 
    \draw [arrow] ($(PhiM.west) +(\vershift)$) -- ($(PsiP.north)-(\horshift)$) node [pos=0.5,anchor=south east,Q node] {\scriptsize $\mathbb{Q}_{\sL},\overline{\mathbb{Q}}_{\sR}$};
    \draw [arrow] ($(PsiP.north)+(\horshift)$) -- ($(PhiM.west) -(\vershift)$) node [pos=0.5,anchor=north west,Q node] {};

	\node [box] at ( -1  , -1cm) {\boxed{\varrho_{\sL}}};

  \end{tikzpicture}
\hspace{0.5cm}
  \begin{tikzpicture}[%
    box/.style={outer sep=1pt},
    Q node/.style={inner sep=1pt,outer sep=0pt},
    arrow/.style={-latex}
    ]%

    \node [box] (PhiM) at ( 0  , 2cm) {\small $\ket{\psi^{\sR}}$};
    \node [box] (PsiP) at (-2cm, 0cm) {\small $\ket{\phi^{\sR}}$};

    \newcommand{\horshift}{0.09cm,0cm}
    \newcommand{\vershift}{0cm,0.10cm}
 
    \draw [arrow] ($(PhiM.west) +(\vershift)$) -- ($(PsiP.north)-(\horshift)$) node [pos=0.5,anchor=south east,Q node] {\scriptsize $\mathbb{Q}_{\sL},\overline{\mathbb{Q}}_{\sR}$};
    \draw [arrow] ($(PsiP.north)+(\horshift)$) -- ($(PhiM.west) -(\vershift)$) node [pos=0.5,anchor=north west,Q node] {};

	\node [box] at ( -1  , -1cm) {\boxed{\varrho_{\sR}}};

  \end{tikzpicture}
\hspace{0.5cm}
  \begin{tikzpicture}[%
    box/.style={outer sep=1pt},
    Q node/.style={inner sep=1pt,outer sep=0pt},
    arrow/.style={-latex}
    ]%

    \node [box] (PhiM) at ( 0  , 2cm) {\small $\ket{\widetilde{\psi}^{\sL}}$};
    \node [box] (PsiP) at (-2cm, 0cm) {\small $\ket{\widetilde{\phi}^{\sL}}$};

    \newcommand{\horshift}{0.09cm,0cm}
    \newcommand{\vershift}{0cm,0.10cm}
 
    \draw [arrow] ($(PhiM.west) +(\vershift)$) -- ($(PsiP.north)-(\horshift)$) node [pos=0.5,anchor=south east,Q node] {\scriptsize $\mathbb{Q}_{\sL},\overline{\mathbb{Q}}_{\sR}$};
    \draw [arrow] ($(PsiP.north)+(\horshift)$) -- ($(PhiM.west) -(\vershift)$) node [pos=0.5,anchor=north west,Q node] {};

	\node [box] at ( -1  , -1cm) {\boxed{\widetilde{\varrho}_{\sL}}};

  \end{tikzpicture}
\hspace{0.5cm}
  \begin{tikzpicture}[%
    box/.style={outer sep=1pt},
    Q node/.style={inner sep=1pt,outer sep=0pt},
    arrow/.style={-latex}
    ]%

    \node [box] (PhiM) at ( 0  , 2cm) {\small $\ket{\widetilde{\phi}^{\sR}}$};
    \node [box] (PsiP) at (-2cm, 0cm) {\small $\ket{\widetilde{\psi}^{\sR}}$};

    \newcommand{\horshift}{0.09cm,0cm}
    \newcommand{\vershift}{0cm,0.10cm}
 
    \draw [arrow] ($(PhiM.west) +(\vershift)$) -- ($(PsiP.north)-(\horshift)$) node [pos=0.5,anchor=south east,Q node] {\scriptsize $\mathbb{Q}_{\sL},\overline{\mathbb{Q}}_{\sR}$};
    \draw [arrow] ($(PsiP.north)+(\horshift)$) -- ($(PhiM.west) -(\vershift)$) node [pos=0.5,anchor=north west,Q node] {};

	\node [box] at ( -1  , -1cm) {\boxed{\widetilde{\varrho}_{\sR}}};

  \end{tikzpicture}
  \caption{Short representations of $\su(1|1)^2_{\ce}$. They differ by the label L or R, and by the grading.}
  \label{fig:su112-repr}
\end{figure}

In Section~\ref{sec:AlgebraTensorProductT4} we showed how it is possible to write the $\psu(1|1)^4_{\ce}$ algebra in terms of $\su(1|1)^2_{\ce}$ generators. In this section we explain how the representations of $\psu(1|1)^4_{\ce}$ that are relevant for {\adsthree} can be understood as proper tensor products of representations of $\su(1|1)^2_{\ce}$.
The representations that we consider in this section are depicted in Figure~\ref{fig:su112-repr}.

We start by considering a possible short representation of $\su(1|1)^2_{\ce}$ that we call $\varrho_{\sL}$. It has dimension two, with one boson denoted by $\phi^{\sL}$ and one fermion denoted by $\psi^{\sL}$. It is defined by the following action of the supercharges that satisfy the commutation relations~\eqref{eq:comm-rel-su112}-\eqref{eq:comm-rel-su112-ce}
\begin{equation}\label{eq:su(1|1)2-repr1}
\boxed{\varrho_{\sL}:} \qquad\qquad
  \begin{aligned}
    \mathbb{Q}_{\smallL} \ket{\phi^{\sL}_p} &= a_p \ket{\psi^{\sL}_p} , \qquad &
    \mathbb{Q}_{\smallL} \ket{\psi^{\sL}_p} &= 0 , \\
    \overline{\mathbb{Q}}{}_{\smallL} \ket{\phi^{\sL}_p} &= 0 , \qquad &
    \overline{\mathbb{Q}}{}_{\smallL} \ket{\psi^{\sL}_p} &= \bar{a}_p \ket{\phi^{\sL}_p} , \\
    \mathbb{Q}_{\smallR} \ket{\phi^{\sL}_p} &= 0 , \qquad &
    \mathbb{Q}_{\smallR} \ket{\psi^{\sL}_p} &= b_p \ket{\phi^{\sL}_p} , \\
    \overline{\mathbb{Q}}{}_{\smallR} \ket{\phi^{\sL}_p} &= \bar{b}_p \ket{\psi^{\sL}_p} , \qquad &
    \overline{\mathbb{Q}}{}_{\smallR} \ket{\psi^{\sL}_p} &= 0 .
  \end{aligned}
\end{equation}
The choice of the coefficients makes sure that the Left and the Right Hamiltonians are positive definite. The above equations identify a \emph{Left} representation, in the sense that on-shell $b_p=\bar{b}_p=0$ the Right charges annihilate the module.
Similarly, one could consider a \emph{Right} representation $\varrho_{\sR}$ where the role of Left and Right charges is exchanged
\begin{equation}\label{eq:su(1|1)2-reprR}
\boxed{\varrho_{\sR}:} \qquad\qquad
  \begin{aligned}
    \mathbb{Q}_{\smallR} \ket{\phi_p^{\sR}} &= a_p \ket{\psi_p^{\sR}} , \qquad &
    \mathbb{Q}_{\smallR} \ket{\psi_p^{\sR}} &= 0 , \\
    \overline{\mathbb{Q}}{}_{\smallR} \ket{\phi_p^{\sR}} &= 0 , \qquad &
    \overline{\mathbb{Q}}{}_{\smallR} \ket{\psi_p^{\sR}} &= \bar{a}_p \ket{\phi_p^{\sR}} , \\
    \mathbb{Q}_{\smallL} \ket{\phi_p^{\sR}} &= 0 , \qquad &
    \mathbb{Q}_{\smallL} \ket{\psi_p^{\sR}} &= b_p \ket{\phi_p^{\sR}} , \\
    \overline{\mathbb{Q}}{}_{\smallL} \ket{\phi_p^{\sR}} &= \bar{b}_p \ket{\psi_p^{\sR}} , \qquad &
    \overline{\mathbb{Q}}{}_{\smallL} \ket{\bar{\psi}_p} &= 0 .
  \end{aligned}
\end{equation}
If for lowering operators we conventionally choose the supercharges $\mathbb{Q}_{\smallL},\overline{\mathbb{Q}}_{\smallR}$---and for raising operators $\mathbb{Q}_{\smallR},\overline{\mathbb{Q}}_{\smallL}$---then the representations $\varrho_{\sL}$ and $\varrho_{\sR}$ are identified by the fact that the highest weight states are respectively $\phi^{\sL}$ and $\psi^{\sR}$. Other two possible representations are found by taking the opposite grading of the representations above, namely by exchanging the role of the boson and the fermion. 
We call $\widetilde{\varrho}_{\sL}$ the Left representation in which $\psi^{\sL}$ is the highest weight state
\begin{equation}\label{eq:su(1|1)2-repr2}
\boxed{\widetilde{\varrho}_{\sL}:} \qquad\qquad
  \begin{aligned}
    \mathbb{Q}_{\smallL} \ket{\widetilde{\psi}_p^{\sL}} &= a_p \ket{\widetilde{\phi}_p^{\sL}} , \qquad &
    \mathbb{Q}_{\smallL} \ket{\widetilde{\phi}_p^{\sL}} &= 0 , \\
    \overline{\mathbb{Q}}{}_{\smallL} \ket{\widetilde{\psi}_p^{\sL}} &= 0 , \qquad &
    \overline{\mathbb{Q}}{}_{\smallL} \ket{\widetilde{\phi}_p^{\sL}} &= \bar{a}_p \ket{\widetilde{\psi}_p^{\sL}} , \\
    \mathbb{Q}_{\smallR} \ket{\widetilde{\psi}_p^{\sL}} &= 0 , \qquad &
    \mathbb{Q}_{\smallR} \ket{\widetilde{\phi}_p^{\sL}} &= b_p \ket{\widetilde{\psi}_p^{\sL}} , \\
    \overline{\mathbb{Q}}{}_{\smallR} \ket{\widetilde{\psi}_p^{\sL}} &= \bar{b}_p \ket{\widetilde{\phi}_p^{\sL}} , \qquad &
    \overline{\mathbb{Q}}{}_{\smallR} \ket{\widetilde{\phi}_p^{\sL}} &= 0 ,
  \end{aligned}
\end{equation}
and similarly $\widetilde{\varrho}_{\sR}$ the one in which $\phi^{\sR}$ is the highest weight state.

\smallskip

The study of short representations of $\su(1|1)^2_{\ce}$ is useful because the exact representations relevant for {\adsthree} are \emph{bi-fundamental representations} of $\su(1|1)^2_{\ce}$. It is easy to check that the Left-massive, the Right-massive and the massless modules correspond to the following tensor products of representations
\begin{equation}
\text{Left}:\  \varrho_{\sL} \otimes \varrho_{\sL},
\qquad\quad
\text{Right}:\  \varrho_{\sR} \otimes \varrho_{\sR},
\qquad\quad
\text{massless}:\  (\varrho_{\sL} \otimes \widetilde{\varrho}_{\sL})^{\oplus 2}.
\end{equation}
For the massless module one has to consider two copies of $\varrho_{\sL} \otimes \widetilde{\varrho}_{\sL}$, hence the symbol $\oplus 2$. These two modules transform one into the other under the fundamental representation of $\su(2)_{\circ}$.
More precisely, one can identify the massive states as
\begin{equation}
\label{eq:mv-tensor}
  \begin{aligned}
    Y^{\sL} = \phi^{\sL} \otimes \phi^{\sL} , \qquad
    \eta^{\sL 1} = \psi^{\sL} \otimes \phi^{\sL} ,\;  \qquad
    \eta^{\sL 2} = \phi^{\sL} \otimes \psi^{\sL} ,\;  \qquad
    Z^{\sL} = \psi^{\sL} \otimes \psi^{\sL} , \\
    Y^{\sR} = {\phi}^{\sR} \otimes {\phi}^{\sR} , \qquad
    \eta^{\sR}_{\ 1} = {\psi}^{\sR} \otimes {\phi}^{\sR} , \qquad
    \eta^{\sR}_{\ 2} = {\phi}^{\sR} \otimes {\psi}^{\sR} , \qquad
    Z^{\sR} = {\psi}^{\sR} \otimes {\psi}^{\sR} ,
  \end{aligned}
\end{equation}
and the massless ones as 
\begin{equation}\label{eq:ml-tensor}
  \begin{aligned}
    T^{1a} = \big(\psi^{\sL} \otimes \tilde{\psi}^{\sL}\big)^{a} ,\quad
    \widetilde{\chi}^{a} = \big(\psi^{\sL} \otimes \tilde{\phi}^{\sL}\big)^{a} , \quad
    \chi^{a} = \big(\phi^{\sL} \otimes \tilde{\psi}^{\sL}\big)^{a} , \quad
     T^{2a} = \big(\phi \otimes \tilde{\phi}^{\sL}\big)^{a} . 
  \end{aligned}
\end{equation}
Together with the identification~\eqref{eq:supercharges-tensor-product} for the $\psu(1|1)^4_{\ce}$ charges in terms of the ones of $\su(1|1)^2_{\ce}$, it is easy to check that we reproduce the action presented in~\eqref{eq:exact-repr-left-massive}-\eqref{eq:exact-repr-right-massive} and~\eqref{eq:exact-repr-massless}.

\subsection{Left-Right symmetry}\label{sec:LR-symmetry}
The labels L and R appearing in the representations for the massive excitations are inherited from the two copies $\psu(1,1|2)_{\sL} \oplus \psu(1,1|2)_{\sR}$ of the symmetry of the string~\cite{Babichenko:2009dk}.
It is clear that exchanging the two labels in this case does not produce any difference.
Considering the commutation relations of $\mathcal{A}$ in~\eqref{eq:cealgebra} we see that they remain invariant under the map
\begin{equation}
 \gen{Q}_{\sL}{}^{\dot{a}} \longleftrightarrow  \gen{Q}_{\sR \dot{a}},
\qquad
\gen{M} \longrightarrow -\gen{M}.
\end{equation}
A Left supercharge with upper $\su(2)_{\bullet}$ index is mapped to a Right supercharge with lower $\su(2)_{\bullet}$ index because they transform under the fundamental and anti-fundamental representations of $\su(2)_{\bullet}$, respectively.
The Left and Right massive modules inherit a similar $\mathbb{Z}_2$ symmetry that we call Left-Right (LR) symmetry. The map here is given by
\begin{equation}\label{eq:LR-massive}
Y^{\sL} \longleftrightarrow {Y}^{\sR}, \qquad Z^{\sL} \longleftrightarrow {Z}^{\sR}, \qquad \eta^{\sL \dot{a}} \longleftrightarrow \eta^{\sR}_{\ \dot{a}}.
\end{equation}
Combining the map for the charges and the one for the states, we find compatibility for the representations~\eqref{eq:exact-repr-left-massive} and~\eqref{eq:exact-repr-right-massive}. This will prove to be extremely useful when constructing the S-matrix, see Section~\ref{sec:LR-symmetry}.
At the level of the bi-fundamental representations, it is clear that the map above is equivalent to exchanging the labels L and R in~\eqref{eq:mv-tensor}.

Let us consider the massless representation in~\eqref{eq:exact-repr-massless}, or its bi-fundamental structure~\eqref{eq:ml-tensor}.
Na\"ively, it seems that the notion of LR symmetry cannot be extended to this module, as only Left representations are used for the construction.
It turns out that LR symmetry is naturally implemented, and the resolution is in the masslessness of these excitations: there exists a momentum-dependent change of basis for the massless states\footnote{Using the parameterisation of the Section~\ref{sec:RepresentationCoefficientsT4} one can check that the rescalings are in fact just a sign, $\frac{a_p}{b_p}=-\text{sign} \big(\sin \frac{p}{2}\big)$.}
\begin{equation}
  \label{eq:mless-rescaling}
  \ket{ \cbchi_p^a } = -\frac{a_p}{b_p} \ket{ \widetilde{\chi}_p^a }, \qquad \ket{ \cchi_p^a } = \frac{b_p}{a_p} \ket{ \chi_p^a },
\end{equation}
under which the action of the supercharges becomes
\begin{equation}\label{eq:repr-massless2}
\boxed{(\varrho_{\sR}\otimes\widetilde{\varrho}_{\sR})^{\oplus 2}}: \qquad
  \begin{aligned}
    \gen{Q}_{\sL}^{\ \dot{a}} \ket{T^{\dot{b}a}_p} &= -\epsilon^{\dot{a}\dot{b}} b_p \ket{\cbchi^\alpha_p},
    \qquad &
    \gen{Q}_{\sL}^{\ \dot{a}} \ket{\cchi^{a}_p}\ &=  b_p \ket{T^{\dot{a}a}_p}, \\
    \overline{\gen{Q}}{}_{\sL \dot{a}} \ket{\cbchi^{a}_p}\ &= \epsilon_{\dot{a}\dot{b}} \bar{b}_p \ket{T^{\dot{b}a}_p},
    \qquad &
    \overline{\gen{Q}}{}_{\sL \dot{a}} \ket{T^{\dot{b}a}_p} &= \delta_{\dot{a}}^{\ \dot{b}} \bar{b}_p \ket{\cchi^a_p}, \\[8pt]
    \gen{Q}_{\sR \dot{a}} \ket{T^{\dot{b}a}_p} &= \delta_{\dot{a}}^{\ \dot{b}} a_p \ket{\cchi^a_p},
    \qquad &
    \gen{Q}_{\sR \dot{a}} \ket{\cbchi^a_p}\ &= \epsilon_{\dot{a}\dot{b}} a_p \ket{T^{\dot{b}a}_p}, \\
    \overline{\gen{Q}}{}_{\sR}^{\ \dot{a}} \ket{\cchi^a_p}\ &= \bar{a}_p \ket{T^{\dot{a}a}_p},  
    \qquad &
    \overline{\gen{Q}}{}_{\sR}^{\ \dot{a}} \ket{T^{\dot{b}a}_p} &= -\epsilon^{\dot{a}\dot{b}} \bar{a}_p \ket{\cbchi^a_p}.
  \end{aligned}
\end{equation}
The bi-fundamental structure in this case corresponds to the identifications
\begin{equation}\label{eq:ml-tensor-R}
  \begin{aligned}
    T_{1a} = \big({\psi}^{\sR} \otimes \widetilde{\psi}^{\sR}\big)_{a} , \quad
    \cbchi_{a} = \big({\phi}^{\sR} \otimes \widetilde{\psi}^{\sR}\big)_{a} , \quad
    \cchi_{a} = \big({\psi}^{\sR} \otimes \widetilde{\phi}^{\sR}\big)_{a} , \quad
    T_{2a} = \big({\phi}^{\sR} \otimes \widetilde{\phi}^{\sR}\big)_{a} .
  \end{aligned}
\end{equation}
This change of basis yields the above representation only when the states are massless, as we need to use explicitly $|a_p|^2=|b_p|^2$.
It is then clear that a notion of LR symmetry is present also for the massless module, where we have the rules
\begin{equation}
\label{eq:LR-massless}
\ket{T^{\dot{a}a}} \longleftrightarrow \ket{T_{\dot{a} a}}, \qquad \ket{\widetilde{\chi}^a} \longleftrightarrow +\frac{b_p}{a_p} \ket{\chi_a}, \qquad \ket{\chi^a} \longleftrightarrow -\frac{a_p}{b_p} \ket{\widetilde{\chi}_a}.
\end{equation}
It is interesting to note that one might perform also a different rescaling
\begin{equation}\label{eq:ml-tensor-LR}
\ket{ \cbchi_p^1 } = \ket{ \widetilde{\chi}_p^1 }, \quad \ket{ \cchi_p^1 } = \ket{ \chi_p^1 },
\qquad
\ket{ \cbchi_p^2 } = -\frac{a_p}{b_p} \ket{ \widetilde{\chi}_p^2 }, \quad \ket{ \cchi_p^2 } = \frac{b_p}{a_p} \ket{ \chi_p^2 }.
\end{equation}
Doing this, one would get a bi-fundamental structure of the form $({\varrho}_{\sL}\otimes\widetilde{\varrho}_{\sL}) \oplus({\varrho}_{\sR}\otimes\widetilde{\varrho}_{\sR})$.
Now both Left and Right representations would be used to construct the massless module, where Left corresponds to the $\su(2)_{\bullet}$ index $\dot{a}=1$ and Right to $\dot{a}=2$.
LR symmetry would be implemented just by swapping the two $\su(2)_{\bullet}$ flavors. 

\subsection{Representation coefficients}\label{sec:RepresentationCoefficientsT4}
In the previous section we presented the action of the odd generators of $\mathcal{A}$ on the massive and massless states. 
It is written in terms of two complex coefficients $a_p,b_p$---depending on the mometum $p$ of the excitation and the eigenvalue $m$ of the central charge $\gen{M}$ on the specific module---and their complex conjugates $\bar{a}_p,\bar{b}_p$.
Computing anti-commutators of supercharges we are able to write the relation between these coefficients and the eigenvalues of the central charges.
These are known at any value of the coupling constant, thanks to the results coming from the explicit worldsheet computation~\eqref{eq:allloop-centralcharges} and the shortening condition~\eqref{eq:short-cond}\footnote{The eigenvalue of the charge $\gen{M}$ is denoted by $m$. In these equations we have the absolute value of $m$ appearing becuase we get $a_p \bar{a}_p - b_p \bar{b}_p=m=+1$ for Left states and $a_p \bar{a}_p - b_p \bar{b}_p=-m=+1$ for Right states. On massless states we have $a_p \bar{a}_p - b_p \bar{b}_p=m=0$.}
\begin{equation}
\begin{aligned}
\gen{M}: & \qquad a_p \bar{a}_p - b_p \bar{b}_p = |m|\,,
\\
\gen{H}: & \qquad a_p \bar{a}_p + b_p \bar{b}_p = \sqrt{ m^2 + 4 h^2 \sin^2 \frac{p}{2}}\,,
\\
\gen{C}: & \qquad \phantom{+ b_p \bar{b}_p\ } a_p b_p  = h\, \frac{i}{2}(e^{ip}-1)\,\zeta\,.
\end{aligned}
\end{equation}
Here $\zeta=e^{2i\, \xi}$ is a function that characterises the representation. On one-particle states it can be taken to be $1$, but in Section~\ref{sec:two-part-repr-T4} we will show that this is not the case when constructing two-particle states.

A way to solve the above equations is to introduce the Zhukovski parameters $x^\pm$, that satisfy
\begin{equation}
\label{eq:zhukovski}
x^+_p +\frac{1}{x^+_p} -x^-_p -\frac{1}{x^-_p} = \frac{2i \, |m|}{h}, \qquad \frac{x^+_p}{x^-_p}=e^{ip}.
\end{equation}
Then we can take the representation coefficients to be
\begin{equation}\label{eq:expl-repr-coeff}
a_p = \eta_p e^{i\xi}, \quad \bar{a}_p = \eta_p \left( \frac{x^+_p}{x^-_p}\right)^{-1/2} e^{-i\xi}, \quad 
b_p = -\frac{\eta_p}{x^-_p} \left( \frac{x^+_p}{x^-_p}\right)^{-1/2} e^{i\xi}, \quad \bar{b}_p = -\frac{\eta_p}{x^+_p} e^{-i\xi},
\end{equation}
where we have introduced the function
\begin{equation}\label{eq:def-eta}
\eta_p = \left( \frac{x^+_p}{x^-_p}\right)^{1/4}  \sqrt{\frac{ih}{2}(x^-_p - x^+_p)}\,.
\end{equation}
This parameterisation coincides with the one of~\cite{Arutyunov:2009ga}.
The constraints on the spectral parameters $x^\pm$ can be solved by taking
\begin{equation}\label{eq:xpm-funct-p}
x^\pm_p = \frac{e^{\pm\frac{i p}{2}} \csc \left(\frac{p}{2}\right) \left(|m|+\sqrt{m^2+4 h^2 \sin ^2\left(\frac{p}{2}\right)}\right)}{2 h}\, ,
\end{equation}
where the branch of the square root has been chosen such that $|x^\pm_p|>1$ for real values of the momentum $p$, when we consider massive states $|m|>0$.
For massless states, we have simply
\begin{equation}
x^\pm_p = \text{sgn}(\sin \tfrac{p}{2})\, e^{\pm\frac{i p}{2}},
\qquad
E_p = 2h \left|\sin \frac{p}{2}\right|.
\end{equation}
In the massless case the spectral parameters lie on the unit circle. Similarly to the situation of massive excitations, the dispersion relation is not relativistic, and now it actually gets the form of a giant magnon dispersion relation at strong coupling~\cite{Hofman:2006xt}.

\section{Summary}
In this chapter we have studied the symmetry algebra that remains after fixing light-cone gauge for strings on {\adsthree}.
We have considered the charges written in terms of worldsheet fields, and to simplify the computations we have truncated them at quadratic order in the expansion in powers of fields.
We actually used a ``hybrid expansion'', in the sense that the dependence on the light-cone coordinate $x^-$ was kept exact.
The coordinate $x^-$ is related to the worldsheet momentum through the Virasoro constraint.

Computing anti-commutators of supercharges we have verified the presence of a central extension when we are off-shell, \ie when we relax the level-matching condition and we consider states whose total worldsheet momentum is not zero.
The hybrid expansion allowed us to derive the exact momentum-dependence of the central charges.

The computation at the near-BMN order revealed that we have four bosonic and four fermionic massive excitations, together with four bosonic and four fermionic massless excitations.
The massive excitations correspond to transverse directions in AdS$_3\times$S$^3$, and they are further divided into two irreducible representations---labelled by Left and Right---of the off-shell symmetry algebra $\mathcal{A}$.
Massless modes correspond to excitations on T$^4$.

We showed that it is possible to deform the near-BMN representations introducing a dependence on the parameter $h$---related to the string tension---that reproduces the exact non-linear momentum dependence of the central charges. We also obtained the ``all-loop dispersion relation'' for the worldsheet excitations.

In the next chapter we will use compatibility with the charges constructed here to bootstrap an all-loop S-matrix.

\chapter{Exact S-matrix for \adsthree}\label{ch:S-matrix-T4}
The Hamiltonian of the gauge-fixed model living on the worldsheet can, in principle, be used to compute the S-matrix, responsible for the scattering of the excitations on the string. To start, the quartic Hamiltonian provides the $2\to 2$ scattering elements at tree-level, and higher corrections may be computed.

In this chapter we want to take a different route. Rather than deriving the \mbox{S-matrix} in perturbation theory, we use a bootstrap procedure to find it at all values of the coupling $h$.
This method relies on a crucial assumption, namely that the theory at hand is \emph{quantum integrable}. 
As anticipated in the introduction of Chapter~\ref{ch:intro}, one can prove \emph{classical Integrability} for strings on {\adsthree}~\cite{Babichenko:2009dk}, meaning that there exists an infinite number of conserved quantities 
in involution with each other.
The assumption is that this property survives at the quantum level, where we find an infinite set of commuting conserved charges labelled by $n_j$
$$
[\gen{J}_{n_1},\gen{J}_{n_2}]=0.
$$
Two of these are the familiar charges that measure momentum and energy of the state. The others are called higher charges, and in relativisitic integrable field theories their eigenvalues typically depend on higher order polynomials in the momenta.
In general, given a state with momentum $p$, each charge acts simply as $\gen{J}_{n_j}\ket{p}=j_{n_j}(p)\ket{p}$.
We should appreciate that the situation is very much constrained, as we have at our disposal an infinite set of independent functions $j_{n_j}(p)$.

The consequences of this are important when we consider the scattering problem. We focus on the in-states prepared at $t=-\infty$ and on the out-states that remain after the collision at $t=+\infty$. We do not try to describe the details of the scattering when the particles are close to each other, as the interactions might be very complicated. We define an object $\mathcal{S}$ that we call S-matrix and that relates the inital and final states
$$
\mathcal{S}\ket{\mathcal{X}^{c_1}(p_1)\ldots\mathcal{X}^{c_{N_{\text{in}}}}(p_{N_{\text{in}}})}=
\mathcal{A}_{\ c'_1\ldots c'_{N_{\text{out}}}}^{c_1\ldots c_{N_{\text{in}}}}(p_1,\ldots p_{N_{\text{in}}};p'_1,\ldots p'_{N_{\text{out}}})
\ket{\mathcal{X}^{c'_1}(p'_1)\ldots\mathcal{X}^{c'_{N_{\text{out}}}}(p'_{N_{\text{out}}})}.
$$
For a generic quantum field theory, the first requirement that we might want to impose on this S-matrix is compatibility with symmetries.
Additionally, we should also impose the unitarity condition, to be sure that no state is missing in the description.

The generic problem is very complicated; in fact, creation and annihilation processes may take place, meaning that interactions might modify the number of particles after the scattering. Following Alexander and Alexei Zamolodchikov~\cite{Zamolodchikov:1978xm}, if in a quantum integrable model we impose conservation for each of the charges $\gen{J}_{n}$
$$
\sum_{k=1}^{N_{\text{in}}} j_n(p_k)
=\sum_{k=1}^{N_{\text{out}}} j_n(p'_k),
\qquad\forall n,
$$
we conclude that the only way to satisfy all these constraints is to conserve under the scattering
\begin{itemize}
\item the number of particles $N_{\text{in}}=N_{\text{out}}$,
\item the set of momenta $\{p_1,\ldots,p_{N_{\text{in}}}\}=\{p'_1,\ldots,p'_{N_{\text{out}}}\}$. 
\end{itemize}
The momenta are allowed to be reshuffled under the scattering, but not to change their values.

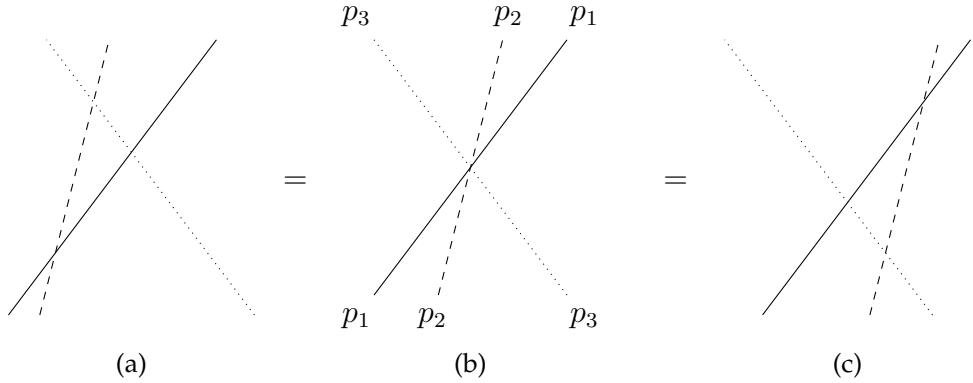
\begin{figure}[t]
  \centering
\hspace{-0.75cm}
 \subfloat[\label{fig:YB-left}]{
  \begin{tikzpicture}[%
    box/.style={outer sep=1pt},
    Q node/.style={inner sep=1pt,outer sep=0pt},
    arrow/.style={-latex}
    ]%
    \node [box] (p1in) at ($(-1.5cm,-2cm)+(0.5cm,0cm)$) {};
    \node [box] (p2in) at (-0.5cm,-2cm) {};
    \node [box] (p3in) at ($(+1.5cm,-2cm)+(1cm,0cm)$) {};

    \node [box] (p1out) at ($(+1.5cm,2cm)+(0.5cm,0cm)$) {};
    \node [box] (p2out) at (+0.5cm,2cm) {};
    \node [box] (p3out) at ($(-1.5cm,2cm)+(1cm,0cm)$) {};

	\draw (p1in) -- (p1out);
	\draw [dashed] (p2in) -- (p2out);
	\draw [dotted] (p3in) -- (p3out);

  \end{tikzpicture}
}
\raisebox{2cm}{$=$}
\hspace{0cm}
 \subfloat[\label{fig:YB-central}]{
  \begin{tikzpicture}[%
    box/.style={outer sep=1pt},
    Q node/.style={inner sep=1pt,outer sep=0pt},
    arrow/.style={-latex}
    ]%
    \node [box] (p1in) at (-1.5cm,-2cm) {$p_1$};
    \node [box] (p2in) at (-0.5cm,-2cm) {$p_2$};
    \node [box] (p3in) at (+1.5cm,-2cm) {$p_3$};

    \node [box] (p1out) at (+1.5cm,2cm) {$p_1$};
    \node [box] (p2out) at (+0.5cm,2cm) {$p_2$};
    \node [box] (p3out) at (-1.5cm,2cm) {$p_3$};

	\draw (p1in) -- (p1out);
	\draw [dashed] (p2in) -- (p2out);
	\draw [dotted] (p3in) -- (p3out);

  \end{tikzpicture}
}
\hspace{0.5cm}
\raisebox{2cm}{$=$}
 \subfloat[\label{fig:YB-right}]{
  \begin{tikzpicture}[%
    box/.style={outer sep=1pt},
    Q node/.style={inner sep=1pt,outer sep=0pt},
    arrow/.style={-latex}
    ]%
    \node [box] (p1in) at ($(-1.5cm,-2cm)+(0cm,0cm)$) {};
    \node [box] (p2in) at ($(-0.5cm,-2cm)+(0.5cm,0cm)$) {};
    \node [box] (p3in) at ($(+1.5cm,-2cm)+(-0.5cm,0cm)$) {};

    \node [box] (p1out) at ($(+1.5cm,2cm)+(0cm,0cm)$) {};
    \node [box] (p2out) at ($(+0.5cm,2cm)+(0.5cm,0cm)$) {};
    \node [box] (p3out) at ($(-1.5cm,2cm)+(-0.5cm,0cm)$) {};

	\draw (p1in) -- (p1out);
	\draw [dashed] (p2in) -- (p2out);
	\draw [dotted] (p3in) -- (p3out);
 
  \end{tikzpicture}
}
 \caption{Parallel lines with the same style correspond to particles having the same momentum. The vertical axis parameterises time, while the horizontal axis space. The action of the higher conserved charges allows us to independently move the wave-packets of the excitations that are scattering. A three-body process like the one in the central figure becomes then equivalent to either the process depicted in the left or the one in the right. In both cases we get a sequence of two-body scatterings. Constistency imposes that these two factorisations should be equivalent. This requirement results in the Yang-Baxter equation, a constraint that the two-body S-matrix should satisfy.}
  \label{fig:Yang-Baxter}
\end{figure}

Already at this stage we find a problem that is much simpler than what is usually considered in a generic quantum field theory.
The fact that we have higher charges gives even more powerful consequences than the ones already mentioned, as one can show that any $N$-particle scattering is \emph{factorisable} into a sequence of two-body processes.
The idea is that the action of the higher charges allows us to move independently the wave packets corresponding to each of the particles that scatter. Thanks to this property, a three-body process like the one in Figure~\ref{fig:YB-central} becomes equivalent to either~\ref{fig:YB-left} or~\ref{fig:YB-right}. 

It is clear that factorisability is possible only if we satisfy the consistency condition stating that the order of factorisation is unimportant.
We then find that the S-matrix has to satisfy the \emph{Yang-Baxter equation} 
$$
\mathcal{S}_{23}\ \mathcal{S}_{13}\ \mathcal{S}_{12}=\mathcal{S}_{12}\ \mathcal{S}_{13}\ \mathcal{S}_{23}\,.
$$
When the above equation is satisfied, the consistency of factorisation of any $N$-body scattering is automatically ensured.
To derive any scattering process it is then enough to know the two-body S-matrix, and this object is indeed the subject of this chapter.

\medskip

In Section~\ref{sec:two-part-repr-T4} we explain how to obtain the action of the charges on two-particle states.
Demanding compatibility with these charges, in Section~\ref{sec:S-mat-T4} we bootstrap the all-loop two-body S-matrix.
The S-matrix is naturally divided into blocks corresponding to the various sectors of scattering---massive, massless, mixed-mass.
In each sector we write the S-matrix as a tensor product of two smaller S-matrices, compatible with the tensor product representations of the previous chapter.
Taking into account the constraints coming from unitarity and LR-symmetry, we show that the S-matrix is fixed completely up to a total of four undetermined scalar function of the momenta, that we call ``dressing factors''.
Further contraints on them are imposed by the crossing equations, that we derive.
Compatibility with the assumption of factorisation of scattering is confirmed by the Yang-Baxter equation, that our S-matrix satisfies.
Section~\ref{sec:Bethe-Yang} is devoted to the derivation of the Bethe-Yang equations.
We first present the procedure and then write explicitly the Bethe-Yang equations that we obtain for {\adsthree}.

\section{Two-particle representations}\label{sec:two-part-repr-T4}
In this section we study the action of the charges on two-particle states. We will show that not all the charges are defined via the standard co-product---for some of them this has to be non-local.
Given a charge $\gen{J}$ acting on a one-particle state $\ket{\mathcal{X}}$ as $\gen{J}\ket{\mathcal{X}}=\ket{\mathcal{Y}}$, the corresponding charge on two-particle states that we get by using the \emph{standard} co-product is 
\begin{equation}
\gen{J}_{12} \equiv \gen{J} \otimes \mathbf{1} + \mathbf{1} \otimes \gen{J},
\quad\implies\quad
\gen{J}_{12} \ket{\mathcal{X}_1\mathcal{X}_2}= \ket{\mathcal{Y}_1\mathcal{X}_2}+\ket{\mathcal{X}_1\mathcal{Y}_2}.
\end{equation}
In case $\gen{J}$ is an odd charge one has to take care of the signs arising when commuting with a fermionic state.
It is easy to check that the standard co-product cannot be used to define the action of the central charge $\gen{C}_{12}$ on two-particle states~\cite{Beisert:2006qh,Arutyunov:2006yd}. Another way to phrase this is to say that we cannot set to zero the parameters $\xi_1$ and $\xi_2$ entering the definition~\eqref{eq:expl-repr-coeff}.
Indeed using $\gen{C}=+\frac{ih}{2}(e^{+i\gen{P}}-1)$ we find
\begin{equation}
\gen{C} \ket{\mathcal{X}_1\mathcal{X}_2} = \frac{ih}{2}(e^{+i(p_1+p_2)}-1) \ket{\mathcal{X}_1\mathcal{X}_2} ,
\end{equation}
while using the combined action on one-particle states we get
\begin{equation}
\gen{C} \ket{\mathcal{X}_1\mathcal{X}_2} = \frac{ih}{2}\left(e^{2i\xi_{1}}(e^{+ip_1}-1)+ e^{2i\xi_{2}}(e^{+ip_2}-1)\right)\ket{\mathcal{X}_1\mathcal{X}_2}.
\end{equation}
In order to have compatibility of the two results, we cannot set $e^{i\xi_{1}}=e^{i\xi_{2}}=1$.
If we require that these factors lie on the unit circle, then we get two possible solutions
\begin{equation}
\{e^{2i\xi_{1}} = 1, e^{2i\xi_{2}} = e^{i\, p_1} \},
\qquad
\qquad
\{e^{2i\xi_{1}} = e^{i\, p_2},e^{2i\xi_{2}} = 1 \}.
\end{equation}
Both these solutions imply that $\gen{C}_{12}$ is defined by a \emph{non-local} product, as the action on one of the two states depends on the momentum of the other.
In the rest of the chapter we will choose the first of the above solutions, namely $\xi_1=0,\ \xi_2=p_1/2$. This choice agrees with the one of~\cite{Arutyunov:2009ga}.

It is obvious that also the action of supercharges on two-particle states is defined by a non-local co-product, and the exact action is found by replacing the value of $\xi_i$ in the definitions of the coefficients $a_p,\bar{a}_p,b_p,\bar{b}_p$ in~\eqref{eq:expl-repr-coeff}.
As an explicit example, when we consider the action of $\gen{Q}_{\sL}^{\ 1}$ on a two-particle state, we find
\begin{equation}
\gen{Q}_{\sL}^{\ 1}(p_1,p_2) = \gen{Q}_{\sL}^{\ 1}(p_1)\otimes\mathbf{1}+e^{i\, p_1/2}\, \Sigma \otimes \gen{Q}_{\sL}^{\ 1}(p_2)\,.
\end{equation}
The matrix $\Sigma$ takes into account the even or odd grading of the states, and is $+1$ on bosons and $-1$ on fermions.
For this reason we can use the ordinary tensor product $\otimes$.

On the other hand, computing the action of the generators corresponding to the Hamiltonian $\gen{H}$ and the angular momentum $\gen{M}$, it is clear that the dependence on the parameters $\xi_i$ cancels, and the action on two-particle states is just given by the standard co-product\footnote{Although we indicate the momentum dependence in both cases, we remind that the eigenvalue of $\gen{M}$ is momentum-independent.}
\begin{equation}
\begin{aligned}
\gen{H}(p_1,p_2)&=\gen{H}(p_1)\otimes\mathbf{1}+\mathbf{1}\otimes\gen{H}(p_2)\,,
\\
\gen{M}(p_1,p_2)&=\gen{M}(p_1)\otimes\mathbf{1}+\mathbf{1}\otimes\gen{M}(p_2)\,.
\end{aligned}
\end{equation}
The same is true for the $\su(2)_{\bullet} \oplus \su(2)_{\circ}$ generators, whose action does not depend on the above coefficients.

We note that a generalisation of this discussion to multi-particle states is possible.
The requirement that $\gen{C}=+\frac{ih}{2}(e^{+i\gen{P}}-1)$ still holds is fullfilled by taking $\xi_1=0$ and $\xi_i=\sum_{j=1}^{i-1}p_j/2$, for $i>1$.

\section{The S-matrix}\label{sec:S-mat-T4}
In this section we present the explicit form of the exact two-body S-matrix for the worldsheet excitations of \adsthree. This is found by fixing invariance of the S-matrix under the symmetry algebra $\mathcal{A}$. 
Depending on convenience, we will use two objects denoted by $\mathcal{S}$ and $\mathbf{S}$. They are related to each other by a simple permutation in the two body space\footnote{The object that here is called $\mathbf{S}$ is denoted by $S$ in~\cite{Arutyunov:2009ga}.}
\begin{equation}
\mathcal{S}=\Pi\, \mathbf{S}\,.
\end{equation}
After constructing the generators on two-particle states as explained in the previous section, we impose compatibility as\footnote{The difference between the two is how we apply the charge after the action of the proper S-matrix.}
\begin{equation}
\begin{aligned}
\mathcal{S}_{12}(p,q)\, \gen{J}_{12}(p,q) - \gen{J}_{12}(q,p)\, \mathcal{S}_{12}(p,q) &=0\,,
\\
\mathbf{S}_{12}(p,q)\, \gen{J}_{12}(p,q) - \gen{J}_{21}(q,p)\, \mathbf{S}_{12}(p,q) &=0\,.
\end{aligned}
\end{equation}
Invariance under the action of the generators $\gen{M}$ and $\gen{H}$ allows us to identify three possible sectors, that we use to divide the S-matrix: 
\begin{itemize}
\item[-] the massive sector $(\bullet\bullet)$, 
\item[-] the massless sector $(\circ\circ)$, 
\item[-] the mixed-mass sector $(\bullet\circ,\circ\bullet)$. 
\end{itemize}
In each of these sectors the set of masses is conserved under the scattering. In the mixed-mass sector we have in addition that the mass is transmitted. In other words, the mass can be thought as a label attached to the momentum of the excitation.

The next generators to consider are the ones of $\su(2)_\bullet \oplus \su(2)_\circ$. Their action is momentum independent, and compatibility of the S-matrix with them allows us to relate or set to zero different scattering elements, in such a way that the $\su(2)$ structures are respected.

Another powerful way to constrain the S-matrix is to consider the $\mathbb{Z}_2$-symmetry introduced in Section~\ref{sec:LR-symmetry}, that we called LR-symmetry. We will then impose that scattering elements that are related by the rules~\eqref{eq:LR-massive} and~\eqref{eq:LR-massless} should be the same.

It is considering compatibility with the supercharges that we see the dependence of the scattering elements on the momenta of the excitations.
In particular, this fixes invariance under the $\psu(1|1)^4_\ce$ subalgebra of $\mathcal{A}$. We will use the fact that fundamental representations of $\psu(1|1)^4_\ce$ can be understood as bi-fundamental representations of $\psu(1|1)^2_\ce$ to rewrite an S-matrix compatible with $\psu(1|1)^4_\ce$-invariance as a proper tensor product of two copies of $\psu(1|1)^2_\ce$-invariant S-matrices.
Let us first construct the relevant S-matrices in this simpler case.

\subsection{The $\su(1|1)^2_{\ce}$-invariant S-matrices}\label{sec:su112-S-matrices}
In Section~\ref{sec:BiFundamentalRepresentationsT4} we presented the possible fundamental representations of $\su(1|1)^2_{\ce}$, that we called $\varrho_{\sL},\varrho_{\sR},\widetilde{\varrho}_{\sL},\widetilde{\varrho}_{\sR}$. They are related by exchanging the labels L and R on the states and on the supercharges, or by exchanging the role of the boson $\phi$ and the fermion $\psi$ composing the short representation. Using the four possible fundamental representations one can construct sixteen different two-particle representations. In this section we discuss only the ones that are relevant for the S-matrix of \adsthree, in particular we start by considering the case in which both particles that scatter belong to the representation $\varrho_{\sL}$. Invariance under the algebra yields an S-matrix $\mathcal{S}^{\sL\sL}$ of the form~\cite{Borsato:2012ud,Borsato:2014exa}
\begin{equation}\label{eq:su(1|1)2-Smat-grad1}
\begin{aligned}
\mathcal{S}^{\sL\sL} \ket{\phi_p^{\sL} \phi_q^{\sL}} &=  A_{pq}^{\sL\sL} \ket{\phi_q^{\sL} \phi_p^{\sL}},
\qquad
&\mathcal{S}^{\sL\sL} \ket{\phi_p^{\sL} \psi_q^{\sL}} &= B_{pq}^{\sL\sL} \ket{\psi_q^{\sL} \phi_p^{\sL}} +  C_{pq}^{\sL\sL} \ket{\phi_q^{\sL} \psi_p^{\sL}}, \\
\mathcal{S}^{\sL\sL} \ket{\psi_p^{\sL} \psi_q^{\sL}} &= F_{pq}^{\sL\sL} \ket{\psi_q^{\sL} \psi_p^{\sL}},\qquad
&\mathcal{S}^{\sL\sL} \ket{\psi_p^{\sL} \phi_q^{\sL}} &= D_{pq}^{\sL\sL} \ket{\phi_q^{\sL} \psi_p^{\sL}} +  E_{pq}^{\sL\sL} \ket{\psi_q^{\sL} \phi_p^{\sL}},
\end{aligned}
\end{equation}
The coefficients appearing are determined up to an overall factor. As a convention we decide to normalise $A_{pq}^{\sL\sL}=1$ and we find
\begin{equation}\label{eq:expl-su112-smat-el}
\begin{aligned}
A^{\sL\sL}_{pq} &= 1, &
\qquad
B^{\sL\sL}_{pq} &= \phantom{-}\left( \frac{x^-_p}{x^+_p}\right)^{1/2} \frac{x^+_p-x^+_q}{x^-_p-x^+_q}, \\
C^{\sL\sL}_{pq} &= \left( \frac{x^-_p}{x^+_p} \frac{x^+_q}{x^-_q}\right)^{1/2} \frac{x^-_q-x^+_q}{x^-_p-x^+_q} \frac{\eta_p}{\eta_q}, 
\qquad &
D^{\sL\sL}_{pq} &= \phantom{-}\left(\frac{x^+_q}{x^-_q}\right)^{1/2}  \frac{x^-_p-x^-_q}{x^-_p-x^+_q}, \\
E^{\sL\sL}_{pq} &= \frac{x^-_p-x^+_p}{x^-_p-x^+_q} \frac{\eta_q}{\eta_p}, 
\qquad &
F^{\sL\sL}_{pq} &= - \left(\frac{x^-_p}{x^+_p} \frac{x^+_q}{x^-_q}\right)^{1/2} \frac{x^+_p-x^-_q}{x^-_p-x^+_q}.
\end{aligned}
\end{equation}
The result is written in terms of the Zhukovski variables introduced in Section~\ref{sec:RepresentationCoefficientsT4}. In particular the result above holds for any value of the masses $|m|$---that appear in the quadratic constraint of~\eqref{eq:zhukovski}---of the two particles, and are valid also for scattering of particles of different masses.

When considering scattering of two $\widetilde{\varrho}_{\sL}$ representations, we find that the result can be rewritten using the above coefficients. Also in this case the overall normalisation is a convention and we decide to write it as 
\begin{equation}\label{eq:su(1|1)2-Smat-grad2}
\begin{aligned}
\mathcal{S}^{\tilde{\sL}\tilde{\sL}} \ket{\tilde{\phi}^{\sL}_p \tilde{\phi}^{\sL}_q} &= -F_{pq}^{\sL\sL} \ket{\tilde{\phi}^{\sL}_q \tilde{\phi}^{\sL}_p}, 
\qquad
&\mathcal{S}^{\tilde{\sL}\tilde{\sL}} \ket{\tilde{\phi}^{\sL}_p \tilde{\psi}^{\sL}_q} &= D_{pq}^{\sL\sL} \ket{\tilde{\psi}^{\sL}_q \tilde{\phi}^{\sL}_p}  -E_{pq}^{\sL\sL} \ket{\tilde{\phi}^{\sL}_q \tilde{\psi}^{\sL}_p}, \\
\mathcal{S}^{\tilde{\sL}\tilde{\sL}} \ket{\tilde{\psi}^{\sL}_p \tilde{\psi}^{\sL}_q} &= -A_{pq}^{\sL\sL} \ket{\tilde{\psi}^{\sL}_q \tilde{\psi}^{\sL}_p}, 
\qquad
&\mathcal{S}^{\tilde{\sL}\tilde{\sL}} \ket{\tilde{\psi}^{\sL}_p \tilde{\phi}^{\sL}_q} &= B_{pq}^{\sL\sL} \ket{\tilde{\phi}^{\sL}_q \tilde{\psi}^{\sL}_p}  -C_{pq}^{\sL\sL} \ket{\tilde{\psi}^{\sL}_q \tilde{\phi}^{\sL}_p}.
\end{aligned}
\end{equation}
The last Left-Left case that we want to consider---as it will be used to construct the S-matrix of {\adsthree}---concerns the scattering of $\varrho_{\sL}$ and $\widetilde{\varrho}_{\sL}$. We write it as 
\begin{equation}\label{eq:su(1|1)2-Smat-grad3}
\begin{aligned}
\mathcal{S}^{{\sL}\tilde{\sL}} \ket{\phi^{\sL}_p \tilde{\phi}^{\sL}_q} &= \phantom{+}B_{pq}^{\sL\sL} \ket{\tilde{\phi}^{\sL}_q \phi^{\sL}_p} -C_{pq}^{\sL\sL} \ket{\tilde{\psi}^{\sL}_q \psi^{\sL}_p},
\qquad
&\mathcal{S}^{{\sL}\tilde{\sL}} \ket{\phi^{\sL}_p \tilde{\psi}^{\sL}_q} &= \phantom{+}A_{pq}^{\sL\sL} \ket{\tilde{\psi}^{\sL}_q \phi^{\sL}_p} , \\
\mathcal{S}^{{\sL}\tilde{\sL}} \ket{\psi^{\sL}_p \tilde{\psi}^{\sL}_q} &= -D_{pq}^{\sL\sL} \ket{\tilde{\psi}^{\sL}_q \psi^{\sL}_p}+E_{pq}^{\sL\sL} \ket{\tilde{\phi}^{\sL}_q \phi^{\sL}_p} ,
\qquad
&\mathcal{S}^{{\sL}\tilde{\sL}} \ket{\psi^{\sL}_p \tilde{\phi}^{\sL}_q} &= -F_{pq}^{\sL\sL} \ket{\tilde{\phi}^{\sL}_q \psi^{\sL}_p} .
\end{aligned}
\end{equation}

In order to complete the discussion and present all the material that is needed to construct the full S-matrix, we now turn to Left-Right scattering.
For the case of two particles with equal masses, requiring just invariance under the symmetry algebra one obtains an S-matrix that is a combination of transmission and reflection, where this terminology should be applied to the LR-flavors.
Imposing LR-symmetry and unitarity one finds that only two solutions are allowed, namely \emph{pure transmission} or \emph{pure reflection}~\cite{Borsato:2012ud}. Compatibility with perturbative results then forces to choose the pure-transmission S-matrix, that is the one presented here. Moreover it is only this S-matrix that satisfies the Yang-Baxter equation.
A process involving the representations $\varrho_{\sL}$ and $\varrho_{\sR}$ yields an S-matrix of the form~\cite{Borsato:2012ud,Borsato:2014exa}
\begin{equation}\label{eq:su(1|1)2-Smat-LRgrad1}
\begin{aligned}
\mathcal{S}^{\sL\sR} \ket{\phi^{\sL}_p \phi^{\sR}_q} &= A^{\sL\sR}_{pq} \ket{\phi^{\sR}_q \phi^{\sL}_p} + B^{\sL\sR}_{pq} \ket{\psi^{\sR}_q \psi^{\sL}_p}, \qquad 
&\mathcal{S}^{\sL\sR} \ket{\phi^{\sL}_p \psi^{\sR}_q} &= C^{\sL\sR}_{pq} \ket{\psi^{\sR}_q \phi^{\sL}_p} , \\
\mathcal{S}^{\sL\sR} \ket{\psi^{\sL}_p \psi^{\sR}_q} &= E^{\sL\sR}_{pq} \ket{\psi^{\sR}_q \psi^{\sL}_p}+F^{\sL\sR}_{pq} \ket{\phi^{\sR}_q \phi^{\sL}_p} ,  \qquad 
& \mathcal{S}^{\sL\sR} \ket{\psi^{\sL}_p \phi^{\sR}_q} &= D^{\sL\sR}_{pq} \ket{\phi^{\sR}_q \psi^{\sL}_p} ,
\end{aligned}
\end{equation}
where the scattering elements can be parametrised explicitly by
\begin{equation}
\begin{aligned}
 A^{\sL\sR}_{pq} &= \zeta_{pq}\, \left(\frac{x^+_p}{x^-_p} \right)^{1/2} \frac{1-\frac{1}{x^+_p x^-_q}}{1-\frac{1}{x^-_p x^-_q}}\,, 
 \qquad &
B^{\sL\sR}_{pq} &= -\frac{2i}{h} \, \left(\frac{x^-_p}{x^+_p}\frac{x^+_q}{x^-_q} \right)^{1/2} \frac{\eta_{p}\eta_{q}}{ x^-_p x^+_q} \frac{\zeta_{pq}}{1-\frac{1}{x^-_p x^-_q}}\,,
\\
 C^{\sL\sR}_{pq} &= \zeta_{pq}\, ,   
\qquad &
D^{\sL\sR}_{pq} &=\zeta_{pq}\, \left(\frac{x^+_p}{x^-_p}\frac{x^+_q}{x^-_q} \right)^{1/2} \frac{1-\frac{1}{x^+_p x^+_q}}{1-\frac{1}{x^-_p x^-_q}}\,, \\
 E^{\sL\sR}_{pq} &= - \zeta_{pq}\, \left(\frac{x^+_q}{x^-_q} \right)^{1/2} \frac{1-\frac{1}{x^-_p x^+_q}}{1-\frac{1}{x^-_p x^-_q}}\,,
\qquad &
F^{\sL\sR}_{pq} &= \frac{2i}{h} \left(\frac{x^+_p}{x^-_p}\frac{x^+_q}{x^-_q} \right)^{1/2}  \frac{\eta_{p}\eta_{q}}{ x^+_p x^+_q} \frac{\zeta_{pq}}{1-\frac{1}{x^-_p x^-_q}}\,,
\end{aligned}
\end{equation}
and we have introduced a convenient factor
\begin{equation}
\zeta_{pq} = \left(\frac{x^+_p}{x^-_p}\right)^{-1/4}\left(\frac{x^+_q}{x^-_q}\right)^{-1/4} \left(\frac{1-\frac{1}{x^-_p x^-_q}}{1-\frac{1}{x^+_p x^+_q}}\right)^{1/2}\,.
\end{equation}
Similarly, an S-matrix $\mathcal{S}^{\sR\sL}$ can be found by swapping the labels L and R in~\eqref{eq:su(1|1)2-Smat-LRgrad1}. Imposing LR-symmetry one has that the explicit parameterisation is the same as in the equations above $ A^{\sR\sL}_{pq}= A^{\sL\sR}_{pq}$, et cetera.
Changing the grading of the first of the two representations, one finds for example an S-matrix
\begin{equation}\label{eq:su(1|1)2-Smat-LRgrad2}
\begin{aligned}
\mathcal{S}^{\tilde{\sL}\sR} \ket{\tilde{\phi}^{\sL}_p \phi^{\sR}_q} &= +D^{\sL\sR}_{pq} \ket{\phi^{\sR}_q \tilde{\phi}^{\sL}_p},  \qquad 
& \mathcal{S}^{\tilde{\sL}\sR} \ket{\tilde{\phi}^{\sL}_p \psi^{\sR}_q} &= -E^{\sL\sR}_{pq} \ket{\psi^{\sR}_q \tilde{\phi}^{\sL}_p} -F^{\sL\sR}_{pq} \ket{\phi^{\sR}_q \tilde{\psi}^{\sL}_p}, \\
\mathcal{S}^{\tilde{\sL}\sR} \ket{\tilde{\psi}^{\sL}_p \psi^{\sR}_q} &= -C^{\sL\sR}_{pq} \ket{\psi^{\sR}_q \tilde{\psi}^{\sL}_p},  \qquad 
& \mathcal{S}^{\tilde{\sL}\sR} \ket{\tilde{\psi}^{\sL}_p \phi^{\sR}_q} &= +A^{\sL\sR}_{pq} \ket{\phi^{\sR}_q \tilde{\psi}^{\sL}_p} +B^{\sL\sR}_{pq} \ket{\psi^{\sR}_q \tilde{\phi}^{\sL}_p}.
\end{aligned}
\end{equation}
To conclude we write down another result that we will need in the following 
\begin{equation}\label{eq:su(1|1)2-Smat-RLgrad2}
\begin{aligned}
\mathcal{S}^{\sR\tilde{\sL}} \ket{\phi^{\sR}_p \tilde{\phi}^{\sL}_q} &= +C^{\sR\sL}_{pq} \ket{\tilde{\phi}^{\sL}_q \phi^{\sR}_p},  \qquad 
& \mathcal{S}^{\sR\tilde{\sL}} \ket{\phi^{\sR}_p \tilde{\psi}^{\sL}_q} &= +A^{\sR\sL}_{pq} \ket{\tilde{\psi}^{\sL}_q \phi^{\sR}_p} - B^{\sR\sL}_{pq} \ket{\tilde{\phi}^{\sL}_q \psi^{\sR}_p}, \\
\mathcal{S}^{\sR\tilde{\sL}} \ket{\psi^{\sR}_p \tilde{\psi}^{\sL}_q} &= -D^{\sR\sL}_{pq} \ket{\tilde{\psi}^{\sL}_q \psi^{\sR}_p},  \qquad 
& \mathcal{S}^{\sR\tilde{\sL}} \ket{\psi^{\sR}_p \tilde{\phi}^{\sL}_q} &= -E^{\sR\sL}_{pq} \ket{\tilde{\phi}^{\sL}_q \psi^{\sR}_p} +  F^{\sR\sL}_{pq} \ket{\tilde{\psi}^{\sL}_q \phi^{\sR}_p}.
\end{aligned}
\end{equation}
The S-matrices presented here are also compatible with braiding and physical unitarity
\begin{equation}
\begin{aligned}
\mathcal{S}^{\sL\sR}\, \mathcal{S}^{\sR\sL}=\mathbf{1}\,,
\qquad
(\mathcal{S}^{\sL\sR})^\dagger\, \mathcal{S}^{\sR\sL}=\mathbf{1}\,,
\\
\mathcal{S}^{\tilde{\sL}\sR}\, \mathcal{S}^{\sR\tilde{\sL}}=\mathbf{1}\,,
\qquad
(\mathcal{S}^{\tilde{\sL}\sR})^\dagger\, \mathcal{S}^{\sR\tilde{\sL}}=\mathbf{1}\,.
\end{aligned}
\end{equation}
We refer to Section~\ref{sec:unitarity-YBe} for a discussion on this.

\subsection{The S-matrix as a tensor product}\label{sec:smat-tensor-prod}
The results of the previous section allow us to rewrite a $\psu(1|1)^4_\ce$-invariant S-matrix as a tensor product of two $\su(1|1)^2_\ce$-invariant S-matrices.
\begin{equation}
\begin{aligned}
\mathcal{S}_{\alg{psu}(1|1)^4}&= S_0 \cdot \mathcal{S}_{\alg{su}(1|1)^2} \;\check{\otimes}\; \mathcal{S}_{\alg{su}(1|1)^2},
\\
\mathbf{S}_{\alg{psu}(1|1)^4}&= S_0 \cdot \mathbf{S}_{\alg{su}(1|1)^2} \;\hat{\otimes}\; \mathbf{S}_{\alg{su}(1|1)^2},
\end{aligned}
\end{equation}
where $S_0$ is a possible prefactor that is not fixed by symmetries.
We introduced the graded tensor products~$\check{\otimes}$ and $\hat{\otimes}$
\begin{equation}
\label{eq:gradedtensorpr}
\begin{aligned}
\left( \mathcal{A}\,\check{\otimes}\,\mathcal{B} \right)_{MM',NN'}^{KK',LL'} &= (-1)^{\epsilon_{M'}\epsilon_{N}+\epsilon_{L}\epsilon_{K'}} \ \mathcal{A}_{MN}^{KL} \  \mathcal{B}_{M'N'}^{K'L'}\,,
\\
\left( \mathbf{A}\,\hat{\otimes}\,\mathbf{B} \right)_{MM',NN'}^{KK',LL'} &= (-1)^{\epsilon_{M'}\epsilon_{N}+\epsilon_{L'}\epsilon_{K}} \ \mathbf{A}_{MN}^{KL} \  \mathbf{B}_{M'N'}^{K'L'}\,,
\end{aligned}
\end{equation}
where the symbol~$\epsilon$ is $1$ for fermions and $0$ for bosons.
Depending on the representations that we want to scatter we have to choose the proper $\alg{su}(1|1)^2$ S-matrices entering the tensor product~\cite{Borsato:2014exa}.
This construction is explained in the rest of this section, while the explicit result for all the scattering elements may be found in Appendix~\ref{app:S-mat-explicit}.


\subsubsection{The massive sector $(\bullet\bullet)$}When considering the massive sector, we can scatter two different irreducible representations $\varrho_{\sL} \otimes \varrho_{\sL}$ and $\varrho_{\sR} \otimes \varrho_{\sR}$, identified by the eigenvalue $m=\pm 1$ of the generator $\gen{M}$. 
We see that this divides the massive sector into four different subsectors: Left-Left (LL), Right-Right (RR), Left-Right (LR) and Right-Left (RL). In each of these subsectors the LR-flavor is transmitted\footnote{As explained in the previous section, one needs to impose also LR-symmetry and unitarity to get pure transmission for the scattering of different flavors~\cite{Borsato:2012ud}.}.
Scattering two Left excitations means that we need to consider the tensor product
\begin{equation}
\text{Left - Left:}\quad
S_0^{\sL\sL} \cdot \mathcal{S}^{\sL\sL}\,\check{\otimes}\,\mathcal{S}^{\sL\sL}\,,
\end{equation}
where the explicit form of $\mathcal{S}^{\sL\sL}$ is given in~\eqref{eq:su(1|1)2-Smat-grad1}.
We need to fix a proper normalisation and we find convenient to do it as 
\begin{equation}\label{eq:norm-LL-massive-sector}
S_0^{\sL\sL} (x^\pm_p,x^\pm_q) = \frac{x^+_p}{x^-_p} \, \frac{x^-_q}{x^+_q} \, \frac{x^-_p - x^+_q}{x^+_p - x^-_q} \, \frac{1-\frac{1}{x^-_p x^+_q}}{1-\frac{1}{x^+_p x^-_q}} \, \frac{1}{\left(\sigma^{\bullet\bullet}_{pq} \right)^2 },
\end{equation}
where $\sigma^{\bullet\bullet}_{pq}$ is called \emph{dressing factor}. Since it cannot be fixed by the symmetries, it will be constrained later by solving the crossing equations derived in Section~\ref{sec:crossing-invar-T4}.
This normalisation is chosen to get for example the following scattering element
\begin{equation}
\bra{Y^{\sL}_q \, Y^{\sL}_p} \mathcal{S} \ket{Y^{\sL}_p \, Y^{\sL}_q}  = \frac{x^+_p}{x^-_p} \, \frac{x^-_q}{x^+_q} \, \frac{x^-_p - x^+_q}{x^+_p - x^-_q} \, \frac{1-\frac{1}{x^-_p x^+_q}}{1-\frac{1}{x^+_p x^-_q}} \, \frac{1}{\left(\sigma^{\bullet\bullet}_{pq} \right)^2 }.
\end{equation}
When we scatter two Right excitations we find an S-matrix
\begin{equation}
\text{Right - Right:}\quad
S_0^{\sR\sR} \cdot \mathcal{S}^{\sR\sR}\,\check{\otimes}\,\mathcal{S}^{\sR\sR}\,,
\end{equation}
and imposing LR-symmetry allows us to relate this result to the previous one, $\mathcal{S}^{\sR\sR}=\mathcal{S}^{\sL\sL}$ and $S_0^{\sR\sR}=S_0^{\sL\sL}$. In particular one does not need to introduce a different dressing factor in this subsector.

On the other hand, scattering a Left excitation with a Right one we get the S-matrix
\begin{equation}
\text{Left - Right:}\quad
S_0^{\sL\sR} \cdot \mathcal{S}^{\sL\sR}\,\check{\otimes}\,\mathcal{S}^{\sL\sR}\,,
\end{equation}
where $\mathcal{S}^{\sL\sR}$ may be found in~\eqref{eq:su(1|1)2-Smat-LRgrad1}.
The preferred normalisation in this case is 
\begin{equation}
S_0^{\sL\sR}(x^\pm_p,x^\pm_q) =\left(\frac{x^+_p}{x^-_p}\right)^{1/2}\left(\frac{x^+_q}{x^-_q}\right)^{-1/2}  \, \frac{1-\frac{1}{x^-_p x^+_q}}{1-\frac{1}{x^+_p x^-_q}}  \, \frac{1}{\left(\tilde{\sigma}^{\bullet\bullet}_{pq}\right)^2}\,,
\end{equation}
where a new dressing factor $\tilde{\sigma}^{\bullet\bullet}_{pq}$ is introduced.
With this normalisation we get for example the following scattering element
\begin{equation}
\bra{Y^{\sR}_q \, Y^{\sL}_p} \mathcal{S} \ket{Y^{\sL}_p \, Y^{\sR}_q}  = \frac{x^+_p}{x^-_p} \, \frac{x^-_q}{x^+_q} \, \frac{1-\frac{1}{x^+_p x^-_q}}{1-\frac{1}{x^+_p x^+_q}}  \, \frac{1-\frac{1}{x^-_p x^+_q}}{1-\frac{1}{x^-_p x^-_q}} \, \frac{1}{\left(\tilde{\sigma}^{\bullet\bullet}_{pq} \right)^2 }.
\end{equation}
To conclude, in the massive sector we need to introduce two unconstrained factors $\sigma^{\bullet\bullet}_{pq},\tilde{\sigma}^{\bullet\bullet}_{pq}$.

\subsubsection{The massless sector $(\circ\circ)$}
Each one-particle massless representation transforms under two copies of $\varrho_{\sL} \otimes \widetilde{\varrho}_{\sL}$, that are further organised in an $\su(2)_{\circ}$ doublet. When scattering two massless modules, we should then consider $16$ copies of $\psu(1|1)^4_\ce$-invariant S-matrices, relating each of the $4$ possible in-states to each of the $4$ possible out-states. Using the $\su(2)_{\circ}$ symmetry one is able to relate all these S-matrices, finding an object that is $\su(2)_{\circ}$-invariant.
More explicitly, the S-matrix in the massless sector can be written as the tensor product of an $\su(2)_{\circ}$-invariant S-matrix and the relevant tensor product realisation of the $\psu(1|1)^4_\ce$-invariant S-matrix
\begin{equation}
(\circ\circ):\quad
S_0^{\circ\circ} \cdot \mathcal{S}_{\alg{su}(2)}\;{\otimes}\;\big(\mathcal{S}^{\sL\sL}\;\check{\otimes}\;\mathcal{S}^{\tilde{\sL}\tilde{\sL}}\big)\,.
\end{equation}
Fixing a preferred normalisation we have
\begin{equation}
\label{eq:su2smat}
\mathcal{S}_{\alg{su}(2)}(p,q)=\frac{1}{1+\varsigma_{pq}}\big(\mathbf{1}+\varsigma_{pq} \Pi\big)
=
\left(\begin{array}{cccc}
1 &	0		&	0		&	0		\\
0		&\frac{1}{1+\varsigma_{pq}} 	&\frac{\varsigma_{pq}}{1+\varsigma_{pq}} &	0		\\
0		&\frac{\varsigma_{pq}}{1+\varsigma_{pq}} &\frac{1}{1+\varsigma_{pq}} &	0		\\
0		&	0		&	0		&	1		\\
\end{array}\right)\,,
\end{equation}
where $\Pi$ is the permutation matrix and $\varsigma_{pq}$ is a function of the two momenta $p,q$ that is not fixed by the $\su(2)_{\circ}$ symmetry. In Sections~\ref{sec:unitarity-YBe} and~\ref{sec:crossing-invar-T4} we will see that further constraints are imposed on that by unitarity, Yang-Baxter equation and crossing invariance.
We choose to fix the overall normalisation as
\begin{equation}
S_0^{\circ\circ} = \left( \frac{x^+_p}{x^-_p} \, \frac{x^-_q}{x^+_q} \right)^{1/2}\, \frac{x^-_p - x^+_q}{x^+_p - x^-_q} \, \frac{1}{\left(\sigma^{\circ\circ}_{pq} \right)^2 }, 
\end{equation}
where we introduced the dressing factor $\sigma^{\circ\circ}_{pq}$ for the massless sector. With this choice, the scattering of two identical bosons coming from the torus is just
\begin{equation}
\begin{aligned}
\bra{T^{\dot{a}a}_q \, T^{\dot{a}a}_p} \mathcal{S} \ket{T^{\dot{a}a}_p \, T^{\dot{a}a}_q} & =  \frac{1}{\left(\sigma^{\circ\circ}_{pq} \right)^2 }.
\end{aligned}
\end{equation}

\subsubsection{The mixed-mass sector $(\bullet\circ),(\circ\bullet)$}
In the mixed-mass sector we may decide to scatter a massive particle with a massless one $(\bullet\circ)$ or vice-versa $(\circ\bullet)$.
Let us focus on the first case. We have also the possibility of choosing the LR-flavor of the first excitation. When this is Left we find the S-matrix
\begin{equation}
\label{eq:mssive-mless1}
 \text{Left (massive) -  massless:}\quad
S_0^{\sL\circ} \cdot \left(\mathcal{S}^{\sL\sL}\,\check{\otimes}\,\mathcal{S}^{\sL\tilde{\sL}} \right)^{\oplus 2}\,.
\end{equation}
The symbol $\oplus 2$ appears because massless excitations are organised in two $\psu(1|1)^4_\ce$-modules, identified by the two $\su(2)_{\circ}$ flavors. Imposing $\su(2)_{\circ}$-invariance one finds that the S-matrix is just the sum of two identical copies. In other words the $\su(2)_{\circ}$ flavor stands as a spectator and it is transmitted under the scattering.

Similarly, for scattering involving Right-massive excitations
\begin{equation}
\label{eq:mssive-mlessR}
 \text{Right (massive) -  massless:}\quad
S_0^{\sR\circ} \cdot \left(\mathcal{S}^{\sR\sL}\,\check{\otimes}\,\mathcal{S}^{\sR\tilde{\sL}} \right)^{\oplus 2}\,.
\end{equation}
This S-matrix is related by LR-symmetry to the previous one. Implementing LR-symmetry as in Section~\ref{sec:LR-symmetry} and using the fact that the second excitation satisfies a massless dispersion relation, one can check that the two S-matrices are mapped one into the other, upon fixing the proper normalisations. This allows us to use just $S_0^{\sL\circ} \cdot \left(\mathcal{S}^{\sL\sL}\,\check{\otimes}\,\mathcal{S}^{\sL\tilde{\sL}} \right)^{\oplus 2}$ for both cases. We also set the overall factor
\begin{equation}
\begin{aligned}
S_0^{\bullet\circ} &\equiv S_0^{\sL\circ} =\left( \frac{x^+_p}{x^-_p} \right)^{-1/2}\, \left( \frac{1-\frac{1}{x^-_p x^+_q}}{1-\frac{1}{x^+_p x^-_q}} \right)^{1/2} \, \left(\frac{1-\frac{1}{x^-_p x^-_q}}{1-\frac{1}{x^+_p x^+_q}} \right)^{1/2} \, \frac{1}{\left(\sigma^{\bullet\circ}_{pq} \right)^2 },
\end{aligned}
\end{equation}
where we have chosen a proper normalisation and introduced the dressing factor $\sigma^{\bullet\circ}_{pq}$ for massive-massless scattering.

Similar considerations apply when considering massless-massive scattering. The second excitation is allowed to take the two different flavors L or R, and in the two cases we find the S-matrices
\begin{equation}
\label{eq:mless-mssive}
\begin{aligned}
 \text{massless - \ \ Left\  (massive):}\quad &
S_0^{\circ \sL} \cdot \left(\mathcal{S}^{\sL\sL}\,\check{\otimes}\,\mathcal{S}^{\tilde{\sL}\sL} \right)^{\oplus 2}\,,
\\
 \text{massless - Right (massive):}\quad &
S_0^{\circ \sR} \cdot \left(\mathcal{S}^{\sL\sR}\,\check{\otimes}\,\mathcal{S}^{\tilde{\sL}\sR} \right)^{\oplus 2}\,.
\end{aligned}
\end{equation}
LR-symmetry allows us to use just $S_0^{\circ \sL} \cdot \left(\mathcal{S}^{\sL\sL}\,\check{\otimes}\,\mathcal{S}^{\tilde{\sL}\sL} \right)^{\oplus 2}$ in both cases. We then introduce the common factor $S_0^{\circ\bullet}$ that we decide to normalise as
\begin{equation}
S_0^{\circ\bullet}\equiv S_0^{\circ \sL}=\left( \frac{x^+_q}{x^-_q} \right)^{1/2}\, \left( \frac{1-\frac{1}{x^-_p x^+_q}}{1-\frac{1}{x^+_p x^-_q}} \right)^{1/2} \, \left(\frac{1-\frac{1}{x^-_p x^-_q}}{1-\frac{1}{x^+_p x^+_q}} \right)^{-1/2} \, \frac{1}{\left(\sigma^{\circ\bullet}_{pq} \right)^2 }.
\end{equation}
The chosen normalisations allow us to write for example the following scattering elements
\begin{equation}
\begin{aligned}
\bra{T^{\dot{a}a}_q \, Y^{\sL}_p} \mathcal{S} \ket{Y^{\sL}_p \, T^{\dot{a}a}_q} 
&=
 \left( \frac{1-\frac{1}{x^+_p x^-_q}}{1-\frac{1}{x^+_p x^+_q}}  \, \frac{1-\frac{1}{x^-_p x^+_q}}{1-\frac{1}{x^-_p x^-_q}} \right)^{1/2} \, \frac{1}{\left(\sigma^{\bullet\circ}_{pq} \right)^2 }, \\
\bra{Y^{\sL}_q \, T^{\dot{a}a}_p} \mathcal{S} \ket{T^{\dot{a}a}_p \, Y^{\sL}_q} 
&=
 \left( \frac{1-\frac{1}{x^+_p x^-_q}}{1-\frac{1}{x^+_p x^+_q}}  \, \frac{1-\frac{1}{x^-_p x^+_q}}{1-\frac{1}{x^-_p x^-_q}} \right)^{1/2} \, \frac{1}{\left(\sigma^{\circ\bullet}_{pq} \right)^2 }.
\end{aligned}
\end{equation}
Later we will discuss how massive-massless and massless-massive scatterings are related in a simple way by unitarity. In particular, this will give a relation between $\sigma^{\bullet\circ}_{pq}$ and $\sigma^{\circ\bullet}_{pq}$, motivating the statement that in the mixed-mass sector we have just one dressing factor.

\subsection{Unitarity and Yang-Baxter equation}\label{sec:unitarity-YBe}
After fixing the S-matrix based on symmetries, one finds that this is determined up to five dressing factors. Two of them belong to the massive sector, describing scattering of excitations with the same or with opposite LR flavors. Other two are responsible for the mixed-mass sector, namely massive-massless and massless-massive scattering. The last one belongs to the massless sector. 

More constraints on those scalar factors come from unitarity. 
One notion of this is the usual \emph{physical} unitarity, that requires the S-matrix to be unitary as a matrix
\be
\mathcal{S}^\dagger_{pq}\mathcal{S}_{pq}=\mathbf{1}\,.
\ee
Another natural constraint for scattering of particles on a line is \emph{braiding} unitarity 
\be
\mathcal{S}_{qp}\mathcal{S}_{pq}=\mathbf{1}\,.
\ee
Its interpretation is that scattering twice two excitations should just bring us back to the initial situation\footnote{We define the S-matrix such that the momentum of the first particle is larger than the one of the second. If the first scattering happens for $p>q$, to evaluate the second process we have to analytically continue the S-matrix to the region where the momentum of the first particle is less than the one of the second excitation.}.
We refer to~\cite{Arutyunov:2006yd} for a justification of this constraint from the point of view of the formalism of the Zamolodchikov-Fadeev algebra applied to worldsheet integrable scattering.

In our case we find the following equations
\begin{equation}
\begin{aligned}
\sigma^{\bullet\bullet}_{qp}=\big(\sigma^{\bullet\bullet}_{pq}\big)^*=\frac{1}{\sigma^{\bullet\bullet}_{pq}}\,,
\qquad
\tilde{\sigma}^{\bullet\bullet}_{qp}=\big(\tilde{\sigma}^{\bullet\bullet}_{pq}\big)^*=\frac{1}{\tilde{\sigma}^{\bullet\bullet}_{pq}}\,,
\qquad
\sigma^{\circ\circ}_{qp}=\big(\sigma^{\circ\circ}_{pq}\big)^*=\frac{1}{\sigma^{\circ\circ}_{pq}}\,,\\
\qquad
\sigma^{\bullet\circ}_{qp}=\big(\sigma^{\circ\bullet}_{pq}\big)^*=\frac{1}{\sigma^{\circ\bullet}_{pq}}\,,
\qquad\qquad
\sigma^{\circ\bullet}_{qp}=\big(\sigma^{\bullet\circ}_{pq}\big)^*=\frac{1}{\sigma^{\bullet\circ}_{pq}}\,.
\qquad\qquad
\end{aligned}
\end{equation}
The first line states that the dressing factors in the massive and massless sectors can be written as exponentials of anti-symmetric functions of the two momenta, and for physical momenta they take values on the unit circle. On the other hand, in the mixed-mass sector unitarity relates massive-massless and massless-massive scattering.
This reduces to four the number of unconstrained dressing factors.

Unitarity imposes also the following constraint on the function $\varsigma_{pq}$ appearing in the $\su(2)_\circ$-invariant S-matrix
\begin{equation}
\varsigma_{qp}=\big(\varsigma_{pq}\big)^*=-\varsigma_{pq}\,,
\end{equation}
meaning that it is a purely imaginary anti-symmetric function of $p$ and $q$.

\medskip

For the integrability of the model, it is necessary for the S-matrix to satisfy the Yang-Baxter equation
\begin{equation}\label{eq:YBe}
\mathcal{S}(q,r)\otimes\mathbf{1} \cdot\mathbf{1}\otimes\mathcal{S}(p,r) \cdot \mathcal{S}(p,q)\otimes\mathbf{1}
=
\mathbf{1} \otimes\mathcal{S}(p,q)\cdot\mathcal{S}(p,r)\otimes \mathbf{1} \cdot \mathbf{1} \otimes\mathcal{S}(q,r)\,.
\end{equation}
This is a crucial requirement to make factorisability of multi-particle scatterings possible.
One may check the Yang-Baxter equation for the full S-matrix, or equivalently for each factor of the tensor product appearing in each sector. Since the $\su(1|1)^2_\ce$ S-matrices of Section~\ref{sec:su112-S-matrices} satisfy the Yang-Baxter equation, it follows that this is true also for the $\psu(1|1)^4_\ce$ S-matrices of the various sectors of our model.

On the S-matrix for the $\su(2)_\circ$ factor, the Yang-Baxter equation yields a further constraint for $\varsigma_{pq}$
\begin{equation}
\varsigma(p,q)-\varsigma(p,r)+\varsigma(q,r)=0\,.
\end{equation}
The above equation is linear thanks to a suitable choice of parameterisation for the $\su(2)_\circ$ S-matrix. 
The solution is a function that is a difference of two rapidities, each depending on just one momentum. Together with the constraints imposed by unitarity we can write
\begin{equation}
\label{eq:varsigma-difference}
\varsigma(p,q)=i\big(w_p-w_q\big),
\end{equation}
where we have introduce a new real function of the momentum $w_p$.

\subsection{Crossing invariance}\label{sec:crossing-invar-T4}
The analytic properties of the dressing factors are revealed after imposing crossing invariance of the S-matrix~\cite{Janik:2006dc}.
A crossing transformation corresponds to an analytic continuation to an unphysical channel, where the energy and the momentum flip sign.

We start the discussion by first considering the massive excitations. Their dispersion relation satisfies
\begin{equation}
E^2 = 1+4h^2\sin^2 \frac{p}{2}\,,
\end{equation}
and we can uniformise it in terms of a complex parameter $z$ with the parameterisation~\cite{Beisert:2006nonlin}
\begin{equation}
p=2 \text{am}z\,,\qquad \sin \frac{p}{2} = \text{sn}(z,k)\,, \qquad E=\text{dn}(z,k)\,,
\end{equation}
where the elliptic modulus is defined as $k=-4h^2<0$.
The curve that we obtain is a torus, and we call $2\omega_1$ and $2\omega_2$ the periods for real and imaginary shifts, respectively. They are obtained by
\begin{equation}
2\omega_1= 4\,\text{K}(k)\,, \qquad 2\omega_2 = 4i\, \text{K}(1-k)-4\, \text{K}(k)\,,
\end{equation}
with $\text{K}(k)$ the complete elliptic integral of the first kind.
Real values of $z$ correspond to real values of the momentum $p$. If we take periodicity into account and we define the physical range of the momentum to be $-\pi\leq p<\pi$, then we may take $-\omega_1/2\leq z<\omega_1/2$.

A crossing transformation corresponds to an analytic continuation to a complex value that we denote by $\bar{z}$, where we flip the signs of the momentum and the energy. We see that this is implemented by shifting $z$ by half of the imaginary period
\begin{equation}
z\to\bar{z}= z+\omega_2\,,
\quad\implies\quad
 p\to \bar{p}=-p\,,\quad E\to \bar{E}=-E\,.
\end{equation}
On the Zhukovski variables and the function $\eta_p$ defined in~\eqref{eq:def-eta}, the crossing transformation implies
\begin{equation}
x^{\pm}(z+\omega_2)= \frac{1}{x^{\pm}(z)}\,,
\qquad
\eta(z+\omega_2) = \frac{i}{x^+(z)}\eta(z)\,.
\end{equation}
We can easily check that for massless excitations the crossing transformation on the parameters $x^\pm$ is implemented in the same way.
It is important to note that for the eigeinvalue of the central charge $\gen{C}$, crossing does not change just its sign
\begin{equation}\label{eq:cross-central-charge}
\frac{i\,h}{2}\left(e^{i\,\bar{p}}-1\right)=-e^{-i\,p}\frac{i\,h}{2}\left(e^{i\,p}-1\right)\,,
\implies
\gen{C}(\bar{z})= -e^{i\,p} \gen{C}(z)\,.
\end{equation}

A crucial step is now to note that we may mimic these transformation laws on the central charges also in a different way, namely by defining a proper  charge conjugation matrix $\mathscr{C}(z)$.
According to the previous discussion, on the central charges $\gen{H}$ and $\gen{M}$ we should impose
\begin{equation}
\label{eq:crossign-u1-charges}
\gen{H}(\bar{z}) = - \mathscr{C}(z)\gen{H}\,\mathscr{C}(z)^{-1}\,,
\qquad
\gen{M} = - \mathscr{C}(z)\gen{M}\,\mathscr{C}(z)^{-1}\,.
\end{equation}
In order to reproduce also the transformation~\eqref{eq:cross-central-charge} on the charge $\gen{C}$,  we impose that on supercharges we have
\begin{equation}
\begin{aligned}
\gen{Q}^{\ \dot{a}}_{\sL}(z+\omega_2)^{\st} &= - e^{-\frac{i}{2}\, p}\ \mathscr{C}(z)\, \gen{Q}^{\ \dot{a}}_{\sL}(z)\, \mathscr{C}^{-1}(z), \\
\gen{Q}_{\sR \dot{a}}(z+\omega_2)^{\st} &= - e^{-\frac{i}{2}\, p}\ \mathscr{C}(z)\, \gen{Q}_{\sR \dot{a}}(z)\, \mathscr{C}^{-1}(z), \\
\overline{\gen{Q}}{}_{\sL  \dot{a}}(z+\omega_2)^{\st} &= -e^{+\frac{i}{2} p}\  \mathscr{C}(z)\, \overline{\gen{Q}}{}_{\sL \dot{a}}(z)\, \mathscr{C}^{-1}(z),\\
\overline{\gen{Q}}{}_{\sR}^{\  \dot{a}}(z+\omega_2)^{\st} &= -e^{+\frac{i}{2} p}\  \mathscr{C}(z)\, \overline{\gen{Q}}{}_{\sR}^{\  \dot{a}}(z)\, \mathscr{C}^{-1}(z).
\end{aligned}
\end{equation}
Here ${}^\st$ denotes supertransposition, that is implemented on supercharges as $\gen{Q}^{\st} = \gen{Q}^{\text{t}} \, \Sigma$. The diagonal matrix $\Sigma$ is the fermion-sign matrix, taking values $+1, -1$ on bosons and fermions respectively.
Compatibility with the $\su(2)_{\bullet}$ generators requires that we exchange the highest and the lowest weight in the doublet representation. We follow the same rule also for $\su(2)_{\circ}$
\begin{equation}
{\gen{J}_{\bullet \dot{b}}}^{\dot{a}}=-\mathscr{C}(z) \, {\gen{J}_{\bullet \dot{a}}}^{\dot{b}}\,\mathscr{C}(z)^{-1}\,,
\qquad
{\gen{J}_{\circ b}}^a=- \mathscr{C}(z) \, {\gen{J}_{\circ a}}^b\,\mathscr{C}(z)^{-1} \,.
\end{equation}
The form of the charge conjugation matrix is not unique, nevertheless all choices yield the same crossing equations.
In the basis 
\begin{equation}
\{ Y^{\sL}, \eta^{\sL 1}, \eta^{\sL 2}, Z^{\sL} \} \oplus \{ Y^{\sR}, \eta^{\sR 1}, \eta^{\sR 2}, Z^{\sR} \} \oplus \{ T^{11}, T^{21}, T^{12}, T^{22} \} \oplus \{ \widetilde{\chi}^1, \chi^1, \widetilde{\chi}^2, \chi^2 \} ,
\end{equation}
we take it to be
\begin{equation}
\newcommand{\0}{\color{black!40}0}
  \renewcommand{\arraystretch}{1.1}
  \setlength{\arraycolsep}{3pt}
  \mathscr{C}_p=\!\left(\!
    \mbox{\footnotesize$
      \begin{array}{cccc|cccc}
        \0 & \0 & \0 & \0 & 1 & \0 & \0 & \0 \\
        \0 & \0 & \0 & \0 & \0 & \0 & -i & \0 \\
        \0 & \0 & \0 & \0 & \0 & i & \0 & \0 \\
        \0 & \0 & \0 & \0 & \0 & \0 & \0 & 1 \\
        \hline
        1 & \0 & \0 & \0 & \0 & \0 & \0 & \0 \\
        \0 & \0 & i & \0 & \0 & \0 & \0 & \0 \\
        \0 & -i & \0 & \0 & \0 & \0 & \0 & \0 \\
        \0 & \0 & \0 & 1 & \0 & \0 & \0 & \0 \\
      \end{array}$}\!
  \right) 
	\oplus
	\!\left(\!
    \mbox{\footnotesize$
      \begin{array}{cccc|cccc}
        \0 & \0 & \0 & 1 & \0 & \0 & \0 & \0 \\
        \0 & \0 & -1 & \0 & \0 & \0 & \0 & \0 \\
        \0 & -1 & \0 & \0 & \0 & \0 & \0 & \0 \\
        1 & \0 & \0 & \0 & \0 & \0 & \0 & \0 \\
        \hline
        \0 & \0 & \0 & \0 & \0 & \0 & \0 & -i \frac{a_p}{b_p} \\
        \0 & \0 & \0 & \0 & \0 & \0 & i \frac{b_p}{a_p} & \0 \\
        \0 & \0 & \0 & \0 & \0 & i \frac{a_p}{b_p} & \0 & \0 \\
        \0 & \0 & \0 & \0 & -i \frac{b_p}{a_p} & \0 & \0 & \0 \\
      \end{array}$}\!
  \right).
\end{equation}
With the explicit form of the charge conjugation matrix we are able to impose crossing invariance on the full S-matrix.
They key is to consider the objects 
\begin{equation}
\mathbf{S}(z_p,z_q)^{-1}
\quad\text{ and }
\quad\mathscr{C}_1(z_p) \cdot \mathbf{S}^{\text{t}_1}(z_p+\omega_2,z_q) \cdot \mathscr{C}_1^{-1}(z_p) \,,
\end{equation}
where we have used the notation $\mathscr{C}_1(z_p) = \mathscr{C} (z_p) \otimes \mathbf{1}$, and ${}^{\text{t}_1}$ denotes transposition on the first space.
We can check that the compatibility condition with the charges for $\mathbf{S}(z_p,z_q)^{-1}$
\begin{equation}
\begin{aligned}
 \gen{J}_{12}(z_p,z_q) \,\mathbf{S}(z_p,z_q)^{-1}- \mathbf{S}(z_p,z_q)^{-1}\,\gen{J}_{21}(z_q,z_p) &=0\,,
\end{aligned}
\end{equation}
is satisfied also by $\mathscr{C}_1(z_p) \cdot \mathbf{S}^{\text{t}_1}(z_p+\omega_2,z_q) \cdot \mathscr{C}_1^{-1}(z_p)$.
They might then differ just by some factors, one for each of the sectors that we identified in Section~\ref{sec:smat-tensor-prod}. Crossing symmetry fixes this freedom and states that these two objects are equal\footnote{A similar statement can be made also for the object $ \mathscr{C}_2^{-1}(z_q) \cdot \mathbf{S}^{\text{t}_2}(z_p,z_q-\omega_2) \cdot   \mathscr{C}_2(z_q)$, where crossing is implemented by shifting the second entry in the opposite direction. This second equation would be related by unitarity to the previous one.}
\begin{equation}\label{eq:cr-matrix-form}
\begin{aligned}
\mathscr{C}_1(z_p) \cdot \mathbf{S}^{\text{t}_1}(z_p+\omega_2,z_q) \cdot \mathscr{C}_1^{-1}(z_p)  \cdot \mathbf{S}(z_p,z_q) &= \mathbf{1} \,.
\end{aligned}
\end{equation}

It is an important fact that the whole set of equations reduces to equations just for the dressing factors 
\begin{equation}\label{eq:cr-massive}
\begin{aligned}
\left(\sigma^{\bullet\bullet}_{pq}\right)^2 \ \left(\tilde{\sigma}^{\bullet\bullet}_{\bar{p}q}\right)^2 &= \left( \frac{x^-_q}{x^+_q} \right)^2 \frac{(x^-_p-x^+_q)^2}{(x^-_p-x^-_q)(x^+_p-x^+_q)} \frac{1-\frac{1}{x^-_px^+_q}}{1-\frac{1}{x^+_px^-_q}}, \\
\left(\sigma^{\bullet\bullet}_{\bar{p}q}\right)^2 \ \left(\tilde{\sigma}^{\bullet\bullet}_{pq}\right)^2 &= \left( \frac{x^-_q}{x^+_q} \right)^2 \frac{\left(1-\frac{1}{x^+_px^+_q}\right)\left(1-\frac{1}{x^-_px^-_q}\right)}{\left(1-\frac{1}{x^+_px^-_q}\right)^2} \frac{x^-_p-x^+_q}{x^+_p-x^-_q},
\end{aligned}
\end{equation}
\begin{equation}\label{eq:cr-mixed}
\begin{aligned}
\left(\sigma^{\bullet \circ}_{\bar{p}q} \right)^2 \ \left( \sigma^{\bullet \circ}_{pq} \right)^2 &= \frac{x^+_p}{x^-_p} \frac{x^-_p-x^+_q}{x^+_p-x^+_q} \frac{1-\frac{1}{x^+_px^+_q}}{1-\frac{1}{x^-_px^+_q}}, \\
\left(\sigma^{\circ \bullet}_{\bar{p}q} \right)^2 \ \left( \sigma^{\circ \bullet}_{pq} \right)^2 &= \frac{x^+_q}{x^-_q} \frac{x^+_p-x^-_q}{x^+_p-x^+_q} \frac{1-\frac{1}{x^+_px^+_q}}{1-\frac{1}{x^+_px^-_q}}
\end{aligned}
\end{equation}
\begin{equation}\label{eq:cr-massless}
\begin{aligned}
\left(\sigma^{\circ\circ}_{\bar{p}q} \right)^2 \ \left(\sigma^{\circ\circ}_{pq} \right)^2 &= \frac{\varsigma_{pq}-1}{\varsigma_{pq}} \, \frac{1-\frac{1}{x^+_p x^+_q}}{1-\frac{1}{x^+_p x^-_q}} \, \frac{1-\frac{1}{x^-_p x^-_q}}{1-\frac{1}{x^-_p x^+_q}},
\end{aligned}
\end{equation}
and for the function $\varsigma_{pq}$ of the $\su(2)_{\circ}$ S-matrix 
\begin{equation}\label{eq:cr-varsigma}
\begin{aligned}
\varsigma_{\bar{p}q} &= \varsigma_{pq}-1\, 
\implies
w(\bar{p})=w(p)+i\,.
\end{aligned}
\end{equation}
In Section~\ref{sec:dressing-factors} we present solutions of the crossing equations for the dressing factors of the massive sector.

\section{Bethe-Yang equations}\label{sec:Bethe-Yang}
The Bethe-Yang equations are quantisation conditions that allow one to solve for the momenta of the excitations of a multi-particle state. With this data one can find the spectrum of the theory in the decompactification limit. In integrable models the Bethe-Yang equations arise when imposing periodicity of the wave-function of an eigenstate of the Hamiltonian.
In our case periodicity is motivated by the fact that we are studying closed strings, and we depict this in Figure~\ref{fig:period-cond}. Instead of looking for eigenstates of the exact quantum Hamiltonian---that is not known---we will construct eigenstates of the exact S-matrix derived in Section~\ref{sec:S-mat-T4}.

\subsection{Deriving the Bethe-Yang equations}
Rather than introducing a toy model to explain how to derive the Bethe-Yang equations, we prefer to guide the reader through the main steps in the case of \adsthree. Indeed, deciding to turning on only certain flavors of the excitations reduces the problem remarkably, and from the operational point of view it makes it conceptually equivalent to other simpler integrable models, such as the Heisenberg spin-chain---see~\cite{1998cond.mat..9162K,Faddeev:1994nk,Faddeev:1996iy} for nice reviews.
The various complications are added gradually, until all the material to construct the full set of Bethe-Yang equations is presented.

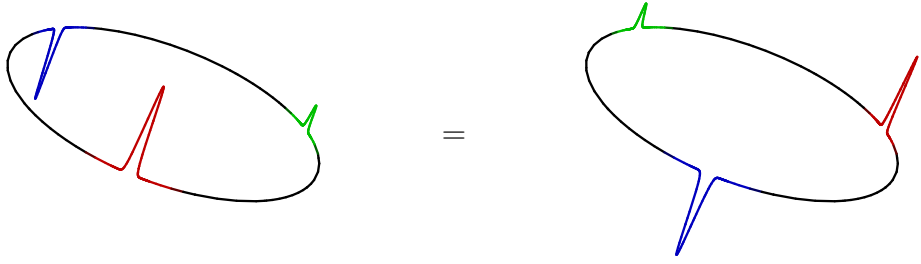
\begin{figure}[t]
  \centering
\includegraphics[width=0.4\textwidth]{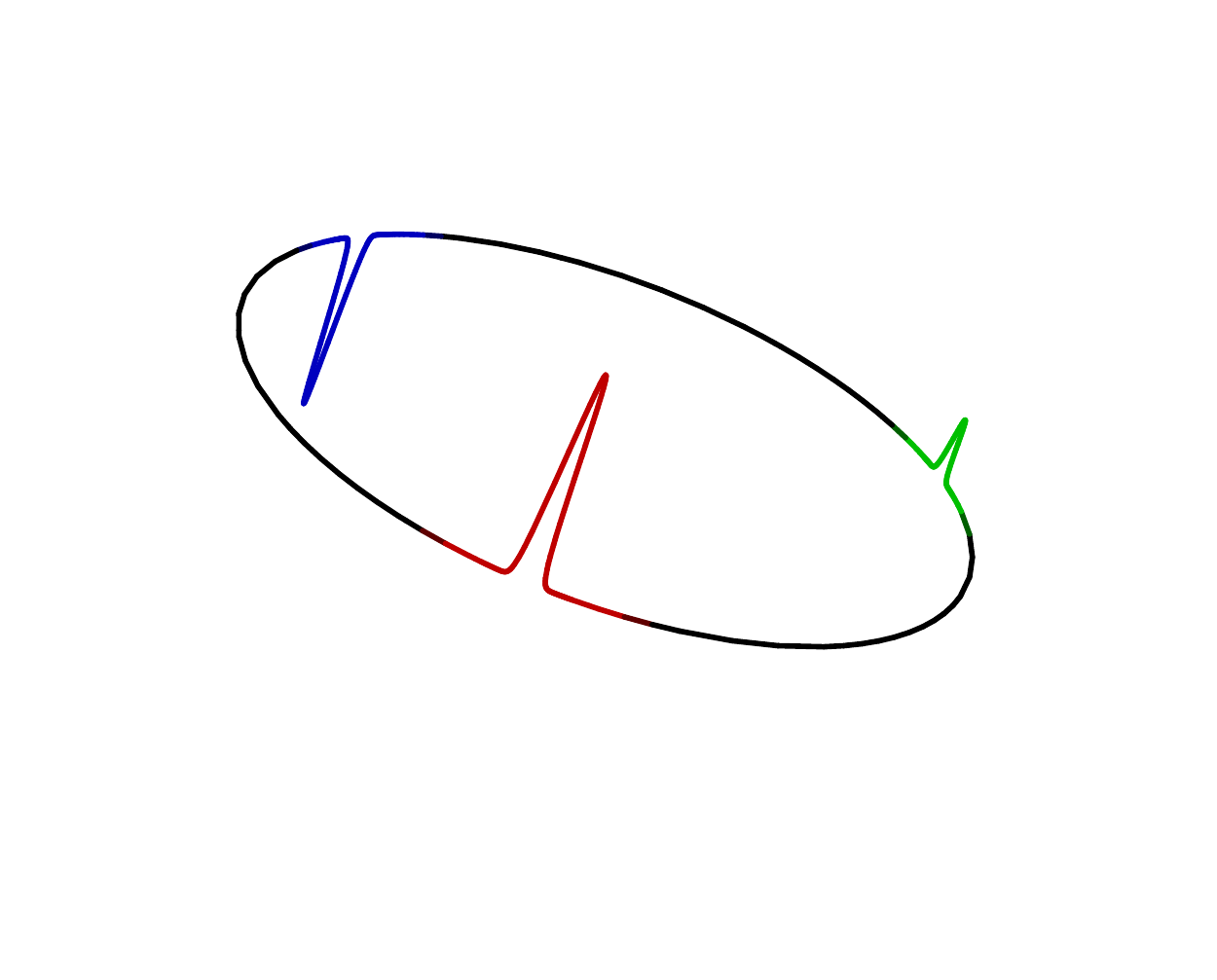}
  \begin{tikzpicture}[%
    box/.style={outer sep=1pt},
    Q node/.style={inner sep=1pt,outer sep=0pt},
    arrow/.style={-latex}
    ]%
	\node [box] at ( 0  , -3cm) {\raisebox{2.5cm}{$=$}};
 \end{tikzpicture}
\includegraphics[width=0.4\textwidth]{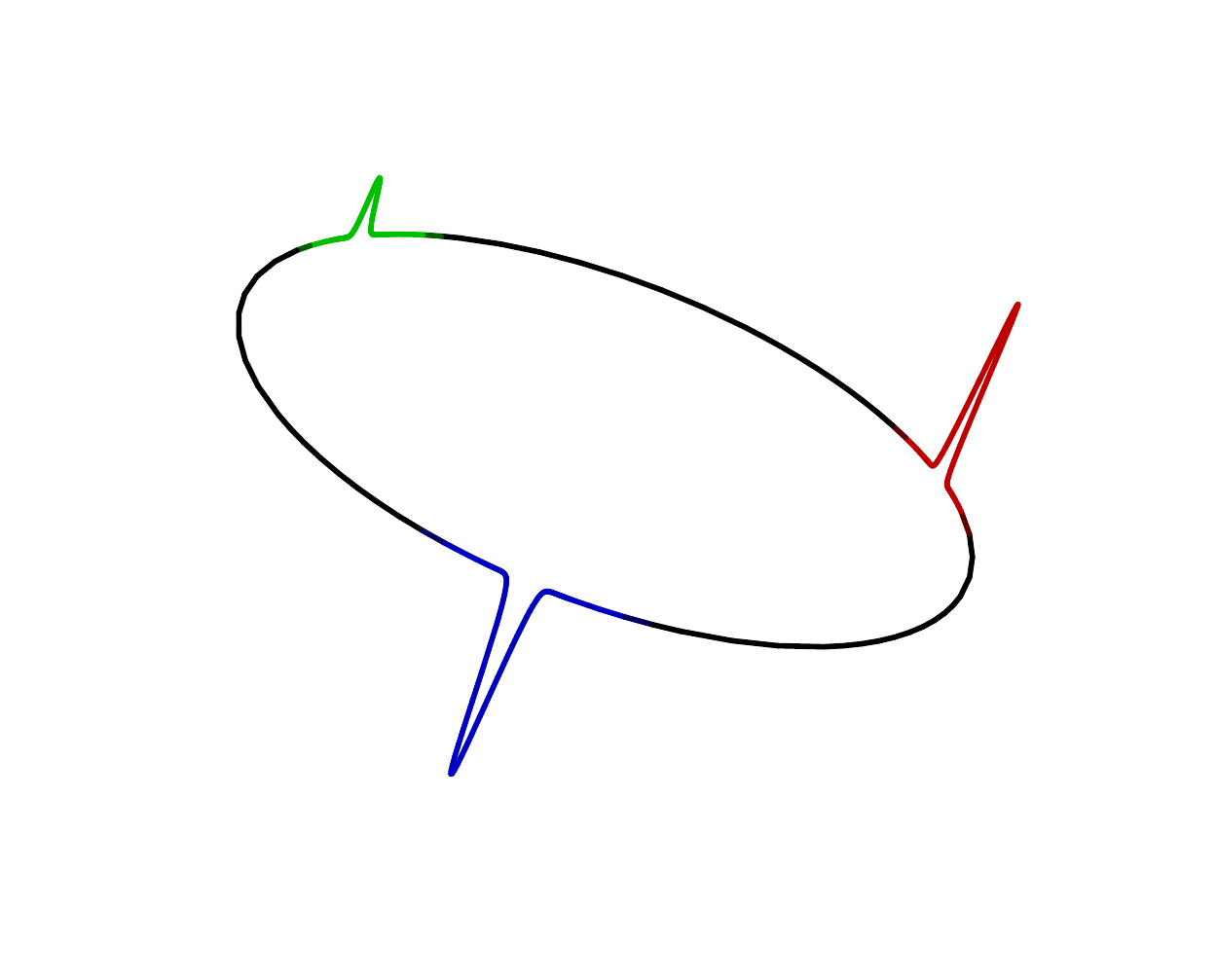}
  \caption{The periodicity condition of the wave-function yields the Bethe-Yang equations. A configuration with excitations localised at given points of the string is equivalent to another one where the excitations are cyclically permuted.}
  \label{fig:period-cond}
\end{figure}

\paragraph{Bethe Ansatz with one flavor.}
Let us start for simplicity with the case in which only excitations of the type $Y^{\sL}$ are present.
We make this choice because the scattering of two $Y^{\sL}$ excitations is very simple
\be
\mathcal{S}\ket{Y^{\sL}_pY^{\sL}_q}= \mathcal{A}_{pq}\ket{Y^{\sL}_qY^{\sL}_p}\,.
\ee
Here $\mathcal{A}_{pq}$ denotes the corresponding scattering element. What is important is that this two-particle state does not mix with others under scattering.
The simplest multiparticle state that we might want to consider is a collection of plane-waves
\begin{equation}
\ket{Y^{\sL}_{p_1}Y^{\sL}_{p_2}\ldots Y^{\sL}_{p_n}}=\sum_{\sigma_1\ll \sigma_2 \ll \ldots \ll \sigma_{n}} \ e^{i \sum_{j=1}^n p_j\sigma_j} \ \ket{Y^{\sL}_{\sigma_1}Y^{\sL}_{\sigma_2}\ldots Y^{\sL}_{\sigma_n}}.
\end{equation}
Here and in the following we always assume that we deal with asymptotic states, meaning that excitations with different momenta are ordered $p_1>\ldots>p_n$ and well separated.
As it is known from the simplest integrable models (\emph{e.g.} the Heisenberg spin-chain), the eigenstates of the Hamiltonian are specific superpositions of plane waves.
Let us focus on the case of just two excitations. We consider a generic state
\begin{equation}
  \begin{aligned}
\ket{\Psi}&=\sum_{\sigma_1\ll \sigma_2} \psi(\sigma_1,\sigma_2) \ket{Y^{\sL}_{\sigma_1}Y^{\sL}_{\sigma_2}}
\\
&=\ket{Y^{\sL}_{p_1}Y^{\sL}_{p_2}}+S(p_1,p_2)\ket{Y^{\sL}_{p_2}Y^{\sL}_{p_1}}.
\end{aligned}
\end{equation}
where by definition we restrict ourselves to the region $\sigma_1\ll \sigma_2$ and we have defined the generic wave-function
\begin{equation}
\psi(\sigma_1,\sigma_2) =e^{i(p_1\sigma_1+p_2\sigma_2)}+S(p_1,p_2)e^{i(p_2\sigma_1+p_1\sigma_2)}.
\end{equation}
The choice $S(p_1,p_2)=1$ would correspond to just the sum of the original plane-waves with the reflected ones.
We choose instead to identify $S(p_1,p_2)=\mathcal{A}_{p_1p_2}$, namely the scattering element of the two excitations.
Thanks to this choice $\ket{\Psi}$ becomes an eigenstate of the S-matrix
\begin{equation}
  \begin{aligned}
\mathcal{S}\ket{\Psi} &=\mathcal{A}_{p_1p_2}\ket{Y^{\sL}_{p_2}Y^{\sL}_{p_1}}+\mathcal{A}_{p_1p_2}\mathcal{A}_{p_2p_1}\ket{Y^{\sL}_{p_1}Y^{\sL}_{p_2}}
=\ket{\Psi}\,,
\end{aligned}
\end{equation}
that is proved using braiding unitarity, \ie $\mathcal{A}_{p_1p_2}\mathcal{A}_{p_2p_1}=1$.
This justifies the choice for the function $S(p_1,p_2)$.

\medskip

The important requirement that we want to impose now is the periodicity of the wave-function, as depicted in Figure~\ref{fig:period-cond}.
An explicit computation gives
\begin{equation}
  \begin{aligned}
\psi(\sigma_2,\sigma_1+L)&=e^{i(p_1\sigma_2+p_2\sigma_1+p_2L)}+\mathcal{A}_{p_1p_2}e^{i(p_2\sigma_2+p_1\sigma_1+p_1L)}
\\
&= e^{ip_1L}\mathcal{A}_{p_1p_2} \left(\left(\mathcal{A}_{p_1p_2}\right)^{-1}e^{i(p_1\sigma_2+p_2\sigma_1+(p_2-p_1)L)}+e^{i(p_2\sigma_2+p_1\sigma_1)}\right)\,.
\end{aligned}
\end{equation}
If we require $\psi(\sigma_2,\sigma_1+L)=\psi(\sigma_1,\sigma_2)$ we find the two equations
\begin{equation}
e^{i p_1L}= \left(\mathcal{A}_{p_1p_2}\right)^{-1},
\qquad
e^{i p_2L}= \left(\mathcal{A}_{p_2p_1}\right)^{-1}.
\end{equation}
These are the Bethe-Yang equations for the particular case at hand.
The generalisation to $N$-particle states is straightforward. We define the wave-function
\begin{equation}
\psi(\sigma_1,\sigma_2,\ldots,\sigma_N) =e^{i\sum_{j=1}^N p_j\sigma_j}+\sum_{\perm} S_{\perm}(p_1,\ldots p_N)e^{i\sum_{j=1}^N p_{\perm(j)}\sigma_j}.
\end{equation}
where we sum over all possible permutations $\perm$.
Once we rewrite a given permutation $\perm$ as a sequence of two-body permutations, we define the function $S_{\perm}(p_1,\ldots p_N)$ as the product of the two-body scattering elements produced by the chosen factorisation.\footnote{For example, given the permutation $1234|3214$ we define 
\begin{equation}
S_{1234|3214} =\mathcal{A}_{p_1p_2} \mathcal{A}_{p_1p_3} \mathcal{A}_{p_2p_3}.
\end{equation}
}
Integrability ensures that different factorisations are equivalent.
Similarly as before, it is possible to check that the state
\begin{equation}
\ket{\Psi}=\sum_{\sigma_1\ll \sigma_2\ll \ldots \sigma_N} \psi(\sigma_1,\sigma_2,\ldots ,\sigma_N) \ket{Y^{\sL}_{\sigma_1}Y^{\sL}_{\sigma_2}\ldots Y^{\sL}_{\sigma_N}}
\end{equation}
is an eigenstate of the S-matrix $\mathcal{S}\ket{\Psi} =\ket{\Psi} $.
Periodicity of the wave-function written as $\psi(\sigma_2,\ldots,\sigma_N,\sigma_1+L)=\psi(\sigma_1,\sigma_2,\ldots,\sigma_N)$ now imposes
\begin{equation}
e^{i p_kL}= \prod_{\substack{j = 1\\j \neq k}}^{N} \left(\mathcal{A}_{p_kp_j}\right)^{-1}\,, \qquad k=1,\ldots,N\,,
\end{equation}
for each of the momenta $p_k$ of the excitations on the worldsheet.
The above result is compatible with the level matching condition. Indeed multiplying all the Bethe equations together we get
\begin{equation}
e^{i \sum_{k=1}^N p_kL}= \prod_{k=1}^N\prod_{\substack{j = 1\\j \neq k}}^{N} \left(\mathcal{A}_{p_kp_j}\right)^{-1}=1\,,
\end{equation}
where we have used that $\mathcal{A}_{p_kp_j}\mathcal{A}_{p_jp_k}=1$, as a consequence of unitarity.
We then recover the quantisation condition on the sum of momenta
\begin{equation}
\sum_{k=1}^N p_k = 2 \pi n\,, \qquad n \in \mathbb{Z},
\end{equation}
that characterises on-shell multi-particle states.

\paragraph{Bethe Ansatz with more flavors.}
It is clear that for the previous construction it was not essential to have only excitations of the same flavor, and we can extend it also to the case in which other types of excitations are present. The only characterising requirement is that the scattering of any of the excitations involved is \emph{diagonal}\footnote{When we write diagonal scattering we mean that other different flavors are not created after the scattering, and that the flavors of the two in-states are \emph{transmitted} to the out-states.}.
In {\adsthree} this situation is realised if for example we allow also for the presence of excitations $Z^{\sR}$.
We denote the relevant scattering elements by
\begin{equation}
\Smat \ket{Z^{\sR}_p Z^{\sR}_q} =\mathcal{C}_{pq} \ket{Z^{\sR}_q Z^{\sR}_p }\,,
\qquad
\Smat \ket{Z^{\sR}_p Y^{\sL}_q}=\widetilde{\mathcal{B}}_{pq}\ket{Y^{\sL}_q Z^{\sR}_p } \,.
\end{equation}
It is clear that unitarity implies that scattering $Y^{\sL}$ and $Z^{\sR}$ in the opposite order yields $\Smat \ket{Y^{\sL}_p Z^{\sR}_q}=(\widetilde{\mathcal{B}}_{qp})^{-1}\ket{Z^{\sR}_q Y^{\sL}_p}$.
The situation to consider---that is new with respect to the case of just one flavor---is when both $\ket{Z^{\sR}}$ and $\ket{Y^{\sL}}$ are present. In the example of a two-particle state we would define
\begin{equation}
\ket{\Psi}=\ket{Y^{\sL}_{p_1}Z^{\sR}_{p_2}}+(\widetilde{\mathcal{B}}_{p_2p_1})^{-1}\ket{Z^{\sR}_{p_2}Y^{\sL}_{p_1}},
\end{equation}
to get an eigenstate of the S-matrix.
The Bethe-Yang equations that we get now after imposing periodicity of the wave-function are
\begin{equation}
e^{i p_1L}= \widetilde{\mathcal{B}}_{p_2p_1},
\qquad
e^{i p_2L}= \left(\widetilde{\mathcal{B}}_{p_2p_1}\right)^{-1}.
\end{equation}
Multiplying these equations we recover again the level-matching condition.
If we had a total number of $N_{\sL}$ excitations of type $Y^{\sL}$ and $N_{\sR}$ excitations of type $Z^{\sR}$ we would generalise the previous construction and find the Bethe-Yang equations
\begin{equation}
\begin{aligned}
e^{i p_kL}&= \prod_{\substack{j = 1\\j \neq k}}^{N_{\sL}} \left(\mathcal{A}_{p_kp_j}\right)^{-1} \ \prod_{j=1}^{N_{\sR}} \widetilde{\mathcal{B}}_{p_jp_k}\,,
\qquad
&&k=1,\ldots,N_{\sL}\,,
\\
e^{i p_kL}&= \prod_{\substack{j = 1\\j \neq k}}^{N_{\sR}} \left(\mathcal{C}_{p_kp_j}\right)^{-1} \ \prod_{j=1}^{N_{\sL}} \left(\widetilde{\mathcal{B}}_{p_kp_j}\right)^{-1}\,,
\qquad
&&k=1,\ldots,N_{\sR}\,.
\end{aligned}
\end{equation}
The first are equations for $p_k$ being the momenta the excitations $Y^{\sL}$, while the second for the excitations $Z^{\sR}$.

In {\adsthree} we may add another type of excitations that scatter diagonally with both  $Y^{\sL}$ and $Z^{\sR}$. They belong to the massless module, and they can be chosen to be of type $\chi^1$. The excitations $Y^{\sL}$, $Z^{\sR}$ and $\chi^1$ that we have chosen here are the highest-weight states in each of the irreducible one-particle representations at our disposal, as it can be checked in Section~\ref{sec:exact-repr-T4}, see also Figure~\ref{fig:massive} and~\ref{fig:massless}.

\paragraph{Non-diagonal scattering: nesting procedure.}
To describe the most generic state we have to allow also for excitations that do not scatter diagonally. We then have to appeal to the nesting procedure to write the corresponding Bethe-Yang equations. The idea is to organise the excitations at our disposal into different levels. Level-I corresponds to the set of excitations that scatter diagonally among each other, as considered in the previous paragraphs. Level-II contains all the excitations that can be created from level-I by the action of lowering operators. Depending on the algebra and representations considered, one might need to go further than level-II, but this will not be our case.

In the following we explain how the nesting procedure works when we choose the lowering operator $\gen{Q}^{\sL 1}$ acting on level-I excitations $Y^{\sL}$. According to the exact representation in Eq.~\eqref{eq:exact-repr-left-massive} this creates a fermionic excitation $\eta^{\sL 1}$. The S-matrix acts on a two-particle state containing both of them as
\begin{equation}
\mathcal{S}\ket{Y^{\sL}_p\eta^{\sL1}_q}=S_0^{\sL\sL}(p,q)\left(A^{\sL\sL}_{pq}B^{\sL\sL}_{pq}\ket{\eta^{\sL1}_qY^{\sL}_p}+A^{\sL\sL}_{pq}C^{\sL\sL}_{pq}\ket{Y^{\sL}_q\eta^{\sL1}_p}\right)\,,
\end{equation}
where the $\su(1|1)^2_{\ce}$ scattering elements may be found in ~\eqref{eq:expl-su112-smat-el}, and the normalisation $S_0^{\sL\sL}(p,q)$ was fixed in~\eqref{eq:norm-LL-massive-sector}.
The S-matrix does not act as simple permutation on the above two-particle scattering elements. The idea is to find a state containing both $Y^{\sL}$ and $\eta^{\sL 1}$ where this is the case.
A way to do it is to consider the linear combination defined by
\begin{equation}
\ket{\mathcal{Y}_y} = f(y,p_1) \ket{\eta^{\sL 1}_{p_1} Y^{\sL}_{p_2}} + f(y,{p_2}) S^{\text{II,I}}(y,p_1) \ket{Y^{\sL}_{p_1} \eta^{\sL 1}_{p_2}}.
\end{equation}
In order to parameterise the state we have introduced a variable $y$. It goes under the name of \emph{auxiliary root} and it is associated to the level-II excitation.
Here $f(y,p_j)$ is interpreted as a normalisation parameter and $S^{\text{II,I}}(y,p_1)$ as the scattering element between the level-II and the level-I excitation.
The diagonal scattering is achieved by imposing the equation 
\begin{equation}\label{eq:compat-cond-level-II}
\mathcal{S} \ket{\mathcal{Y}_y} = \mathcal{A}_{p_1p_2} \ket{\mathcal{Y}_y}_\perm\,,
\end{equation}
where $\ket{\mathcal{Y}_y}_\perm$ is the permuted state that is found from $\ket{\mathcal{Y}_y}$ after exchanging the momenta $p_1$ and $p_2$.
This equation is motivated by the fact that we want to interpret level-II excitations as created on top of the ones of level-I.
Thanks to the compatibility condition~\eqref{eq:compat-cond-level-II}, it is enough to define 
\begin{equation}
\ket{\Psi}=\ket{\mathcal{Y}_y}+\mathcal{A}_{p_1p_2}\ket{\mathcal{Y}_y}_\perm,
\end{equation}
to get an eigenstate of the S-matrix.
Because of the above definitions the wave-function looks more involved
\begin{equation}
\begin{aligned}
\psi(\sigma_1,\sigma_2)&=\left[ f(y,p_1)+f(y,p_2) S^{\text{II,I}}(y,p_1)\right]e^{i(p_1\sigma_1+p_2\sigma_2)}\\
&+\mathcal{A}_{p_1p_2}  \left[ f(y,p_2)+f(y,p_1) S^{\text{II,I}}(y,p_2)\right]e^{i(p_2\sigma_1+p_1\sigma_2)}\,.
\end{aligned}
\end{equation}
Imposing the periodicity condition $\psi(\sigma_2,\sigma_1+L)=\psi(\sigma_1,\sigma_2)$ and matching the coefficients for $f(y,p_1)$ and $f(y,p_2)$ we find the Bethe equations
\begin{equation}
\begin{aligned}
e^{i p_1L}&= \left(\mathcal{A}_{p_1p_2}\right)^{-1} \ S^{\text{II,I}}(y,p_1),
\qquad
1=S^{\text{II,I}}(y,p_1)\ S^{\text{II,I}}(y,p_2)\,,
\\
e^{i p_2L}&= \left(\mathcal{A}_{p_2p_1}\right)^{-1} \ S^{\text{II,I}}(y,p_2)\,.
\end{aligned}
\end{equation}
The introduction of level-II excitations then has the consequence of producing factors of $S^{\text{II,I}}(y,p_j)$ in the Bethe-Yang equations, confirming the interpretation of these terms as scattering elements between excitations of different levels.
We also find a new equation for the auxiliary root $y$, conceptually similar to the equations for $p_1$ and $p_2$. The variable $y$ carries no momentum, and the left-hand-side of its equation is just $1$.

In the particular case we considered, when we impose~\eqref{eq:compat-cond-level-II} we find
\begin{equation}
f(y,p_j)=  \frac{g(y) \eta_{p_j}}{y - x^+_{p_j}}\,,
\qquad
S^{\text{II,I}}(y,p_j)= \left(\frac{x^+_{p_j}}{x^-_{p_j}}\right)^{1/2} \frac{y - x^-_{p_j}}{y - x^+_{p_j}}  \,.
\end{equation}

To derive the Bethe-Yang equations for a state with a number of $N^{\text{I}}_{\sL}$ excitations $Y^{\sL}$ and $N^{\text{II}}_{\sL}$ excitations $\eta^{\sL 1}$ we repeat the above procedure, generalising it to multiparticle states.
One should also take into account the possibility of a non-trivial scattering of level-II excitations among each other.
To check whether such a scattering element $S^{\text{II,II}}(y_1,y_2)$ exists, for two such excitations parameterised by the auxiliary roots $y_1$ and $y_2$, we consider the state
\begin{equation}
  \begin{aligned}
    \ket{\mathcal{Y}_{y_1} \mathcal{Y}_{y_2}} &= f(y_1,p_1) f(y_2,p_2) S^{\text{II},\text{I}}(y_2,p_1) \ket{\eta^{\sL 1}_{p_1} \eta^{\sL 1}_{p_2}} \\
    &\qquad + f(y_2,p_1) f(y_1,p_2) S^{\text{II},\text{I}}(y_1,p_1) S^{\text{II},\text{II}}(y_1,y_2) \ket{\eta^{\sL 1}_{p_1} \eta^{\sL 1}_{p_2}} \,,
  \end{aligned}
\end{equation}
where the functions $f(y_j,p_k)$ and $S^{\text{II},\text{I}}(y_j,p_k)$ are the ones calculated previously.
Imposing the compatibility condition $\mathcal{S} \ket{\mathcal{Y}_{y_1} \mathcal{Y}_{y_2}} = \mathcal{A}_{p_1p_2} \ket{\mathcal{Y}_{y_1} \mathcal{Y}_{y_2}}_\perm$---where the permuted state is found by exchanging the momenta $p_1$ and $p_2$---we find 
\begin{equation}
S^{\text{II},\text{II}}(y_1,y_2) = -1.
\end{equation}
The scattering element is \emph{trivial} and there is no contribution to the Bethe-Yang equations. The minus sign is present here because we are permuting two fermionic states.
The periodicity condition is then written as
\begin{equation}
\begin{aligned}
e^{i p_kL}&= \prod_{\substack{j = 1\\j \neq k}}^{N_{\sL}} \left(\mathcal{A}_{p_kp_j}\right)^{-1} \ \prod_{j=1}^{N_{\sL}^{\text{II}}} S^{\text{II},\text{I}}(y_j,p_k)\,,
\qquad
&&k=1,\ldots,N_{\sL}\,,
\\
1&=  \prod_{j=1}^{N_{\sL}} S^{\text{II},\text{I}}(y_k,p_j)\,,
\qquad
&&k=1,\ldots,N_{\sL}^{\text{II}}\,,
\end{aligned}
\end{equation}
where we have defined the total number of excitations $N_{\sL}=N_{\sL}^{\text{I}}+N_{\sL}^{\text{II}}$.

Similar computations may be done to consider scattering of different level-II excitations with other level-I states.
It is clear that we need to associate one auxiliary root to each lowering operator of the algebra $\mathcal{A}$, creating the corresponding level-II excitation.

\subsection{Bethe-Yang equations for \adsthree}
\label{sec:BAE}
Using the procedure explained in the previous section, we can derive the whole set of \mbox{Bethe-Yang} equations for \adsthree, when we allow for a generic number of excitations of each type.
We write them explicitly\footnote{Here we write the Bethe-Yang equations following the convention of~\cite{Borsato:2014hja} for the normalisation of the mixed-mass sector. This differs from the normalisation of~\cite{Borsato:2016kbm}.} in Equations~\eqref{eq:BA-1}-\eqref{eq:BA-su(2)}.
In the following we use the shorthand notation $\nu_k \equiv e^{i p_k}.$
When restricting to the massive sector, the factors of $\nu$ have also the meaning of \emph{frame-factors}, and they allow us to relate the string frame to the spin-chain frame, see Section~\ref{sec:S-matr-spin-chain-T4}.

\begin{align}
    \label{eq:BA-1}
    1 &= 
    \prod_{j=1}^{K_2} \frac{y_{1,k} - x_j^+}{y_{1,k} - x_j^-} \nu_j^{-\frac{1}{2}}
    \prod_{j=1}^{K_{\bar{2}}} \frac{1 - \frac{1}{y_{1,k} \bar{x}_j^-}}{1- \frac{1}{y_{1,k} \bar{x}_j^+}} \nu_j^{-\frac{1}{2}}
    \prod_{j=1}^{K_0} \frac{y_{1,k} - z_j^+ }{y_{1,k} - z_j^- } \nu_j^{-\frac{1}{2}} , \\
\nonumber \\
    \begin{split}
    \label{eq:BA-2}
      \left(\frac{x_k^+}{x_k^-}\right)^L &=
      \prod_{\substack{j = 1\\j \neq k}}^{K_2} \nu_k^{-1}\nu_j \frac{x_k^+ - x_j^-}{x_k^- - x_j^+} \frac{1- \frac{1}{x_k^+ x_j^-}}{1- \frac{1}{x_k^- x_j^+}} (\sigma^{\bullet\bullet}_{kj})^2
      \prod_{j=1}^{K_1}  \nu_k^{\frac{1}{2}} \, \frac{x_k^- - y_{1,j}}{x_k^+ - y_{1,j}}
      \prod_{j=1}^{K_3} \nu_k^{\frac{1}{2}} \, \frac{x_k^- - y_{3,j}}{x_k^+ - y_{3,j}}
      \\ &\times
      \prod_{j=1}^{K_{\bar{2}}} \nu_j  \frac{1- \frac{1}{x_k^+ \bar{x}_j^+}}{1- \frac{1}{x_k^- \bar{x}_j^-}} \frac{1- \frac{1}{x_k^+ \bar{x}_j^-}}{1- \frac{1}{x_k^- \bar{x}_j^+}} (\widetilde{\sigma}^{\bullet\bullet}_{kj})^2
      \prod_{j=1}^{K_{\bar{1}}} \nu_k^{-\frac{1}{2}} \, \frac{1 - \frac{1}{x_k^- y_{\bar{1},j}}}{1- \frac{1}{x_k^+ y_{\bar{1},j}}}
      \prod_{j=1}^{K_{\bar{3}}} \nu_k^{-\frac{1}{2}} \, \frac{1 - \frac{1}{x_k^- y_{\bar{3},j}}}{1- \frac{1}{x_k^+ y_{\bar{3},j}}} 
     \\ &\times
      \prod_{j=1}^{K_{0}} \nu_k^{\frac{1}{2}} \left(\frac{1- \frac{1}{x_k^+ z_j^+}}{1- \frac{1}{x_k^- z_j^-}}\right)^{\frac{1}{2}} \left(\frac{1- \frac{1}{x_k^+ z_j^-}}{1- \frac{1}{x_k^- z_j^+}}\right)^{\frac{1}{2}} (\sigma^{\bullet\circ}_{kj})^2,
    \end{split} \\
\nonumber \\
    \label{eq:BA-3}
    1 &= 
    \prod_{j=1}^{K_2} \frac{y_{3,k} - x_j^+}{y_{3,k} - x_j^-} \nu_j^{-\frac{1}{2}}
    \prod_{j=1}^{K_{\bar{2}}} \frac{1 - \frac{1}{y_{3,k} \bar{x}_j^-}}{1- \frac{1}{y_{3,k} \bar{x}_j^+}} \nu_j^{-\frac{1}{2}}
    \prod_{j=1}^{K_0} \frac{y_{3,k} - z_j^+ }{y_{3,k} - z_j^- } \nu_j^{-\frac{1}{2}} , \\
\nonumber \\ 
\nonumber \\ 
\nonumber \\
    \label{eq:BA-1b}
    1 &= 
    \prod_{j=1}^{K_{\bar{2}}} \frac{y_{\bar{1},k} - \bar{x}_j^-}{y_{\bar{1},k} - \bar{x}_j^+} \nu_j^{\frac{1}{2}}
    \prod_{j=1}^{K_2} \frac{1 - \frac{1}{y_{\bar{1},k} x_j^+}}{1- \frac{1}{y_{\bar{1},k} x_j^-}} \nu_j^{\frac{1}{2}}
    \prod_{j=1}^{K_{0}}   \frac{y_{\bar{1},k} - z_j^-}{y_{\bar{1},k} - z_j^+} \nu_j^{\frac{1}{2}}, \\
\nonumber \\
    \begin{split}
    \label{eq:BA-2b}
      \left(\frac{\bar{x}_k^+}{\bar{x}_k^-}\right)^L &=
      \prod_{\substack{j = 1\\j \neq k}}^{K_{\bar{2}}} \frac{\bar{x}_k^- - \bar{x}_j^+}{\bar{x}_k^+ - \bar{x}_j^-} \frac{1- \frac{1}{\bar{x}_k^+ \bar{x}_j^-}}{1- \frac{1}{\bar{x}_k^- \bar{x}_j^+}}  (\sigma^{\bullet\bullet}_{kj})^2
      \prod_{j=1}^{K_{\bar{1}}} \nu_k^{-\frac{1}{2}} \frac{\bar{x}_k^+ - y_{\bar{1},j}}{\bar{x}_k^- - y_{\bar{1},j}}
      \prod_{j=1}^{K_{\bar{3}}} \nu_k^{-\frac{1}{2}} \frac{\bar{x}_k^+ - y_{\bar{3},j}}{\bar{x}_k^- - y_{\bar{3},j}}
      \\ &\times
      \prod_{j=1}^{K_2} \nu_k^{-1} \frac{1- \frac{1}{\bar{x}_k^- x_j^-}}{1- \frac{1}{\bar{x}_k^+ x_j^+}} \frac{1- \frac{1}{\bar{x}_k^+ x_j^-}}{1- \frac{1}{\bar{x}_k^- x_j^+}}(\widetilde{\sigma}^{\bullet\bullet}_{kj})^2
      \prod_{j=1}^{K_{1}} \nu_k^{\frac{1}{2}} \frac{1 - \frac{1}{\bar{x}_k^+ y_{1,j}}}{1- \frac{1}{\bar{x}_k^- y_{1,j}}}
      \prod_{j=1}^{K_{3}} \nu_k^{\frac{1}{2}}\frac{1 - \frac{1}{\bar{x}_k^+ y_{3,j}}}{1- \frac{1}{\bar{x}_k^- y_{3,j}}} 
      \\ &\times
      \prod_{j=1}^{K_{0}} \nu_k^{-\frac{1}{2}}\nu_j^{-1}  \left(\frac{1- \frac{1}{\bar{x}_k^- z_j^-}}{1- \frac{1}{\bar{x}_k^+ z_j^+}}\right)^{\frac{3}{2}} \left(\frac{1- \frac{1}{\bar{x}_k^+ z_j^-}}{1- \frac{1}{\bar{x}_k^- z_j^+}}\right)^{\frac{1}{2}} (\sigma^{\bullet\circ}_{kj})^2,
    \end{split} \\
\nonumber \\
    \label{eq:BA-3b}
    1 &= 
    \prod_{j=1}^{K_{\bar{2}}} \frac{y_{\bar{3},k} - \bar{x}_j^-}{y_{\bar{3},k} - \bar{x}_j^+} \nu_j^{\frac{1}{2}}
    \prod_{j=1}^{K_2} \frac{1 - \frac{1}{y_{\bar{3},k} x_j^+}}{1- \frac{1}{y_{\bar{3},k} x_j^-}}\nu_j^{\frac{1}{2}}
    \prod_{j=1}^{K_{0}}   \frac{y_{\bar{3},k} - z_j^-}{y_{\bar{3},k} - z_j^+} \nu_j^{\frac{1}{2}}, 
\end{align}
\begin{align}
\begin{split}
    \label{eq:BA-0}
      \left(\frac{z_k^+}{z_k^-}\right)^L &=
      \prod_{\substack{j = 1\\j \neq k}}^{K_0}\nu_k^{-\frac{1}{2}}\nu_j^{\frac{1}{2}}  \frac{z_k^+ - z_j^-}{z_k^- - z_j^+}(\sigma^{\circ\circ}_{kj})^2
	\prod_{j=1}^{K_{4}}  \frac{w_k - y_{4,j} - i/2}{w_k -y_{4,j} + i/2} 
      \\ &
     \hspace{-8pt} \prod_{j=1}^{K_{2}} \nu_j^{-\frac{1}{2}} \left( \frac{1-\frac{1}{z_k^- x_j^-}}{1-\frac{1}{z_k^+ x_j^+}} \right)^{\frac{1}{2}} \left(\frac{1-\frac{1}{z_k^+ x_j^-}}{1-\frac{1}{z_k^- x_j^+}} \right)^{\frac{1}{2}}
 (\sigma^{\circ\bullet}_{kj})^2 \
	\prod_{j=1}^{K_1}\nu_k^{\frac{1}{2}} \frac{z_k^- - y_{1,j}}{z_k^+ - y_{1,j}}
      \prod_{j=1}^{K_3} \nu_k^{\frac{1}{2}}  \frac{z_k^- - y_{3,j}}{z_k^+ - y_{3,j}}
     \\ &
     \hspace{-8pt}      \prod_{j=1}^{K_{\bar{2}}} \nu_k\nu_j^{\frac{1}{2}}  \left(\frac{1-\frac{1}{z_k^+ \bar{x}_j^+}}{1-\frac{1}{z_k^- \bar{x}_j^-}}\right)^{\frac{3}{2}} \left( \frac{1-\frac{1}{z_k^+ \bar{x}_j^-}}{1-\frac{1}{z_k^- \bar{x}_j^+}} \right)^{\frac{1}{2}} (\sigma^{\circ\bullet}_{kj})^2 	
      \prod_{j=1}^{K_{\bar{1}}}  \nu_k^{-\frac{1}{2}} \frac{z_k^+ - y_{\bar{1},j}}{z_k^- - y_{\bar{1},j}}
      \prod_{j=1}^{K_{\bar{3}}} \nu_k^{-\frac{1}{2}} \frac{z_k^+ - y_{\bar{3},j}}{z_k^- - y_{\bar{3},j}},
    \end{split} \\
\nonumber \\
\begin{split}
    \label{eq:BA-su(2)}
   1  & 
   = \prod_{\substack{j = 1\\j \neq k}}^{K_{4}} \frac{y_{4,k} - y_{4,j} + i}{y_{4,k} - y_{4,j} - i}
    \prod_{j=1}^{K_{0}} \frac{y_{4,k} - w_j - i/2}{y_{4,k} - w_j + i/2}  
    \end{split}
\end{align}

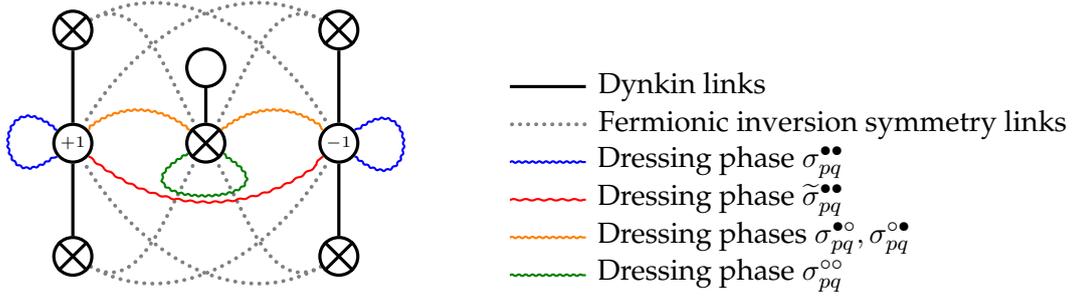
\begin{figure}
  \centering
  
\begin{tikzpicture}
  \begin{scope}
    \coordinate (m) at (0cm,0cm);

    \node (v1L) at (-1.75cm, 1.5cm) [dynkin node] {};
    \node (v2L) at (-1.75cm, 0cm) [dynkin node] {$\scriptscriptstyle +1$};
    \node (v3L) at (-1.75cm,-1.5cm) [dynkin node] {};

    \draw [dynkin line] (v1L) -- (v2L);
    \draw [dynkin line] (v2L) -- (v3L);
    
    \node (v1R) at (+1.75cm, 1.5cm) [dynkin node] {};
    \node (v2R) at (+1.75cm, 0cm) [dynkin node] {$\scriptscriptstyle -1$};
    \node (v3R) at (+1.75cm,-1.5cm) [dynkin node] {};

    \draw [dynkin line] (v1L.south west) -- (v1L.north east);
    \draw [dynkin line] (v1L.north west) -- (v1L.south east);

    \draw [dynkin line] (v3L.south west) -- (v3L.north east);
    \draw [dynkin line] (v3L.north west) -- (v3L.south east);
    
    \draw [dynkin line] (v1R.south west) -- (v1R.north east);
    \draw [dynkin line] (v1R.north west) -- (v1R.south east);

    \draw [dynkin line] (v3R.south west) -- (v3R.north east);
    \draw [dynkin line] (v3R.north west) -- (v3R.south east);

    \draw [dynkin line] (v1R) -- (v2R);
    \draw [dynkin line] (v2R) -- (v3R);

    \node (v0) at (0cm, 0cm) [dynkin node] {};
    \draw [dynkin line] (v0.south west) -- (v0.north east);
    \draw [dynkin line] (v0.north west) -- (v0.south east);

    \node (vsu2) at (0cm, +1cm) [dynkin node] {};
    \draw [dynkin line] (vsu2.south) -- (v0.north);

    \begin{pgfonlayer}{background}
      \draw [inverse line] [out=  0+30,in= 120] (v1L) to (v2R);
      \draw [inverse line] [out=  0-30,in=240] (v3L) to (v2R);
      \draw [inverse line] [out=180-30,in= 60] (v1R) to (v2L);
      \draw [inverse line] [out=180+30,in=300] (v3R) to (v2L);

      \draw [inverse line] [out=  0+30,in= 120] (v1L) to (v0);
      \draw [inverse line] [out=  0-30,in=240] (v3L) to (v0);
      \draw [inverse line] [out=180-30,in= 60] (v1R) to (v0);
      \draw [inverse line] [out=180+30,in=300] (v3R) to (v0);
    \end{pgfonlayer}

    \draw [red phase] [out=-40,in=180+40] (v2L) to (v2R);

    \draw [blue phase] [out=180-40,in= 180+40,loop] (v2L) to (v2L);
    \draw [blue phase] [out=-40,in=40,loop] (v2R) to (v2R);

    \draw [green phase] [out=180+30,in= -30,loop] (v0) to (v0);
    \draw [brown phase] [out=180-40,in= +40] (v0) to (v2L);
    \draw [brown phase] [out=+40,in=180-40] (v0) to (v2R);
  \end{scope}

  \begin{scope}[xshift=+4cm,yshift=-0.75cm]
    \draw [dynkin line]  (0cm,1.5cm) -- (1cm,1.5cm) node [anchor=west,black] {\small Dynkin links};
    \draw [inverse line] (0cm+0.75pt,1.0cm) -- (1cm,1.0cm) node [anchor=west,black] {\small Fermionic inversion symmetry links};
    \draw [blue phase]   (0cm,0.5cm) -- (1cm,0.5cm) node [anchor=west,black] {\small Dressing phase $\sigma^{\bullet\bullet}_{pq}$};
    \draw [red phase]    (0cm,0.0cm) -- (1cm,0.0cm) node [anchor=west,black] {\small Dressing phase $\widetilde{\sigma}^{\bullet\bullet}_{pq}$};
    \draw [brown phase]   (0cm,-0.5cm) -- (1cm,-0.5cm) node [anchor=west,black] {\small Dressing phases $\sigma^{\bullet\circ}_{pq},\sigma^{\circ\bullet}_{pq}$};
   \draw [green phase]    (0cm,-1cm) -- (1cm,-1cm) node [anchor=west,black] {\small Dressing phase $\sigma^{\circ\circ}_{pq}$};
  \end{scope}
\end{tikzpicture}

  \caption{The Dynkin diagram for $\alg{psu}(1,1|2)^2$ plus the root for massless modes, with the various interaction terms appearing in the Bethe ansatz indicated.}
  \label{fig:bethe-equations}
\end{figure}

A nice way to visualise the equations is done by considering Figure~\ref{fig:bethe-equations}. We recognise two copies of the Dynkin diagram of $\alg{psu}(1,1|2)$, corresponding to L and R. The crossed nodes denote fermionic roots, the other bosonic ones. The diagrams are in the $\su(2)$ grading for the Left copy and in the $\sls(2)$ grading for the Right copy. We refer to Section~\ref{sec:psu112-algebra} for a discussion on the possible different gradings of $\alg{psu}(1,1|2)$.
Between these two Dynkin diagrams we find two additional nodes, one fermionic and one bosonic. The latter corresponds to the $\su(2)_{\circ}$ lowering operator\footnote{In~\cite{Borsato:2016kbm} the S-matrix scattering the $\su(2)_\circ$ flavors of massless excitations was conjectured to be trivial at all orders in $h$ to match with the perturbative results of~\cite{Sundin:2016gqe}. This was obtained by sending the rapidity $w_p \to \infty$. As a result this node and the auxiliary root $y_4$ are missing in the Bethe-Yang equations of~\cite{Borsato:2016kbm}.}, and we associate to it the auxiliary root $y_4$, whose Bethe-Yang equation is~\eqref{eq:BA-su(2)}.
The external fermionic nodes in the diagram correspond to the four lowering fermionic operators of $\mathcal{A}$. We associate to them the auxiliary roots $y_1,y_3$ for the Left part of the diagram, and $y_{\bar{1}},y_{\bar{3}}$ for the Right part of the diagram. The corresponding Bethe-Yang equations may be found in~\eqref{eq:BA-1},\eqref{eq:BA-3} and~\eqref{eq:BA-1b},\eqref{eq:BA-3b}.
The three nodes aligned horizontally in Figure~\ref{fig:bethe-equations} are the momentum carrying nodes. The one on the left corresponds to Left massive excitations, for which we use parameters $x^\pm$. To distinguish the Right massive excitations that correspond to the node on the right, we use the notation $\bar{x}^\pm$ for the spectral parameters. The two equations for L and R are found in~\eqref{eq:BA-2} and~\eqref{eq:BA-2b}.
The node in the middle of the diagram is associated to massless excitations. For them we use the notation $z^\pm$, and their equation is~\eqref{eq:BA-0}.
The table below recaps our notation for the spectral parameters and the number of excitations in each case.
We also write our choice of level I excitations.
$$
\begin{aligned}
\hline
& & 
&\text{Left massive} & \qquad
&\text{Right massive} & \qquad
&\text{Massless}& \\
\hline
& \text{spectral parameters}& 
&\ \ \ \ \ \ x^\pm & 
&\ \ \ \ \ \ \bar{x}^\pm &
&\ \ \ \ z^\pm& \\
& \text{number of excitations}&
&\ \ \ \ \ \ K_2 & 
&\ \ \ \ \ \ K_{\bar{2}} &
&\ \ \ \ K_0& \\
\hline
& \text{level I excitations}&
&\ \ \ \ \ \ Y^{\smallL} & 
&\ \ \ \ \ \ Z^{\smallR} & 
&\ \ \ \ \chi^1&\\
\hline
\end{aligned}
$$
These level I excitations have been chosen because they scatter diagonally among each other.
They are highest weight states for the raising operators $\bar{\gen{Q}}_{\smallL1},\bar{\gen{Q}}_{\smallL2},\gen{Q}_{\smallR1},\gen{Q}_{\smallR2}$ and $\gen{J}_{\circ 2}^{ \ \ 1} $.

The table below shows the notation for the auxiliary root and number of excitations associated to each lowering operator
$$
\begin{aligned}
& &
&\gen{Q}^{\smallL1} & \qquad
&\gen{Q}^{\smallL2} & \qquad
&\bar{\gen{Q}}^{\smallR1} & \qquad
&\bar{\gen{Q}}^{\smallR2} & \qquad
& \gen{J}_{\circ 1}^{ \ \ 2} & 
 \\
\hline
& \text{auxiliary root}& 
& y_1& 
&y_3& 
& y_{\bar{1}}& 
& y_{\bar{3}}&
& y_{4}& \\
& \text{number of excitations}&
& K_1& 
&K_3& 
& K_{\bar{1}}& 
& K_{\bar{3}}& 
&K_4& \\
\hline
\end{aligned}
$$
Solutions of the Bethe-Yang equations allow to compute the spectrum only up to wrapping corrections. In the model we are studying virtual particles wrapping the cylinder can be either massive or massless, and in~\cite{Abbott:2015pps} it was shown that the latter contributions cannot be discarded if one wants to reproduce the energy of semiclassical spinning strings as computed in~\cite{Beccaria:2012kb}.

\vspace{12pt}
\smallskip

\section{Summary}
In this chapter we have constructed the action of the charges on multi-particle states, using the results of the previous chapter.
In particular, we have used a non-local co-product to write supercharges acting on two-particle representations.
This was needed in order to reproduce the exact eigenvalues of the charges appearing in the central extension.

Compatibility with the bosonic and fermionic generators allowed us to fix the all-loop S-matrix almost completely.
We found a total of four ``dressing factors'' that are not fixed by symmetries, and that are further constrained by unitarity, LR-symmetry and crossing invariance.
We have also checked that the S-matrix that we have derived satisfies the Yang-Baxter equation, confirming compatibility with the assumption of factorisation of scattering.

We have imposed periodicity of the wave-function, motivated by the fact that we are describing closed strings.
Using the ``nesting procedure'' we have derived the complete set of Bethe-Yang equations, which should encode the spectrum of strings on the background {\adsthree} up to wrapping corrections.

\chapter{The massive sector of \adsthree}\label{ch:massive-sector-T4}
In this chapter we concentrate on the massive sector\footnote{The massive sector of {\adsthree} has been discussed in detail also in the thesis of A. Sfondrini~\cite{Sfondrini:2014via}, to which we refer for an alternative presentation.} of {\adsthree}.
The massive sector corresponds to strings moving only in the AdS$_3\times$S$^3$ subspace of the background. 
From the point of view of worldsheet scattering, the results of the previous chapter show that we indeed identify a sector if we consider only massive excitations for the incoming states. In fact, Integrability ensures that if we scatter two massive excitations, then massless particles never appear in the asymptotic out-states.

Focusing on smaller sectors of the theory is a good method to better understand the results obtained.
Moreover, the study of the massive sector allows us to compare to the case of {\adsfive}, where only massive excitations are present.

We start by explaining how we can encode the integrable model found from the point of view of the string into a spin-chain description.
As reviewed in Chapter~\ref{ch:intro}, in AdS$_5$/CFT$_4$ integrable spin-chains emerge when considering the spectrum of the dilatation operator in the gauge theory~\cite{Minahan:2002ve}. The idea here is to construct a spin-chain from which we can derive essentially the same all-loop S-matrix and Bethe-Yang equations valid for the massive sector of the string.

We will also consider the crossing equations of Section~\ref{sec:crossing-invar-T4} for the dressing factors governing massive scattering, and derive solutions for them. 
The solution of these equations is not unique, and we will motivate our choice by commenting on the analytical structure of these functions.
We will also take a proper limit of the Bethe-Yang equations, to obtain the ``finite-gap equations''. We conclude with a discussion and with a collection of the references to the perturbative calculations that succesfully tested our findings.

\section{Spin-chain description}
The spin-chain that we construct in this section shares the $\psu(1,1|2)^2$ symmetry of the massive sector of strings in {\adsthree}~\cite{Babichenko:2009dk}.
We start by presenting this superalgebra.
\subsection{$\psu(1,1|2)^2$ algebra}\label{sec:psu112-algebra}
The bosonic subalgrebra of $\psu(1,1|2)$ is $\su(1,1)\oplus\su(2)$. In what follows it is more convenient to change the real form of the non-compact subalgebra and consider instead $\sls(2)\oplus\su(2)$. We denote the corresponding generators by $\gen{S}_0,\gen{S}_\pm$ and $\gen{L}_5,\gen{L}_\pm$ respectively.
In addition to those we have also eight supercharges, that we denote with the help of three indices $\gen{Q}_{a\alpha\dot{\alpha}}$, each of them taking values $\pm$.
The commutation relations of $\psu(1,1|2)$ then read as 
\begin{equation}
  \begin{aligned}
    \comm{\gen{S}_0}{\gen{S}_\pm} &= \pm \gen{S}_\pm ,\qquad &
    \comm{\gen{S}_+}{\gen{S}_-} &= 2 \gen{S}_0 , \\
    \comm{\gen{L}_5}{\gen{L}_\pm} &= \pm \gen{L}_\pm , \qquad &
    \comm{\gen{L}_+}{\gen{L}_-} &= 2 \gen{L}_5 , 
  \end{aligned}
\end{equation}
\begin{equation}
  \begin{aligned}
    \comm{\gen{S}_0}{\gen{Q}_{\pm\beta\dot{\beta}}} &= \pm\frac{1}{2} \gen{Q}_{\pm\beta\dot{\beta}} , \qquad &
    \comm{\gen{S}_\pm}{\gen{Q}_{\mp\beta\dot{\beta}}} &= \gen{Q}_{\pm\beta\dot{\beta}} , \\
    \comm{\gen{L}_5}{\gen{Q}_{b\pm\dot{\beta}}} &= \pm\frac{1}{2} \gen{Q}_{b\pm\dot{\beta}} ,\qquad &
    \comm{\gen{L}_\pm}{\gen{Q}_{b\mp\dot{\beta}}} &= \gen{Q}_{b\pm\dot{\beta}} , \\
  \end{aligned}
\end{equation}
\begin{equation}
  \begin{aligned}
    \acomm{\gen{Q}_{\pm++}}{\gen{Q}_{\pm--}} &= \pm \gen{S}_{\pm} , \qquad &
    \acomm{\gen{Q}_{\pm+-}}{\gen{Q}_{\pm-+}} &= \mp \gen{S}_{\pm}  , \\
    \acomm{\gen{Q}_{+\pm+}}{\gen{Q}_{-\pm-}} &= \mp \gen{L}_{\pm} , \qquad &
    \acomm{\gen{Q}_{+\pm-}}{\gen{Q}_{-\pm+}} &= \pm \gen{L}_{\pm} ,
\\
    \acomm{\gen{Q}_{+\pm\pm}}{\gen{Q}_{-\mp\mp}} &= - \gen{S}_0 \pm \gen{L}_5 , \qquad &
    \acomm{\gen{Q}_{+\pm\mp}}{\gen{Q}_{-\mp\pm}} &= + \gen{S}_0 \mp \gen{L}_5 .
  \end{aligned}
\end{equation}
As it is clear from the equations above, the first index of a supercharge spans an $\sls(2)$ doublet, while the second index an $\su(2)$ one.
The last index is not associated to an $\su(2)$ doublet\footnote{The third index on a supercharge spans an $\su(2)$ doublet in the case of the $\alg{d}(2,1,\alpha)$ superalgebra, of which $\psu(1,1|2)$ can be seen as a particular contraction---a proper $\alpha \to 1$ or $\alpha \to 0$ limit. For generic $\alpha$ the generator $\gen{R}_8$ is the Cartan element of the additional $\su(2)$.}.
The superalgebra admits a $\alg{u}(1)$ automorphism generated by the charge $\gen{R}_8$, that acts on the supercharges as 
\begin{equation}
  \comm{\gen{R}_8}{\gen{Q}_{b\beta\pm}} = \pm \frac{1}{2} \gen{Q}_{b\beta\pm} ,
\end{equation}
and that commutes with all bosonic generators.

Let us present the possible choices of Serre-Chevalley basis. For superalgebras the inequivalent possibilities are associated to different Dynkin diagrams. Each of them corresponds to the choice of Cartan generators $\gen{h}_i$ and the corresponding raising and lowering operators $\gen{e}_i,\gen{f}_i$. The index $i$ runs from $1$ to the rank of the superalgebra, that is $3$ in the case of $\psu(1,1|2)$.
In this basis the commutation relations acquire the form\footnote{If both $\gen{e}_i$ and $\gen{f}_j$ are fermionic, then the commmutator $[,]$ should be replaced by the anti-commutator $\{,\}$.}
\begin{equation}
  \comm{\gen{h}_i}{\gen{h}_j} = 0 , \qquad
  \comm{\gen{e}_i}{\gen{f}_j} = \delta_{ij} \gen{h}_j , \qquad
  \comm{\gen{h}_i}{\gen{e}_j} = + A_{ij} \gen{e}_j , \qquad
  \comm{\gen{h}_i}{\gen{f}_j} = - A_{ij} \gen{f}_j ,
\end{equation}
where $A_{ij}$ is the Cartan matrix.

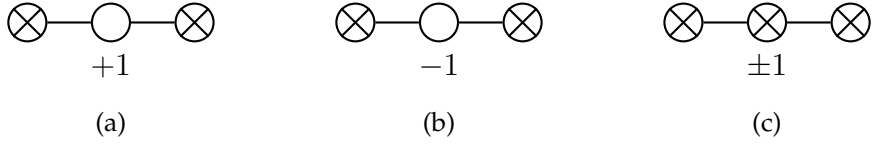
\begin{figure}
  \centering

  \subfloat[\label{fig:dynkin-su22-su}]{
    \begin{tikzpicture}
      [
      thick,
      node/.style={shape=circle,draw,thick,inner sep=0pt,minimum size=5mm}
      ]

      \useasboundingbox (-1.5cm,-1cm) rectangle (1.5cm,1cm);

      \node (v1) at (-1.1cm, 0cm) [node] {};
      \node (v2) at (  0.0cm, 0cm) [node] {};
      \node (v3) at (  1.1cm, 0cm) [node] {};

      \draw (v1.south west) -- (v1.north east);
      \draw (v1.north west) -- (v1.south east);

      \draw (v3.south west) -- (v3.north east);
      \draw (v3.north west) -- (v3.south east);

      \draw (v1) -- (v2);
      \draw (v2) -- (v3);

      \node at (v2.south) [anchor=north] {$+1$};
    \end{tikzpicture}
  }
  \hspace{1cm}
  \subfloat[\label{fig:dynkin-su22-sl}]{
    \begin{tikzpicture}
      [
      thick,
      node/.style={shape=circle,draw,thick,inner sep=0pt,minimum size=5mm}
      ]

      \useasboundingbox (-1.5cm,-1cm) rectangle (1.5cm,1cm);

      \node (v1) at (-1.1cm, 0cm) [node] {};
      \node (v2) at (  0.0cm, 0cm) [node] {};
      \node (v3) at (  1.1cm, 0cm) [node] {};

      \draw (v1.south west) -- (v1.north east);
      \draw (v1.north west) -- (v1.south east);

      \draw (v3.south west) -- (v3.north east);
      \draw (v3.north west) -- (v3.south east);

      \draw (v1) -- (v2);
      \draw (v2) -- (v3);

      \node at (v2.south) [anchor=north] {$-1$};
    \end{tikzpicture}
  }
  \hspace{1cm}
  \subfloat[\label{fig:dynkin-su22-fff}]{
    \begin{tikzpicture}
      [
      thick,
      node/.style={shape=circle,draw,thick,inner sep=0pt,minimum size=5mm}
      ]

      \useasboundingbox (-1.5cm,-1cm) rectangle (1.5cm,1cm);

      \node (v1) at (-1.1cm, 0cm) [node] {};
      \node (v2) at (  0.0cm, 0cm) [node] {};
      \node (v3) at (  1.1cm, 0cm) [node] {};

      \draw (v1.south west) -- (v1.north east);
      \draw (v1.north west) -- (v1.south east);

      \draw (v2.south west) -- (v2.north east);
      \draw (v2.north west) -- (v2.south east);

      \draw (v3.south west) -- (v3.north east);
      \draw (v3.north west) -- (v3.south east);

      \draw (v1) -- (v2);
      \draw (v2) -- (v3);

      \node at (v2.south) [anchor=north] {$\pm 1$};
    \end{tikzpicture}
  }
  
  \caption{Three Dynkin diagrams for $\psu(1,1|2)$. A cross denotes a fermionic root.}
  \label{fig:dynkin-su22}
\end{figure}

The superalgebra $\psu(1,1|2)$ admits three inequivalent gradings, see Figure~\ref{fig:dynkin-su22} for the corresponding Dynkin diagrams.
The Dynkin diagram in Figure~\ref{fig:dynkin-su22}~\subref{fig:dynkin-su22-su} corresponds to the $\su(2)$ grading. The choice for the Cartan generators and the simple roots is
\begin{equation}\label{eq:SC-basis-su2}
  \begin{aligned}
    \gen{h}_1 &= -\gen{S}_0 - \gen{L}_5 , \qquad &
    \gen{e}_1 &= +\gen{Q}_{+--} , \qquad &
    \gen{f}_1 &= +\gen{Q}_{-++} , \\
    \gen{h}_2 &= +2\gen{L}_5 , \qquad &
    \gen{e}_2 &= +\gen{L}_+ , \qquad &
    \gen{f}_2 &= +\gen{L}_- , \\
    \gen{h}_3 &= -\gen{S}_0 - \gen{L}_5 , \qquad &
    \gen{e}_3 &= +\gen{Q}_{+-+} , \qquad &
    \gen{f}_3 &= -\gen{Q}_{-+-} .
  \end{aligned}
\end{equation}
This leads to the Cartan matrix
\begin{equation}\label{eq:Cartan-su2}
  \begin{pmatrix}
     0 & -1 &  0 \\
    -1 & +2 & -1 \\
     0 & -1 &  0
  \end{pmatrix}.
\end{equation}
In Figure~\ref{fig:dynkin-su22}~\subref{fig:dynkin-su22-sl} we find the Dynkin diagram in the $\sls(2)$ grading. For Cartan generators and simple roots we take
\begin{equation}\label{eq:SC-basis-sl2}
  \begin{aligned}
    \hat{\gen{h}}_1 &= +\gen{S}_0 + \gen{L}_5 , \qquad &
    \hat{\gen{e}}_1 &= -\gen{Q}_{-++} , \qquad &
    \hat{\gen{f}}_1 &= +\gen{Q}_{+--} , \\
    \hat{\gen{h}}_2 &= -2\gen{S}_0 , \qquad &
    \hat{\gen{e}}_2 &= +\gen{S}_+ , \qquad &
    \hat{\gen{f}}_2 &= -\gen{S}_- , \\
    \hat{\gen{h}}_3 &= +\gen{S}_0 + \gen{L}_5 , \qquad &
    \hat{\gen{e}}_3 &= -\gen{Q}_{-+-} , \qquad &
    \hat{\gen{f}}_3 &= -\gen{Q}_{+-+} ,
  \end{aligned}
\end{equation}
with the Cartan matrix
\begin{equation}\label{eq:Cartan-sl2}
  \begin{pmatrix}
     0 & +1 &  0 \\
    +1 & -2 & +1 \\
     0 & +1 &  0
  \end{pmatrix}.
\end{equation}
The two above are the gradings that will be more relevant for us. For completeness we write down also the choice corresponding to Figure~\ref{fig:dynkin-su22}~\subref{fig:dynkin-su22-fff}, where all simple roots are fermionic. The choice for the raising operators $\gen{e}_i$ can be either
\begin{equation}
  \gen{Q}_{+-+}, \quad \gen{Q}_{++-}, \quad \gen{Q}_{-++} \,, \qquad 
  \text{or} \qquad
  \gen{Q}_{-+-}, \quad \gen{Q}_{--+}, \quad \gen{Q}_{+--} \,.\,
\end{equation}
This leads to the Cartan matrices
\begin{equation}\label{eq:Cartan-ferm}
    \begin{pmatrix}
     0 & +1 &  0 \\
    +1 & 0 & -1 \\
     0 & -1 &  0
  \end{pmatrix} , \qquad
  \text{and} \qquad
    \begin{pmatrix}
     0 & -1 &  0 \\
    -1 & 0 & +1 \\
     0 & +1 &  0
  \end{pmatrix}.
\end{equation}

\subsection{Spin-chain representation}
To construct a spin-chain that transforms under \emph{one} copy of $\psu(1,1|2)$, we put at each site an infinite dimensional representation denoted by $(-\tfrac{1}{2};\tfrac{1}{2})$~\cite{OhlssonSax:2011ms,Borsato:2013qpa}. This consists of the bosonic $\algSU(2)$ doublet $\phi^{(n)}_{\pm}$, where the indices $\pm$ label the two $\su(2)$ states, and two $\algSU(2)$ singlets $\psi^{(n)}_{\pm}$. The index $n$ indicates the $\algSL(2)$ quantum number. Explicitly, the action of the bosonic generators is
\begin{equation}\label{eq:su112-bosonicgen-representation}
  \begin{gathered}
      \gen{L}_5 \ket{\phi_{\pm}^{(n)}} = \pm \frac{1}{2} \ket{\phi_{\pm}^{(n)}} , \qquad
      \gen{L}_+ \ket{\phi_{-}^{(n)}} = \ket{\phi_{+}^{(n)}} , \qquad
      \gen{L}_- \ket{\phi_{+}^{(n)}} = \ket{\phi_-^{(n)}} , \\
    \begin{aligned}
      \gen{S}_0 \ket{\phi_{\beta}^{(n)}} &= - \left( \tfrac{1}{2} + n \right) \ket{\phi_{\beta}^{(n)}} , &
      \gen{S}_0 \ket{\psi_{\dot\beta}^{(n)}} &= - \left( 1 + n \right) \ket{\psi_{\dot\beta}^{(n)}} , \\
      \gen{S}_+ \ket{\phi_{\beta}^{(n)}} &= +n \ket{\phi_{\beta}^{(n-1)}} , &
      \gen{S}_+ \ket{\psi_{\dot\beta}^{(n)}} &= +\sqrt{(n + 1)n} \ket{\psi_{\dot\beta}^{(n-1)}} , \\
      \gen{S}_- \ket{\phi_{\beta}^{(n)}} &= -(n+1)\ket{\phi_{\beta}^{(n+1)}} , &
      \gen{S}_- \ket{\psi_{\dot\beta}^{(n)}} &= -\sqrt{(n + 2) (n + 1)} \ket{\psi_{\dot\beta}^{(n+1)}} .
    \end{aligned} 
  \end{gathered}
\end{equation}
The supercharges relate the bosons and the fermions as
\begin{equation}\label{eq:su112-fermionicgen-representation}
    \begin{aligned}
      \gen{Q}_{-\pm\dot\beta} \ket{\phi_{\mp}^{(n)}} &= \pm \sqrt{n+1} \ket{\psi_{\dot\beta}^{(n)}} , &
      \gen{Q}_{+\pm\dot\beta} \ket{\phi_{\mp}^{(n)}} &= \pm \sqrt{n} \ket{\psi_{\dot\beta}^{(n-1)}} , \\
      \gen{Q}_{-\beta\pm} \ket{\psi_{\mp}^{(n)}} &= \mp \sqrt{n+1} \ket{\phi_{\beta}^{(n+1)}} , &
      \gen{Q}_{+\beta\pm} \ket{\psi_{\mp}^{(n)}} &= \mp \sqrt{n+1} \ket{\phi_{\beta}^{(n)}} .
    \end{aligned}
\end{equation}
In the $\su(2)$ grading the highest weight state is $\phi_{+}^{(0)}$, since it is annihilated by the raising operators $\gen{L}_+,\gen{Q}_{+--}\gen{Q}_{+-+}$.
Nevertheless one can check that also the positive roots $\gen{Q}_{-+\pm}$ annihilate this state. For this reason the representation is \emph{short}, and one has the identity
\begin{equation}
  \acomm{\gen{Q}_{+-\mp}}{\gen{Q}_{-+\pm}} \ket{\phi^{(0)}_+} = \mp (\gen{S}_0 + \gen{L}_5 ) \ket{\phi^{(0)}_+} = 0.
\end{equation}
When constructing a spin-chain of length $L$, we need to consider the $L$-fold tensor product of the representation presented above. Since the symmetry algebra consists of two copies of $\psu(1,1|2)$ labelled by L and R, we actually need to take the tensor product of two such spin-chains. In particular, we define the ground state of the $\psu(1,1|2)_{\sL}\oplus \psu(1,1|2)_{\sR}$ spin-chain as
\begin{equation}
  \ket{\vacuum}_L = \Ket{(\phi^{(0)}_+)^L} \otimes \Ket{(\phi^{(0)}_+)^L}.
\end{equation}
This is the heighest weight state of the representation $(-\tfrac{L}{2};\tfrac{L}{2}) \otimes (-\tfrac{L}{2};\tfrac{L}{2})$, where by definition charges with label L act on the first $L$-fold product, while charges with label R on the second one.
The shortening condition is also inherited, giving a total of eight supercharges preserving the ground state. We denote them as
\begin{equation}
  \begin{aligned}
    \gen{Q}_{\sL}^{\ 1} = +\gen{Q}_{-++}^{\sL} , \quad
    \gen{Q}_{\sL}^{\ 2} = -\gen{Q}_{-+-}^{\sL} , \quad
    \overline{\gen{Q}}_{\sL 1} = +\gen{Q}_{+--}^{\sL} , \quad
    \overline{\gen{Q}}_{\sL 2} = +\gen{Q}_{+-+}^{\sL} , \\
    \gen{Q}_{\sR 1} = +\gen{Q}_{-++}^{\sR} , \quad
    \gen{Q}_{\sR 2} = -\gen{Q}_{-+-}^{\sR} , \quad
    \overline{\gen{Q}}_{\sR}^{\ 1} = +\gen{Q}_{+--}^{\sR} , \quad
    \overline{\gen{Q}}_{\sR}^{\ 2} = +\gen{Q}_{+-+}^{\sR} , 
  \end{aligned}
\end{equation}
where we use the same notation as for the charges derived from the string theory. The ground state is preserved also by the central charges $\gen{H}_{\sL},\gen{H}_{\sR}$ defined as
\begin{equation}
 \gen{H}_{\sL,\sR} = -\gen{S}_0^{\sL,\sR}- \gen{L}_5^{\sL,\sR}.
\end{equation}
Using the $\psu(1,1|2)$ commutation relations one can check that these generators close into four copies of $\su(1|1)^2$
\begin{equation}\label{eq:su11-su11-algebra}
  \acomm{\gen{Q}_{\sI}^{\ \dot{a}}}{\overline{\gen{Q}}_{\sJ \dot{b}}} = \delta^{\dot{a}}_{\ \dot{b}} \delta_{\sI\sJ} \gen{H}_{\sI}\,, \qquad\quad \sI,\sJ=\sL,\sR.
\end{equation}
It is clear that this algebra coincides with the one found from the string theory~\eqref{eq:cealgebra}, once the central extension is turned off $\gen{C}=\overline{\gen{C}}=0$ and we identify
\begin{equation}
  \gen{H} = \gen{H}_{\sL} + \gen{H}_{\sR} , \qquad
  \gen{M} = \gen{H}_{\sL} - \gen{H}_{\sR} .
\end{equation}
To introduce excited states of the spin-chain we just need to replace the highest weight $\phi_+^{(0)}$ with another state of the same module.
A nice way to organise excited states is to look at the eigenvalues of the charges $\gen{H}_{\sL,\sR}$. Considering the states $\phi_+^{(n)}$ or $\phi_-^{(n)}$ sitting on a site of one of the two copies of the spin-chain increases the eigenvalue of the corresponding Hamiltonian by $n$ and $n+1$ respectively. For the states $\psi_\pm^{(n)}$ this is increased by $n+1$.
The lightest states---the ones increasing the Hamiltonian just by $1$---are then
\begin{equation}
  \phi_-^{(0)} , \qquad \psi_+^{(0)} , \qquad \psi_-^{(0)} , \qquad \text{and} \qquad \phi_+^{(1)}.
\end{equation}
They transform in the familiar fundamental representation of $\psu(1|1)^4$---equivalently in the bi-fundamental representation of $\su(1|1)^2$, see \emph{e.g.} Figure~\ref{fig:massive}. We decide to introduce a notation for the excited states of the spin-chain that makes clear the bi-fundamental nature of the representation. We write
\begin{equation}\label{eq:bi-fund-fields}
  \Phi^{\sI++} = +\phi_-^{\sI(0)} , \qquad
  \Phi^{\sI--} = +\phi_+^{\sI(1)} , \qquad
  \Phi^{\sI-+} = +\psi_+^{\sI(0)} , \qquad
  \Phi^{\sI+-} = -\psi_-^{\sI(0)} ,
\end{equation}
where we introduced a label I=L,R to distinguish the two types of excitations on the spin-chain.
It is easy to check that these states transform under the two irreducible representations $\varrho_{\sL} \otimes \varrho_{\sL} $ and $\varrho_{\sR} \otimes \varrho_{\sR} $ of Section~\ref{sec:BiFundamentalRepresentationsT4}, once the supercharges  are rewritten in terms of $\su(1|1)^2$ supercharges as in Section~\ref{sec:AlgebraTensorProductT4}. 
To match with Equations~\eqref{eq:su112-fermionicgen-representation} one should set $a=\bar{a}=1$ and $b=\bar{b}=0$ in the exact short representations of $\psu(1,1)^4_{\ce}$. We conclude that the lightest excitations in~\eqref{eq:su112-fermionicgen-representation} correspond to an on-shell representation at zero coupling.
In the next section we discuss how the central extension and the coupling dependence are implemented in the spin-chain description.

\subsection{Central extension}
In order to make the Hamiltonian $\gen{H}$ dependent on the coupling constant and the momenta of the excitations, we need to deform the above representation. 
We give a momentum to a one-particle excitation by writing the plane-wave
\begin{equation}
  \ket{\mathcal{X}_p} = \sum_{n=1}^{L} e^{ipn} \ket{ \vacuum^{n-1} \mathcal{X} \vacuum^{L-n} } ,
\end{equation}
where $\vacuum$ denotes a vacuum site and $L$ is the length of the spin-chain.
When we consider multi-particle states we write a similar expression, where we always assume that the spin-chain length is very large $L\gg 1$ and the excitations are well separated. In other words we consider only \emph{asymptotic} states, and we make use of the S-matrix to relate in- and out-states.

In order to get the central extension and find non-vanishing eigenvalues for the central charges $\gen{C},\overline{\gen{C}}$ like in Equation~\eqref{eq:cealgebra}, we have to allow for a non-trivial action of the Right supercharges on Left excitations and vice versa. We impose that this action produces \emph{length-changing} effects on the spin-chain, by removing or adding vacuum sites.
The addition and the removal of vacuum sites is denoted by $\vacuum^{\pm}$ and it produces new momentum-dependent phase factors once these symbols are commuted to the far left of the excitations
\begin{equation}
\ket{\mathcal{X}_p\, \vacuum^\pm } = e^{\pm i\, p}\ket{\vacuum^\pm \, \mathcal{X}_p},
\end{equation}
as it can be checked from the plane-wave Ansatz.

Once we consider a spin-chain invariant under $\su(1|1)^2$, a way to centrally-extend it is to take~\cite{Borsato:2012ud}
\begin{equation}\label{eq:chiral-rep}
  \begin{aligned}
    \gen{Q}_{\smallL} \ket{\phi_p^{\sL}} &= a_p \ket{\psi_p^{\sL}} , \qquad &
    \gen{Q}_{\smallL} \ket{\psi_p^{\sL}} &= 0 , \\
    \overline{\gen{Q}}_{\smallL} \ket{\phi_p^{\sL}} &= 0 , \qquad &
    \overline{\gen{Q}}_{\smallL} \ket{\psi_p^{\sL}} &= \bar{a}_p \ket{\phi_p^{\sL}} , \\
    \gen{Q}_{\smallR} \ket{\phi_p^{\sL}} &= 0 , \qquad &
    \gen{Q}_{\smallR} \ket{\psi_p^{\sL}} &= b_p \ket{\vacuum^+\, \phi_p^{\sL}} , \\
    \overline{\gen{Q}}_{\smallR} \ket{\phi_p^{\sL}} &= \bar{b}_p \ket{\vacuum^-\, \psi_p^{\sL}} , \qquad &
    \overline{\gen{Q}}_{\smallR} \ket{\psi_p^{\sL}} &= 0 ,
  \end{aligned}
\end{equation}
and similarly for the Right module, after we exchange the labels L and R.
For the bi-fundamental representations of the spin-chain excitations that we want to consider we then get
\begin{equation}\label{eq:representation-LL}
  \begin{aligned}
    \gen{Q}_{\smallL}^{\ 1} \ket{\Phi_p^{\sL++}} &= +a_p \ket{\Phi_p^{\sL-+}} , \qquad &
    \gen{Q}_{\smallL}^{\ 1} \ket{\Phi_p^{\sL+-}} &= +a_p \ket{\Phi_p^{\sL--}} , \\
    \overline{\gen{Q}}_{\smallL 1} \ket{\Phi_p^{\sL-+}} &= +b_p \ket{\Phi_p^{\sL++}} , &
    \overline{\gen{Q}}_{\smallL 1} \ket{\Phi_p^{\sL--}} &= +b_p \ket{\Phi_p^{\sL+-}} , \\
    \gen{Q}_{\smallL}^{\ 2} \ket{\Phi_p^{\sL++}} &= +a_p \ket{\Phi_p^{\sL+-}} , &
    \gen{Q}_{\smallL}^{\ 2} \ket{\Phi_p^{\sL-+}} &= -a_p \ket{\Phi_p^{\sL--}} , \\
    \overline{\gen{Q}}_{\smallL 2} \ket{\Phi_p^{\sL+-}} &= +b_p \ket{\Phi_p^{\sL++}} , &
    \overline{\gen{Q}}_{\smallL 2} \ket{\Phi_p^{\sL--}} &= -b_p \ket{\Phi_p^{\sL-+}} ,
  \end{aligned}
\end{equation}
\begin{equation}
  \begin{aligned}
    \gen{Q}_{\smallR 1} \ket{\Phi_p^{\sL--}} &= +c_p \ket{\vacuum^+\Phi_p^{\sL+-} } , \qquad &
    \gen{Q}_{\smallR 1} \ket{\Phi_p^{\sL-+}} &= +c_p \ket{\vacuum^+\Phi_p^{\sL++} } , \\
    \overline{\gen{Q}}_{\smallR}^{\ 1} \ket{\Phi_p^{\sL++}} &= +d_p \ket{\vacuum^-\Phi_p^{\sL-+} } , \qquad &
    \overline{\gen{Q}}_{\smallR}^{\ 1} \ket{\Phi_p^{\sL+-}} &= +d_p \ket{\vacuum^-\Phi_p^{\sL--} } , \\
    \gen{Q}_{\smallR 2} \ket{\Phi_p^{\sL--}} &= -c_p \ket{\vacuum^+\Phi_p^{\sL-+} } , \qquad &
    \gen{Q}_{\smallR 2} \ket{\Phi_p^{\sL+-}} &= +c_p \ket{\vacuum^+\Phi_p^{\sL++} } , \\
    \overline{\gen{Q}}_{\smallR}^{\ 2} \ket{\Phi_p^{\sL++}} &= +d_p \ket{\vacuum^-\Phi_p^{\sL+-} } , \qquad &
    \overline{\gen{Q}}_{\smallR}^{\ 2} \ket{\Phi_p^{\sL-+}} &= -d_p \ket{\vacuum^-\Phi_p^{\sL--} } .
  \end{aligned}
\end{equation}
Using the commutation relations we find the actions of the central charges
\begin{equation}
  \begin{aligned}
    \gen{H}_{\sL} \ket{\Phi_p^{\sL\pm\pm}} &= a_p \bar{a}_p \ket{\Phi_p^{\sL\pm\pm}} , \qquad &
    \gen{C} \ket{\Phi_p^{\sL\pm\pm}} &= a_p b_p \ket{\vacuum^+\Phi_p^{\sL\pm\pm} } , \\
    \gen{H}_{\sR} \ket{\Phi_p^{\sL\pm\pm}} &= b_p \bar{b}_p \ket{\Phi_p^{\sL\pm\pm}} , \qquad &
    \overline{\gen{C}} \ket{\Phi_p^{\sL\pm\pm}} &= \bar{a}_p \bar{b}_p \ket{\vacuum^-\Phi_p^{\sL\pm\pm}} .
  \end{aligned}
\end{equation}
We stress that the length-changing effects are crucial if we want the central charges $\gen{C},\overline{\gen{C}}$ to have the correct eigenvalues also on multi-particle states. On two-particle states we find\footnote{Differently from the original paper~\cite{Borsato:2012ud,Borsato:2013qpa}, we modify the construction by moving the added or removed vacuum sites $\vacuum^\pm$ to the left of the excitations. This allows us to get the central charge $\gen{C}=\frac{ih}{2} (e^{i \gen{P}} - 1)$ that matches the one derived from the worldsheet computation, rather than $\gen{C}=\frac{ih}{2} (e^{-i \gen{P}} - 1)$. Moreover, the relation between the S-matrix for excitations on the string discussed in Section~\ref{sec:S-mat-T4} and the one for excitations on the spin-chain can be related in a simple way, see~\eqref{eq:S-mat-sp-ch-S-mat-str}.}
\begin{equation}
\begin{aligned}
\gen{C}\ket{\Phi_p^{\sL\pm\pm}\Phi_q^{\sL\pm\pm}} &= a_p b_p \ket{\vacuum^+\Phi_p^{\sL\pm\pm}\Phi_q^{\sL\pm\pm}}+a_q b_q \ket{\Phi_p^{\sL\pm\pm}\vacuum^+\Phi_q^{\sL\pm\pm}}
\\
&= (a_p b_p +e^{+i\, p}a_q b_q)\ket{\vacuum^+\Phi_p^{\sL\pm\pm}\Phi_q^{\sL\pm\pm}}.
\end{aligned}
\end{equation}
Setting
\begin{equation}
  a_p b_p = \frac{ih}{2} (e^{ip} - 1) ,
\end{equation}
we find the eigenvalue
\begin{equation}
  a_p b_p +e^{+i\, p}a_q b_q = \frac{ih}{2} \left( e^{i(p+q)} - 1 \right) .
\end{equation}
One can repeat the discussion in Section~\ref{sec:RepresentationCoefficientsT4} to find the expressions of the representation coefficients that reproduce the exact eigenvalues of the central charges. Even though other choices are allowed, we prefer to keep the same parameterisation~\eqref{eq:expl-repr-coeff} used for the description of the string.

In the spin-chain description one does not need to introduce the parameter $\xi$ for the coefficients in~\eqref{eq:expl-repr-coeff}. This was introduced in the string picture to get a non-local action of the supercharges and reproduce the correct eigenvalue of the central charges on multiparticle states. In the context of the dynamical spin-chain, the same role is played by the length-changing effects, allowing us to set  $\xi=0$ also for multiparticle states.

When considering one-particle states we can identify the representation of the string with the one presented here for the spin-chain as follow
\begin{equation}\label{eq:identif-string-sp-ch-states}
\begin{aligned}
\ket{\Phi^{\sL++}}=\ket{Y^{\sL}},\qquad\ket{\Phi^{\sL-+}}=\ket{\eta^{\sL 1}},\qquad\ket{\Phi^{\sL+-}}=\ket{\eta^{\sL 2}},\qquad\ket{\Phi^{\sL--}}=\ket{Z^{\sL}},
\\
\ket{\Phi^{\sR++}}=\ket{Y^{\sR}},\qquad\ket{\Phi^{\sR-+}}=\ket{\eta^{\sR}_{\ 1}},\qquad\ket{\Phi^{\sR+-}}=\ket{\eta^{\sR}_{\ 2}},\qquad\ket{\Phi^{\sR--}}=\ket{Z^{\sR}}.
\end{aligned}
\end{equation}
This identification is possible by comparing the action of the supercharges on one-particle states. 

\subsection{The S-matrix for the spin-chain}\label{sec:S-matr-spin-chain-T4}
Repeating the derivation of Section~\ref{sec:S-mat-T4} one may find the exact S-matrix governing the scattering of the spin-chain excitations.
This is essentially the same object as the one found from the string description, with the only exception that it is written in a different basis.
Indeed, although the action of the charges on one-particle representations agrees on the two side---yielding the identification~\eqref{eq:identif-string-sp-ch-states}---one can check that it is different on two-particle states. This is just a consequence of the fact that the basis for the two-particle representations on the two sides are related by a matrix that acts non-locally on the states.

To be precise, the S-matrix $\mathcal{S}^{\text{sp-ch}}$ of the spin-chain picture is related to the one found from the string theory $\mathcal{S}^{\text{str}}$ as\footnote{This relation may be found by realising that we can map the supercharges on the two sides by $\gen{Q}^{\text{str}}_{pq} = (\mathbf{1} \otimes \mathbf{U}^\dagger_p) \, \cdot \, \gen{Q}^{\text{sp-ch}}_{pq} \, \cdot \, (\mathbf{1} \otimes \mathbf{U}_p)$.}
 \begin{equation}\label{eq:S-mat-sp-ch-S-mat-str}
  \mathcal{S}^{\text{sp-ch}}_{pq} = (\mathbf{1} \otimes \mathbf{U}_q) \, \cdot \, \mathcal{S}^{\text{str}}_{pq} \, \cdot \, (\mathbf{1} \otimes \mathbf{U}^\dagger_p) .
\end{equation}
In the basis
\begin{equation}\label{eq:1partbasis}
\left(\Phi^{\sL++},\,\Phi^{\sL+-},\,\Phi^{\sL-+},\,\Phi^{\sL--},\,\Phi^{\sR++},\,\Phi^{\sR+-},\,\Phi^{\sR-+},\,\Phi^{\sR--},\right) ,
\end{equation}
the matrix that we need is
\begin{equation}
 \mathbf{U}_p=\text{diag} \left(e^{i\, p},e^{i\, p/2},e^{i\, p/2},1,e^{i\, p},e^{i\, p/2},e^{i\, p/2},1\right) .
\end{equation}
While the string-frame S-matrix satisfies the standard Yang-Baxter equation~\eqref{eq:YBe}, it is easy to check that the above redefinition implies that for the S-matrix of the spin-chain we must have a \emph{twisted} Yang-Baxter equation 
 \begin{equation}\label{eq:YB-mat-twist}
  \left(\mathbf{F}_p^{\phantom{1}}\mathcal{S}_{qr}\mathbf{F}_p^{{-1}}\right) \otimes \mathbf{1} \, \cdot \,
  \mathbf{1}\otimes\mathcal{S}_{pr} \, \cdot \,
  \left(\mathbf{F}_r^{\phantom{1}} \mathcal{S}_{pq}\mathbf{F}_r^{{-1}}\right) \otimes \mathbf{1}
  =
   \mathbf{1}\otimes\mathcal{S}_{pq} \, \cdot \,
  \left(\mathbf{F}_q^{\phantom{1}}\mathcal{S}_{pr}\mathbf{F}_q^{-1}\right) \otimes \mathbf{1} \, \cdot \,
  \mathbf{1}\otimes\mathcal{S}_{qr},
\end{equation}
where $\mathbf{F}_p\equiv\mathbf{U}_p\otimes \mathbf{U}_p$.
The same result is actually found after carefully considering the length changing effects on the three-particle states on which we want to check Yang-Baxter.

It is possible to repeat the derivation of the previous chapter and write down the Bethe-Yang equations for the spin-chain.
Doing so one finds the same six equations for the massive excitations~\eqref{eq:BA-1}-\eqref{eq:BA-3b}, where the interaction terms with massless excitations are obviously missing.
Moreover, because of the change of basis~\eqref{eq:S-mat-sp-ch-S-mat-str}, the factors $\nu_j$ are absent in the Bethe-Yang equations for the spin-chain.
These factors can actually be reabsorbed in the definition of the length, allowing us to relate the length of the string to the length of the spin-chain.

It would be very interesting to construct a long-range spin-chain that describes the CFT$_2$ dual to {\adsthree} in the spirit of the succesful program carried on in AdS$_5$/CFT$_4$, and compare it to our construction.
We refer to~\cite{Pakman:2009mi,Sax:2014mea} for papers taking some preliminary steps in this direction.

\section{Dressing factors}\label{sec:dressing-factors}
In Section~\ref{sec:smat-tensor-prod} we determined the S-matrix up to a total of four unconstrained dressing factors.
In this section we present a solution to the crossing equations~\eqref{eq:cr-massive} for the dressing factors in the massive sector~\cite{Borsato:2013hoa}.

\subsection{Solution of the crossing equations}
As explained in Section~\ref{sec:unitarity-YBe}, thanks to the unitarity conditions the dressing factors are written as
\begin{equation}
\sigma^{\bullet\bullet}_{pq} = \text{exp}(i \ \theta^{\bullet\bullet}_{pq}),
\qquad
\tilde{\sigma}^{\bullet\bullet}_{pq} = \text{exp}(i \ \tilde{\theta}^{\bullet\bullet}_{pq}),
\end{equation}
where $\theta^{\bullet\bullet}_{pq},\tilde{\theta}^{\bullet\bullet}_{pq}$ are real anti-symmetric functions of the physical momenta.
In both cases we will assume that it is possible to rewrite them as~\cite{Arutyunov:2006iu}
\begin{equation}
\label{eq:thchi}
\theta(p,q) = \chi(x_p^+,x_q^+) +\chi(x_p^-,x_q^-) -\chi(x_p^+,x_q^-) -\chi(x_p^-,x_q^+)\,,
\end{equation}
with $\chi$ anti-symmetric, to respect braiding unitarity.
Instead of solving the crossing equations
\begin{equation}
\begin{aligned}
\left(\sigma^{\bullet\bullet}_{pq}\right)^2 \ \left(\tilde{\sigma}^{\bullet\bullet}_{\bar{p}q}\right)^2 &= \tilde{c}_{pq}=\left( \frac{x^-_q}{x^+_q} \right)^2 \frac{(x^-_p-x^+_q)^2}{(x^-_p-x^-_q)(x^+_p-x^+_q)} \frac{1-\frac{1}{x^-_px^+_q}}{1-\frac{1}{x^+_px^-_q}}, \\
\left(\sigma^{\bullet\bullet}_{\bar{p}q}\right)^2 \ \left(\tilde{\sigma}^{\bullet\bullet}_{pq}\right)^2 &= c_{pq}=\left( \frac{x^-_q}{x^+_q} \right)^2 \frac{\left(1-\frac{1}{x^+_px^+_q}\right)\left(1-\frac{1}{x^-_px^-_q}\right)}{\left(1-\frac{1}{x^+_px^-_q}\right)^2} \frac{x^-_p-x^+_q}{x^+_p-x^-_q},
\end{aligned}
\end{equation}
we prefer to study the ones obtained by taking the product and the ratio of these two
\begin{equation}
\begin{aligned}
\left(\sigma^{+}_{pq}\sigma^{+}_{\bar{p}q}\right)^2&=\left(\sigma^{\bullet\bullet}_{pq}\tilde{\sigma}^{\bullet\bullet}_{pq}\right)^2 \ \left(\sigma^{\bullet\bullet}_{\bar{p}q}\tilde{\sigma}^{\bullet\bullet}_{\bar{p}q}\right)^2 =c_{pq}\tilde{c}_{pq},\\
\frac{\left(\sigma^{-}_{pq}\right)^2}{\left(\sigma^{-}_{\bar{p}q}\right)^2}&=\left(\frac{\sigma^{\bullet\bullet}_{pq}}{\tilde{\sigma}^{\bullet\bullet}_{pq}}\right)^2 \ \left(\frac{\sigma^{\bullet\bullet}_{\bar{p}q}}{\tilde{\sigma}^{\bullet\bullet}_{\bar{p}q}}\right)^{-2} =\frac{\tilde{c}_{pq}}{c_{pq}},\\
\end{aligned}
\end{equation}
where the symbols $+$ and $-$ are introduced to remind that the corresponding phases are the sum and the difference of the original ones
\begin{equation}
\theta^{+}_{pq}=\theta^{\bullet\bullet}_{pq}+\tilde{\theta}^{\bullet\bullet}_{pq},
\qquad
\theta^{-}_{pq}=\theta^{\bullet\bullet}_{pq}-\tilde{\theta}^{\bullet\bullet}_{pq}.
\end{equation}
This rewriting turns out to be very convenient, one reason being that the solution for $\theta^{+}_{pq}$ can be found by using the results valid for describing the integrable model of \adsfive.

\paragraph{Solution for the sum of the phases}
The right-hand-side of the crossing equation for $\sigma^{+}_{pq}$ can be rewritten as 
\begin{equation}\label{eq:rhs-sum-phases}
c_{pq}\tilde{c}_{pq} = \frac{(c^{\BES}_{pq})^3}{(c^{\BES}_{pq})^*},
\end{equation}
where $*$ denotes complex conjugation and $c^{\BES}_{pq}$ is the right-hand-side of the crossing equation of {\adsfive} satisfied by the Beisert-Eden-Staudacher (BES)  dressing factor~\cite{Beisert:2006ez}
\begin{equation}\label{eq:cr-BES}
\sigma^\BES_{pq} \sigma^\BES_{\bar{p}q} = c^{\BES}_{pq}=
\frac{x_q^-}{x_q^+}\frac{x_p^- - x_q^+}{x_p^- - x_q^-} \frac{1-\frac{1}{x_p^+x_q^+}}{1-\frac{1}{x_p^+x_q^-}}.
\end{equation}
A useful representation of this solution in the physical region was given by Dorey, Hofman and Maldacena (DHM)~\cite{Dorey:2007xn} in terms of a double integral on unit circles
\begin{equation}
\chi^\BES(x,y)= i \ointc \frac{dw}{2 \pi i} \ointc \frac{dw'}{2 \pi i} \, \frac{1}{x-w}\frac{1}{y-w'} \log{\frac{\Gamma[1+i \frac{h}{2}(w+1/w-w'-1/w')]}{\Gamma[1-i \frac{h}{2}(w+1/w-w'-1/w')]}}\,. 
\label{eq:besdhmrep}
\end{equation}
For later convenience, we note that taking the strong coupling limit $h\to \infty$ of this solution one recovers the Arutyunov-Frolov-Staudacher (AFS) phase~\cite{Arutyunov:2004vx}, whose factor may be written in terms of the spectral parameters as 
\begin{equation}
\label{eq:AFS-xpxm}
\sigma^\AFS_{pq} = \left(  \frac{1-\frac{1}{x_p^-x_q^+}}{1-\frac{1}{x_p^+x_q^-}} \right) \left(  \frac{1-\frac{1}{x_p^+x_q^-}}{1-\frac{1}{x_p^+x_q^+}}  \frac{1-\frac{1}{x_p^-x_q^+}}{1-\frac{1}{x_p^-x_q^-}} \right)^{i \frac{h}{2} (x_p+1/x_p-x_q-1/x_q)}\,.
\end{equation}
Pushing the expansion at strong coupling to the next-to-leading order one finds the Hern\'andez-L\'opez (HL) factor~\cite{Hernandez:2006tk}, that solves the crossing equation
\begin{equation}\label{eq:cr-HL}
\sigma^\HL_{pq} \sigma^\HL_{\bar{p}q} = \sqrt{ \frac{c^{\BES}_{pq}}{c^{\BES}_{\bar{p}q}} }=\sqrt{c^{\BES}_{pq}\,(c^{\BES}_{pq})^*}\,.
\end{equation}
A possible representation of this phase may be obtained by expanding the one for BES, giving
\begin{equation}\label{eq:DHM-HL}
\chi^\HL(x,y)= \frac{\pi}{2} \ointc \frac{dw}{2 \pi i} \ointc \frac{dw'}{2 \pi i} \, \frac{1}{x-w}\frac{1}{y-w'} \, \text{sign}(w'+1/w'-w-1/w)\,.
\end{equation}
The BES and HL phases can then be used as building blocks to construct the solution for the sum of our phases. Using the identity~\eqref{eq:rhs-sum-phases} we see that we can solve the crossing equation for $\sigma^{+}_{pq}$ if we define it as
\begin{equation}
\label{eq:sumsolution}
\sigma^{+}_{pq}=\frac{(\sigma^{\BES}_{pq})^2}{\sigma^{\HL}_{pq}}\,, \qquad\qquad \theta^{+}_{pq}=2\theta^{\BES}_{pq}-\theta^{\HL}_{pq}\,.
\end{equation}
We now present the solution for the factor $\sigma^{-}_{pq}$.

\paragraph{Solution for the difference of the phases}
The crossing equation for $\sigma^{-}_{pq}$ is
\begin{equation}\label{eq:cr-ratio}
\frac{(\sigma^{-}_{pq})^2}{(\sigma^{-}_{\bar{p}q})^2}=\frac{\ell^-(x_p^+,x_q^-)\ \ell^-(x_p^-,x_q^+)}{\ell^-(x_p^+,x_q^+)\ \ell^-(x_p^-,x_q^-)}, \qquad \ell^-(x,y)\equiv(x-y)\left(1-\frac{1}{xy}\right).
\end{equation}
A solution of this equation is given by
\begin{equation}\label{eq:chi-}
\begin{aligned}
\chi^-(x,y) &=\ointc \, \frac{dw}{8\pi} \frac{\text{sign}((w-1/w)/i)}{x-w} \log{\ell^-(y,w)}  \ - x \leftrightarrow y \\
&=\left( \inturl - \intdlr \right)\frac{dw}{8\pi} \frac{1}{x-w} 
\log{\ell^-(y,w)} \ - x \leftrightarrow y\,,
\end{aligned}
\end{equation}
where in the second line we have split the integration along the upper and the lower semicircles.
The proof that $\chi^-$ satisfies the crossing equation may be found in Appendix~\ref{app:crossing-AdS3}.

\paragraph{Recap of the solutions}
The above results allow us to write the following solution for the dressing phases of the massive sector
\begin{equation}
\label{eq:solution}
\begin{aligned}
  \chi^{\bullet\bullet}(x,y) &= \chi^{\text{BES}}(x,y)+\frac{1}{2}\left(-\chi^{\text{HL}}(x,y)+\chi^{-}(x,y)\right) \,, \\
  \tilde{\chi}^{\bullet\bullet}(x,y) &= \chi^{\text{BES}}(x,y)+\frac{1}{2}\left(-\chi^{\text{HL}}(x,y)-\chi^{-}(x,y)\right) \,.
\end{aligned}
\end{equation}
A consequence of this result is that both factors $\sigma^{\bullet\bullet}$ and $\tilde{\sigma}^{\bullet\bullet}$ reduce to the AFS dressing factor at strong coupling. At the next order they are not just the HL dressing factor: its contribution to the phases is just half\footnote{Remember that BES contains one power of HL in the expansion.} of what one has in the case of \adsfive, and we discover a novel piece produced by $\chi^-$.

\subsection{Bound states}
In this section we discuss the possibility of bound states arising in the scattering processes.
This proves to be a good way to validate the proposed solutions of the crossing equations.

Let us consider a two-particle state with excitations of momenta $p,q$ described by the wave-function\footnote{Here $\sigma_1$ and $\sigma_2$ denote the worldsheet spatial coordinate. We trust it does not create confusion with the notation for the dressing factors.}
\begin{equation}
  \psi(\sigma_1,\sigma_2) = e^{i(p \sigma_1 + q \sigma_2)} + S(p,q) e^{i(p \sigma_2 + q \sigma_1)} .
\end{equation}
Here we are considering just the region $\sigma_1 \ll \sigma_2$. The first and second terms correspond to the in-coming and out-going waves, respectively. After the scattering one picks up a phase-shift $S(p,q)$.
A bound state may arise when the S-matrix exibits a pole, and this can happen for complex values of the two momenta 
\begin{equation}\label{eq:mom-imag-bound-state}
  p = \frac{p'}{2} + iv , \qquad
  q = \frac{p'}{2} - iv.
\end{equation}
The relevant behaviour of the wave-function is then $\psi(\sigma_1,\sigma_2) \sim e^{-v(\sigma_2 - \sigma_1)}$, and the normalisability condition implies that we should impose $v>0$.

This condition must be checked when studying the possible bound states that are allowed by representation theory. These are found by studying when a generic multi-particle representation becomes short. A feauture of $\psu(1|1)^4_\ce$ is that all \emph{short} representations\footnote{Similarly, for $\su(1|1)^2_\ce$, short representations have dimension 2, and long ones dimension 4.} have dimension 4, while all \emph{long} representations have dimension 16. 
When considering a two-particle representation obtained as the tensor product of two Left massive modules, we find that in general it is a long representation. However, there exist particular values for the momenta $p,q$ of the excitations such that the representation becomes short. This happens for $x^+_p=x^-_q$ or for $x^-_p=x^+_q$.
In the first case we find that the bosonic state $\ket{Y^{\sL}_pY^{\sL}_q}$ survives in the module, while in the second case $\ket{Z^{\sL}_p Z^{\sL}_q}$. We will refer to the two cases as $\su(2)$ and $\sls(2)$ bound states respectively.
We see that in these situations the momenta of the excitations develop a non-zero imaginary part, as in~\eqref{eq:mom-imag-bound-state}. Nevertheless, only the $\su(2)$ bound state satisfies the condition $v>0$, while for the $\sls(2)$ case the imaginary part of $p$ is negative.
The former is considered to be a bound state in the spectrum, while the latter should not appear.
Later we will check that indeed  the S-matrix exibits a pole when scattering two $Y^{\sL}$ excitations, while it is regular when scattering two $Z^{\sL}$ excitations.
The case of two Right massive excitations is equivalent, thanks to LR symmetry.

The situation is different when we consider two massive excitations with different LR flavor. In that case the representation becomes short for $x^+_p=1/x^+_q$ or $x^-_p=1/x^-_q$. Neither of these cases satisfy $|x^\pm_p|>1$ and $|x^\pm_q|>1$, necessary to remain in the physical region. For this reason there are no supersymmetric bound states in the LR-sector.

\medskip

As anticipated, the above results must be checked at the level of the S-matrix of Section~\ref{sec:smat-tensor-prod} derived from symmetries, including the solutions~\eqref{eq:solution} for the dressing factors that satisfy the crossing equations.
This provides a non-trivial check of the validity of the solutions.
The first process to consider is 
\begin{equation}
\mathcal{A}_{pq}=\bra{Y^{\sL}_q \, Y^{\sL}_p} \mathcal{S} \ket{Y^{\sL}_p \, Y^{\sL}_q}  = \frac{x^+_p}{x^-_p} \, \frac{x^-_q}{x^+_q} \, \frac{x^-_p - x^+_q}{x^+_p - x^-_q} \, \frac{1-\frac{1}{x^-_p x^+_q}}{1-\frac{1}{x^+_p x^-_q}} \, \frac{1}{\left(\sigma^{\bullet\bullet}_{pq} \right)^2 }.
\end{equation}
The dressing factor $1/\left(\sigma^{\bullet\bullet}_{pq} \right)^2$ is regular at $x^+_p=x^-_q$, as it is shown in~\ref{sec:sing-dressing}. The element $\mathcal{A}_{pq}$ has then a single pole at this point, confirming the presence of the expected $\su(2)$ bound state.

Similarly we can check that in the LR sector there is no pole in the physical region. We consider the scattering element
\begin{equation}
\widetilde{\mathcal{B}}_{pq}=\bra{Y^{\sL}_q Z^{\sR}_p }\Smat \ket{Z^{\sR}_p Y^{\sL}_q}  = \frac{x^+_p}{x^-_p}\, \frac{1-\frac{1}{x_p^- x_q^+}}{1-\frac{1}{x_p^+ x_q^-}} \,  \frac{1-\frac{1}{x_p^+ x_q^+}}{1-\frac{1}{x_p^- x_q^-}} \frac{1}{\left(\tilde{\sigma}^{\bullet\bullet}_{pq} \right)^2 },
\end{equation}
and we see that both the rational factors and $\tilde{\sigma}^{\bullet\bullet}_{pq}$ are regular in the physical region, in particular at the point $x^+_p=x^-_q$.

It is interesting to see how these processes in the $s$-channel automatically give constraints for processes in the $t$-channel, as a consequence of crossing symmetry.
The crossing equations~\eqref{eq:cr-massive} can be written in the simple form
\begin{equation}
\begin{aligned}
\mathcal{A}_{pq} \widetilde{\mathcal{A}}_{\bar{p}q}=1,& \qquad \text{where}\quad \widetilde{\mathcal{A}}_{pq}=\bra{Y^{\sL}_q Y^{\sR}_p }\Smat \ket{Y^{\sR}_p Y^{\sL}_q}\,,
\\
\widetilde{\mathcal{B}}_{pq} \mathcal{B}_{\bar{p}q}=1,& \qquad \text{where}\quad \mathcal{B}_{pq}=\bra{Y^{\sL}_q Z^{\sL}_p }\Smat \ket{Z^{\sL}_p Y^{\sL}_q}\,,
\end{aligned}
\end{equation}
involving explicit scattering elements.
It is clear that the presence of a single pole for $\mathcal{A}_{pq}$ at $x^+_p=x^-_q$ implies a zero for $\widetilde{\mathcal{A}}_{\bar{p}q}$ at the point $1/x^+_{\bar{p}}=x^-_q$. This is then responsible for a process in the $t$-channel.
Similarly, regularity of $\widetilde{\mathcal{B}}_{pq}$ implies regularity of $\mathcal{B}_{\bar{p}q}$, and consequently no corresponding process in the $t$-channel\footnote{When checking regularity of $\mathcal{B}_{\bar{p}q}$ one has to carefully analytically continue the dressing factor for crossed values of the momentum $p$. Doing that one discovers that $\left(\sigma^{\bullet\bullet}_{\bar{p}q} \right)^{-2}$ has a zero at $1/x^+_{\bar{p}}=x^-_q$ that cancels the apparent pole coming from the rational terms.}.

\medskip 

The discussion on the pole structure of the S-matrix is important to justify the validity of the solution to the crossing equations.
It is indeed always possible to multiply the solutions that we proposed by the so-called CDD factors, that solve the homogeneous crossing equations
\begin{equation}
\sigma^{\CDD}_{pq}\,\widetilde{\sigma}^{\CDD}_{\bar{p}q}=1\,,\qquad
\sigma^{\CDD}_{\bar{p}q}\,\widetilde{\sigma}^{\CDD}_{pq}=1\,.
\end{equation}
Usually these are meromorphic functions of the spectral parameters, obtained by taking 
\begin{equation}
\chi^{\CDD}_{pq}=\frac{i}{2}\log\frac{(x-y)^{c_1}}{(1-xy)^{c_2}}\,,\qquad
\widetilde{\chi}^{\CDD}_{pq}=\frac{i}{2}\log\frac{(x-y)^{c_2}}{(1-xy)^{c_1}}\,.
\end{equation}
It is clear that such solutions introduce new zeros and poles that modify the analytical structure of the S-matrix elements, spoiling the bound state interpretation.
These considerations allow us to rule out the possibility of CDD factors of this form
\begin{equation}
\sigma^{\CDD}_{pq}=1\,,\qquad\widetilde{\sigma}^{\CDD}_{pq}=1\,.
\end{equation}
A possibility that might still be valid is to introduce factors that satisfy the homogeneous crossing equation and that have no poles or zeros in the physical region.
Nevertheless, further independent validations of the phases proposed here appeared in the literature, and we refer to Section~\ref{sec:perturbative-results} for a collection of them.

\section{Finite-gap equations}\label{sec:strong-limit-T4}
Taking the limit of large string tension, one can make contact with solutions of rigid strings that are constructed explicitly by solving the classical equations of motion.
On the other hand, the formulation in terms of a Lax connection allows one to write down the so-called \emph{finite-gap equations}, from which one can find the spectrum of the classical integrable model.
We refer to~\cite{SchaferNameki:2010jy} for a review on this in the context of AdS/CFT.
Here we take a proper thermodynamic limit of the Bethe-Yang equations of Section~\ref{sec:BAE} in the limit of large tension, to recover the finite-gap equations for the massive sector of {\adsthree}.

To start, we expand $x^\pm$ for large values of the string tension.
We consider the Zhukovski parameters $x^\pm_p$ expressed in terms of the momentum $p$ as in Eq.~\eqref{eq:xpm-funct-p}, that solves the constraints~\eqref{eq:zhukovski}. 
The large-tension limit of these parameters is obtained by first rescaling the momentum $p= {\rm p}/h$ and then expanding the expressions at large $h$, obtaining\footnote{Since we are considering massive excitations, we obtain the same result for left or right movers on the worldsheet. One should instead distinguish between these two cases when considering massless excitations.} 
\begin{equation}\label{eq:resc-p-xpm}
x^\pm_p =\frac{\sqrt{m^2+{\rm p}^2}+|m|}{{\rm p}} \pm \frac{i \left(\sqrt{m^2+{\rm p}^2}+|m|\right)}{2 h} +\mathcal{O}(1/h^2).
\end{equation}
We parameterise the leading contribution in the expansion with a spectral parameter $x$, obtaining
\begin{equation}
x=\frac{\sqrt{m^2+{\rm p}^2}+|m|}{{\rm p}}
\implies
{\rm p}=\frac{2 |m| x}{x^2-1},
\end{equation}
\begin{equation}
x^\pm_p =x \pm \frac{i \, |m| \, x^2}{h(x^2-1)} +\mathcal{O}(1/h^2).
\end{equation}
Notice that it was important to assume that $m\neq 0$ when solving for ${\rm p}$ in terms of $x$.
For a single excitation, momentum and energy (difference) are given by
\begin{equation}
\begin{aligned}
{\rm p} &= h\, p = -i \, h \log \frac{x^+_p}{x^-_p}  
= \frac{2 |m|\, x}{x^2 - 1}+\mathcal{O}(1/h),
\\
\Delta E_p &= -|m| + \sqrt{m^2 +4 h^2 \sin^2 \frac{p}{2}} = -i \, h \left( \frac{1}{x^-_p} - \frac{1}{x^+_p} \right) 
= \frac{2 |m|}{x^2 - 1}  +\mathcal{O}(1/h).
\end{aligned}
\end{equation}
To take the finite-gap limit we consider a large number $K_i$ of excitations, where the index $i$ denotes the possible types of massive excitations and it takes values $i=1,2,3,\bar{1},\bar{2},\bar{3}$. More precisely, we take the number of excitations to scale like the string tension $K_i \sim h$, and we define densities as
\begin{equation}
\rho_i(x) \equiv \frac{2}{h} \sum_{k=1}^{K_i}  \frac{x^2}{x^2-1} \delta(x-x_{i,k})\,.
\end{equation}
Momentum and energy (difference) for the collection of the excitations are then expressed as the integrals
\begin{equation}
\mathcal{P}_i \equiv \int {\rm d}x \ \frac{\rho_i(x)}{x} ,
\qquad
\epsilon_i \equiv \int {\rm d}x \ \frac{\rho_i(x)}{x^2} .
\end{equation}

The finite-gap limit of the Bethe-Yang equations is taken by considering each factor\footnote{Here $S_{pq}$ stands for any product of rational expressions of $x^\pm$ and dressing phases that appear in our Bethe-Yang equations.} $S_{pq}$ and by computing $-i\, \log S_{pq}$ in the large $h$ limit, using the formulas above for the expansion of $x^\pm$ and keeping the auxiliary roots finite.
For the massive sector of {\adsthree} this yields the following equations\footnote{To avoid confusion we remind that when writing the finite-gap equations one uses a convention in notation that is a bit different from the one of the Bethe-Yang equations. For finite-gap we use the letter $x$ for the variable that solves the given equation, while $y$ is used for any variable on which we integrate. There is no distinction anymore in the notation for momentum carrying nodes and auxiliary roots.}
\begin{equation}
\begin{aligned}
2\pi n_1 &= - \int\frac{\rho_2(y)}{x-y}dy- \int\frac{\rho_{\bar{2}}(y)}{x-1/y}\frac{dy}{y^2} 
{ -\frac{1}{2} \left( \mathcal{P}_{2} + \mathcal{P}_{\bar{2}} \right)}\,,\\
2\pi n_2&=-\frac{x}{x^2-1}2\pi\mathcal{E}-\int\frac{\rho_1(y)}{x-y}dy + 2 \pint\frac{\rho_2(y)}{x-y}dy -\int\frac{\rho_3(y)}{x-y}dy\\
&\phantom{={}} +\int\frac{\rho_{\bar{1}}(y)}{x-1/y}\frac{dy}{y^2} + \int\frac{\rho_{\bar{3}}(y)}{x-1/y}\frac{dy}{y^2}+\frac{1}{x^2-1}\mathcal{M}
{ + \left( \mathcal{P}_{2} + \mathcal{P}_{\bar{2}} \right)}\,, \\
2\pi n_3&= - \int\frac{\rho_2(y)}{x-y}dy- \int\frac{\rho_{\bar{2}}(y)}{x-1/y}\frac{dy}{y^2} 
{ -\frac{1}{2} \left( \mathcal{P}_{2} + \mathcal{P}_{\bar{2}} \right)}\,,\\
\label{eq:FGlimit}
2\pi n_{\bar{1}}&= \int\frac{\rho_{2}(y)}{x-1/y}\frac{dy}{y^2}+\int\frac{\rho_{\bar{2}}(y)}{x-y}dy 
{ +\frac{1}{2} \left( \mathcal{P}_{2} + \mathcal{P}_{\bar{2}} \right)} \,,\\
2\pi n_{\bar{2}}&=-\frac{x}{x^2-1}2\pi\mathcal{E}-\int\frac{\rho_1(y)}{x-1/y}\frac{dy}{y^2}  -\int\frac{\rho_3(y)}{x-1/y}\frac{dy}{y^2}\\
&\phantom{={}} +\int\frac{\rho_{\bar{1}}(y)}{x-y}dy -2 \pint\frac{\rho_{\bar{2}}(y)}{x-y}dy + \int\frac{\rho_{\bar{3}}(y)}{x-y}dy+\frac{1}{x^2-1}\mathcal{M}\,, \\
2\pi n_{\bar{3}}&= \int\frac{\rho_{2}(y)}{x-1/y}\frac{dy}{y^2}+\int\frac{\rho_{\bar{2}}(y)}{x-y}dy 
{ +\frac{1}{2} \left( \mathcal{P}_{2} + \mathcal{P}_{\bar{2}} \right)}\,. \\ 
\end{aligned}
\end{equation}
The explicit factors containing $\mathcal{P}_{2} + \mathcal{P}_{\bar{2}}$ are frame-dependent, and would not be present if we took the finite-gap limit of the Bethe-Yang equations written in the spin-chain frame. 
The limit allows us also to read off the residue of the quasi-momentum $\mathcal{E}$, that is the same for the node $2$ and $\bar{2}$.
\begin{equation}
\mathcal{E} = \frac{1}{2 \pi} \left(\frac{2}{h}L -\epsilon_1 +2 \epsilon_2 -\epsilon_3 +\epsilon_{\bar{1}} +\epsilon_{\bar{3}}
{ -\frac{2}{h} \left(\frac{1}{2}K_{1} -K_2 +\frac{1}{2}K_{3} -\frac{1}{2}K_{\bar{1}}-\frac{1}{2}K_{\bar{3}} \right) } \right),
\end{equation}
The factor $1/h$ is consistent with the fact that we have taken the length $L$ and the excitation numbers to be large, and only the ratios $L/h, K_i/h$ remain finite.
The quantity $\mathcal{M}$ reads as
\begin{equation}\label{eq:winding-finite-gap}
\mathcal{M}=\mathcal{P}_1+\mathcal{P}_3-\mathcal{P}_{\bar{1}}+2 \mathcal{P}_{\bar{2}}- \mathcal{P}_{\bar{3}}
.
\end{equation}
The finite-gap equations that we have obtained here are equivalent to the ones constructed in~\cite{Babichenko:2009dk} with the help of the Lax connection.

\section{Concluding remarks}\label{sec:perturbative-results}
In this chapter we have focused on the massive sector of {\adsthree}.
First we showed that it is possible to construct a dynamic spin-chain with $\psu(1,1|2)_{\sL}\oplus \psu(1,1|2)_{\sR}$ symmetry.
We have then derived the S-matrix governing the scattering of the spin-chain excitations, showing that it is related to the worldsheet S-matrix of the previous chapter by a change of basis for the two-particle states.

We have also solved the crossing equations for the two ``dressing factors'' of the massive sector\footnote{Solutions to the crossing equations in the massless and mixed-mass sectors were proposed in~\cite{Borsato:2016kbm}} that are not fixed by compatibility of symmetries.
Commenting on the analytical properties of our factors, we have motivated the choice of the solutions of the crossing equations.

We have also made contact with the ``finite-gap equations''---that are obtained from the Lax formulation of the classical integrable model---by taking a proper limit of the Bethe-Yang equations derived in the previous chapter.

\bigskip

Let us conclude the chapters devoted to AdS$_3$/CFT$_2$ by referring to the independent perturbative tests of the all-loop results presented here.

Tree-level scattering elements involving excitations of the massive sector of the background {\adsthree} were computed in~\cite{Hoare:2013pma}.
There the more general case in which a $B$-field is present was actually considered.
For the pure RR case, the same tree-level results had appeared in~\cite{Sundin:2013ypa}, where also certain one-loop processes in the ``near-flat-space'' limit\footnote{The near-flat-space limit is achieved by having a momentum that scales like $p\sim \lambda^{-1/4}$~\cite{Maldacena:2006rv} and at leading order it can be seen as a further expansion on top of the near-BMN limit. It was used also to eliminate apparent ultra-violet divergences arising in near-BMN worldsheet computations, that were finally resolved in~\cite{Roiban:2014cia}.} and interactions involving massless excitations were produced.
Agreement with these perturbative results was shown in~\cite{Borsato:2013qpa}.

The ``Hern\'andez-L\'opez order'' of the dressing phases in the massive sector was addressed with semiclassical methods in~\cite{Beccaria:2012kb} and~\cite{Abbott:2013ixa}. Of these two results, only the latter agrees with the findings presented in~\cite{Borsato:2013hoa} and reviewed here. The resolution of this mismatch was explained in~\cite{Abbott:2015pps}, where it was shown that the procedure of~\cite{Beccaria:2012kb} for deriving the phases must be modified by taking into account wrapping corrections due to massless virtual particles, as these are not suppressed.

The S-matrix including the proposed all-loop dressing phases was shown to agree with two-loop worldsheet calculations obtained with unitarity techniques in~\cite{Engelund:2013fja}. These are actually able to probe just the log-dependence of the scattering processes.
Different unitarity techniques that are able to account also for the rational terms were performed in~\cite{Bianchi:2014rfa}, where it was shown that the full momentum-dependence of the scattering elements in the massive sector matches at one loop.

Certain one-loop processes obtained with standard near-BMN computations confirmed again agreement with the large-tension expansion of the all-loop scattering elements~\cite{Sundin:2014sfa}.
This result was later extended to the full theory---including the massless and mixed-mass sectors---in~\cite{Sundin:2016gqe}, finding again match at one-loop with the proposed S-matrix.

In~\cite{Sundin:2014ema,Sundin:2015uva} the two-loop correction to the two-point function was computed. While for massive excitations this agrees with the expansion of the exact dispersion relation, a mismatch is found for the massless ones. At present this is still an unresolved problem, which might be explained by unexpected quantum corrections to the central charges $\gen{C},\overline{\gen{C}}$ or by ambiguities in treating massless modes in perturbation theory.

\chapter{Bosonic {\etaadsfive}}\label{ch:qAdS5Bos}
This is the first of two chapters devoted to the investigation of another integrable $\sigma$-model motivated by the AdS/CFT correspondence.
It corresponds to a particular deformation of the $\sigma$-model for strings on {\adsfive}.

Beisert and Koroteev first constructed an R-matrix invariant under a $q$-deformation of the $\su(2|2)_{\ce}$ superalgebra~\cite{Beisert:2008tw}.
After solving the crossing equation for the factor that was not fixed by the symmetries, Hoare, Hollowood and Miramontes~\cite{Hoare:2011wr} proposed an S-matrix that was conjectured to correspond to a quantum integrable model realising the $q$-deformation of the model for the AdS$_5$/CFT$_4$ dual pair.
Up to now, no explicit realisation of a $q$-deformation of $\mathcal{N}=4$ Super Yang-Mills has been constructed.

A deformation of the string $\sigma$-model on {\adsfive} was proposed by Delduc, Magro and Vicedo in~\cite{Delduc:2013qra}, building on previous results for bosonic cosets~\cite{Delduc:2013fga}.
It preserves the classical integrability of the original model, and replaces the original $\psu(2,2|4)$ symmetry with the quantum group $U_q(\psu(2,2|4))$~\cite{Delduc:2014kha}.
The parameter that is used to deform the theory was called $\eta$, and the procedure is often referred to as ``$\eta$-deformation''. We adopt this terminology here.
The deformation is of the type of the Yang-Baxter $\sigma$-model constructed by Klim\v{c}\'{i}k~\cite{Klimcik:2002zj,Klimcik:2008eq}, that generalises the work of Cherednik~\cite{Cherednik:1981df}.

In this chapter we focus on the bosonic sector of the deformed model.
For convenience, we start by reviewing the undeformed case, then we study the effects of the deformation and explain how to match with the large-tension limit of the proposed S-matrix invariant under the $q$-deformed algebra.

\section{Undeformed model}
{\adsfive} is the product of the five-dimensional Anti-de Sitter and the five-dimensional sphere.
Let us start with the compact space. We use six coordinates $Y_A, A=1,\ldots,6$ to parameterise the Euclidean space $\mathbb{R}^6$. 
The five-dimensional sphere is identified by the constraint $Y_AY_B\delta^{AB}=1$. A convenient parameterisation of these coordinates is 
\be\label{eq:sph-coord-S5}
\begin{aligned}
Y_1+iY_2 = r\cos\xi\,e^{i\p_1}\,,\quad Y_3+iY_4 = r\sin\xi\,e^{i\p_2}\,,\quad Y_5+iY_6 = \sqrt{1-r^2}\,e^{i\p_3}\,,
\end{aligned}
\ee
where $0<r<1$ is the radius of the three-sphere, and for the angles we have the ranges $0<\xi<\pi/2,\ 0<\phi_i<2\pi$.
From now on we rename $\phi_3=\phi$, as this will be the angle that we will use to fix light-cone gauge, see Section~\ref{sec:Bos-string-lcg} for a generic treatment and Section~\ref{sec:quartic-action-lcg-etaads} for the case at hand.
The metric on $\mathbb{R}^6$ ${\rm d}s^2_{\mathbb{R}^6}={\rm d}Y_A{\rm d}Y_B \delta^{AB}$ then induces the metric on the sphere
\be\label{eq:metrc-S5-sph-coord}
\begin{aligned}
{\rm d}s^2_{\text{S}^5}&=\left(1-r^2\right){\rm d}\phi^2
  +\frac{{\rm d}r^2}{ \left(1-r^2\right)}
  + r^2\left( {\rm d}\xi^2+\cos ^2\xi \, {\rm d}\phi_1^2+ \sin^2\xi\, {\rm d}\phi_2^2\right) \,.
\end{aligned}
\ee
Let us comment also on another convenient parameterisation, that will be useful in Section~\ref{sec:pert-bos-S-mat-eta-ads5s5} for implementing perturbation theory on the worldsheet.
The above constraint may be satisfied also by\footnote{For $y_i$ and also for the coordinates $z_i$ introduced later, we do not distinguish between upper or lower indices $y^i=y_i,\ z^i=z_i$.}
\be\label{eq:embed-eucl-coord-S5}
Y_1+iY_2= \frac{y_1+iy_2}{1+\frac{|y|^2}{4}}\,,
\qquad
Y_3+iY_4= \frac{y_3+iy_4}{1+\frac{|y|^2}{4}}\,,
\qquad
Y_5+iY_6= \frac{1-\frac{|y|^2}{4}}{1+\frac{|y|^2}{4}}e^{i\phi}\,,
\ee
where we have defined $|y|^2\equiv y_iy_i$ and we have $-2<y_i<2$.
The metric of the sphere in these coordinates reads as
\be\label{eq:metrc-S5-eucl-coord}
{\rm d}s^2_{\text{S}^5}=\left(\frac{1-\frac{|y|^2}{4}}{1+\frac{|y|^2}{4}}\right)^2 {\rm d}\phi^2
+\frac{{\rm d}y_i{\rm d}y_i}{\left(1+\frac{|y|^2}{4}\right)^2}\,.
\ee
The discussion for five-dimensional Anti-de Sitter follows a similar route. We embed it into $\mathbb{R}^{2,4}$ spanned by $Z_A, A=0,\ldots,5$, and we identify it with the constraint $Z^AZ^B\eta_{AB}=-1$, where $\eta_{AB}=\text{diag}(-1,1,1,1,1,-1)$.
A parameterisation---reminiscent of the one for the sphere---for which the AdS constraint is satisfied is
\be\label{eq:sph-coord-AdS5}
Z_1+iZ_2 = \r\cos\z\,e^{i\psi_1}\,,\quad Z_3+iZ_4 = \r\sin\z\,e^{i\psi_2}\,,\quad Z_0+iZ_5 = \sqrt{1+\r^2}\,e^{it}\,,
\ee
where $0<\r<\infty$, and for the angles we have the ranges $0<\zeta<\pi/2,\ 0<\psi_i<2\pi$. We take the universal cover of AdS$_5$, where $t$ is the non-compact time coordinate.
Using these local coordinates the metric for Anti-de Sitter is
\be
\begin{aligned}\label{eq:metrc-AdS5-sph-coord}
{\rm d}s^2_{\text{AdS}_5}&=-\left(1+\rho^2\right){\rm d}t^2
  +\frac{{\rm d}\rho^2}{ \left(1+\rho^2\right)}
  + \rho^2\left( {\rm d}\zeta^2+\cos ^2\zeta \, {\rm d}\psi_1^2+ \sin^2\zeta\, {\rm d}\psi_2^2\right) \,.
\end{aligned}
\ee
Also in this case we mention an alternative parameterisation that will be useful for perturbation theory
\be\label{eq:embed-eucl-coord-AdS5}
Z_1+iZ_2= \frac{z_1+iz_2}{1-\frac{|z|^2}{4}}\,,
\qquad
Z_3+iZ_4= \frac{z_3+iz_4}{1-\frac{|z|^2}{4}}\,,
\qquad
Z_5+iZ_6= \frac{1+\frac{|z|^2}{4}}{1-\frac{|z|^2}{4}}e^{it}\,,
\ee
where $|z|^2\equiv z_iz_i$ and the space is covered by $-2<z_i<2$. The metric in these coordinates is 
\be\label{eq:metrc-AdS5-eucl-coord}
{\rm d}s^2_{\text{AdS}^5}=-\left(\frac{1+\frac{|z|^2}{4}}{1-\frac{|z|^2}{4}}\right)^2 {\rm d}t^2
+\frac{{\rm d}z_i{\rm d}z_i}{\left(1-\frac{|z|^2}{4}\right)^2}\,.
\ee
These two spaces are also realised as the following cosets
\be
\text{AdS}_5: \quad \frac{\text{SU}(2,2)}{\text{SO}(4,1)} \,,
\qquad\qquad
\text{S}^5: \quad \frac{\text{SU}(4)}{\text{SO}(5)}\,.
\ee
Then the action of the string may be written in the form of a non-linear $\sigma$-model, where the base space is the worldsheet and the target space is {\adsfive}.
We do that by considering coset elements $\alg{g_a}$ and $\alg{g_s}$ that depend on the local coordinates parameterising Anti-de Sitter and the sphere.
It is natural to represent these elements in terms of $4\times 4$ matrices that satisfy a reality condition compatible with $SU(2,2)$ and $SU(4)$.
We refer to Appendix~\ref{app:bos-eta-def} for possible parameterisations.
The two group elements may be considered at the same time by defining the $8\times 8$ matrix
\be\label{eq:8x8-bos-el}
{\alg{g_b}}=
\left(
\begin{array}{cc}
\alg{g_a} & 0 \\
0 & \alg{g_s}
\end{array}
\right)\,.
\ee
In Section~\ref{sec:algebra-basis} we will realise the $\su(2,2|4)\supset \su(2,2)\oplus \su(4)$ algebra in terms of $8\times 8$ matrices, making the above definition naturally motivated.
After constructing the current $A\equiv -{\alg{g_b}}^{-1}{\rm d}{\alg{g_b}}$ that is an element of the algebra $\su(2,2)\oplus\su(4)$, we have to decompose it into $A=A^++A^-$, where $A^+$ belongs to the denominator of the coset, while $A^-$ to its complement\footnote{The subspaces with labels $+$ and $-$ appearing in this chapter correspond to the subspaces of grading $0$ and $2$ respectively of Section~\ref{sec:algebra-basis}}. In particular 
\be
\begin{aligned}
&A^{\alg{a}+}\in \so(4,1), \qquad &&A^{\alg{a}-}\in \su(2,2)\setminus \so(4,1),\\
&A^{\alg{s}+}\in \so(5), \qquad &&A^{\alg{s}-}\in \su(4)\setminus \so(5)\,.
\end{aligned}
\ee
Then the action for the bosonic string may be written as
\be
S^\bos=-\frac{g}{2}\int {\rm d}\tau{\rm d}\sigma\, (\g^{\a\b}-\eps^{\a\b}) \Str\left(A^{-}_{\a}A^{-}_{\b}\right)\,,
\ee
where we need to define a graded trace that we call supertrace\footnote{The minus sign in front of the trace for the sphere contribution is motivated by the fact that we want the correct signature for this space. It becomes natural when we think of the full $\psu(2,2|4)$ algebra, see Section~\ref{sec:algebra-basis}.} $\Str\equiv \tr_{\alg{a}}-\tr_{\alg{s}}$.
It is easy to check that the contribution with $\eps^{\a\b}$ vanishes.
Therefore, after choosing an explicit coset representative and rewriting the action in the Polyakov form~\eqref{eq:bos-str-action} of Section~\ref{sec:Bos-string-lcg}, we find that the $B$-field is zero
\begin{equation}\label{eq:bos-act-undef-adsfive}
\begin{aligned}
S^{\bos}&= -\frac{g}{2}\int \, {\rm d}\sigma {\rm d} \tau\  \gamma^{\alpha\beta} \partial_\alpha X^M \partial_\beta X^N G_{MN}\,.
\end{aligned}
\end{equation}
If we use coordinates $X^0,\ldots,X^4$ to parameterise AdS$_5$ and $X^5,\ldots, X^9$ for S$^5$, the metric $G_{MN}$ is in block form, with the upper-left block containing the AdS$_5$ metric and the lower-right block the S$^5$ metric.
If we use the coset representatives of Eq.~\eqref{basiccoset} we find the metrics in the form~\eqref{eq:metrc-S5-sph-coord} and~\eqref{eq:metrc-AdS5-sph-coord}, while~\eqref{eq:eucl-bos-coset-el} yields the metrics~\eqref{eq:metrc-S5-eucl-coord} and~\eqref{eq:metrc-AdS5-eucl-coord}.

\section{Deformed model}\label{sec:def-bos-model}
The deformed model is obtained by inserting a linear operator acting on one of the two currents in the action of the non-linear $\sigma$-model~\cite{Delduc:2013fga,Delduc:2013qra}.
For the deformation of the full supercoset $\sigma$-model we refer to Section~\ref{sec:def-lagr-supercos}. Here it will be enough to notice that when restricted to the bosonic model, the deformed action may be written as
\be
\tilde{S}^{\bos}=-\frac{\tilde{g}}{2}\int{\rm d}\sigma {\rm d} \tau \ \left(\gamma^{\a\b}-\eps^{\a\b}\right)\, \Str\left(A^{-}_{\a}\cdot\op^{-1}_{\bos}(A^{-}_{\beta})\right) \, ,
\ee
where $\op^{-1}_{\bos}$ is the inverse of the linear operator
\be\label{eq:def-op-def-bos-mod}
\op_{\bos}=\gen{1}-\frac{2\eta}{1-\eta^2} R_{\alg{g_b}} \circ P^{(-)}\,,
\ee
mapping the algebra $\su(2,2)\oplus\su(4)$ to itself.

The deformation parameter is $\eta\in]-1,1[$, where the range is chosen to have invertibility for $\op_{\bos}$. Setting $\eta=0$ we recover the undeformed model.
The definition of $\op_{\bos}$ depends on the composition of the operators $P^{(-)}$ and $R_{\alg{g_b}}$. The former is the projector onto the component ``$-$'' of the algebra, while the latter is defined as
\be\label{eq:defin-Rg-bos}
R_{\alg{g_b}} = \text{Adj}_{{\alg{g_b}}^{-1}} \circ R \circ \text{Adj}_{{\alg{g_b}}}\,,
\ee
meaning that its action on a matrix $M$ is $R_{\alg{g_b}}(M) = {\alg{g_b}}^{-1}R({\alg{g_b}} M{\alg{g_b}}^{-1}){\alg{g_b}}$.
The linear operator $R$ satisfies the \emph{modified classical Yang-Baxter equation}
\be\label{eq:mod-cl-YBeq-R}
[R(M),R(N)]-R([R(M),N]+[M,R(N)])=[M,N]\,.
\ee
According to the definition given in~\cite{Delduc:2013fga,Delduc:2013qra}, it multiplies by $-i$ and $+i$ generators associated with positive and negative roots respectively, and by $0$ Cartan generators. Strictly speaking it is defined on the complexified algebra, and what we will use is its restriction to $\su(2,2)\oplus\su(4)$. On elements of the algebra written as $8\times 8$ matrices, we may write its action as
\be
R(M)_{ij} = -i\, \eps_{ij} M_{ij}\,,\quad \eps_{ij} = \left\{\begin{array}{ccc} 1& \rm if & i<j \\
0&\rm if& i=j \\
-1 &\rm if& i>j \end{array} \right.\,.
\ee
The action for the deformed model is multiplied by $\tilde{g}$ that plays the role of the effective string tension, related to the one of the undeformed theory $g$ by\footnote{Our $\e$-dependent prefactor  differs from the one in \cite{Delduc:2013qra}. Our choice is necessary to match the perturbative worldsheet scattering matrix with the $q$-deformed one.}
\be
\label{eq:def-g-tilde}
\tilde{g}=\frac{1+\eta^2}{1-\eta^2}\, g\,.
\ee
To obtain better expressions we also introduce a new deformation parameter related to $\eta$ as
\be
\varkappa=\frac{2\eta}{1-\eta^2}\,,
\qquad\qquad
0<\vk<\infty\,,
\ee
which as we will see is a convenient choice.
In order to compute the action for the deformed theory one has to first study the operator $\op_\bos$ and invert it.
From its definition it is clear that $\op_\bos$ acts as the identity operator on the $10+10$ generators of $\so(4,1)\oplus\so(5)\subset\su(2,2)\oplus\su(4)$.
When acting on the $5+5$ generators of the coset $\su(2,2)\oplus\su(4)\setminus\so(4,1)\oplus\so(5)$ we see that we never mix generators of Anti-de Sitter and of the sphere. In Appendix~\ref{app:bosonic-op-and-inverse} we provide the explicit results for the inverse operator $\op^{-1}_{\bos}$.

When we put the action of the deformed model in the form presented in~\eqref{eq:bos-str-action}
\begin{equation}\label{eq:bos-lagr-eta-def-Pol}
S^{\bos}=-\frac{\tilde{g}}{2} \int \, {\rm d}\sigma {\rm d} \tau\  \left( \, \gamma^{\alpha\beta} \partial_\alpha X^M \partial_\beta X^N \widetilde{G}_{MN} -\epsilon^{\alpha\beta} \partial_\alpha X^M \partial_\beta X^N \widetilde{B}_{MN} \right)\,,
\end{equation}
we find that the metric is deformed and that a $B$-field is generated.
The result is particularly simple when expressed in terms of the coordinates~\eqref{eq:metrc-S5-sph-coord} and~\eqref{eq:metrc-AdS5-sph-coord}, related to the coset representative~\eqref{basiccoset}. 
\begin{figure}[t]
  \centering
\includegraphics[width=0.4\textwidth]{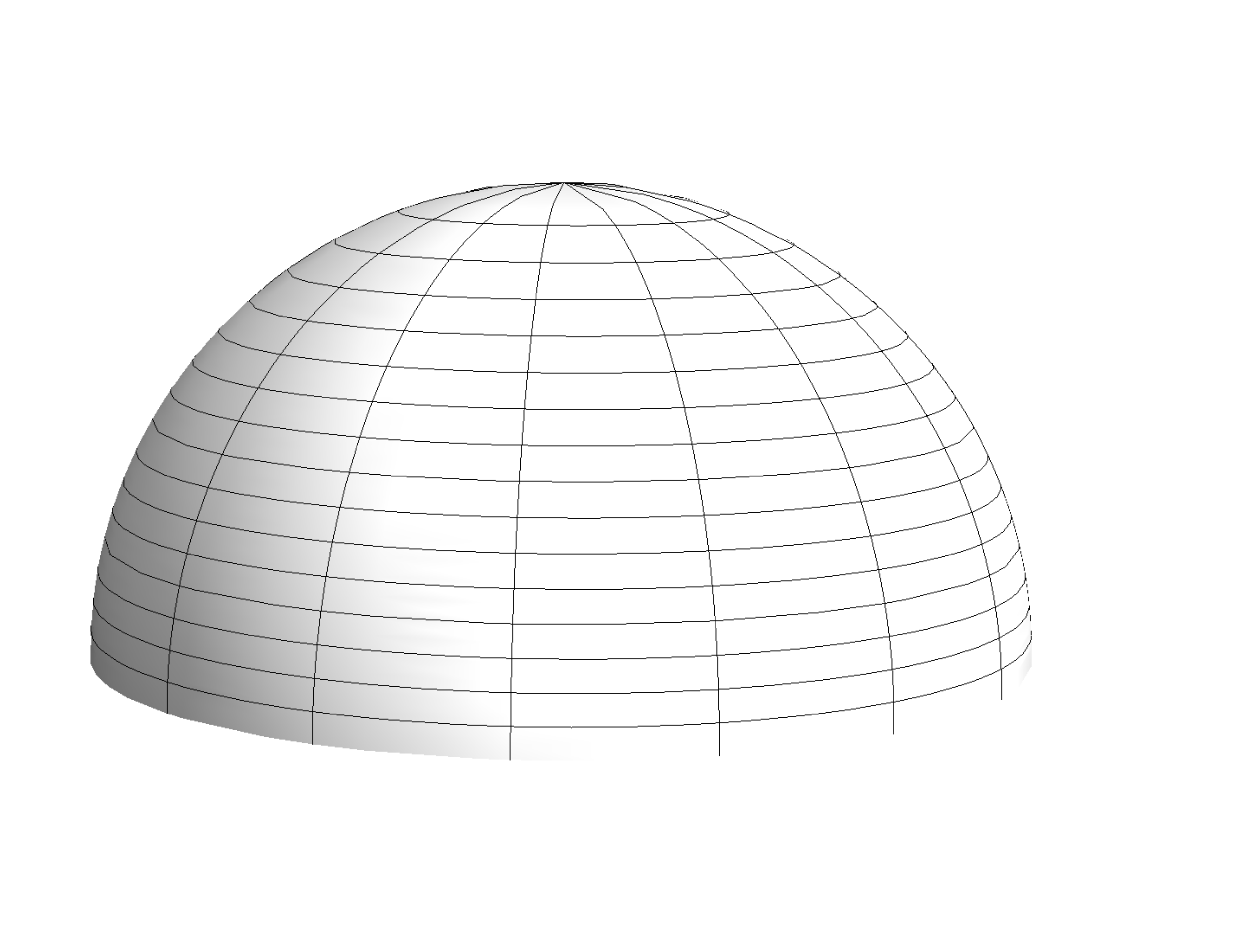}
\raisebox{-0.4cm}{
\includegraphics[width=0.4\textwidth]{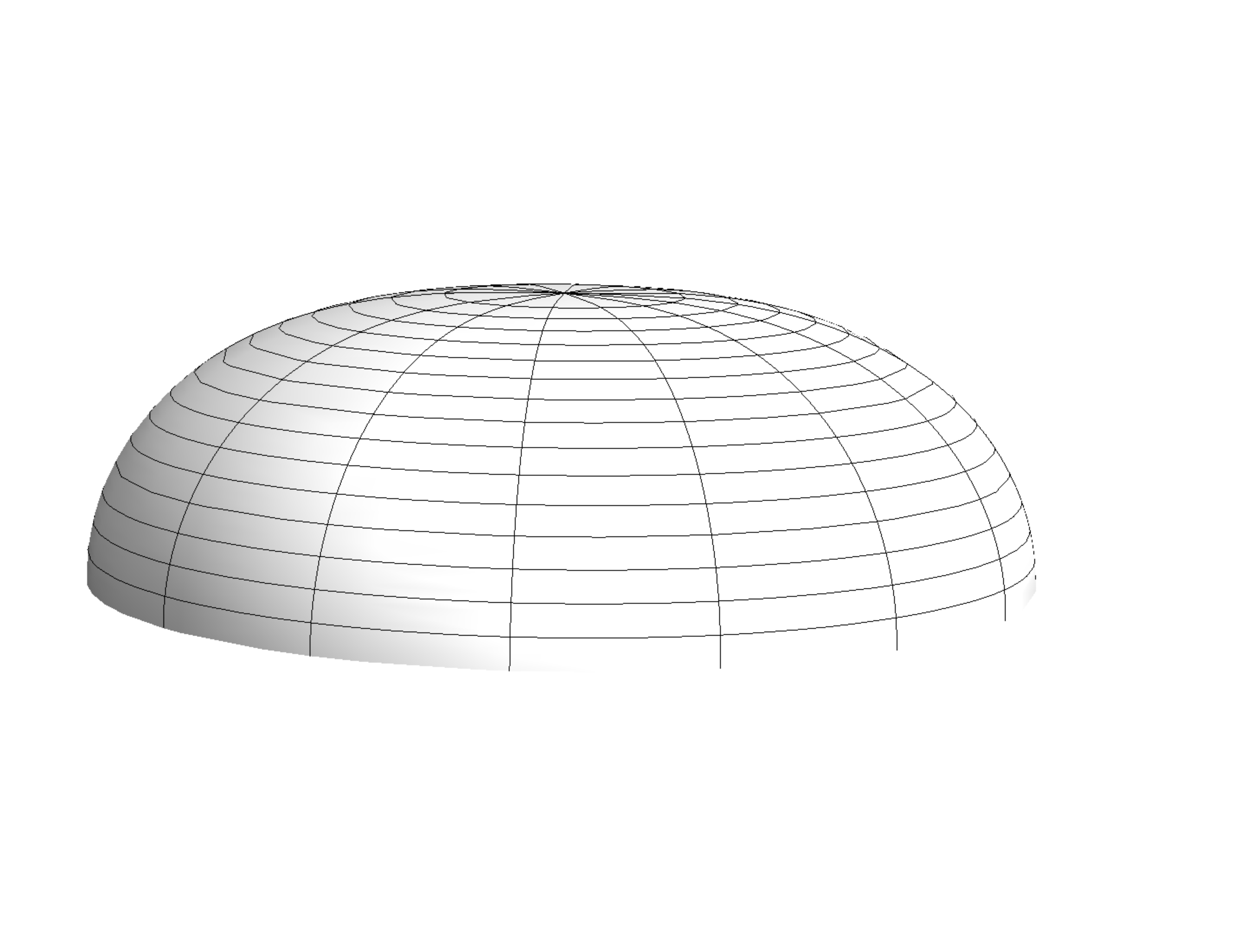}
}
  \caption{The left figure represents the hemisphere parameterised by the coordinates $r\in[0,1]$ and $\phi\in[0,2\pi]$. On the right we draw the squashed hemisphere that we find when we turn on the deformation. The figure was generated with deformation parameter $\vk=1$.}
  \label{fig:eta-def-sphere}
\end{figure}
\begin{figure}[t]
  \centering
\raisebox{2.5cm}{
\includegraphics[width=0.2\textwidth]{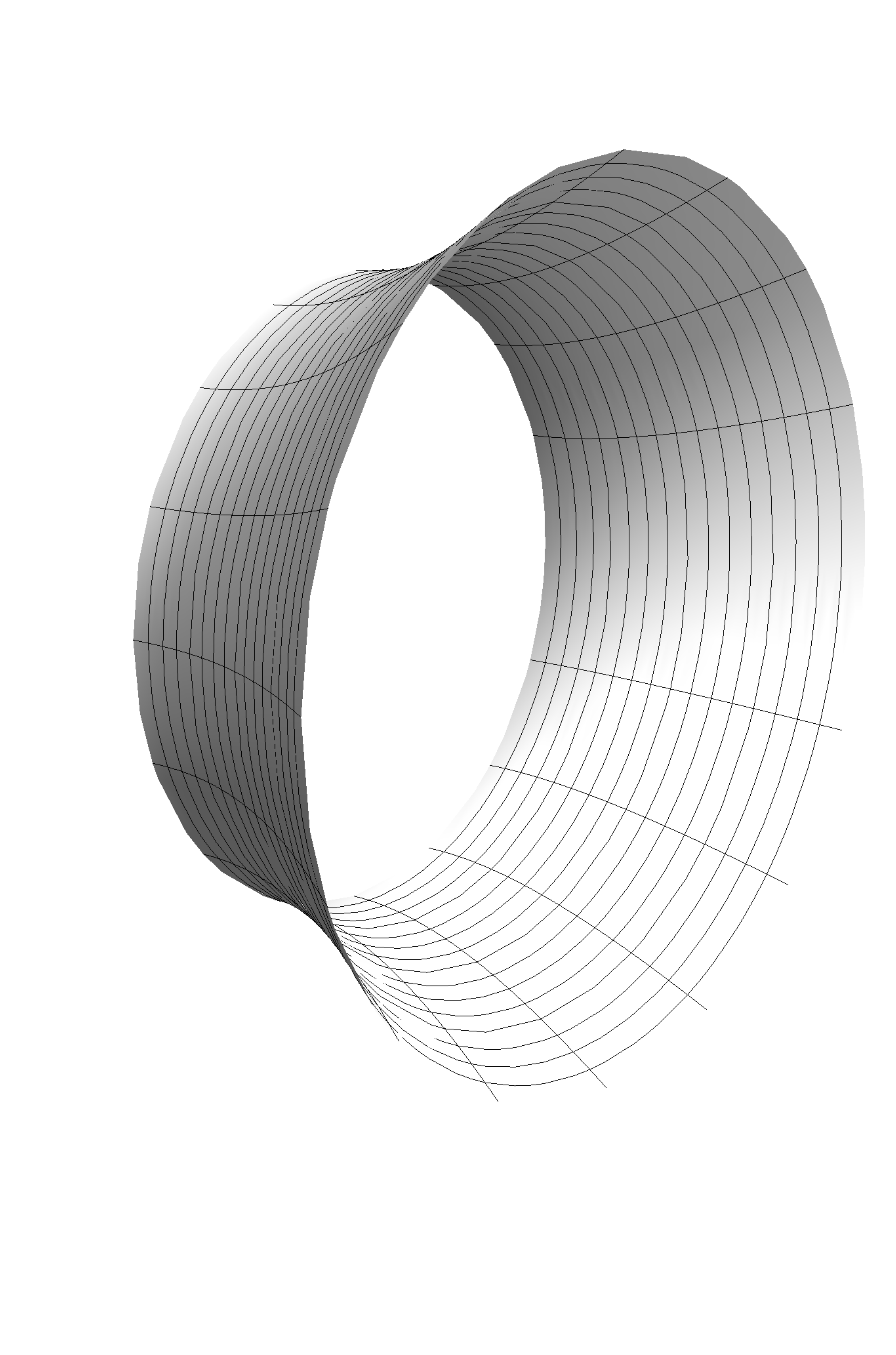}
}
\includegraphics[width=0.4\textwidth]{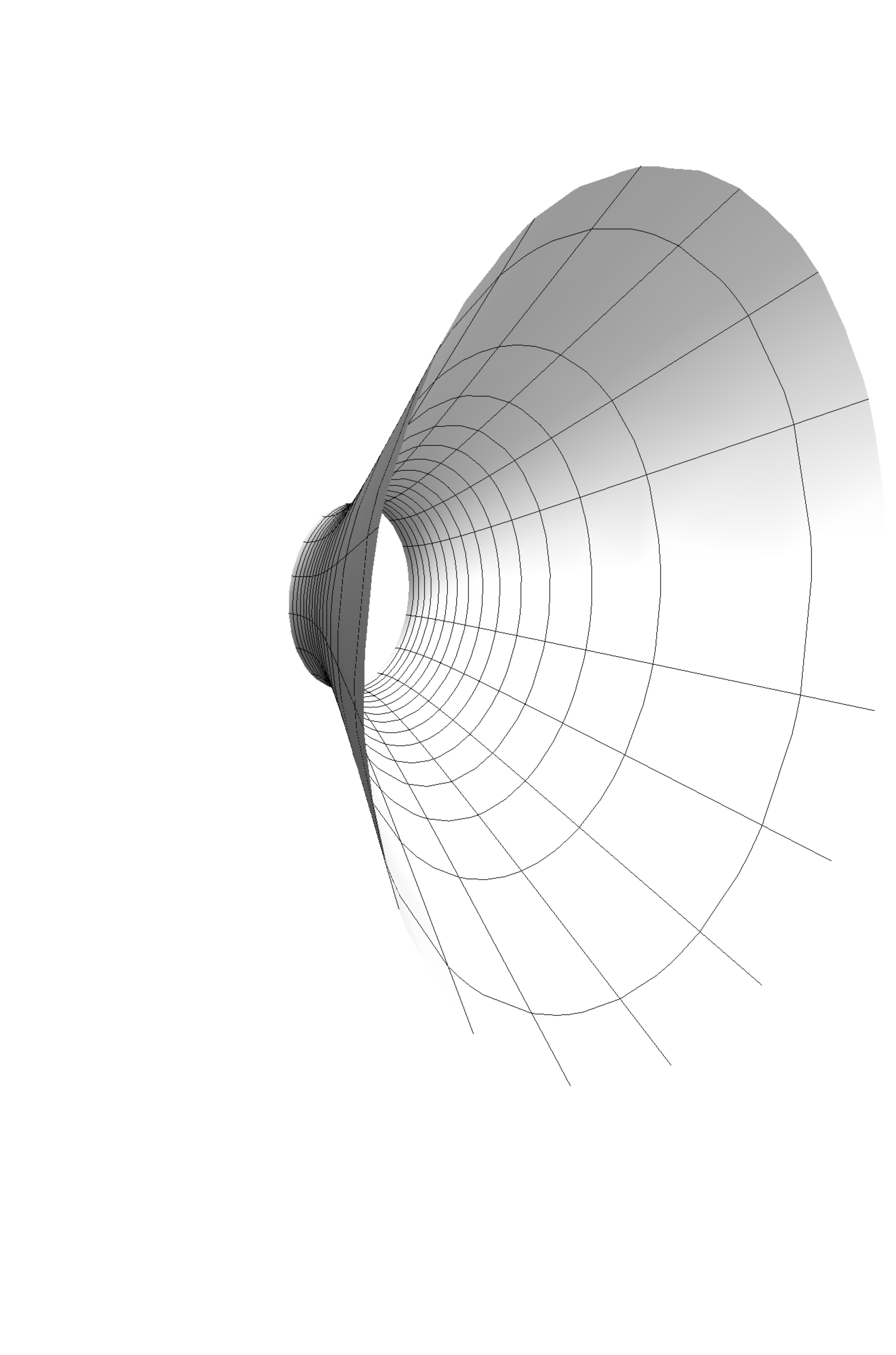}
  \caption{In the left figure we draw the space parameterised by the coordinates $\rho\in[0,\infty[$---here we stop the range of $\rho$ at the value $1$---and $t\in[0,2\pi]$. On the right we draw its deformation, generated with $\vk=1$. The right figure has actually been rescaled to fit the page: the circles at $\rho=0$ have the same radius in the two cases.}
  \label{fig:eta-def-AdS}
\end{figure}
In particular, we find that the metrics for the deformed AdS$_5$ and the deformed S$^5$ are~\cite{Arutyunov:2013ega}
\be\label{eq:metrc-etaAdS5S5-sph-coord}
\begin{aligned}
{\rm d}s^2_{(\text{AdS}_5)_{\eta}}=&-\frac{1+\rho^2}{1-\vk^2\rho^2}{\rm d}t^2
  +\frac{{\rm d}\rho^2}{ \left(1+\rho^2\right)(1-\vk^2\rho^2)}\\
&  + \frac{\rho^2}{1+\vk^2\rho^4\sin^2\zeta}\left( {\rm d}\zeta^2+\cos ^2\zeta \, {\rm d}\psi_1^2\right) 
+\rho^2 \sin^2\zeta\, {\rm d}\psi_2^2\,,
\\
\\
{\rm d}s^2_{(\text{S}^5)_{\eta}}=&\frac{1-r^2}{1+\vk^2 r^2}{\rm d}\phi^2
  +\frac{{\rm d}r^2}{ \left(1-r^2\right)(1+\vk^2 r^2)}\\
&  + \frac{r^2}{1+\vk^2r^4\sin^2\xi}\left( {\rm d}\xi^2+\cos ^2\xi \, {\rm d}\phi_1^2\right) 
+r^2 \sin^2\xi\, {\rm d}\phi_2^2\,.
\end{aligned}
\ee
Figure~\ref{fig:eta-def-sphere} and~\ref{fig:eta-def-AdS} represent the effect of the deformation on the sphere and on AdS.
We find the $B$-field $B=\frac{1}{2} B_{MN}\ {\rm d}X^M\wedge {\rm d}X^N$~\cite{Arutyunov:2013ega}
\be\label{eq:B-field-etaAdS5S5-sph-coord}
\begin{aligned}
\widetilde{B}_{(\text{AdS}_5)_{\eta}} &= +\frac{\vk}{2} \left( \frac{\rho^4 \sin (2\zeta)}{1+\vk^2 \rho^4\sin^2 \zeta} {\rm d}\psi_1\wedge{\rm d}\zeta + \frac{2 \rho}{1-\vk^2 \rho^2}{\rm d}t\wedge{\rm d}\rho\right),
\\
\widetilde{B}_{(\text{S}_5)_{\eta}} &= -\frac{\vk}{2} \left( \frac{r^4 \sin (2\xi)}{1+\vk^2 r^4\sin^2 \xi}{\rm d}\phi_1\wedge{\rm d}\xi + \frac{2r}{1+\vk^2 r^2}{\rm d}\phi\wedge{\rm d}r\right).
\end{aligned}
\ee
It is easy to see that the contributions of the components $B_{t\rho}$ and $B_{\phi r}$ to the Lagrangian are total derivatives, meaning that they can be ignored.

We refer to Appendix~\ref{app:bos-lagr-eta-def} for the Lagrangian written in the coordinates $(t,z_i)$ and $(\phi,y_i)$.
Let us note that in the undeformed case the action is invariant with respect to two copies of ${\rm SO}(4)$, one of them corresponding to rotations of $z_i$ and the other copy of $y_i$. 
In the above action this symmetry is broken down to four copies of ${\rm SO}(2)\sim {\rm U}(1)$, corresponding to shifts of the angles $\psi_i$ and $\phi_i$. Thus, together with the two ${\rm U}(1)$ isometries acting on 
$t$ and $\phi$, the deformed action is invariant under ${\rm U}(1)^3\times {\rm U}(1)^{3}$.
We also find that the range of $\r$ is reduced under the deformation $0\le \r\le 1/\vk$, to preserve the time-like nature of $t$.
The (string frame) metric of the deformed AdS is singular at $\r=1/\vk$. 
This is not just a coordinate singularity, as the Ricci scalar has a pole there.
Without knowing the dilaton it is unclear whether the Einstein-frame metric exhibits the same singularity.

\section{Perturbative bosonic worldsheet S-matrix}\label{sec:pert-bos-S-mat-eta-ads5s5}
In this Section we want to compute the perturbative S-matrix governing worldsheet scattering between two bosonic excitations.
We will then compare it with the large-tension limit of the $q$-deformed S-matrix proposed in~\cite{Hoare:2011wr} and find agreement.

\subsection{Quartic action in light-cone gauge}\label{sec:quartic-action-lcg-etaads}
Since we are interested in the perturbative expansion in powers of fields around $\rho=0,\ r=0$, we first expand the full bosonic Lagrangian up to quartic order in $\r$, $r$ and their derivatives. To simplify the result we also make the shifts of $\rho$ and $r$ as described in Appendix \ref{app:bos-lagr-eta-def}, {\it c.f.}~\eqref{shift}.
We then change the spherical coordinates to the Euclidean coordinates $(z_i,y_i)_{i=1,\ldots,4}$ introduced in~\eqref{eq:embed-eucl-coord-S5} and~\eqref{eq:embed-eucl-coord-AdS5}---as they are the preferred ones for perturbation theory---and we further expand the resulting action up to the quartic order in $z$ and $y$ fields.  
In this way we find the Lagrangian up to quartic order $\lagr=\lagr^{G,\alg{a}}+\lagr^{B,\alg{a}}+\lagr^{G,\alg{s}}+\lagr^{B,\alg{s}}$, where we have separated the contributions of AdS$_5$ from the ones of $S^5$, and the contributions of the metric $G_{MN}$ from the ones of the $B$-field
\begin{equation}\la{Lquart}
\begin{aligned}
\lagr^{G,\alg{a}} &= -\frac{\tilde{g}}{2}  \,  \gamma^{\alpha \beta} \Bigg[ -\left( 1 + (1+\varkappa^2) |z|^2 +\frac{1}{2}(1+\varkappa^2)^2|z|^4\right) \pa_{\a}t \pa_{\b}t \\
&\qquad\qquad + \left(1+(1-\varkappa^2)\frac{|z|^2}{2}\right) \pa_{\a}z_i\pa_{\beta}z_i \Bigg] \, , \\
\lagr^{B,\alg{a}} &= +2\tilde{g} \,   \varkappa (z_3^2+z_4^2) \epsilon^{\a\b} \pa_{\a}z_1 \pa_{\b}z_2 \, , \\
\lagr^{G,\alg{s}}&= -\frac{\tilde{g}}{2}\,  \gamma^{\alpha \beta} \Bigg[ \left( 1 - (1+\varkappa^2) |y|^2 +\frac{1}{2}(1+\varkappa^2)^2|y|^4\right) \pa_{\a}\phi \pa_{\b}\phi \\
&\qquad\qquad+ \left(1-\frac{1}{2}(1-\varkappa^2)|y|^2 \right) \pa_{\a}y_i\pa_{\beta}y_i  \Bigg] \, ,\\
\lagr^{B,\alg{s}}&= - 2\tilde{g}\,    \varkappa (y_3^2+y_4^2) \epsilon^{\a\b} \pa_{\a}y_1 \pa_{\b}y_2\, .
\end{aligned}
\end{equation}
Here we use the notation $|z|\equiv(z_iz_i)^{1/2},\ |y|\equiv(y_iy_i)^{1/2}$.
The ``metric part'' of this Lagrangian has a manifest ${\rm SO}(4)\times {\rm SO(4)}$ symmetry at quartic order, which is however broken by the Wess-Zumino terms. 

\medskip

We first need to impose the uniform light-cone gauge, as explained more generally in Section~\ref{sec:Bos-string-lcg}.
We follow exactly the same notation and conventions.
After that we take the decompactification limit and perform the large-tension expansion presented in Section~\ref{sec:decomp-limit}
The gauge-fixed action is organised in the form
\begin{equation}\label{eq:large-tens-exp-eta-def}
S= \int {\rm d}\tau {\rm d} \sigma \, \left( p_\mu \dot{x}^\mu - \mathcal{H}_2 - \frac{1}{g} \mathcal{H}_4 - \ldots \right),
\end{equation}
where we find the quadratic Hamiltonian 
\begin{equation}\label{eq:quadr-hamilt-eta-def}
\mathcal{H}_2 = \frac{1}{2} p_\mu^2 + \frac{1}{2} (1+\varkappa^2) (X^\mu)^2 + \frac{1}{2} (1+\varkappa^2) (X'^\mu)^2.
\end{equation}
The quartic Hamiltonian in a general $a$-gauge is 
\begin{equation}
\begin{aligned}
\mathcal{H}_4 &= \frac{1}{4} \Bigg( (2 \varkappa^2 |z|^2 -(1+\varkappa^2) |y|^2 ) |p_z|^2 - (2 \varkappa^2 |y|^2 -(1+\varkappa^2) |z|^2 ) |p_y|^2 \\
&+\left(1+\varkappa ^2\right) \left(\left(2 |z|^2-\left(1+\varkappa ^2\right) |y|^2\right)|z'|^2 + \left(\left(1+\varkappa ^2\right) |z|^2-2
   |y|^2\right)|y'|^2\right)\Bigg) \\
&- 2 \varkappa  \left(1+\varkappa ^2\right)^{1\ov2} \left(\left(z_3^2+z_4^2\right) \left(p_{z_1} z_2'-p_{z_2} z_1'\right) -  \left(y_3^2+y_4^2\right) \left(p_{y_1} y_2'-p_{y_2} y_1'\right) \right) \\
&+\frac{(2a-1)}{8} \Bigg( (|p_y|^2+|p_z|^2)^2 -(1+\varkappa^2)^2 (|y|^2+|z|^2)^2  \\
&+2 (1+\varkappa^2)(|p_y|^2+|p_z|^2)(|y'|^2+|z'|^2)+(1+\varkappa^2)^2 (|y'|^2+|z'|^2)^2 -4 (1+\varkappa^2) (x_-')^2\Bigg).
\end{aligned}
\end{equation}
Here we use the notation $|p_z|\equiv  (p_{z_i}p_{z_i})^{1/2}, \ |p_y|\equiv ( p_{y_i} p_{y_i})^{1/2}$. 
 
To simplify the quartic piece, we  
can remove the terms of the form $|p_z|^2|y|^2$ and $|p_y|^2|z|^2$ by performing a canonical transformation generated by
\begin{equation}
V= \frac{(1+\varkappa^2)}{4} \int {\rm d}\sigma \Big( p_{y_i} y_{i} |z|^2 -p_{z_i} z_{i} |y|^2  \Big) .
\end{equation}
After this is done, the quartic Hamiltonian reads as
\begin{equation}
\begin{aligned}
\mathcal{H}_4 &= \frac{(1+\varkappa^2)}{2} ( |z|^2 |z'|^2- |y|^2 |y'|^2 ) + \frac{(1+\varkappa^2)^{2}}{2} (|z|^2 |y'|^2-|y|^2 |z'|^2) \\
&+\frac{\varkappa^2}{2} ( |z|^2  |p_z|^2 - |y|^2  |p_y|^2 )  \\
&- 2\varkappa(1+\varkappa^2)^{1\ov2}  \left[\left(z_3^2+z_4^2\right) \left(p_{z_1} z_2'-p_{z_2} z_1'\right) -  \left(y_3^2+y_4^2\right) \left(p_{y_1} y_2'-p_{y_2} y_1'\right) \right]  \\
&+\frac{(2a-1)}{8}\Bigg( (|p_y|^2+|p_z|^2)^2 -  (1+\varkappa^2)^2 (|y|^2+|z|^2)^2  \\
&+2  (1+\varkappa^2) (|p_y|^2+|p_z|^2)(|y'|^2+|z'|^2)+ (1+\varkappa^2)^2 (|y'|^2+|z'|^2)^2 -4  (1+\varkappa^2) (x_-')^2\Bigg). 
\end{aligned}
\end{equation}

We recall that in the undeformed case the full theory---including both the bosons and the fermions---is invariant with respect to the two copies of the centrally extended superalgebra $\psu(2|2)$, each containing two
$\su(2)$ subalgebras.  To render invariance under $\su(2)$ subalgebras manifest, one can introduce the two-index notation for the worldsheet fields. It is convenient to adopt the same 
notation also for the deformed case\footnote{This parameterisation is different from the one used in \cite{Arutyunov:2009ga}, as we exchange the definitions for $Y^{1\dot{1}}$ and $Y^{2\dot{2}}$ and the definitions for $Y^{1\dot{2}}$ and $Y^{2\dot{1}}$. This does not matter in the undeformed case but is needed here in order to correctly match the perturbative S-matrix with the $q$-deformed one computed from symmetries.}
\begin{equation}
\begin{aligned}
&Z^{3\dot{4}} =\tfrac{1}{2} (z_3-i z_4),  \qquad &Z^{3\dot{3}} =\tfrac{1}{2} (z_1-i z_2), \\
& Z^{4\dot{3}}=-\tfrac{1}{2} (z_3+i z_4),    \qquad &Z^{4\dot{4}}=\tfrac{1}{2} (z_1+i z_2), 
\end{aligned}
\end{equation}   
\begin{equation}
\begin{aligned}
&Y^{1\dot{2}}=-\tfrac{1}{2} (y_3+i y_4), \qquad   &Y^{1\dot{1}}=\tfrac{1}{2} (y_1+i y_2), \\
&Y^{2\dot{1}}=\tfrac{1}{2} (y_3-i y_4), \qquad &Y^{2\dot{2}}=\tfrac{1}{2} (y_1-i y_2)\, .
\end{aligned}
\end{equation}
In terms of two-index fields the quartic Hamiltonian becomes $\mathcal{H}_4 = \mathcal{H}^G_4 + \mathcal{H}^{B}_4$,
where $\mathcal{H}^G_4$ is the contribution coming from the spacetime metric and $\mathcal{H}^{B}_4 $ from the $B$-field
{\small
\bea\label{eq:quartic-hamilt-eta-def}
\mathcal{H}^G_4 &=&2(1+\varkappa^2) \left( Z_{\alpha\dot{\alpha}} Z^{\alpha\dot{\alpha}} Z'_{\beta\dot{\beta}} Z'^{\beta\dot{\beta}} -Y_{a\dot{a}}Y^{a\dot{a}} Y'_{b\dot{b}}Y'^{b\dot{b}} \right) \nonumber \\
&+& 2(1+\varkappa^2)^{2} \left( Z_{\alpha\dot{\alpha}} Z^{\alpha\dot{\alpha}} Y'_{b\dot{b}}Y'^{b\dot{b}} - Y_{a\dot{a}}Y^{a\dot{a}} Z'_{\beta\dot{\beta}} Z'^{\beta\dot{\beta}} \right) \nonumber \\
& +&\frac{\varkappa^2}{2} \left( Z_{\alpha\dot{\alpha}} Z^{\alpha\dot{\alpha}} P_{\beta\dot{\beta}} P^{\beta\dot{\beta}} - Y_{a\dot{a}}Y^{a\dot{a}} P_{b\dot{b}}P^{b\dot{b}} \right) \\
&+&\frac{(2a-1)}{8} \Bigg( \frac{1}{4}(P_{a\dot{a}}P^{a\dot{a}}+P_{\alpha\dot{\alpha}}P^{\alpha\dot{\alpha}})^2 -4 (1+\varkappa^2)^2 (Y_{a\dot{a}}Y^{a\dot{a}}+Z_{\alpha\dot{\alpha}}Z^{\alpha\dot{\alpha}})^2 \nonumber  \\
&+&2 (1+\varkappa^2) (P_{a\dot{a}}P^{a\dot{a}}+P_{\alpha\dot{\alpha}}P^{\alpha\dot{\alpha}})(Y'_{a\dot{a}}Y'^{a\dot{a}}+Z'_{\alpha\dot{\alpha}}Z'^{\alpha\dot{\alpha}})+4 (1+\varkappa^2)^2 (Y'_{a\dot{a}}Y'^{a\dot{a}}+Z'_{\alpha\dot{\alpha}}Z'^{\alpha\dot{\alpha}})^2 \nonumber \\
&-&4 (1+\varkappa^2) (P_{a\dot{a}}Y'^{a\dot{a}} +P_{\alpha\dot{\alpha}}Z'^{\alpha\dot{\alpha}})^2\Bigg), \nonumber \\
\mathcal{H}^{B}_4 &=& 8 i\varkappa(1+\varkappa^2)^{1\ov2} \left( Z^{3\dot{4}} Z^{4\dot{3}} ( P_{3\dot{3}} Z'^{3\dot{3}} -P_{4\dot{4}} Z'^{4\dot{4}} ) + Y^{1\dot{2}} Y^{2\dot{1}} ( P_{1\dot{1}} Y'^{1\dot{1}} -P_{2\dot{2}} Y'^{2\dot{2}} ) \right)\, . \nonumber
\eea
}
Here indices are raised and lowered with the $\epsilon$-tensors, where $\epsilon^{12}=-\epsilon_{12}=\epsilon^{34}=-\epsilon^{34}=+1$, and similarly for dotted indices.
Note that we have used the Virasoro constraint $C_1=0$ given in~\eqref{eq:Vira-constr-bos} in order to express $x'_-$ in terms of the two index fields.
The gauge dependent terms multiplying $(2a-1)$ are invariant under SO(8) as in the underformed case.

\subsection{Tree-level bosonic S-matrix}
The computation of the tree level bosonic S-matrix follows the route reviewed in Section~\ref{sec:large-tens-exp}, see~\cite{Arutyunov:2009ga} for more details.
We first quantise the theory by introducing creation and annihilation operators as
\be
\begin{aligned}
Z^{\a\dot{\a}}(\sigma,\tau)&=\frac{1}{\sqrt{2\pi}}\int {\rm d}p\frac{1}{2\sqrt{\omega_p}}\left( 
e^{ip\sigma}a^{\a\dot{\a}}(p,\tau)+e^{-ip\sigma}\eps^{\a\b}\eps^{\dot{\a}\dot{\b}}a^\dagger_{\b\dot{\b}}(p,\tau)\right)\,,
\\
Y^{a\dot{a}}(\sigma,\tau)&=\frac{1}{\sqrt{2\pi}}\int {\rm d}p\frac{1}{2\sqrt{\omega_p}}\left( 
e^{ip\sigma}a^{a\dot{a}}(p,\tau)+e^{-ip\sigma}\eps^{ab}\eps^{\dot{a}\dot{b}}a^\dagger_{b\dot{b}}(p,\tau)\right)\,,
\\
P_{\a\dot{\a}}(\sigma,\tau)&=\frac{1}{\sqrt{2\pi}}\int {\rm d}p\, i\, \sqrt{\omega_p}\left( 
e^{-ip\sigma}a^\dagger_{\a\dot{\a}}(p,\tau)-e^{ip\sigma}\eps_{\a\b}\eps_{\dot{\a}\dot{\b}}a^{\b\dot{\b}}(p,\tau)\right)\,,
\\
P_{a\dot{a}}(\sigma,\tau)&=\frac{1}{\sqrt{2\pi}}\int {\rm d}p\, i\, \sqrt{\omega_p}\left( 
e^{-ip\sigma}a^\dagger_{a\dot{a}}(p,\tau)-e^{ip\sigma}\eps_{ab}\eps_{\dot{a}\dot{b}}a^{b\dot{b}}(p,\tau)\right)\,,
\end{aligned}
\ee
where the frequency $\omega_p$ is related to the momentum $p$ as 
\be\la{omega}
\omega_p=(1+\varkappa^2)^{1\ov2} \sqrt{1+p^2} = \sqrt{1+p^2\ov 1-\dpnu^2}\,,
\ee
and we have introduced a new convenient parameterisation of the deformation as
\be
\dpnu = {\varkappa\ov (1+\varkappa^2)^{1\ov2}}={2\e\ov 1+\e^2}\,.
\ee
We compute the T-matrix defined by~\eqref{eq:def-Tmat} using Equation~\eqref{eq:pert-Tmat}.
The free Hamiltonian governing the dynamics of in- and out-states is found by rewriting the quadratic Hamiltonian~\eqref{eq:quadr-hamilt-eta-def} in terms of the oscillators $a^\dagger,a$. At leading order in the large-tension expansion the potential $\gen{V}$ is essentially the quartic Hamiltonian~\eqref{eq:quartic-hamilt-eta-def} written for $a^\dagger,a$
\be
\gen{V}=\frac{1}{g}\ \gen{H}_4+\mathcal{O}(1/g^2)\,.
\ee
The additional power of $1/g$ comes from the expansion in powers of fields~\eqref{eq:field-expansion}, as it is seen also in~\eqref{eq:large-tens-exp-eta-def}.

It is convenient to rewrite the tree-level S-matrix as a sum of two terms $\mathbb{T}=\mathbb{T}^G + \mathbb{T}^{B}$, coming from $\mathcal{H}^G_4$ and $\mathcal{H}^{B}_4$ respectevely. The reason is that $\mathbb{T}^G$ preserves the $\alg{so}(4)\oplus \alg{so}(4)$ symmetry, while $\mathbb{T}^{B}$ breaks it.
To write the results we consider states with momenta $p,p'$---and corresponding frequencies $\omega,\omega'$---and we always assume that $p>p'$.
To have a nicer notation, we denote the states found by acting with the creation operators on the vacuum by $\ket{Z_{\a\dot{\a}}}\equiv a^\dagger_{\a\dot{\a}}\ket{\vacuum}, \ \ket{Y_{a\dot{a}}}\equiv a^\dagger_{a\dot{a}}\ket{\vacuum}$.
The action of $\mathbb{T}^G$ on the two-particle states is given by\footnote{Here a $'$ on a state is used when the corresponding momentum is $p'$.}
\begin{equation}
\label{Tmatrix}
\begin{aligned}
\mathbb{T}^G \, \ket{Y_{a\dot{c}} Y_{b\dot{d}}'} &= \left[ \frac{1-2a}{2}(p \omega' - p' \omega) +\frac{1}{2} \frac{ (p-p')^2 +\dpnu^2 (\omega-\omega')^2}{p \omega' - p' \omega}  \right] \, \ket{Y_{a\dot{c}} Y_{b\dot{d}}'}  \\ 
& + \frac{p p' + \dpnu^2 \omega \omega'  }{p \omega' - p' \omega} \left( \ket{Y_{a\dot{d}} Y_{b\dot{c}}'} + \ket{Y_{b\dot{c}} Y_{a\dot{d}}'} \right), \\ \\
\mathbb{T}^G \, \ket{Z_{\alpha\dot{\gamma}} Z_{\beta\dot{\delta}}'} &=  \left[ \frac{1-2a}{2}(p \omega' - p' \omega) -\frac{1}{2} \frac{ (p-p')^2 +\dpnu^2 (\omega-\omega')^2  }{p \omega' - p' \omega} \right] \, \ket{Z_{\alpha\dot{\gamma}} Z_{\beta\dot{\delta}}'} \\
& - \frac{ p p' + \dpnu^2 \omega \omega'  }{p \omega' - p' \omega} \left( \ket{Z_{\alpha\dot{\delta}} Z_{\beta\dot{\gamma}}'} + \ket{Z_{\beta\dot{\gamma}} Z_{\alpha\dot{\delta}}'} \right), \\ \\
\mathbb{T}^G \, \ket{Y_{a\dot{b}} Z_{\alpha\dot{\beta}}'} &=  \left[ \frac{1-2a}{2}(p \omega' - p' \omega) -\frac{1}{2}\frac{\om^2-\om'^2}{p \omega' - p' \omega} \right] \, \ket{Y_{a\dot{b}} Z_{\alpha\dot{\beta}}'}, \\ \\
\mathbb{T}^G \, \ket{Z_{\alpha\dot{\beta}} Y_{a\dot{b}}'} &=  \left[ \frac{1-2a}{2}(p \omega' - p' \omega)  +\frac{1}{2}\frac{\om^2-\om'^2}{p \omega' - p' \omega} \right] \, \ket{Z_{\alpha\dot{\beta}} Y_{a\dot{b}}'},
\end{aligned}
\end{equation}
and the  action of $\mathbb{T}^{B}$ on the two-particle states is
\begin{equation}
\begin{aligned}
\mathbb{T}^{B} \, \ket{Y_{a\dot{c}} Y_{b\dot{d}}'} &
 = i \dpnu\left(\epsilon_{ab} \ket{Y_{b\dot{c}} Y_{a\dot{d}}'} 
+\epsilon_{\dot{c}\dot{d}} \ket{Y_{a\dot{d}} Y_{b\dot{c}}'}\right)
, \\
\mathbb{T}^{B} \, \ket{Z_{\alpha\dot{\gamma}} Z_{\beta\dot{\delta}}'} &
 = i \dpnu \left( \epsilon_{\alpha\beta} \ket{Z_{\beta\dot{\gamma}} Z_{\alpha\dot{\delta}}'} 
+ \epsilon_{\dot{\gamma}\dot{\delta}} \ket{Z_{\alpha\dot{\delta}} Z_{\beta\dot{\gamma}}'}\right)\,,
\end{aligned}
\end{equation}
where on the r.h.s. we obviously do not sum over the repeated indices.  

\medskip 

In the undeformed case, the S-matrix ${\mathbb S}$ computed in perturbation theory is factorised into the product of two S-matrices, each of them invariant under one copy of the centrally extended superalgebra $\psu(2|2)$~\cite{Beisert:2005tm,Arutyunov:2006yd}
\be
\mathbb{S}_{\psu(2|2)^2_{\ce}} = \mathbb{S}_{\psu(2|2)_{\ce}}\, \hat{\otimes}\, \mathbb{S}_{\psu(2|2)_{\ce}}\,.
\ee
Using~\eqref{eq:def-Tmat} one finds the corresponding factorisation rule for the T-matrix 
\bea\label{eq:factoris-rule-T-mat-AdS5}
{\mathbb T}^{P\dot{P},Q\dot{Q}}_{M\dot{M},N\dot{N}}=(-1)^{\eps_{\dot M}(\eps_{N}+\eps_{Q})}{\cal T}_{MN}^{PQ}\delta_{\dot{M}}^{\dot{P}}\delta_{\dot{N}}^{\dot{Q}}
+(-1)^{\eps_Q(\eps_{\dot{M}}+\eps_{\dot{P}})}\delta_{M}^{P}\delta_{N}^{Q} {\cal T}_{\dot{M}\dot{N}}^{\dot{P}\dot{Q}}\, .
\eea 
Here $M=(a, \alpha)$ and $\dot{M}=(\dot{a},\dot{\alpha})$, and  dotted and undotted indices are referred to two copies of $\psu(2|2)$, respectively, while 
$\eps_{M}$ and $\eps_{\dot{M}}$ describe statistics of the corresponding indices, {\it i.e.} they  are zero for bosonic (Latin) indices and equal to one for fermionic (Greek) ones.
For the bosonic model the factor ${\mathcal T}$ can be regarded as a $16\times 16$ matrix.

\smallskip

It is not difficult to see that the same type of factorisation persists in the deformed case as well. Indeed, from \eqref{Tmatrix} we extract the following elements for the ${\mathcal T}$-matrix
\bea\la{cTmatr}
\begin{aligned}
&\cT_{ab}^{cd}=
A\,\de_a^c\de_b^d+B\,\de_a^d\de_b^c+W\, \eps_{ab}\de_a^d\de_b^c\, ,   \\ 
&\cT_{\a\b}^{\g\de}=
D\,\de_\a^\g\de_\b^\de+E\,\de_\a^\de\de_\b^\g+W\, \eps_{\a\b}\, \de_{\a}^{\delta}\de_{\b}^{\gamma}\,,  \\
&\cT_{a\b}^{c\de}= G\,\de_a^c\de_\b^\de\,,\qquad
~\cT_{\a b}^{\g d}= L\,\de_\a^\g\de_b^d\,,
\end{aligned}
\eea where the
coefficients are given by 
\bea
\begin{aligned}
\la{Tmatrcoef}
&A(p,p')= \frac{1-2a}{4}(p \omega' - p' \omega) +\frac{1}{4} \frac{ (p-p')^2 +\dpnu^2 (\omega-\omega')^2}{p \omega' - p' \omega} \,,\\
&B(p,p')=-E(p,p')= \frac{p p' + \dpnu^2 \omega \omega'  }{p \omega' - p' \omega} \,, \\
&D(p,p')=\frac{1-2a}{4} (p \omega' - p' \omega) -\frac{1}{4} \frac{ (p-p')^2 +\dpnu^2 (\omega-\omega')^2  }{p \omega' - p' \omega} \,,\\
&G(p,p')=-L(p',p)=\frac{1-2a}{4}(p \omega' - p' \omega) -\frac{1}{4} \frac{\om^2-\om'^2}{p \omega' - p' \omega} \,,\\
&W(p,p')= i\dpnu   \, .
\end{aligned}
\eea
Here $W$ corresponds to the contribution of the Wess-Zumino term and it does not actually depend on the particle momenta.
All the four remaining coefficients $\cT_{ab}^{\g\de},\cT_{\a\b}^{cd},\cT_{a\b}^{\g d},\cT_{\a b}^{\g d}$ vanish in the bosonic case but will be switched on once fermions are
taken into account. The matrix ${\mathcal T}$ is recovered from its matrix elements as follows
\bea
\nonumber
{\mathcal T}={\cal T}_{MN}^{PQ}\, E_P^M\otimes E_Q^N=\cT_{ab}^{cd}\, E_c^a\otimes E_d^b+\cT_{\a\b}^{\g\de}\, E_\g^\a\otimes E_\delta^\b+
\cT_{a\b}^{c\de}\, E_c^a\otimes E_\de^\b+\cT_{\a b}^{\g d}\, E_\g^\a\otimes E_d^b\, ,
\eea
where $E_M^N$ are the standard matrix unities. For the reader convenience we present ${\mathcal T}$ as an explicit $16\times 16$ matrix\footnote{See Appendix 8.5 of 
\cite{Arutyunov:2006yd}
for the corresponding matrix in the undeformed case.}

{\scriptsize
\begin{eqnarray}
\newcommand{\0}{\color{black!40}0}
{\mathcal T}\equiv \left( \begin{array}{ccccccccccccccccccc}
{\cal A}_1&\0&\0&\0&|&\0&\0&\0&\0&|&\0&\0&\0&\0&|&\0&\0&\0&\0\\
\0&{\cal A}_2&\0&\0&|&{\cal A}_4&\0&\0&\0&|&\0&\0&\0&\0&|&\0&\0&\0 &\0\\
\0&\0&{\cal A}_3&\0&|&\0&\0&\0&\0&|&\0&\0&\0&\0&|&\0&\0&\0&\0\\
\0&\0&\0&{\cal A}_3&|&\0&\0&\0&\0&|&\0&\0&\0&\0&|&\0&\0&\0&\0\\
-&-&-&-&-&-&-&-&-&-&-&-&-&-&-&-&-&-&-\\
\0&{\cal A}_5&\0&\0&|& {\cal A}_2&\0&\0&\0&|&\0&\0&\0&\0&|&\0&\0& \0 &\0\\
\0&\0&\0&\0&|&\0&{\cal A}_1&\0&\0&|&\0&\0&\0&\0&|&\0&\0&\0&\0\\
\0&\0&\0&\0&|&\0&\0&{\cal A}_3&\0&|&\0&\0&\0&\0&|&\0&\0&\0&\0\\
\0&\0&\0&\0&|&\0&\0&\0&{\cal A}_3&|&\0&\0&\0&\0&|&\0&\0&\0&\0\\
-&-&-&-&-&-&-&-&-&-&-&-&-&-&-&-&-&-&-\\
\0&\0&\0&\0&|&\0&\0&\0&\0&|&{\cal A}_8&\0&\0&\0&|&\0&\0&\0&\0\\
\0&\0&\0&\0&|&\0&\0&\0&\0&|&\0&{\cal A}_8&\0&\0&|&\0&\0&\0&\0\\
\0&\0&\0&\0&|&\0&\0&\0&\0&|&\0&\0& {\cal A}_6 &\0&|&\0&\0&\0&\0\\
\0&\0&\0&\0&|& \0&\0&\0&\0&|&\0&\0&\0&{\cal A}_7&|&\0&\0&{\cal A}_9&\0\\
-&-&-&-&-&-&-&-&-&-&-&-&-&-&-&-&-&-&-\\
\0&\0&\0&\0&|&\0&\0&\0&\0&|&\0&\0&\0&\0&|&{\cal A}_8&\0&\0&\0\\
\0&\0&\0&\0&|&\0&\0&\0&\0&|&\0&\0&\0&\0&|&\0&{\cal A}_8&\0&\0\\
\0&\0&\0&\0&|&\0&\0&\0&\0&|&\0&\0&\0&{\cal A}_{10}&|&\0&\0&{\cal A}_7&\0\\
\0&\0&\0&\0&|&\0&\0&\0&\0&|&\0&\0&\0&\0&|&\0&\0&\0&{\cal A}_6\\
\end{array} \right) \, .\nonumber
\end{eqnarray}
 }
Here the non-trivial matrix elements of ${\mathcal T}$ are given by
 \bea
 &&{\cal A}_1=A+B\, ,\quad
 {\cal A}_2=A\, ,\quad
 {\cal A}_4=B-W\, ,\quad
 {\cal A}_5=B+W\,, \quad{\cal A}_6=D+E\, ,
 \\\nonumber
 &&{\cal A}_6=D+E\, ,\quad
 {\cal A}_7=D\, , \quad
 {\cal A}_8=L\, ,\quad
 {\cal A}_9=E-W=-{\cal A}_5 \, ,\quad
 {\cal A}_{10}=E+W=-{\cal A}_4\, .
 \eea

 We conclude this section by pointing out that the matrix ${\mathcal T}$ that we have found satisfies the classical Yang-Baxter equation
\bea
[\cT_{12}(p_1,p_2),\cT_{13}(p_1,p_3)+\cT_{23}(p_2,p_3)]+[\cT_{13}(p_1,p_3),\cT_{23}(p_2,p_3)]=0\, 
\eea  
for any value of the deformation parameter $\dpnu$.

\subsection{Comparison with the $q$-deformed S-matrix}

In this subsection we show that the perturbative bosonic worldsheet S-matrix coincides with the first nontrivial term in the large-$g$ expansion of the $q$-deformed \ads S-matrix\footnote{The difference with the expansion performed in \cite{Beisert:2010kk} is that we include the dressing factor in the definition of the S-matrix.}.

 Let us recall that---up to an overall factor---the $q$-deformed \ads S-matrix is given by a tensor product of two copies of the $\psu(2|2)_q$-invariant S-matrix~\cite{Beisert:2008tw} which we denote just by $\mathbf{S}$, to avoid a heavy notation. The matrix may be found in~\eqref{Sqmat} of Appendix~\ref{app:matrixSmatrix}. 
We also need to multiply it by the overall factor $S_{\su(2)}$~\cite{Hoare:2011wr}
\be
\begin{aligned}
&S_{\su(2)}(p_1,p_2) \ \mathbf{S}_{12}\, \hat{\otimes} \, \mathbf{S}_{12}\, , \\
&S_{\su(2)}(p_1,p_2)=\frac{e^{i a(p_2{\cal E}_1-p_1{\cal E}_2)}}{\sigma_{12}^2}\frac{x_1^++\xi}{x_1^-+\xi}\frac{x_2^-+\xi}{x_2^++\xi}\cdot
\frac{x_1^--x_2^+}{x_1^+-x_2^-}\frac{1-\frac{1}{x_1^-x_2^+}}{1-\frac{1}{x_1^+x_2^-}}\, .
\end{aligned}
\ee
Here  $\hat{\otimes} $ stands for the graded tensor product, $a$ is the parameter of the light-cone gauge---see Eq.~\eqref{eq:lc-coord}---$\sigma$ is the dressing factor, and ${\cal E}$ is the $q$-deformed dispersion relation \eqref{qdisp} whose large $g$ expansion starts with $\om$.  The dressing factor can be found by solving the corresponding crossing equation,  and  it is given by \cite{Hoare:2011wr}
\begin{equation}\label{eq:def-theta}
\sigma_{12}=e^{i\t_{12}}\,,\quad \theta_{12} = \chi(x^+_1,x^+_2) + \chi(x^-_1,x^-_2)- \chi(x^+_1,x^-_2) - \chi(x^-_1,x^+_2),
\end{equation}
where
\begin{equation}\la{chi12}
\chi(x_1,x_2) = i \oint \frac{dz}{2 \pi i} \frac{1}{z-x_1} \oint \frac{dz'}{2 \pi i} \frac{1}{z'-x_2} \log\frac{\Gamma_{q^2}(1+\frac{i g}{2} (u(z)-u(z')))}{\Gamma_{q^2}(1-\frac{i g}{2} (u(z)-u(z')))}.
\end{equation}
Here $\Gamma_{q}(x)$ is the $q$-deformed Gamma function which for complex $q$
 admits an integral representation  \eqref{lnGovG} \cite{Hoare:2011wr}.

To develop the large $g$ expansion of the $q$-deformed \ads S-matrix, one  has to assume that $q=e^{-\qdp/g}$ where $\qdp$ is a deformation parameter which is kept fixed in the limit $g\to\infty$, and should be related to $\dpnu$. Then, due to the factorisation of the perturbative bosonic worldsheet S-matrix and of the $q$-deformed \ads S-matrix, it is sufficient to compare the ${\cal T}$-matrix \eqref{cTmatr} with the ${\mathbf T}$-matrix appearing in the expansion of one copy $\mathbf S$ with the proper factor
\be\la{qTpert}
(S_{\su(2)})^{1/2}\, \gen{1}_g\, \mathbf{S}=\gen{1} +\frac{i}{g}{\mathbf T}\, ,
\ee
where $\gen{1}_g$ is the graded identity which is introduced so that the expansion starts with $\gen{1}$. 
To check whether $\mathbf{T}=\mathcal{T}$ at leading order, the only term which is not straightforward to expand is the $S_{\su(2)}$ scalar factor because it contains the dressing phase $\t_{12}$.  
It is clear that it will contribute only to the part of the ${\mathbf T}$-matrix proportional to the identity matrix. 
If we study the expansion of just $\gen{1}_g\, \mathbf{S}$ without the $S_{\su(2)}$ factor, we find that it is indeed related to the matrix $\mathcal{T}$ computed in perturbation theory by
\be
\gen{1}_g \mathbf{S}= \gen{1} + \frac{i}{g} (\mathcal{T}-{\cal A}_1 \gen{1})\,,
\ee
provided we identify the parameters $q$ and $\nu$, or $q$ and $\vk$, as
\be
q=e^{-\nu/g}=e^{-\vk/\tilde{g}}\,,
\ee
showing that $q$ is real.
What is left to check is then the overall normalisation, namely that the $1/g$ term in the expansion of $S_{\su(2)}^{1/2}$ is equal to $ {\cal A}_1$. 
To this end one should find the large $g$ expansion of the dressing phase $\t_{12}$. This is done by first expanding the ratio of $\Gamma_{q^2}$-functions in \eqref{chi12} with $u(z)$ and $u(z')$ being kept fixed, using Eq.~\eqref{qGamma1}.
Next, one combines it with the expansion of the $\frac{1}{z-x_1^\pm} \frac{1}{z'-x_2^\pm} $ terms which appear in the integrand of \eqref{eq:def-theta}. As a result one finds that the dressing phase is of order $1/g$ just as it was in the undeformed case~\cite{Arutyunov:2004vx}.  
One may check numerically that the element ${\cal A}_1$ is indeed equal to the $1/g$ term in the expansion of $S_{\su(2)}^{1/2}$
.   In fact it is not difficult to extract from  ${\cal A}_1$ the leading term in the large $g$ expansion of the dressing phase which appears to be very simple
 \be
\t_{12}= \frac{\dpnu^2 \left(\omega _1-\omega _2\right)+p_2^2 \left(\omega _1-1\right)-p_1^2 \left(\omega _2-1\right)}{2g
   \left(p_1+p_2\right)} +\mathcal{O}(1/g^2)\,.
\ee
It would be curious to derive this expression directly from the double integral representation. Note that, doing this double integral, one could also get the full AFS order of the phase, which would be a deformation of the one in~\eqref{eq:AFS-xpxm}.

\section{Concluding remarks}
In this chapter we have studied the bosonic sector of the string on $\eta$-deformed {\adsfive}.
We have derived the deformed metric and the $B$-field that is generated.
After computing the tree-level scattering processes involving bosonic excitations on the worldsheet, we were able to succesfully match with the large-tension expansion of the all-loop S-matrix found by imposing the $\psu_q(2|2)_{\ce}$ symmetry.

The bosonic background that we have derived was further studied in a series of papers.
Giant magnons were studied in~\cite{Arutynov:2014ota,Khouchen:2014kaa,Ahn:2014aqa} and other classical solutions in~\cite{Kameyama:2014vma,Panigrahi:2014sia,Banerjee:2015nha}. Minimal surfaces were considered in~\cite{Kameyama:2014via,Bai:2014pya} and three-point correlation functions in~\cite{Ahn:2014iia,Bozhilov:2015kya}. For deformations of classical integrable models corresponding to subsectors of the bosonic theory we refer to~\cite{Kameyama:2014bua,Arutyunov:2014cda}.
The pertubative S-matrix that we have computed was studied at one and two loops using unitarity techniques in~\cite{Engelund:2014pla}.

In~\cite{Hoare:2014pna} truncations to lower dimensional models and special limits were considered. In particular, one way was provided to prove that the limit of maximal deformation is related by double T-duality to dS$_5\times$H$^5$, namely the product of five-dimensional de Sitter and five-dimensional hyperboloid.
A similar and more physical ($\eta\to1,\ \vk\to \infty$) limit was studied in~\cite{Arutyunov:2014cra}, where it was shown agreement with the background of the mirror model---first introduced to develop the Thermodynamic Bethe Ansatz~\cite{Ambjorn:2005wa,Arutyunov:2007tc,Arutyunov:2009zu}---obtained by performing a double-Wick rotation on the light-cone gauge-fixed string.
The exact spectrum was actually considered in~\cite{Arutynov:2014ota}, where the notion of ``mirror duality'' was introduced, after observing that original and mirror models are related by small/large values of the deformation parameter.

A proposal on how to deform the $\sigma$-model on {\adsfive} to obtain the $q$-deformation in the case of $q$ being root of unity is the $\lambda$-deformation of~\cite{Hollowood:2014qma}.
We refer also to~\cite{Vicedo:2015pna,Hoare:2015gda} for a relation between the $\eta$-deformation and the $\lambda$-deformation.
Generalisations of the deformation procedure were studied in~\cite{Kawaguchi:2014fca,Matsumoto:2014cja,vanTongeren:2015soa,vanTongeren:2015uha} were Jordanian and other deformations based on $R$-matrices satisfying the classical Yang-Baxter equation were considered.

In~\cite{Lunin:2014tsa} it was shown that in the two cases of (AdS$_2\times$S$^2)_\eta$ and (AdS$_3\times$S$^3)_\eta$ it is possible to add to the deformed metric the missing NSNS scalar and RR fields, to obtain a background satisfying the supergravity equations of motion. Of these two cases, only for (AdS$_2\times$S$^2)_\eta$ it  was conjectured that the solution\footnote{A one-parameter family of solutions was actually found, and the conjecture proposes one specific point to correspond to the deformed model.} corresponds to the $\eta$-deformation. The explicit check of this is still missing.
The case of {\etaadsfive} is technically more complicated and a supergravity solution has not been found. 
One of the main motivations of the next chapter is to compute the Lagrangian of the superstring in the deformed model up to quadratic order in fermions. This will allow us to read off the couplings to the unknown RR fields.

\chapter{{\etaadsfive} at order $\theta^2$}\label{ch:qAdS5Fer}
In this chapter we push the computation of the Lagrangian for the deformed model up to quadratic order in fermions.
The motivation for doing that is to discover the couplings to the unknown RR fields and the dilaton, which should complete the deformed metric and $B$-field to a full type IIB background.

We start in Section~\ref{sec:algebra-basis} by presenting a convenient realisation of $\psu(2,2|4)$ and in~\ref{sec:psu224-current} by computing the current for this algebra.
Using the results collected in Section~\ref{sec:inverse-op} regarding the inverse operator that is used to define the deformed action, we compute the Lagrangian in Section~\ref{sec:-eta-def-lagr-quad-theta} and we show how to recast it in the standard form of Green-Schwarz.
In~\ref{sec:kappa-symm-eta-def-quad-theta} we compute the kappa-variations of the bosonic and fermionic fields, and of the worldsheet metric, to confirm the results for the background fields obtained in the previous section.
Section~\ref{sec:discuss-eta-def-background} is devoted to a discussion on the results that we have found.
We show that the background fields that we have derived are not compatible with the equations of motion of type IIB supergravity.
We comment on this result and on particular limits of the $\sigma$-model action.

\section{The $\psu(2,2|4)$ algebra}\label{sec:algebra-basis}
The subject of this section is the $\psu(2,2|4)$ algebra, which plays a central role in the construction of the action for the superstring on {\adsfive} and its deformation.
We start from the algebra $\alg{sl}(4|4)$, one particular element of which may be written as a $8\times 8$ matrix
\be
M=
\left(
\begin{array}{c|c}
 m_{11} & m_{12} 
\\
\hline
m_{21} & m_{22}
\end{array}
\right)\,,
\ee
where each $m_{ij}$ above is a $4\times 4$ block. The matrix $M$ is required to have vanishing supertrace, defined as $\Str M=\tr m_{11}-\tr m_{22}$.
The $\mathbb{Z}_2$ structure identifies the diagonal blocks $m_{11}, m_{22}$ as even, while the off-diagonal blocks $m_{12},m_{21}$ as odd.
Later we will multiply the former by Grassmann-even (bosonic) variables, while the latter by Grassmann-odd (fermionic) ones.

We find the algebra $\su(2,2|4)$ by imposing a proper reality condition
\be\label{eq:real-cond-su224}
M^\dagger H+HM=0\,,
\ee
where we have defined the matrix $H$ as
\be
H=\left(
\begin{array}{cc}
 \Sigma & 0  \\
 0 &  \mathbf{1}_4  \\
\end{array}
\right)\,,
\ee
and the diagonal matrix $\Sigma=\text{diag}(1,1,-1,-1)$.
We will present an explicit realisation of this algebra in terms of $8\times 8$ matrices. Since $\su(2,2|4)$ is non-compact, the above representation is non-unitary. 
The algebra $\psu(2,2|4)$ is then found by projecting away the generator proportional to the identity operator.

To construct our $8\times 8$ matrices we will use $4\times 4$ gamma-matrices~\cite{Arutyunov:2009ga}.
In Appendix~\ref{app:su224-algebra} we write our preferred choice for the $4\times 4$ gamma-matrices.
 They are all Hermitian and satisfy the $\text{SO}(5)$ Clifford algebra 
\be
\{ \gamma_m, \gamma_n\} = 2 \delta_{mn}\,, \qquad m=0,\ldots,4\,.
\ee
We need two copies of these matrices, one for Anti-de Sitter and one for the sphere.
In the first case we will denote them with a check $\check{\g}_m$, in the second with a hat $\hat{\g}_m$
\be\label{eq:gamma-AdS5-S5}
\begin{aligned}
\text{AdS}_5:\quad&\check{\gamma}_0 = i \gamma_0, \qquad \check{\gamma}_m = \gamma_m, \quad m=1,\cdots,4, \\
\text{S}^5:\quad&\hat{\gamma}_{m+5} = -\gamma_m, \quad m=0,\cdots,4.
\end{aligned}
\ee
We have chosen to enumerate the gamma-matrices for AdS$_5$ from $0$ to $4$ and the ones for S$^5$ from $5$ to $9$ to have a better notation when we want to write ten-dimensional expressions.
The $i$ in the definition of $\check{\g_0}$ is needed to reproduce the signature of the metric.
We will not use the notation of~\cite{Arutyunov:2009ga} for the generators.
In the following we provide explicitly our preferred basis.


\paragraph{Even generators}

We denote 10 (5 for AdS$_5$ + 5 for S$^5$) of the even generators by $\gen{P}$ and the remaining 20 (10 for AdS$_5$ + 10 for S$^5$) by $\gen{J}$.
The generators $\check{\gen{P}}_m$ for  AdS$_5$ and the generators $\hat{\gen{P}}_{m}$ for S$^5$ are defined as
\be
\check{\gen{P}}_m = 
\left(
\begin{array}{cc}
 -\frac{1}{2} \check{\gamma}_m & \mathbf{0}_4  \\
 \mathbf{0}_4 & \mathbf{0}_4 \\
\end{array}
\right), \quad m=0,\ldots4,
\qquad
\hat{\gen{P}}_{m} = 
\left(
\begin{array}{cc}
 \mathbf{0}_4 & \mathbf{0}_4  \\
 \mathbf{0}_4 & \frac{i}{2} \hat{\gamma}_m \\
\end{array}
\right), \quad m=5,\ldots9.
\ee
After defining $\check{\gamma}_{mn} \equiv \frac{1}{2} [\check{\gamma}_m,\check{\gamma}_n]$ and $\hat{\gamma}_{mn} \equiv \frac{1}{2} [\hat{\gamma}_m,\hat{\gamma}_n]$ we also write the generators $\check{\gen{J}}_{mn}$ and $\hat{\gen{J}}_{mn}$ for  AdS$_5$ and  for S$^5$
\be
\check{\gen{J}}_{mn} = 
\left(
\begin{array}{cc}
 \frac{1}{2} \check{\gamma}_{mn} & \mathbf{0}_4  \\
 \mathbf{0}_4 & \mathbf{0}_4 \\
\end{array}
\right), \quad m,n=0,\ldots4,
\qquad
\hat{\gen{J}}_{mn} = 
\left(
\begin{array}{cc}
 \mathbf{0}_4 & \mathbf{0}_4  \\
 \mathbf{0}_4 & \frac{1}{2} \hat{\gamma}_{mn} \\
\end{array}
\right), \quad m,n=5,\ldots9.
\ee
All the generators satisfy Equation~\eqref{eq:real-cond-su224} and hence belong to $\su(2,2|4)$.

\paragraph{Odd generators}
To span all the 32 odd generators of $\alg{su}(2,2|4)$ we use a label $I=1,2$ and two spinor indices $\ul{\a},\ul{a}=1,2,3,4$. Greek spinor indices are used for AdS$_5$, Latin ones for S$^5$.
Our preferred basis for the odd generators is
\be\label{eq:def-odd-el-psu224}
\begin{aligned}
\genQind{I}{\a}{a} &=e^{+i\pi/4}
\left(
\begin{array}{cc}
 \mathbf{0}_4 & m^{\ \, \ul{\a}}_{I \, \ul{a}}  \\
 K \left(m^{\ \, \ul{\a}}_{I \, \ul{a}}\right)^\dagger K & \mathbf{0}_4 \\
\end{array}
\right),
\\
{\left(m^{\ \, \ul{\a}}_{1 \, \ul{a}}\right)_j}^k &= e^{+i\pi/4+i\phi_{\gen{Q}}} \, \delta^{\ul{\a}}_j \delta_{\ul{a}}^k,
\qquad\qquad
{\left(m^{\ \, \ul{\a}}_{2 \, \ul{a}}\right)_j}^k = -e^{-i\pi/4+i\phi_{\gen{Q}}} \,  \delta^{\ul{\a}}_j \delta_{\ul{a}}^k.
\end{aligned}
\ee
Here $m^{\ \, \ul{\a}}_{I \, \ul{a}}$ are $4\times 4$ matrices, and $K$ is defined in~\eqref{eq:SKC-gm}. The phase $\phi_{\gen{Q}}$ corresponds to the $U(1)$ automorphism of $\su(2,2|4)$, and we set $\phi_{\gen{Q}}=0$. 
These supermatrices are constructed in such a way that they do \emph{not} satisfy Eq.~\eqref{eq:real-cond-su224} but rather $\gen{Q}^\dagger i\,  \widetilde{H}+\widetilde{H}\gen{Q}=0$ where we have defined
\be
 \widetilde{H}\equiv
\left(
\begin{array}{cc}
 K & 0  \\
 0 &  K  \\
\end{array}
\right).
\ee
The supermatrices $\gen{Q}$ can be seen as complex combinations of supermatrices $\mathcal{Q}$ satisfying~\eqref{eq:real-cond-su224}
\be
\mathcal{Q}= e^{+i\pi/4} \,
\left(
\begin{array}{cc}
 C & 0  \\
 0 &  K  \\
\end{array}
\right)
\gen{Q},
\qquad
\gen{Q}= -e^{-i\pi/4} \,
\left(
\begin{array}{cc}
 C & 0  \\
 0 &  K  \\
\end{array}
\right)
\mathcal{Q}.
\ee
The matrix $C$ is defined in Eq.~\eqref{eq:SKC-gm}.
On the one hand, taking linear combinations of $\mathcal{Q}$'s with Grassmann variables and imposing that $\ferm{\vartheta}{I}{\a}{a} \mathcal{Q}^{I \, \ul{\a}}_{\ \, \ul{a}}$ belongs to $\alg{su}(2,2|4)$ would translate into the fact that the fermions $\ferm{\vartheta}{I}{\a}{a}$ are real.
On the other hand, imposing that $\ferm{\theta}{I}{\a}{a} \genQind{I}{\a}{a}$ belongs to $\alg{su}(2,2,|4)$ $(\ferm{\theta}{I}{\a}{a} \genQind{I}{\a}{a})^\dagger= -H(\ferm{\theta}{I}{\a}{a} \genQind{I}{\a}{a})H^{-1}$ gives
\be
\ferm{{\theta^\dagger}}{I}{a}{\a} =-i\, \ferm{\theta}{I}{\nu}{b} \ C^{\ul{\nu}\ul{\a}} K_{\ul{b}\ul{a}} .
\ee
Defining the barred version of a fermion we find the Majorana condition in the form
\be
\ferm{\bar{\theta}}{I}{a}{\a} \equiv \ferm{{\theta^\dagger}}{I}{a}{\nu} {(\check{\gamma}^0)_{\ul{\nu}}}^{\ul{\a}} = - \ferm{\theta}{I}{\nu}{b} \ K^{\ul{\nu}\ul{\a}} K_{\ul{b}\ul{a}} .
\ee
Later on we will decide to write the fermions $\ferm{\theta}{I}{\a a}{}$ with both spinor indices lowered and $\ferm{\bar{\theta}}{I}{}{\a a}$ with both spinor indices raised, so the above equation reads as
\be\label{eq:Majorana-cond-ferm-sp-ind}
\ferm{\bar{\theta}}{I}{}{\a a}  = + \ferm{\theta}{I}{\nu b}{} \ K^{\ul{\nu}\ul{\a}} K^{\ul{b}\ul{a}} ,
\ee
matching with~\cite{Metsaev:1998it}.
Let us comment on the fact that the matrix $K$ is the charge conjugation matrix for the $\g$-matrices. We call it $K$ to keep the same notation of~\cite{Arutyunov:2009ga}.
We refer to Appendix~\ref{app:su224-algebra} for our conventions with spinors.
A more compact notation is achieved by actually omitting the spinor indices. The above equation then reads as
\be\label{eqMajorana-cond-compact-not}
\bar{\theta}_I = \theta_I^\dagger \bg^0=+ \theta_I^t \, (K\otimes K)\,,
\ee
where $\bg^0\equiv \check{\g}^0\otimes \gen{1}_4$, and Hermitian conjugation and transposition are implemented only in the space spanned by the spinor indices, where the matrices $\bg^0$ and $K\otimes K$ are acting.


\paragraph{Commutation relations}
It is convenient to rewrite the commutation relations when considering the Grassmann enveloping algebra. In this way we may suppress the spinor indices to obtain more compact expressions. We define $\gen{Q}^{I} \theta_I\equiv \genQind{I}{\a a}{} \ferm{\theta}{I}{\a a}{}$ and we introduce the $16\times 16$ matrices
\be\label{eq:def16x16-gamma}
\begin{aligned}
& \bg_m \equiv \check{\g}_m \otimes \mathbf{1}_4,
\quad m=0, \cdots, 4,
\qquad
& \bg_m \equiv  \mathbf{1}_4 \otimes i\hat{\g}_m,
\quad m=5, \cdots, 9, \\
& \bg_{mn} \equiv \check{\g}_{mn} \otimes \mathbf{1}_4,
\quad m,n=0, \cdots, 4,
\qquad
& \bg_{mn} \equiv  \mathbf{1}_4 \otimes \hat{\g}_{mn},
\quad m,n=5, \cdots, 9. \\
\end{aligned}
\ee
The first space in the tensor product is spanned by the AdS spinor indices, the second by the sphere spinor indices.
To understand the 10-dimensional origin of these objects see appendix~\ref{sec:10-dim-gamma}.
In the context of type IIB, one usually continues to refer to them as gamma-matrices even though they do not satisfy Clifford algebra relations.

In our basis the commutation relations involving only bosonic elements read as
\be
\begin{aligned}
\text{AdS}_5 : \quad &[ \check{\gen{P}}_m, \check{\gen{P}}_n ] = \check{\gen{J}}_{mn}, \ 
&\ \  \text{S}^5 : \quad &[ \hat{\gen{P}}_m, \hat{\gen{P}}_n ] = -\hat{\gen{J}}_{mn}, \\
&[ \check{\gen{P}}_{m}, \check{\gen{J}}_{np} ] = \eta_{mn} \check{\gen{P}}_p - \ _{n \leftrightarrow p}, \ 
& &[ \hat{\gen{P}}_{m}, \hat{\gen{J}}_{np} ] = \eta_{mn} \hat{\gen{P}}_p - \ _{n \leftrightarrow p},\\
&[ \check{\gen{J}}_{mn}, \check{\gen{J}}_{pq} ] = (\eta_{np} \check{\gen{J}}_{mq} - _{m \leftrightarrow n} ) - _{p \leftrightarrow q} \ 
& &[ \hat{\gen{J}}_{mn}, \hat{\gen{J}}_{pq} ] = (\eta_{np} \hat{\gen{J}}_{mq} - _{m \leftrightarrow n} ) - _{p \leftrightarrow q},
\end{aligned}
\ee
where $\eta_{mn}= \text{diag}(-1,1,1,1,1,1,1,1,1,1)$.
Generators of the two different spaces commute with each other.
The generators $\gen{J}$ identify the bosonic subalgebra $\alg{so}(4,1)\oplus\alg{so}(5)$.

With the above definitions, the commutation relations of $\su(2,2|4)$ involving odd generators are\footnote{For commutators of two odd elements we need to multiply by two \emph{different} fermions $\lambda_I,\psi_I$, otherwise the right hand side vanishes.}
\be
\begin{aligned}
& [\gen{Q}^{I} \theta_I, \gen{P}_m] = - \frac{i}{2} \epsilon^{IJ} \gen{Q}^{J} \bg_m  \theta_I, & \qquad
& [\gen{Q}^{I} \theta_I, \gen{J}_{mn}] =  -\frac{1}{2} \delta^{IJ} \gen{Q}^{J} \bg_{mn}  \theta_I, & 
\end{aligned}
\ee
\be\label{eq:comm-rel-QQ-su224}
\begin{aligned}
[ \gen{Q}^{I} \lambda_I, \gen{Q}^{J} \psi_J ] =& \, i \, \delta^{IJ} \bar{\lambda}_I \bg^m  \psi_J \ \gen{P}_m  %
-  \frac{1}{2} \epsilon^{IJ} \bar{\lambda}_I (\bg^{mn} \check{\gen{J}}_{mn} -\bg^{mn}  \hat{\gen{J}}_{mn}) \psi_J \ 
 - \frac{i}{2} \delta^{IJ} \bar{\lambda}_I \psi_J \mathbf{1}_8.
\end{aligned}
\ee
Here we have also used the Majorana condition to rewrite the result in terms of the fermions $\bar{\lambda}_I$, and for completeness we indicate also the generator proportional to the identity operator.
We refer to Appendix~\ref{app:su224-algebra} for the commutation relations with explicit spinor indices.

\paragraph{Supertraces}
In the computation for the Lagrangian we will need to take the supertrace of products of two generators of the algebra. For the non-vanishing ones we find
\be
\begin{aligned}
\Str[\gen{P}_m\gen{P}_n]&=\eta_{mn}, \\
\text{AdS}_5 : \quad \Str[\check{\gen{J}}_{mn}\check{\gen{J}}_{pq}]&= - (\eta_{mp}\eta_{nq}-\eta_{mq}\eta_{np}),  \\
\text{S}^5 : \quad \Str[\hat{\gen{J}}_{mn}\hat{\gen{J}}_{pq}]&= + (\eta_{mp}\eta_{nq}-\eta_{mq}\eta_{np}),  \\
\Str[\gen{Q}^I \lambda_I \, \gen{Q}^J \psi_J ]&= -2 \epsilon^{IJ} \bar{\lambda}_I \psi_J = -2 \epsilon^{JI} \bar{\psi}_J \lambda_I \,.
\end{aligned}
\ee

\paragraph{$\mathbb{Z}_4$ decomposition}
The $\su(2,2|4)$ algebra admits a $\mathbb{Z}_4$ decomposition, compatible with the commutation relations.
We call $\Omega$ the outer automorphism, that acts on elements of the algebra as
\be
\Omega(M)=i^k\, M\,, \qquad k=0,\ldots, 4\,,
\ee
identifying four different subspaces of $\su(2,2|4)$ labelled by $k$.
We define it as~\cite{Arutyunov:2009ga}
\be
\Omega(M) = - \mathcal{K} M^{st} \mathcal{K}^{-1},
\ee
with $\mathcal{K}=\text{diag}(K,K)$ and $^{st}$ denoting the supertranspose
\be
M^{st}\equiv
\left(
\begin{array}{c|c}
 m_{11}^t & -m_{21}^t
\\
\hline
m_{12}^t & m_{22}^t
\end{array}
\right)\,.
\ee
If we consider bosonic generators, it is easy to see that  $\gen{J}$ and $\gen{P}$ belong to the subspaces of grading 0 and 2 respectively\footnote{The subspaces of grading $0$ and $2$ of this chapter correspond to the subspaces with label $+$ and $-$ respectively of Chapter~\ref{ch:qAdS5Bos}.}
\be
\Omega(\gen{J})=+\gen{J}\,,
\qquad
\Omega(\gen{P})=-\gen{P}\,.
\ee
In our basis, the action on odd generators is also very simple
\be
\Omega(\genQind{I}{\a a}{})=\sigma_3^{II}\, i\, \genQind{I}{\a a}{}\,,
\ee
meaning that odd elements with $I=1$ have grading 1, and with $I=2$ have grading 3.
It is natural to introduce projectors $P^{(k)}$ on each subspace, whose action may be found by
\be\label{eq:def-proj-Z4-grad}
P^{(k)}(M)=\frac{1}{4}\left( M+i^{3k}\Omega(M)+i^{2k}\Omega^2(M) +i^{k}\Omega^{3}(M)\right)\,.
\ee
Then $P^{(0)}$ will project on generators $\gen{J}$, $P^{(2)}$ on generators $\gen{P}$, and $P^{(1)},P^{(3)}$ on odd elements with labels $I=1,2$
\be
P^{(1)}(\genQind{I}{\a a}{}) = \frac{1}{2} (\delta^{IJ}+\sigma_3^{IJ}) \genQind{J}{\a a}{}, \qquad 
P^{(3)}(\genQind{I}{\a a}{}) = \frac{1}{2} (\delta^{IJ}-\sigma_3^{IJ}) \genQind{J}{\a a}{}. 
\ee
The definition of the coset uses the $\mathbb{Z}_4$ grading, as the generators that are removed are the $\gen{J}$'s spanning the $\alg{so}(4,1) \oplus \alg{so}(5)$ subalgebra, that conincides with the subspace of grading $0$.

\section{The current}\label{sec:psu224-current}
In this section we compute the current that enters the definition of the Lagrangian up to quadratic order in fermions.
We start by defining a coset element of $\text{PSU}(2,2|4)/\text{SO}(4,1)\times \text{SO}(5)$, that we choose to write as 
\be\label{eq:choice-full-coset-el}
\alg{g}=\gb \cdot \gf.
\ee 
Here $\gb$ is a bosonic group element. We choose the same representative used in Chapter~\ref{ch:qAdS5Bos}, see Equation~\eqref{basiccoset}. The fermionic group element is denoted by $\gf$ and we define it simply through the exponential map
\be
\gf=\text{exp} \chi\,,\qquad\qquad \chi \equiv \genQind{I}{}{} \ferm{\theta}{I}{}{}\,.
\ee
One may prefer different choices, \eg $\gf= \chi + \sqrt{1+\chi^2}$ turns out to be more convenient when we want to expand up to fourth order~\cite{Arutyunov:2009ga}, since it generates no cubic term in the expansion. At quadratic order the two parameterisations are equivalent.

Let us comment on the fact that other choices for $\alg{g}$ are also possible, \eg we could put the fermions to the left and use an element of the form $\gf \cdot \gb$.
In~\cite{Frolov:2006cc,Arutyunov:2009ga} yet another choice was made, namely $\Lambda(t,\phi)\cdot \gf\cdot{\alg{g}}_{\text{X}}$, where $\Lambda(t,\phi)$ is the group element for shifts of $t$ and $\phi$, while ${\alg{g}}_{\text{X}}$ contains the remaining bosonic isometries.
We prefer to use~\eqref{eq:choice-full-coset-el} as in~\cite{Metsaev:1998it} because its expansion in powers of fermions is simpler.
This choice corresponds to fermions that are not charged under global bosonic isometries.

The current is defined as $A=-\alg{g}^{-1}{\rm d}\alg{g}$, and being an element of the algebra we decompose it in terms of linear combinations of the generators
\be
A= L^m \gen{P}_m + \frac{1}{2} L^{mn} \gen{J}_{mn} +  \genQind{I}{\a a}{}\ferm{L}{I}{\a a}{}.
\ee
It is useful to look at the purely bosonic and purely fermionic currents separately, that are found after switching off the fermions and the bosons respectively. The purely bosonic current is a combination of even generators only
\be
\Ab= -\gb^{-1} {\rm d} \gb = e^m \gen{P}_m + \frac{1}{2} \omega^{mn} \gen{J}_{mn} .
\ee
The coefficients in front of the generators $\gen{P}_m$ are the components of the vielbein, while the ones in front of the generators $\gen{J}_{mn}$ are the components of the spin-connection for the ten-dimensional metric.
To write them explicitly, let us choose to enumerate the ten spacetime coordinates as
\be
\begin{aligned}
& X^0 = t, & \quad X^1= \psi_2,& \quad X^2= \psi_1, &\quad X^3=\zeta, &\quad X^4= \rho,\\
& X^5 = \phi, & \quad X^6= \phi_2,& \quad X^7= \phi_1, &\quad X^8=\xi, &\quad X^9= r.
\end{aligned}
\ee
We find that in our parameterisation the vielbein $e^m = e^m_M {\rm d}X^M$ is diagonal and given by\footnote{To avoid confusion with tangent indices, we write curved indices with the explicit names of the spacetime coordinates.}
\be
\begin{aligned}
e^0_t = \sqrt{1+\rho^2}, \quad & e^1_{\psi_2} = -\rho \sin \zeta , \quad &e^2_{\psi_1} =-\rho \cos \zeta, \quad &e^3_{\zeta} = -\rho, \quad& e^4_{\rho} = -\frac{1}{\sqrt{1+\rho^2}}, \\
e^5_{\phi} = \sqrt{1-r^2}, \quad & e^6_{\phi_2} = -r \sin \xi , \quad &e^7_{\phi_1} =-r \cos \xi, \quad &e^8_\xi = -r, \quad& e^9_r = -\frac{1}{\sqrt{1-r^2}}. 
\end{aligned}
\ee
The non-vanishing components of the spin connection $\omega^{mn} = \omega^{mn}_M {\rm d}X^M$ are
\be
\begin{aligned}
& \omega^{04}_t = \rho,  \quad & \omega^{34}_\zeta =  -\sqrt{1+\rho^2}, \quad & \quad & \\
& \omega^{13}_{\psi_2} = -\cos \zeta, \quad & \omega^{14}_{\psi_2} = -\sqrt{1+\rho^2} \sin \zeta , \quad & \omega^{23}_{\psi_1} = \sin \zeta, \quad & \omega^{24}_{\psi_1} = -\sqrt{1+\rho^2} \cos \zeta, \\
\\
& \omega^{59}_\phi = -r,  \quad & \omega^{89}_\xi =  -\sqrt{1-r^2}, \quad & \quad & \\
& \omega^{68}_{\phi_2} = -\cos \xi, \quad & \omega^{69}_{\phi_2} = -\sqrt{1-r^2} \sin \xi , \quad & \omega^{78}_{\phi_1} = \sin \xi, \quad & \omega^{79}_{\phi_1} = -\sqrt{1-r^2} \cos \xi, 
\end{aligned}
\ee
and it can be checked that $\omega_M^{mn}$ satisfies the correct equation for the spin-connection~\eqref{eq:spi-conn-vielb}.

The purely fermionic current is decomposed in terms of even and odd generators
\be
\Af= -\gf^{-1} {\rm d} \gf = \Omega^m \gen{P}_m + \frac{1}{2} \Omega^{mn} \gen{J}_{mn} + \ferm{\Omega}{I}{\a a}{} \genQind{I}{\a a}{}
\ee
where we have defined the to-be-determined quantities $\Omega^m, \Omega^{mn}, \ferm{\Omega}{I}{\a a}{} $.
After expanding $\gf$ in powers of $\theta$, at quadratic order in the fermions we find
\be
\begin{aligned}
\Af=& -\gf^{-1} {\rm d} \gf \\
=& - \gen{Q}^{I} \, {\rm d} \theta_I 
+ \frac{1}{2}[ \gen{Q}^{I} \theta_I , \gen{Q}^{J} \, {\rm d}\theta_J  ]+\mathcal{O}(\theta^3)\\
=& - \gen{Q}^{I} \, {\rm d} \theta_I 
+ \frac{i}{2} \delta^{IJ} \bar{\theta}_I  \bg^m  {\rm d} \theta_J \ \gen{P}_m 
- \frac{1}{4} \epsilon^{IJ} \bar{\theta}_I  \bg^{mn}  {\rm d} \theta_J \ \check{\gen{J}}_{mn} 
+ \frac{1}{4} \epsilon^{IJ} \bar{\theta}_I   \bg^{mn}  {\rm d} \theta_J \ \hat{\gen{J}}_{mn}  +\mathcal{O}(\theta^3),
\end{aligned}
\ee
where we make use of the commutation relations~\eqref{eq:comm-rel-QQ-su224} of $\psu(2,2|4)$---meaning that we also project out the generator proportional to the identity operator.

When we repeat the computation for the full current we see that the computation is similar to the one of the fermionic current, upon replacing ${\rm d} \to ({\rm d}-\Ab)$~\cite{Metsaev:1998it}
\be
\begin{aligned}
A=& -\alg{g}^{-1}{\rm d}\alg{g}= -\gf^{-1} ({\rm d}-\Ab) \gf \\
=& \Ab- 
\gen{Q}^{I} {\rm d} \theta_I - [ \gen{Q}^{I} \theta_I , \Ab]\\
& +\frac{1}{2} \left[\gen{Q}^{I} \theta_I , \left(\gen{Q}^{J} {\rm d} \theta_J - [ \gen{Q}^{J} \theta_J , \Ab] \right)\right]
+\mathcal{O}(\theta^3) \\
=& \left( e^m +\frac{i}{2}  \bar{\theta}_I  \bg^m  D^{IJ} \theta_J \right) \gen{P}_m 
 - \gen{Q}^{I} \, D^{IJ} \theta_J \\
& +\frac{1}{2} \omega^{mn} {\gen{J}}_{mn} - \frac{1}{4} \epsilon^{IJ} \bar{\theta}_I \left( \bg^{mn} \check{\gen{J}}_{mn} - \bg^{mn} \hat{\gen{J}}_{mn}  \right)  D^{JK} \theta_K 
+\mathcal{O}(\theta^3)
\end{aligned}
\ee
where the operator $D^{IJ}$ on fermions $\theta$ is
\be\label{eq:op-DIJ-psu224-curr}
D^{IJ} = \delta^{IJ} \left( {\rm d} - \frac{1}{4} \omega^{mn} \bg_{mn}  \right)
+ \frac{i}{2} \epsilon^{IJ} e^m  \bg_m  .
\ee
Sometimes it is useful to write it as
\be
D^{IJ} = \mathcal{D}^{IJ}
+ \frac{i}{2} \epsilon^{IJ} e^m  \bg_m  ,
\qquad
\mathcal{D}^{IJ} \equiv\delta^{IJ} \left( d - \frac{1}{4} \omega^{mn} \bg_{mn}  \right),
\ee
where $\mathcal{D}^{IJ}$ is the covariant derivative on the fermions.

The contribution of the generators $\gen{J}$ to the current will be irrelevant for the computation of the Lagrangian, since they are projected out when defining the coset.

Imposing the flatness condition on the current
\be
\epsilon^{\a\b} (\pa_\a A_\b -\frac{1}{2} [A_\a,A_\b])=0
\ee
and projecting on the bosonic generators we find the following equations for the vielbein and the spin connection
\be\label{eq:d-veilbein}
\begin{aligned}
\epsilon^{\a\b} (\pa_\a e^m_\b - \omega^{mq}_\a e_{q\b}) &= 0,
\end{aligned}
\ee
\be\label{eq:d-spin-conn}
\begin{aligned}
\epsilon^{\a\b} (\pa_\a \check{\omega}^{mn}_\b - \check{\omega}^{m}_{\ p\a} \check{\omega}^{pn}_\b - \check{e}^m_\a \check{e}^n_\b) &= 0,
\qquad
\epsilon^{\a\b} (\pa_\a \hat{\omega}^{mn}_\b - \hat{\omega}^{m}_{\ p\a} \hat{\omega}^{pn}_\b + \hat{e}^m_\a \hat{e}^n_\b) &= 0.
\end{aligned}
\ee

\section{Expansion in fermions of the inverse operator $\op^{-1}$}\label{sec:inverse-op}
In this section we collect the relevant ingredients to construct the Lagrangian of the deformed model, once we include also the fermionic degrees of freedom.
Following~\cite{Delduc:2013qra} we define linear combinations of the projectors introduced in Section~\ref{sec:algebra-basis}
\be\label{eq:defin-op-d-dtilde}
d=P^{(1)}+\frac{2}{1-\eta^2}P^{(2)}-P^{(3)},
\qquad
\tilde{d}=-P^{(1)}+\frac{2}{1-\eta^2}P^{(2)}+P^{(3)},
\ee
that are understood as being one the transpose of the other $\Str[Md(N)]=\Str[\tilde{d}(M)N]$.
Here $\eta$ is the deformation parameter already introduced in Chapter~\ref{ch:qAdS5Bos}.
The above definitions imply
\be
\begin{aligned}
d(\gen{J}_{mn}) &=\tilde{d}(\gen{J}_{mn})= \gen{0}, \\
d(\gen{P}_m) &=\tilde{d}(\gen{P}_m)= \frac{2}{1-\eta^2} \gen{P}_m, \\
d(\gen{Q}^{I}) &=-\tilde{d}(\gen{Q}^{I}) = (\sigma_3)^{IJ} \gen{Q}^{J}\,.
\end{aligned}
\ee
We define the operator $R_\ag$
\be\label{Rgop}
R_\ag = \text{Adj}_{\alg{g}^{-1}} \circ R \circ \text{Adj}_{\alg{g}}\,,
\ee
that differ from~\eqref{eq:defin-Rg-bos} because now the group element $\ag$ given in~\eqref{eq:choice-full-coset-el} contains also the fermions.
For the operator $R$ we use again the definition
\be\label{Rop}
R(M)_{ij} = -i\, \eps_{ij} M_{ij}\,,\quad \eps_{ij} = \left\{\begin{array}{ccc} 1& \rm if & i<j \\
0&\rm if& i=j \\
-1 &\rm if& i>j \end{array} \right.\,,
\ee
that now becomes relevant also on odd roots.
Even when we consider the full $\psu(2,2|4)$, the operator $R$ multiplies by $-i$ and $+i$ generators associated with positive and negative roots respectively, and by $0$ Cartan generators. The operator $R$ still satisfies the modified classical Yang-Baxter equation~\eqref{eq:mod-cl-YBeq-R}.
The action of $R_{\gb}$ defined through the \emph{bosonic} coset element was studied already in Chapter~\ref{ch:qAdS5Bos}. In the basis of generators used in this chapter we write its action as
\be\label{eq:Rgb-action-lambda}
\begin{aligned}
R_{\gb}(\gen{P}_m) &= {\lambda_m}^n \gen{P}_n + \frac{1}{2} \lambda_m^{np} \gen{J}_{np} , \\
R_{\gb}(\gen{J}_{mn}) &=  \lambda_{mn}^{p} \gen{P}_p +\frac{1}{2} \lambda_{mn}^{pq} \gen{J}_{pq}, \\
R_{\gb}(\gen{Q}^{I}) &= R (\gen{Q}^{I}) = -\epsilon^{IJ} \gen{Q}^{J} ,\\
\end{aligned}
\ee
where the coefficients ${\lambda_m}^n, \lambda_m^{np}, \lambda_{mn}^{p}, \lambda_{mn}^{pq}$ for our particular parameterisation are collected in Appendix~\ref{sec:useful-results-eta-def}, see~\eqref{eq:lambda11}-\eqref{eq:lambda22a}.
They satisfy the properties
\be\label{eq:swap-lambda}
{\lambda_m}^n = - \, \eta_{mm'} \eta^{nn'} {\lambda_{n'}}^{m'},
\qquad
\check{\lambda}_m^{np} = \eta_{mm'} \eta^{nn'} \eta^{pp'} \check{\lambda}^{m'}_{n'p'},
\qquad
\hat{\lambda}_m^{np} = -\, \eta_{mm'} \eta^{nn'} \eta^{pp'} \hat{\lambda}^{m'}_{n'p'},
\ee
that are used to simplify some terms in the Lagrangian.

The operator used to deformed the model is defined as
\be\label{eq:defin-op-def-supercoset}
\op=1-\eta R_\ag \circ d\,,
\ee
and we find convenient to expand it in powers of the fermions $\theta$ as
\be
\op=\op_{(0)}+\op_{(1)}+\op_{(2)}+\cdots\,,
\ee
where $\op_{(k)}$ is the contribution at order $\theta^k$. When restricting the action of $\op_{(0)}$ to bosonic generators only, we recover the operator $\op_{\bos}$ defined in~\eqref{eq:def-op-def-bos-mod} and used to deform the purely bosonic model in Chapter~\ref{ch:qAdS5Bos}.
The action of $\op_{(0)}$ is defined also on odd elements.
The fermionic corrections that we will need read explicitly as
\be\label{eq:expans-ferm-op}
\begin{aligned}
\op_{(1)} (M) &= \eta [ \chi,R_{\gb} \circ d (M)] -\eta R_{\gb} ([\chi , d (M)] ), \\
\op_{(2)} (M) &= \eta [\chi , R_{\gb} ([\chi,d(M)])] - \frac{1}{2} \eta  R_{\gb} ( [\chi,[\chi,d(M)]])- \frac{1}{2} \eta  ( [\chi,[\chi,R_{\gb} \circ d(M)]]) \\
& = \frac{1}{2} \eta \left(  [\chi , [\chi , R_{\gb} \circ d(M)]] -R_{\gb} [\chi, [\chi,d(M)]] \right) - [\chi, \op_{(1)}(M)],
\end{aligned}
\ee
where we use again the notation $\chi \equiv \genQind{I}{}{}\ferm{\theta}{I}{}{} $.

It is actually the inverse operator $\op^{-1}$ that enters the definition of the deformed Lagrangian.
Its action is trivial only on generators $\gen{J}$ of grading 0, on which it acts as the the identity, at any order in fermions.
To find its action also on the other generators, we invert it perturbatively in powers of fermions.
We write it as
\be
\op^{-1}=\opinv_{(0)}+\opinv_{(1)}+\opinv_{(2)}+\cdots\,,
\ee
where $\opinv_{(k)}$ is the contribution at order $\theta^k$.
Demanding that $\op\cdot\op^{-1}=\op^{-1}\cdot\op=1$ we find
\begin{equation}\label{eq:expans-ferm-inv-op}
\begin{aligned}
\opinv_{(1)} & = - \opinv_{(0)} \circ \op_{(1)} \circ \opinv_{(0)} , \\
\opinv_{(2)} & = - \opinv_{(0)} \circ \op_{(2)} \circ \opinv_{(0)} - \opinv_{(1)} \circ \op_{(1)} \circ \opinv_{(0)}.
\end{aligned}
\end{equation}
We will not need higher order contributions.

\paragraph{Order $\theta^0$}
When we switch off the fermions in $\op^{-1}$ we recover the results of Chapter~\ref{ch:qAdS5Bos}. In particular, using the results of Appendix~\ref{app:bosonic-op-and-inverse}
rewritten for our basis of the generators we find that on $\gen{P}_m$ it gives
\be\label{eq:action-Oinv0-P}
\opinv_{(0)} (\gen{P}_{m})= {k_m}^n \gen{P}_n + \frac{1}{2} w_m^{np} \gen{J}_{np},
\ee
where we have
\be\label{eq:k-res1}
\begin{aligned}
& k_0^{\ 0} = k_4^{\ 4} = \frac{1}{1-\vk^2 \rho^2} , \qquad 
& k_1^{\ 1} = 1, \qquad 
& k_2^{\ 2} = k_3^{\ 3} = \frac{1}{1+\vk^2 \rho^4 \sin^2 \zeta}, \\
& k_5^{\ 5} =k_9^{\ 9}= \frac{1}{1+\vk^2 r^2}, \qquad 
& k_6^{\ 6} = 1, \qquad 
& k_7^{\ 7} = k_8^{\ 8}= \frac{1}{1+\vk^2 r^4 \sin^2 \xi},
\end{aligned}
\ee
\be\label{eq:k-res2}
\begin{aligned}
& k_0^{\ 4} = +k_4^{\ 0}= \frac{\vk \rho}{1-\vk^2 \rho^2}, \qquad 
& k_2^{\ 3}=-k_3^{\ 2}=- \frac{\vk \rho^2 \sin \zeta}{1+\vk^2 \rho^4 \sin^2 \zeta},  \\
& k_5^{\ 9} = - k_9^{\ 5}=\frac{\vk r}{1+\vk^2 r^2}, \qquad 
& k_7^{\ 8}= -k_8^{\ 7}=\frac{\vk r^2 \sin \xi}{1+\vk^2 r^4 \sin^2 \xi}.
\end{aligned}
\ee
The coefficients $w_m^{np}$ do not contribute to the Lagrangian, because the generators $\gen{J}$ are projected out by the operators $d,\tilde{d}$.

When acting on odd elements, the inverse operator rotates only the labels $I,J$ without modifying the spinor indices
\be
\opinv_{(0)} (\gen{Q}^{I})= \frac{1}{2} (1+\sqrt{1+\vk^2}) \, \gen{Q}^{I}- \frac{\vk}{2} {\sigma_1}^{IJ} \, \gen{Q}^{J}. 
\ee

\paragraph{Order $\theta^1$}
We use~\eqref{eq:expans-ferm-op} and~\eqref{eq:expans-ferm-inv-op} to compute the action of $\op_{(1)}$ and $\opinv_{(1)}$ on $\gen{P}_m$ and $\gen{Q}^I$.
First we find
\be
\op_{(1)}(\gen{P}_m) = 
\frac{\vk}{2} \gen{Q}^I \left[ 
 \delta^{IJ} \left(i \bg_m - \frac{1}{2} \lambda_m^{np} \bg_{np} \right) 
+ i \epsilon^{IJ} {\lambda_m}^n \bg_n  
\right] \theta_J\,,
\ee
and we use this result to get
\be
\begin{aligned}
\opinv_{(1)}(e^m\gen{P}_m) = 
-\frac{\vk}{4} \gen{Q}^I \ e^m {k_m}^n \ \Bigg[ 
& \left((1+\sqrt{1+\vk^2})\delta^{IJ} -\vk \sigma_1^{IJ}\right) \left(i \bg_n - \frac{1}{2} \lambda_n^{pq} \bg_{pq} \right) \\
 & + i \left((1+\sqrt{1+\vk^2}) \epsilon^{IJ} + \vk \sigma_3^{IJ}\right) {\lambda_n}^p \bg_p 
\Bigg] \theta_J\,.
\end{aligned}
\ee
For later convenience we rewrite this as
\be
\begin{aligned}
\opinv_{(1)}(e^m\gen{P}_m) = 
-\frac{\vk}{4} \gen{Q}^I \ e^m {k_m}^n \ \Bigg[ 
& \left((1+\sqrt{1+\vk^2})\delta^{IJ} -\vk \sigma_1^{IJ}\right) \Delta^1_n \\
 & + \left((1+\sqrt{1+\vk^2}) \epsilon^{IJ} + \vk \sigma_3^{IJ}\right) \Delta^3_n
\Bigg] \theta_J\,,
\end{aligned}
\ee
where $\Delta^1_n\equiv\left(i \bg_n- \frac{1}{2} \lambda_n^{pq} \bg_{pq} \right)$, $\Delta^3_n\equiv i{\lambda_n}^p \bg_p $.
On odd generators we find
\be
\begin{aligned}
\op_{(1)}(\gen{Q}^I\psi_I) =
\frac{1-\sqrt{1+\vk^2}}{\vk} \ \bar{\theta}_J  \Bigg[ 
\sigma_1^{JI} \left( i\bg_p +\frac{1}{2} \lambda^{mn}_{ p}  \bg_{mn}   \right) 
 - i \, \sigma_3^{JI} {\lambda_p}^n \bg_n 
\Bigg] \psi_I \ \eta^{pq} \gen{P}_q  + \cdots \,,
\end{aligned}
\ee
that helps to calculate
\be
\begin{aligned}
\opinv_{(1)}(\gen{Q}^I\psi_I) =
- \frac{1}{2} \ \bar{\theta}_K  \Bigg[ &
 (-\vk \sigma_1^{KI} +(-1+\sqrt{1+\vk^2})\delta^{KI}) \left( i\bg_p +\frac{1}{2}  \lambda^{mn}_{ p} \bg_{mn}  \right) \\
& + i \, (\vk \sigma_3^{KI} -(-1+\sqrt{1+\vk^2})\epsilon^{KI})  {\lambda_p}^n \bg_n
\Bigg] \psi_I \ k^{pq} \ \gen{P}_q  + \cdots \,.
\end{aligned}
\ee
In these formulae we have omitted the terms proportional to $\gen{J}_{mn}$ and replaced them by dots, since they do not contribute to the computation of the Lagrangian.
It is interesting to note that the last result can be rewritten as
\be
\begin{aligned}
\opinv_{(1)}(\gen{Q}^I\psi_I) =
- \frac{1}{2} \ \bar{\theta}_K  \Bigg[ &
 (-\vk \sigma_1^{KI} +(-1+\sqrt{1+\vk^2})\delta^{KI}) \bar{\Delta}^{1}_{p} \\
& + (\vk \sigma_3^{KI} -(-1+\sqrt{1+\vk^2})\epsilon^{KI}) \bar{\Delta}^{3}_{p}
\Bigg] \psi_I \  k^{pq} \ \gen{P}_q  + \cdots
\end{aligned}
\ee
where one needs to use \eqref{eq:swap-lambda}.
The quantities $\bar{\Delta}^{3}_{p'},\bar{\Delta}^{1}_{p'}$ are defined by $(\Delta^{3}_{p'} \theta_K)^\dagger \check{\gamma}^0 = \bar{\theta}_K \bar{\Delta}^{3}_{p'}$ and $(\Delta^{1}_{p'} \theta_K)^\dagger \check{\gamma}^0 = \bar{\theta}_K \bar{\Delta}^{1}_{p'}$.

\paragraph{Order $\theta^2$}
We need to compute the action of $\op$ and $\op^{-1}$ at order $\theta^2$ just on generators $\gen{P}_m$. Indeed the operators $\op_{(2)}$ and $\opinv_{(2)}$ acting on generators $\gen{Q}^I$ contribute only at quartic order in the Lagrangian.
First we find
\be
\begin{aligned}
\op_{(2)}({\gen{P}}_m) = 
- \frac{\vk}{2} \bar{\theta}_K \Bigg[ 
& \delta^{KI} \left(  -\bg_q \left(\bg_m +\frac{i}{4} \lambda_m^{np} \bg_{np} \right) 
+ \frac{i}{4} \lambda^{np}_{q} \bg_{np}  \bg_m \right) \\
& -\frac{1}{2} \epsilon^{KI} \left(  \bg_q \, {\lambda_m}^n \bg_n 
- {\lambda_q}^p \bg_p \bg_m  \right) 
\Bigg] \theta_I \ \eta^{qr} \gen{P}_r
+\cdots\,,
\end{aligned}
\ee
that gives
\be
\begin{aligned}
 -\opinv_{(0)} \circ \op_{(2)} \circ \opinv_{(0)}(e^m\gen{P}_m) = 
&- \frac{\vk}{2} \bar{\theta}_K \ e^m {k_m}^n \ \Bigg[ 
 \delta^{KI} \left(  \bg_u \left(\bg_n +\frac{i}{4} \lambda_n^{pq} \bg_{pq} \right)
- \frac{i}{4} \lambda^{pq}_{\ u} \bg_{pq}  \bg_n \right) \\
 & +\frac{1}{2} \epsilon^{KI} \left(  \bg_u {\lambda_n}^p \bg_p 
- {\lambda_u}^p \bg_p \bg_n  \right) 
\Bigg] \theta_I {k}^{uv} \ \gen{P}_v 
+\cdots\,.
\end{aligned}
\ee
Also here the dots stand for contributions proportional to $\gen{J}_{mn}$ that we are omitting.
The last formula that we will need is
\be
\begin{aligned}
&  -\opinv_{(1)} \circ \op_{(1)} \circ \opinv_{(0)}(e^m\gen{P}_m) =  - \frac{\vk}{4} \bar{\theta}_K \ e^m {k_m}^n \times\\
& \times \Bigg[ 
(-1+\sqrt{1+\vk^2}) \delta^{KJ} \bigg( \left( \bg_u -\frac{i}{2} \lambda_u^{pq}\bg_{pq}   \right) \left(\bg_n +\frac{i}{2} \lambda_n^{rs} \bg_{rs}\right) 
+ {\lambda_u}^p\bg_p   {\lambda_n}^r \bg_r \bigg) \\
 &+(-1+\sqrt{1+\vk^2}) \epsilon^{KJ} \bigg(-{\lambda_u}^p \bg_p \left(\bg_n +\frac{i}{2} \lambda_n^{rs} \bg_{rs}\right) 
+ \left( \bg_u -\frac{i}{2} \lambda_u^{pq}\bg_{pq}   \right) {\lambda_n}^r \bg_r \bigg) \\
 & -\vk \sigma_1^{KJ} \bigg( \left( \bg_u -\frac{i}{2}\lambda_u^{pq} \bg_{pq}   \right) \left(\bg_n +\frac{i}{2} \lambda_n^{rs} \bg_{rs}\right) 
- {\lambda_u}^p\bg_p   {\lambda_n}^r \bg_r \bigg) \\
 &+\vk \sigma_3^{KJ} \bigg( {\lambda_u}^p\bg_p \left(\bg_n +\frac{i}{2} \lambda_n^{rs} \bg_{rs}\right) 
+ \left( \bg_u -\frac{i}{2} \lambda_u^{pq} \bg_{pq}  \right) {\lambda_n}^r \bg_r \bigg)
\Bigg] \theta_J {k}^{uv} \ \gen{P}_v 
+\cdots\,,
\end{aligned}
\ee
where we have rewritten the result using~\eqref{eq:swap-lambda}

\section{The Lagrangian}\label{sec:-eta-def-lagr-quad-theta}
We first repeat the exercise of computing the Lagrangian in the undeformed case, as done in~\cite{Metsaev:1998it}, and then we derive the results for the $\eta$-deformed model.

\subsection{Undeformed case}
When we send $\eta\to 0$ we recover the Lagrangian for the superstring on {\adsfive}
\be
\lagr = - \frac{g}{2}  \left( \gamma^{\a\b}\Str[A^{(2)}_\a A^{(2)}_\b]+\epsilon^{\a\b}\Str[A^{(1)}_\a A^{(3)}_\b] \right)\,,
\ee
where $A^{(k)}=P^{(k)}A$.
The purely bosonic Lagrangian is easily found by setting the fermions to zero and one obtains
\be
\lagr_{\{00\}} = - \frac{g}{2} \gamma^{\a\b} e^m_\a e^n_\b \, \eta_{mn}.
\ee
We are using the notation $\{00\}$ to remind that we are considering both currents $A_\a$ and $A_\b$ entering the definition of the Lagrangian at order 0 in the fermions.
This Lagrangian matches with the one presented in~\eqref{eq:bos-act-undef-adsfive}.

If we want to look at the Lagrangian that is quadratic in fermions, we have to compute three terms, that according to our notation we call $\{02\},\{20\},\{11\}$.
It is convenient to consider the contributions $\{02\},\{20\}$ together. In fact---using the properties of the supertrace---it is easy to show that their sum is symmetric in $\a,\b$, meaning that what we get is multiplied just by $\g^{\a\b}$
\be
\lagr_{\{02\}} + \lagr_{\{20\}} = - \frac{g}{2} \gamma^{\a\b} \, i \bar{\theta}_I e^m_\a \bg_m  D^{IJ}_\b \theta_J .
\ee
By similar means one also shows that the contribution $\{11\}$ is antisymmetric in $\a,\b$ and thus yields the quadratic order of the Wess-Zumino term 
\be
\begin{aligned}
\lagr_{\{11\}} 
&=- \frac{{g}}{2}  \epsilon^{\a\b} \bar{\theta}_I  \sigma_3^{IJ}   \, i \, e^m_\a \bg_m D^{JK}_\b \theta_K +\text{tot. der.}
\end{aligned}
\ee
For the details of the computation we refer to the discussion for the deformed case after Eq.~\eqref{eq:orig-lagr-101}.
The sum of the contributions at quadratic order in $\theta$ gives
\be
\lagr^{\fer^2}=- \frac{{g}}{2} \, i \, \bar{\theta}_I \left(\gamma^{\a\b} \delta^{IJ}+ \epsilon^{\a\b}  \sigma_3^{IJ} \right)  e^m_\a \bg_m D^{JK}_\b \theta_K ,
\ee
that matches with the correct Lagrangian expected for type IIB~\eqref{eq:IIB-action-theta2}.
In particular one finds a five-form~\cite{Metsaev:1998it}
\be
\slashed{F}^{(5)}=F^{(5)}_{m_1m_2m_3m_4m_5}\G^{m_1m_2m_3m_4m_5}=4e^{-\varphi_0}(\G^{01234}-\G^{56789})\,,
\ee
originated by the term multiplied by $\eps^{IJ}$ in the definition~\eqref{eq:op-DIJ-psu224-curr} of $D^{IJ}$,
and a constant dilaton $\varphi=\varphi_0$.

\subsection{Deformed case}\label{sec:def-lagr-supercos}
In the deformed case the Lagrangian is defined as~\cite{Delduc:2013qra}
\be\label{eq:def-deformed-lagr-full}
\begin{aligned}
\lagr &= - \frac{g}{4} (1+\eta^2) (\gamma^{\a\b} - \epsilon^{\a\b}) \Str[\tilde{d}(A_\a) \, \op^{-1}(A_\b)] \\
& = - \frac{g}{2} \frac{\sqrt{1+\vk^2}}{1+\sqrt{1+\vk^2}} (\gamma^{\a\b} - \epsilon^{\a\b}) \Str[\tilde{d}(A_\a) \, \op^{-1}(A_\b)].
\end{aligned}
\ee
In the notation introduced in the previous section, the bosonic Lagrangian already obtained in Section~\ref{sec:def-bos-model} is
\be
\lagr_{\{000\}}  = - \frac{\tilde{g}}{2}  (\gamma^{\a\b} - \epsilon^{\a\b}) \ e^m_\a e^n_\b {k_n}^p \eta_{mp},
\qquad
\tilde{g} \equiv g\sqrt{1+\vk^2}.
\ee
Here we need three numbers to label the contribution to the Lagrangian: we indicate the order in powers of fermions for the current $A_\a$, for the inverse operator $\op^{-1}$ and for the current $A_\b$ respectively.
When we rewrite this result in the usual form~\eqref{eq:bos-str-action} of the Polyakov action we recover the deformed metric and the $B$-field of Section~\ref{sec:def-bos-model}, see Eq.~\eqref{eq:metrc-etaAdS5S5-sph-coord} and~\eqref{eq:B-field-etaAdS5S5-sph-coord}.
We may rewrite the deformed metric in terms of a vielbein $\widetilde{G}_{MN}=\widetilde{e}^m_M\widetilde{e}^n_N \eta_{mn}$, that we choose to be diagonal
\be\label{eq:def-vielb-comp}
\begin{aligned}
\widetilde{e}^0_t=\frac{\sqrt{1+\rho ^2}}{\sqrt{1-\vk ^2 \rho ^2}},
\quad
\widetilde{e}^1_{\psi_2}=-\rho  \sin \zeta,
\quad
\widetilde{e}^2_{\psi_1}=-\frac{\rho  \cos \zeta}{\sqrt{1+\vk ^2 \rho ^4 \sin ^2\zeta}},
\\
\widetilde{e}^3_\zeta=-\frac{\rho }{\sqrt{1+\vk ^2   \rho ^4 \sin ^2\zeta}},
\quad
\widetilde{e}^4_\rho=-\frac{1}{\sqrt{1+\rho ^2} \sqrt{1-\vk ^2 \rho ^2}},
\\
\widetilde{e}^5_\phi=\frac{\sqrt{1-r^2}}{\sqrt{1+\vk ^2 r^2}},
\quad
\widetilde{e}^6_{\phi_2}=-r \sin \xi ,
\quad
\widetilde{e}^7_{\phi_1}=-\frac{r \cos \xi }{\sqrt{1+\vk ^2   r^4 \sin ^2\xi}},
\\
\widetilde{e}^8_\xi=-\frac{r}{\sqrt{1+\vk ^2 r^4 \sin ^2\xi}},
\quad
\widetilde{e}^9_r=-\frac{1}{\sqrt{1-r^2} \sqrt{1+\vk ^2 r^2}}.
\end{aligned}
\ee
The Lagrangian quadratic in fermions is now divided into six terms: three of them when we choose $\op^{-1}$ at order 0 in fermions $(\{002\},\{200\},\{101\})$, two when it is at order 1 $(\{011\},\{110\})$ and one when it is at order 2 $(\{020\})$.
We start by considering the following two contributions
\be
\begin{aligned}
\lagr_{\{002\}} & = - \frac{\tilde{g}}{2} (\gamma^{\a\b} - \epsilon^{\a\b}) \, \frac{i}{2}\bar{\theta}_I (e^m_\a {k^n}_{m}\bg_n ) D^{IJ}_\b \theta_J , \\
\lagr_{\{200\}} & = - \frac{\tilde{g}}{2} (\gamma^{\a\b} - \epsilon^{\a\b}) \, \frac{i}{2}\bar{\theta}_I (e^m_\b {k_{m}}^{n}\bg_n ) D^{IJ}_\a \theta_J .
\end{aligned}
\ee
where ${k^n}_{m} = {k_q}^p \eta^{nq}\eta_{mp}$. Now the sum of $\lagr_{\{002\}}+\lagr_{\{200\}}$ gives a non-trivial contribution also to the Wess-Zumino term, since the matrix $k_{mn}$ has a non-vanishing anti-symmetric part.

Considering the case $\{101\}$, it is easy to see that the insertion of $\opinv_{(0)}$ between two odd currents does not change the fact that the expression is anti-symmetric in $\a,\b$.
In Appendix~\ref{app:der-Lagr-101} we show the steps needed to rewrite the original result~\eqref{eq:orig-lagr-101} in the standard form
\be\label{eq:Lagr-101}
\begin{aligned}
\lagr_{\{101\}} &= - \frac{\tilde{g}}{2}  \epsilon^{\a\b} \bar{\theta}_L \, i \, e^m_\a \bg_m \left( \sigma_3^{LK}  D^{KJ}_\b \theta_J
-\frac{\vk}{1+\sqrt{1+\vk^2}} \ \epsilon^{LK}  \mathcal{D}^{KJ}_\b \theta_J \right) \\
&= - \frac{\tilde{g}}{2}  \epsilon^{\a\b} \bar{\theta}_I \left( \sigma_3^{IJ}  
 -\frac{\vk}{1+\sqrt{1+\vk^2}} \ \epsilon^{IJ}  \right) \, i \, e^m_\a \bg_m \mathcal{D}_\b \theta_J  + \frac{\tilde{g}}{4}  \epsilon^{\a\b} \bar{\theta}_I  \sigma_1^{IJ}  e^m_\a \bg_m e^n_\b \bg_n  \theta_J ,
\end{aligned}
\ee
up to a total derivative.

Let us now consider the inverse operator at first order in the $\theta$ expansion.
The two contributions $\{011\},\{110\}$ can be naturally considered together\footnote{The result can be put in this form thanks to the properties~\eqref{eq:swap-lambda}.}
\be
\begin{aligned}
\lagr_{\{011\}+\{110\}}  
= & - \frac{\tilde{g}}{4}  (\gamma^{\a\b} - \epsilon^{\a\b}) 
 \bar{\theta}_K 
\Bigg[ - (\vk \sigma_1^{KI}-(-1+\sqrt{1+\vk^2})\delta^{KI}) \left( i\bg_p +\frac{1}{2}\bg_{mn}  \lambda_{p}^{mn} \right) \\
& + (\vk \sigma_3^{KI} - (-1+\sqrt{1+\vk^2})\epsilon^{KI}) \ i\bg_n {\lambda_p}^n  \Bigg] 
 (k^p_{\ q}e^q_{\a} D^{IJ}_\b +{k_{q}}^{p}e^q_{\b} D^{IJ}_\a )\theta_J.
\end{aligned}
\ee
To conclude, the last contribution to the Lagrangian that we should consider is the one in which the inverse operator is at order $\theta^2$. We find
\be
\begin{aligned}
\lagr_{\{020\}}  = & - \frac{\tilde{g}}{2} (\gamma^{\a\b} - \epsilon^{\a\b}) \  \frac{\vk}{4} e^v_\a e^m_\b \, {k^{u}}_v {k_m}^n \, \bar{\theta}_K  \\
\Bigg[  
&- 2 \delta^{KI} \left(  \bg_u \left(\bg_n +\frac{i}{4} \lambda_n^{pq} \bg_{pq} \right)
- \frac{i}{4} \bg_{pq}  \bg_n \lambda^{pq}_{\ u}\right) 
  - \epsilon^{KI} \left(  \bg_u {\lambda_n}^p \bg_p 
-  \bg_p \bg_n {\lambda_u}^p \right)  \\
& -(-1+\sqrt{1+\vk^2}) \delta^{KI} \bigg( \left( \bg_u -\frac{i}{2} \bg_{pq}  \lambda_u^{pq} \right) \left(\bg_n +\frac{i}{2} \lambda_n^{rs} \bg_{rs} \right) 
 + \bg_p  {\lambda_u}^p {\lambda_n}^r \bg_r \bigg) \\
& -(-1+\sqrt{1+\vk^2}) \epsilon^{KI} \bigg(- \bg_p {\lambda_u}^p \left(\bg_n +\frac{i}{2} \lambda_n^{rs} \bg_{rs}\right) 
 + \left( \bg_u -\frac{i}{2} \bg_{pq}  \lambda_u^{pq} \right) {\lambda_n}^r \bg_r \bigg) \\
& + \vk \sigma_1^{KI} \bigg( \left( \bg_u -\frac{i}{2} \bg_{pq}  \lambda_u^{pq} \right) \left(\bg_n +\frac{i}{2} \lambda_n^{rs} \bg_{rs}\right) 
 - \bg_p  {\lambda_u}^p {\lambda_n}^r \bg_r \bigg) \\
& -\vk \sigma_3^{KI} \bigg( \bg_p{\lambda_u}^p \left(\bg_n +\frac{i}{2} \lambda_n^{rs} \bg_{rs}\right) 
 + \left( \bg_u -\frac{i}{2} \bg_{pq}  \lambda_u^{pq} \right) {\lambda_n}^r \bg_r \bigg)
\Bigg] \theta_I .
\end{aligned}
\ee
Summing up all the above contributions we discover that the result is \emph{not} written in the standard form of the Green-Schwarz action for type IIB superstring~\eqref{eq:IIB-action-theta2}. This issue is addressed in the next section.

\subsection{Canonical form}\label{sec:canonical-form}
The Lagrangian for the deformed model that we obtain from the definition~\eqref{eq:def-deformed-lagr-full} is not in the standard form of the Green-Schwarz action for type IIB superstring. 
It is clear that a field redefinition of the bosonic and fermionic coordinates will in general modify the form of the action.
The strategy of this section is to find a field redefinition that recasts the result that we have obtained in the desired form~\eqref{eq:IIB-action-theta2}.

Let us focus for the moment just on the contributions involving derivatives on fermions, whose expression is not canonical.
For convenience we collect these terms here. We write separately the contributions contracted with $\g^{\a\b}$ and $\epsilon^{\a\b}$
\be\label{eq:non-can-lagr-kin}
\begin{aligned}
\lagr^{\g,\pa} = -\frac{\tilde{g}}{2}  \ \g^{\a\b} \bar{\theta}_I 
\Bigg[ &
 \frac{i}{2} (\sqrt{1+\vk^2} \delta^{IJ} - \vk \sigma_1^{IJ} ) \bg_n 
 -\frac{1}{4} (\vk \sigma_1^{IJ} -(-1+ \sqrt{1+ \vk^2}) \delta^{IJ} )  \lambda^{pq}_{n} \bg_{pq} 
\\
& +\frac{i}{2} (\vk \sigma_3^{IJ} -(-1+ \sqrt{1+ \vk^2}) \epsilon^{IJ} ) {\lambda_{n}}^{p} \bg_{p} 
\Bigg] 
 ({k^n}_{m}+{k_{m}}^n) e^m_\a \pa_\b \theta_J,
\end{aligned}
\ee
\be\label{eq:non-can-lagr-WZ}
\begin{aligned}
\lagr^{\epsilon,\pa} =  -\frac{\tilde{g}}{2}  \ \epsilon^{\a\b} \bar{\theta}_I 
\Bigg[ &
\Bigg(
-\frac{i}{2} (\sqrt{1+\vk^2} \delta^{IJ}  - \vk \sigma_1^{IJ} ) \bg_n 
  +\frac{1}{4} (\vk \sigma_1^{IJ} -(-1+ \sqrt{1+ \vk^2}) \delta^{IJ} )  \lambda^{pq}_{n} \bg_{pq} 
\\
&  -\frac{i}{2} (\vk \sigma_3^{IJ} -(-1+ \sqrt{1+ \vk^2}) \epsilon^{IJ} )  {\lambda_{n}}^{p} \bg_{p} 
\Bigg) ({k^n}_{m}-{k_{m}}^n) \\
& +i  \left( \sigma_3^{IJ} - \frac{-1+ \sqrt{1+ \vk^2}}{\vk} \epsilon^{IJ} \right) \bg_m 
\Bigg]  e^m_\a \pa_\b \theta_J.
\end{aligned}
\ee
To simplify the result we first redefine our fermions as
\be\label{eq:red-fer-2x2-sp}
\theta_I \to \frac{\sqrt{1+\sqrt{1+\vk^2}}}{\sqrt{2}} \left(\delta^{IJ} + \frac{\vk}{1+\sqrt{1+\vk^2}} \sigma_1^{IJ} \right) \theta_J.
\ee
The contributions to the Lagrangian that we are considering are then transformed as
\be
\begin{aligned}
\lagr^{\g,\pa} \to \lagr^{\g,\pa} = \lagr^{\g,\pa}_1 + \lagr^{\g,\pa}_2,
\\
\lagr^{\g,\pa}_1 = -\frac{\tilde{g}}{2}  \ \g^{\a\b} \bar{\theta}_I 
\Bigg[ &
 \frac{i}{2}  \delta^{IJ}  \bg_n 
+\frac{i}{2} \vk \sigma_3^{IJ}  {\lambda_{n}}^{p} \bg_{p} 
\Bigg] 
 ({k^n}_{m}+{k_{m}}^n) e^m_\a \pa_\b \theta_J,
\\
\lagr^{\g,\pa}_2 = -\frac{\tilde{g}}{2}  \ \g^{\a\b} \bar{\theta}_I 
\Bigg[ &
 -\frac{1}{4} (\vk \sigma_1^{IJ} +(-1+ \sqrt{1+ \vk^2}) \delta^{IJ} )  \lambda^{pq}_{n} \bg_{pq} 
\\
& -\frac{i}{2} (-1+ \sqrt{1+ \vk^2}) \epsilon^{IJ}  {\lambda_{n}}^{p} \bg_{p} 
\Bigg] 
 ({k^n}_{m}+{k_{m}}^n) e^m_\a \pa_\b \theta_J.
\end{aligned}
\ee
\be
\begin{aligned}
\lagr^{\epsilon,\pa} \to \lagr^{\epsilon,\pa} &= \lagr^{\epsilon,\pa}_1 + \lagr^{\epsilon,\pa}_2,
\\
\lagr^{\epsilon,\pa}_1 =  -\frac{\tilde{g}}{2}  \ \epsilon^{\a\b} \bar{\theta}_I 
\Bigg[ &
-\Bigg(
\frac{i}{2}  \delta^{IJ}  \bg_n 
  +\frac{i}{2} \vk \sigma_3^{IJ}   {\lambda_{n}}^{p} \bg_{p} 
\Bigg) ({k^n}_{m}-{k_{m}}^n) 
 +i   \sigma_3^{IJ} \bg_m 
\Bigg]  e^m_\a \pa_\b \theta_J ,
\\
\lagr^{\epsilon,\pa}_2 =  -\frac{\tilde{g}}{2}  \ \epsilon^{\a\b} \bar{\theta}_I 
\Bigg[ &
\Bigg(
  \frac{1}{4} (\vk \sigma_1^{IJ} +(-1+ \sqrt{1+ \vk^2}) \delta^{IJ} )   \lambda^{pq}_{n} \bg_{pq} 
\\
& +\frac{i}{2} (-1+ \sqrt{1+ \vk^2}) \epsilon^{IJ}   {\lambda_{n}}^{p} \bg_{p} 
\Bigg) ({k^n}_{m}-{k_{m}}^n) \\
& -i   \frac{-1+ \sqrt{1+ \vk^2}}{\vk} \epsilon^{IJ} \bg_m 
\Bigg]  e^m_\a \pa_\b \theta_J.
\end{aligned}
\ee
The various terms have been divided into $\lagr_1^\pa$ and $\lagr_2^\pa$ according to the symmetry properties of the objects involved. 
In particular, given an expression of the form $\theta_I M^{IJ} \pa\theta_J$, we decide to organise the terms according to
\be
\begin{aligned}
\theta_I M^{IJ} \pa\theta_J = -\pa\theta_I M^{IJ} \theta_J &\quad\implies\quad \lagr_1^\pa\,,
\\
\theta_I M^{IJ} \pa\theta_J = +\pa\theta_I M^{IJ} \theta_J &\quad\implies\quad \lagr_2^\pa\,.
\end{aligned}
\ee
The symmetry properties are dictated by purely algebraic manipulations---we are not integrating by parts---based on the symmetry properties of the gamma matrices contained in $M^{IJ}$, and on the symmetry properties of $M^{IJ}$ under the exchange of $I,J$.
We also use the ``Majorana-flip''relations of Eq.~\eqref{eq:symm-gamma-otimes-gamma}. 

We make this distinction because we can show that we can remove $\mathcal{L}^{\g,\pa}_2$ by shifting the bosonic coordinates with $\vk$-dependent corrections that are quadratic in fermions.
Let us consider the redefinition
\be\label{eq:red-bos}
X^M \longrightarrow X^M + \bar{\theta}_I \ f^M_{IJ} (X) \ \theta_J,
\ee
where $f^M_{KI} (X)$ is a function of the bosonic coordinates that for the moment is not fixed.
Requiring that the shift is non-vanishing--we use~\eqref{eq:Maj-flip}---shows that the quantity $f^M_{IJ} (X)$ has the same symmetry properties of the terms that we collected in $\lagr_2^\pa$.
This shift produces contributions to the \emph{fermionic} Lagrangian originating from the \emph{bosonic} one~\eqref{eq:bos-lagr-eta-def-Pol}.
We find that it is modified as $\lagr^{\alg{b}} \to \lagr^{\alg{b}} +\delta \lagr^{\alg{b},\g}_m +\delta \lagr^{\alg{b},\g}_2 +\delta \lagr^{\alg{b},\epsilon}_m +\delta \lagr^{\alg{b},\epsilon}_2+\mathcal{O}(\theta^4)$ where
\be\label{eq:red-bos-lagr}
\begin{aligned}
\delta \lagr^{\alg{b},\g}_m &= + \tilde{g} \g^{\a\b} \left(  -  \pa_\a X^M \ \bar{\theta}_I \ \widetilde{G}_{MN} \left( \pa_\b f^N_{IJ} \right) \  \theta_J 
 - \frac{1}{2}  \partial_\alpha X^M \partial_\beta X^N \pa_P \widetilde{G}_{MN} \ \bar{\theta}_I \, f^P_{IJ} \theta_J  \right),\\
\delta \lagr^{\alg{b},\g}_2 &= + \tilde{g} \g^{\a\b} \left( - 2  \pa_\a X^M \ \bar{\theta}_I \ \widetilde{G}_{MN} f^N_{IJ} \ \pa_\b \theta_J  \right),\\
\delta \lagr^{\alg{b},\epsilon}_m &=+ \tilde{g} \epsilon^{\a\b}   \left( +  \pa_\a X^M \ \bar{\theta}_I \ \widetilde{B}_{MN} \left( \pa_\b f^N_{IJ} \right) \  \theta_J 
 + \frac{1}{2}   \partial_\alpha X^M \partial_\beta X^N \pa_P \widetilde{B}_{MN} \ \bar{\theta}_I \, f^P_{IJ} \theta_J \right), \\
\delta \lagr^{\alg{b},\epsilon}_2 &=+ \tilde{g} \epsilon^{\a\b}   \left( 2  \pa_\a X^M \ \bar{\theta}_I \ \widetilde{B}_{MN} f^N_{IJ} \ \pa_\b \theta_J \right).
\end{aligned}
\ee
Here we have used $\pa \bar{\theta}_I \ f^M_{IJ} (X) \ \theta_J = + \bar{\theta}_I \ f^M_{IJ} (X) \ \pa \theta_J$, consequence of the symmetry properties of $f^M_{IJ} (X)$, and we have stopped at quadratic order in fermions.

It is now easy to see that if we define the function
\be\label{eq:def-shift-bos-f}
\begin{aligned}
f^M_{IJ}(X) &=  e^{Mp} \Bigg[ \frac{1}{8}  \left( \vk \sigma_1^{IJ} - (1-\sqrt{1+\vk^2}) \delta^{IJ} \right)   \lambda_{p}^{mn}   \bg_{mn}  
-\frac{i}{4} (1-\sqrt{1+\vk^2}) \epsilon^{IJ}   {\lambda_p}^n  \bg_n \Bigg],
\end{aligned}
\ee
then we are able to remove completely the contribution $\lagr^{\g,\pa}_2$ from the Lagrangian
\be
\lagr^{\g,\pa}_2 + \delta \lagr^{\alg{b},\g}_2 =0.
\ee
On the other hand this shift of the bosonic coordinates is not able to remove completely $\mathcal{L}^{\epsilon,\pa}_2$: there is actually cancellation of the terms with\footnote{This statement is true if one includes also the components $B_{t\rho},B_{\phi r}$ of the $B$-field in the bosonic Lagrangian. Clearly these will contribute giving also new terms with no derivatives on fermions contained in $\delta \lagr_m^{\bos,\eps}$ of~\eqref{eq:red-bos-lagr}. If these components are not included, cancellation of terms with $\delta^{IJ}, \sigma_1^{IJ}$ is not complete, but what is left may be rewritten as a term with no derivatives on fermions, up to a total derivative. The two ways of proceeding are equivalent.} $\delta^{IJ}, \sigma_1^{IJ}$, but the ones with $\epsilon^{IJ}$ are not removed. However, in the Wess-Zumino term we are allowed to perform partial integration\footnote{Performing partial integration in the Lagrangian with $\g^{\a\b}$ would generate derivatives of the worldsheet metric and also of $\pa_\a X^M$.} to rewrite the result such that---up to a total derivative---the partial derivative acts on the bosons and not on the fermions
\be\label{eq:WZ-shift-bos-eps}
\begin{aligned}
\lagr^{\epsilon,\pa}_2 + \delta \lagr^{\alg{b},\epsilon}_2 = &
\frac{\tilde{g}}{2}  \epsilon^{\a\b} \bar{\theta}_I \frac{-1+\sqrt{1+\vk^2}}{\vk} \epsilon^{IJ}  \\
& e^m_\a  \left(
i  \delta^q_m - \frac{i}{2} \vk (k^n_{\ m} - {k_{m}}^{n} ) \lambda_n^{\ q}  + \frac{i}{2} \vk \tilde{B}_{mn} (k^{pn} + k^{np} ) \lambda_p^{\ q}
\right) \bg_q \pa_\b \theta_J \\
= &
\frac{\tilde{g}}{2}  \epsilon^{\a\b} \bar{\theta}_I \frac{-1+\sqrt{1+\vk^2}}{\vk} \epsilon^{IJ}   e^m_\a  i  \bg_m \pa_\b \theta_J \\
= &
-\frac{\tilde{g}}{4}  \epsilon^{\a\b} \bar{\theta}_I \frac{-1+\sqrt{1+\vk^2}}{\vk} \epsilon^{IJ} 
  \pa_\a X^M  \left(\pa_\b  e^m_M   \right)  i\bg_m \theta_J 
 + \text{tot. der.}
\end{aligned}
\ee
This method works thanks to the symmetry properties of $f^M_{IJ}(X)$. We have also used the identity
\be
k^p_{\ m} - {k_{m}}^{p}  - \tilde{B}_{mn} (k^{pn} + k^{np} )  =0.
\ee
After the shift of the bosonic coordinates, the only terms containing derivatives on fermions are $\lagr^{\g,\pa}_1$ and $\lagr^{\epsilon,\pa}_1$.
Let us stress again that the shift will also introduce new couplings without derivatives on fermions, as showed in~\eqref{eq:red-bos-lagr}.
We collect in Eq.~\eqref{eq:lagr-gamma-no-F-red} and~\eqref{eq:lagr-epsilon-no-F-red} the expression for the total Lagrangian at this point.

\medskip

In order to put the remaining terms in canonical form we redefine the fermions as $\theta_I \to U_{IJ}\theta_J$, where the matrix $U_{IJ}$ acts both on the $2\times 2$ space spanned by the labels $I,J$ and on the space of spinor indices---that we are omitting here.
We actually write the matrix $U_{IJ}$ as factorised in the AdS and sphere spinor indices parts
\be\label{eq:red-ferm-Lor-as}
\begin{aligned}
\theta_I &\to( U_{IJ}^{\alg{a}}\otimes  U_{IJ}^{\alg{s}})\theta_J\,,
\\
\theta_{I\ul{\a}\ul{a}} &\to( U_{IJ}^{\alg{a}})_{\ul{\a}}^{\ \ul{\nu}}  (U_{IJ}^{\alg{s}})_{\ul{a}}^{\ \ul{b}}\theta_{J\ul{\nu}\ul{b}}\,.
\end{aligned}
\ee
This is not the most generic redefinition, but it will turn out to be enough.
Each of the matrices $U_{IJ}^{\alg{a}}$ and $U_{IJ}^{\alg{s}}$ may be expanded in terms of the tensors spanning the $2\times 2$ space
\be
U_{IJ}^{\alg{a},\alg{s}}=\delta_{IJ}\, U_{\delta}^{\alg{a},\alg{s}}+\sigma_{1\, IJ}\, U_{\sigma_1}^{\alg{a},\alg{s}}+\epsilon_{IJ}\, U_{\epsilon}^{\alg{a},\alg{s}}+\sigma_{3\, IJ}\, U_{\sigma_3}^{\alg{a},\alg{s}}\,.
\ee
The objects $U_{\mu}^{\alg{a},\alg{s}}$ with $\mu=\delta,\sigma_1,\epsilon,\sigma_3$ are $4\times 4$ matrices that may be written in the convenient basis of $4\times 4$ gamma matrices.
From the Majorana condition~\eqref{eqMajorana-cond-compact-not} we find that in order to preserve $\theta_I^\dagger \bg^0=+ \theta_I^t \, (K\otimes K)$ under the field redefinition, we have to impose 
\be
\bg^0\, ((U_{\mu}^{\alg{a}})^\dagger \otimes (U_{\mu}^{\alg{s}})^\dagger) \bg^0= -(K\otimes K) ((U_{\mu}^{\alg{a}})^t \otimes (U_{\mu}^{\alg{s}})^t) (K\otimes K)\,.
\ee
We impose $\check{\g}^0\, (U_{\mu}^{\alg{a}})^\dagger  \check{\g}^0= K (U_{\mu}^{\alg{a}})^t K$ and $(U_{\mu}^{\alg{s}})^\dagger=- K (U_{\mu}^{\alg{s}})^t K$ and we find that they are solved by
\be
\begin{aligned}
& U^{\alg{a}}_\mu \equiv f^{\alg{a}}_{\mu} \mathbf{1} + i f^p_\mu \check{\g}_p  + \frac{1}{2} f^{pq}_\mu \check{\g}_{pq} ,
\qquad
U^{\alg{s}}_\mu \equiv f^{\alg{s}}_{\mu} \mathbf{1} - f^p_\mu  \hat{\g}_p - \frac{1}{2} f^{pq}_\mu \hat{\g}_{pq} ,
\end{aligned}
\ee
where the coefficients $f$ are \emph{real} functions of the bosonic coordinates.
In other words, what we have fixed in the above equation are the factors of $i$ in front of these coefficients, using~\eqref{eq:symm-prop-5dim-gamma} and~\eqref{eq:herm-conj-prop-5dim-gamma}.

On the other hand the barred version of the fermions will be redefined as $\bar{\theta}_I\to \bar{\theta}_J \bar{U}_{IJ}$ with a matrix $\bar{U}_{IJ}=( \bar{U}_{IJ}^{\alg{a}}\otimes  \bar{U}_{IJ}^{\alg{s}})$, that we expand again in the tensors of the $2\times 2$ space. To preserve $\bar{\theta}_I=\theta_I^t (K\otimes K)$ we have to impose 
\be
( \bar{U}_{\mu}^{\alg{a}}\otimes  \bar{U}_{\mu}^{\alg{s}})=(K\otimes K)((U_{\mu}^{\alg{a}})^t \otimes (U_{\mu}^{\alg{s}})^t) (K\otimes K),
\ee
that allows us to define
\be
\begin{aligned}
& \bar{U}^{\alg{a}}_\mu\equiv f^{\alg{a}}_{\mu} \mathbf{1} + i f^p_\mu \check{\g}_p  - \frac{1}{2} f^{pq}_\mu \check{\g}_{pq}  ,
\qquad
\bar{U}^{\alg{s}}_\mu\equiv f^{\alg{s}}_{\mu} \mathbf{1} - f^p_\mu  \hat{\g}_p + \frac{1}{2} f^{pq}_\mu \hat{\g}_{pq}  .
\end{aligned}
\ee
Here the coefficients $f$ are the same entering the definition of $U^{\alg{a},\alg{s}}_\mu $.

In order to get a canonical expression for the terms containing derivatives on fermions
\be
\begin{aligned}
\lagr^{\g,\pa}_1 \to &
 -\frac{\tilde{g}}{2} \g^{\a\b} \, i  \, \bar{\theta}_I \, \delta^{IJ} \, \widetilde{e}^m_\a \bg_m \pa_\b \theta_J, \\
\lagr^{\epsilon,\pa}_1 \to &
 -\frac{\tilde{g}}{2} \epsilon^{\a\b} \, i  \, \bar{\theta}_I \, \sigma^{IJ}_3 \, \widetilde{e}^m_\a \bg_m \pa_\b \theta_J ,
\end{aligned}
\ee
where $\widetilde{e}^m_\a$ is the deformed vielbein given in~\eqref{eq:def-vielb-comp}, we set all coefficients $f$ for the field redefinition to $0$, except for the redefinition ${U}_{\mu}^{\alg{a}}$ of the AdS factor
\be
\begin{aligned}
f^{\alg{a}}_{\delta} &=
\frac{1}{2} \sqrt{\frac{\left(1+\sqrt{1-\vk ^2 \rho ^2}\right) \left(1+\sqrt{1+\vk ^2 \rho ^4 \sin ^2\zeta }\right)}{\sqrt{1-\vk ^2 \rho ^2} \sqrt{1+\vk ^2 \rho ^4 \sin
   ^2\zeta }}}, \\
f^1_\delta &=
-\frac{\vk ^2 \rho ^3 \sin \zeta }{f^{\alg{a}}_{\text{den}}}, \\
f^{04}_{\sigma_3} &=
\frac{\vk  \rho  \left(1+\sqrt{1+\vk ^2 \rho ^4 \sin ^2\zeta }\right)}{f^{\alg{a}}_{\text{den}}}, \\
f^{23}_{\sigma_3} &=
\frac{\vk  \rho ^2 \sin \zeta  \left(1+\sqrt{1-\vk ^2 \rho ^2}\right)}{f^{\alg{a}}_{\text{den}}}, \\
f^{\alg{a}}_{\text{den}} &\equiv 2 (1-\vk ^2 \rho ^2)^{\frac{1}{4}} (1+\vk ^2 \rho ^4 \sin ^2\zeta )^{\frac{1}{4}} \sqrt{1+\sqrt{1-\vk ^2 \rho ^2}}  \sqrt{1+\sqrt{1+\vk ^2 \rho ^4
   \sin ^2\zeta }},
\end{aligned}
\ee
 and for the redefinition ${U}_{\mu}^{\alg{s}}$ of the sphere factor
\be
\begin{aligned}
f^{\alg{s}}_{\delta} &=
\frac{1}{2} \sqrt{\frac{\left(1+\sqrt{1+\vk ^2 r ^2}\right) \left(1+\sqrt{1+\vk ^2 r ^4 \sin ^2\xi }\right)}{\sqrt{1+\vk ^2 r ^2} \sqrt{1+\vk ^2 r ^4 \sin
   ^2\xi }}}, \\
f^6_{\delta} &=
\frac{\vk ^2 r ^3 \sin \xi }{f^{\alg{s}}_{\text{den}}}, \\
f^{59}_{\sigma_3} &=
\frac{\vk  r  \left(1+\sqrt{1+\vk ^2 r ^4 \sin ^2\xi }\right)}{f^{\alg{s}}_{\text{den}}}, \\
f^{78}_{\sigma_3} &=
\frac{\vk  r ^2 \sin \xi  \left(1+\sqrt{1+\vk ^2 r ^2}\right)}{f^{\alg{s}}_{\text{den}}}, \\
f^{\alg{s}}_{\text{den}} &\equiv 2 (1+\vk ^2 r ^2)^{\frac{1}{4}} (1+\vk ^2 r ^4 \sin ^2\xi )^{\frac{1}{4}} \sqrt{1+\sqrt{1+\vk ^2 r ^2}}  \sqrt{1+\sqrt{1+\vk ^2 r ^4
   \sin ^2\xi }}.
\end{aligned}
\ee
Since the particular redefinition that we have chosen is diagonal in the labels $I,J$---it involves just the tensors $\delta$ and $\sigma_3$---it is interesting to look at the transformation rules for the two sets of Majorana-Weyl fermions separately.
We define
\be
U_{(1)} \equiv U_\delta+U_{\sigma_3}\,,
\qquad
U_{(2)} \equiv U_\delta-U_{\sigma_3}\,,
\implies
\theta_I\to U_{(I)} \theta_I\quad I=1,2.
\ee
These matrices satisfy 
\be\label{eq:transf-rule-gamma-ferm-rot}
\begin{aligned}
& \bar{U}_{(I)} U_{(I)}  = \gen{1}_4,
\qquad
&& \bar{U}_{(I)} \bg_m U_{(I)} =  (\Lambda_{(I)})_m^{\ n} \bg_n ,
\\
&U_{(I)} \bar{U}_{(I)} = \gen{1}_4,
\qquad
 &&\bar{U}_{(I)} \bg_{mn} U_{(I)}  =  (\Lambda_{(I)})_m^{\ p}  (\Lambda_{(I)})_n^{\ q} \bg_{pq} ,
\end{aligned}
\ee
where we do not sum over $I$.
The matrices  $\Lambda_{(I)}$ look very simple
\be\label{eq:Lambda-res1}
\begin{aligned}
& (\Lambda_{(I)})_0^{\ 0} = (\Lambda_{(I)})_4^{\ 4} = \frac{1}{\sqrt{1-\vk^2 \rho^2}} , \quad
&& (\Lambda_{(I)})_5^{\ 5} =(\Lambda_{(I)})_9^{\ 9}= \frac{1}{\sqrt{1+\vk^2 r^2}}, \\  
& (\Lambda_{(I)})_1^{\ 1} = 1, \quad
&& (\Lambda_{(I)})_6^{\ 6} = 1, \\  
& (\Lambda_{(I)})_2^{\ 2} = (\Lambda_{(I)})_3^{\ 3} = \frac{1}{\sqrt{1+\vk^2 \rho^4 \sin^2 \zeta}}, \quad
&& (\Lambda_{(I)})_7^{\ 7} =(\Lambda_{(I)})_8^{\ 8}= \frac{1}{\sqrt{1+\vk^2 r^4 \sin^2 \xi}}, 
\end{aligned}
\ee
\be\label{eq:Lambda-res2}
\begin{aligned}
& (\Lambda_{(I)})_0^{\ 4} = +(\Lambda_{(I)})_4^{\ 0}= -\frac{\sigma_{3II}\, \vk \rho}{\sqrt{1-\vk^2 \rho^2}}, \quad 
&&(\Lambda_{(I)})_5^{\ 9} = - (\Lambda_{(I)})_9^{\ 5}=-\frac{\sigma_{3II}\, \vk r}{\sqrt{1+\vk^2 r^2}},\\
& (\Lambda_{(I)})_2^{\ 3}=-(\Lambda_{(I)})_3^{\ 2}= \frac{\sigma_{3II}\, \vk \rho^2 \sin \zeta}{\sqrt{1+\vk^2 \rho^4 \sin^2 \zeta}},  \quad
&& (\Lambda_{(I)})_7^{\ 8}= -(\Lambda_{(I)})_8^{\ 7}=-\frac{\sigma_{3II}\, \vk r^2 \sin \xi}{\sqrt{1+\vk^2 r^4 \sin^2 \xi}},
\end{aligned}
\ee
and they satisfy the remarkable property of being ten-dimensional Lorentz transformations
\be
(\Lambda_{(I)})_m^{\ p}\ (\Lambda_{(I)})_n^{\ q}\  \eta_{pq}=\eta_{mn}\,, \qquad I=1,2\,.
\ee
We refer to Appendix~\ref{app:total-lagr-field-red} for some comments on how to efficiently implement this field redefinition of the fermions in the Lagrangian.
\subsection{The quadratic Lagrangian}
In this section we show that the field redefinition that was found to put the terms with derivatives acting on fermions into canonical form is actually able to put the whole action in the standard form of Green-Schwarz for type IIB superstring~\eqref{eq:IIB-action-theta2}.

In order to identify the bakground fields that are coupled to the fermions, we can do the computation separately for the part contracted with $\g^{\a\b}$ and the one with $\epsilon^{\a\b}$, and then check that they yield the same results.
It is convenient to consider separately the terms that are diagonal and the ones that are off-diagonal in the labels $I,J$.
The correct identification of the fields is achieved by looking at the tensor structure \emph{after} the rotation of the fermions~\eqref{eq:red-ferm-Lor-as} is implemented.
In the contribution contracted with $\g^{\a\b}$, the terms without derivatives on fermions that are multiplied by $\delta^{IJ}$ will then correspond to the coupling to the spin connection. The terms multiplied by $\sigma_3^{IJ}$ contain the coupling to the field strength of the $B$-field.
The RR-fields are identified by looking at the contributions to the Lagrangian off-diagonal in $IJ$, and by selecting the appropriate Gamma-matrix structure.
Taking into account just the anti-symmetry in the indices, the number of different components for the form $F^{(r)}$ is given by $\sum_{n_1 = 0}^9 \sum_{n_2 =n_1 +1}^9 \cdots \sum_{n_r =n_{r-1} +1}^9$, meaning
\be
F^{(1)}:
\quad
10,
\qquad
F^{(3)}:
\quad
120,
\qquad
F^{(5)}:
\quad
252.
\ee
If we consider also self-duality for $F^{(5)}$ this gives a total number of $10+120+252/2=256$ different components. It is then possible to identify uniquely the RR fields, since the matrices $\bg$ of rank $1,3,5$ are all linearly independent and a $16\times 16$-matrix has indeed $256$ entries.
To impose automatically the self-duality condition for the 5-form, we will rewrite---when necessary---the components in terms of the components $F^{(5)}_{0qrst}$ (there are 126 of them), using
\be
F_{m_1m_2m_3m_4m_5}=+\frac{1}{5!}\eps_{m_1\ldots m_{10}}F^{m_6m_7m_8m_9m_{10}}\,,
\ee
where $\eps^{0\ldots 9}=1$ and $\eps_{0\ldots 9}=-1$.
One should remember that for the Wess-Zumino contribution with $\eps^{\a\b}$ there is an additional $\sigma_3^{IJ}$ as in~\eqref{eq:IIB-action-theta2}.

We find that the Lagrangian quadratic in fermions
 is written in the standard form
\be\label{eq:lagr-quad-ferm}
\begin{aligned}
\lagr^{\fer^2} &= - \frac{\tilde{g}}{2}  \, i \, \bar{\Theta}_I \,(\g^{\a\b} \delta^{IJ}+\epsilon^{\a\b} \sigma_3^{IJ}) \widetilde{e}^m_\a \G_m \, \widetilde{D}^{JK}_\b \Theta_K  , 
\end{aligned}
\ee
where the $32\times 32$ ten-dimensional $\G$-matrices are constructed in~\eqref{eq:def-10-dim-Gamma} and the $32$-dimensional fermions $\T$ in~\eqref{eq:def-32-dim-Theta}.
The operator $\widetilde{D}^{IJ}_\a$ acting on the fermions has the desired form 
\be\label{eq:deform-D-op}
\begin{aligned}
\widetilde{D}^{IJ}_\a  = &
\delta^{IJ} \left( \pa_\a  -\frac{1}{4} \widetilde{\omega}^{mn}_\a \G_{mn}  \right)
+\frac{1}{8} \sigma_3^{IJ} \widetilde{e}^m_\a \widetilde{H}_{mnp} \G^{np}
\\
&-\frac{1}{8} e^{\varphi} \left( \epsilon^{IJ} \G^p \widetilde{F}^{(1)}_p + \frac{1}{3!}\sigma_1^{IJ} \G^{pqr} \widetilde{F}^{(3)}_{pqr} + \frac{1}{2\cdot5!}\epsilon^{IJ} \G^{pqrst} \widetilde{F}^{(5)}_{pqrst}   \right) \widetilde{e}^m_\a \G_m.
\end{aligned}
\ee
We use the tilde on all quantities to remind that we are discussing the deformed model.
The deformed spin connection satisfies the expected equation
\be
\widetilde{\omega}_M^{mn}=
- \widetilde{e}^{N \, [m} \left( \pa_M \widetilde{e}^{n]}_N - \pa_N \widetilde{e}^{n]}_M + \widetilde{e}^{n] \, P} \widetilde{e}_M^p \pa_P \widetilde{e}_{Np} \right),
\ee
where tangent indices $m,n$ are raised and lowered with $\eta_{mn}$, while curved indices $M,N$ with the deformed metric $\widetilde{G}_{MN}$.
From the computation of the deformed Lagrangian we find a field $\widetilde{H}^{(3)}$ with the following non-vanishing components
\be
\widetilde{H}_{234} = - 4 \vk \rho \frac{\sqrt{1+\rho^2}\sqrt{1-\vk^2\rho^2}\sin \zeta}{1+\vk^2 \rho^4 \sin^2 \zeta},
\qquad
\widetilde{H}_{789} = + 4 \vk r \frac{\sqrt{1-r^2}\sqrt{1+\vk^2r^2}\sin \xi}{1+\vk^2 r^4 \sin^2 \xi},
\ee
where we have specified \emph{tangent} indices.
Translating this into \emph{curved} indices we find agreement with the expected result
\be
\begin{aligned}
\widetilde{H}_{\psi_1\zeta\rho} &= \frac{2 \vk  \rho ^3 \sin (2 \zeta )}{\left(1+\vk ^2 \rho ^4 \sin ^2 \zeta\right)^2} 
& = \pa_\rho B_{\psi_1\zeta}, 
\qquad
\widetilde{H}_{\phi_1\xi r} &= -\frac{2 \vk  r^3 \sin (2 \xi )}{\left(1+\vk ^2 r^4 \sin ^2\xi\right)^2}
& = \pa_r B_{\phi_1\xi}.
\end{aligned}
\ee
The new results that can be obtained from the Lagrangian quadratic in fermions are the components of the RR-fields. When we specify \emph{tangent} indices we get
\be\label{eq:flat-comp-F1}
\begin{aligned}
&e^{\varphi} \widetilde{F}_1 =-4 \vk ^2  \ c_{F}^{-1} \ \rho ^3 \sin \zeta , \qquad
&&e^{\varphi} \widetilde{F}_6&= +4 \vk ^2  \ c_{F}^{-1} \ r^3 \sin\xi ,
\end{aligned}
\ee
\be\label{eq:flat-comp-F3}
\begin{aligned}
&e^{\varphi} \widetilde{F}_{014} = + 4 \vk   \ c_{F}^{-1} \ \rho ^2 \sin\zeta, \qquad
&&e^{\varphi} \widetilde{F}_{123} = -4 \vk   \ c_{F}^{-1} \ \rho  , \\
&e^{\varphi} \widetilde{F}_{569}= + 4 \vk   \ c_{F}^{-1} \ r^2 \sin\xi, \qquad
&&e^{\varphi} \widetilde{F}_{678} = -4 \vk   \ c_{F}^{-1} \ r, \\
&e^{\varphi} \widetilde{F}_{046} = +4 \vk^3   \ c_{F}^{-1} \ \rho  r^3 \sin \xi, \qquad
&&e^{\varphi} \widetilde{F}_{236} = -4 \vk^3   \ c_{F}^{-1} \ \rho ^2 r^3 \sin \zeta   \sin\xi, \\
&e^{\varphi} \widetilde{F}_{159} = - 4 \vk^3   \ c_{F}^{-1} \ \rho ^3 r \sin\zeta, \qquad
&&e^{\varphi} \widetilde{F}_{178} = -4 \vk^3   \ c_{F}^{-1} \  \rho ^3 r^2 \sin\zeta \sin \xi, \\
\end{aligned}
\ee
\be\label{eq:flat-comp-F5}
\begin{aligned}
&e^{\varphi} \widetilde{F}_{01234} = + 4    \ c_{F}^{-1} , \qquad
&&e^{\varphi} \widetilde{F}_{02346} = -4 \vk ^4   \ c_{F}^{-1}\rho ^3 r^3 \sin \zeta  \sin\xi , 
\\
&e^{\varphi} \widetilde{F}_{01459} = +4 \vk ^2 \ c_{F}^{-1} \rho ^2 r \sin\zeta, \qquad
&&e^{\varphi} \widetilde{F}_{01478} = +4 \vk ^2 \ c_{F}^{-1} \rho ^2 r^2   \sin\zeta \sin \xi , 
\\
&e^{\varphi} \widetilde{F}_{04569}= +4 \vk ^2 \ c_{F}^{-1} \rho  r^2 \sin \xi, \qquad
&&e^{\varphi} \widetilde{F}_{04678} = -4 \vk ^2 \ c_{F}^{-1} \rho  r.
\end{aligned}
\ee
For simplicity we have defined the common coefficient 
\be
c_{F} = \frac{1}{\sqrt{1+\vk ^2}}\sqrt{1-\vk ^2 \rho^2} \sqrt{1+\vk ^2 \rho ^4 \sin ^2\zeta} \sqrt{1+\vk ^2 r^2} \sqrt{1+\vk ^2 r^4 \sin ^2\xi}.
\ee
The same results written in \emph{curved} indices are
\be\label{eq:curved-comp-F1}
\begin{aligned}
e^{\varphi} \widetilde{F}_{\psi_2} &=4 \vk ^2 \ c_{F}^{-1} \ \rho ^4 \sin ^2 \zeta , \qquad
e^{\varphi} \widetilde{F}_{\phi_2} &= -4 \vk ^2  \ c_{F}^{-1} \ r^4 \sin ^2\xi ,
\end{aligned}
\ee
\be\label{eq:curved-comp-F3}
\begin{aligned}
e^{\varphi} \widetilde{F}_{t\psi_2 \rho } &= + 4 \vk   \ c_{F}^{-1} \ \frac{ \rho ^3 \sin ^2\zeta}{1-\vk ^2 \rho ^2}, \qquad
&e^{\varphi} \widetilde{F}_{\psi_2 \psi_1 \zeta } &= +4 \vk   \ c_{F}^{-1} \ \frac{ \rho ^4 \sin \zeta  \cos \zeta }{1+\vk ^2 \rho ^4 \sin^2\zeta }, \\
e^{\varphi} \widetilde{F}_{\phi \phi_2 r } &= + 4 \vk  \ c_{F}^{-1} \ \frac{ r^3 \sin ^2\xi}{1+\vk ^2 r^2}, \qquad
&e^{\varphi} \widetilde{F}_{\phi_2 \phi_1 \xi } &= +4 \vk  \ c_{F}^{-1} \ \frac{ r^4 \sin \xi \cos\xi}{1+\vk ^2 r^4 \sin ^2\xi}, \\
e^{\varphi} \widetilde{F}_{t \rho \phi_2 } &= + 4 \vk^3  \ c_{F}^{-1} \ \frac{\rho  r^4 \sin ^2\xi }{1-\vk ^2 \rho ^2}, \qquad
&e^{\varphi} \widetilde{F}_{\psi_1 \zeta \phi_2 } &= +4 \vk^3   \ c_{F}^{-1} \ \frac{ \rho ^4 r^4 \sin \zeta  \cos \zeta  \sin ^2\xi}{1+\vk ^2 \rho ^4 \sin ^2\zeta}, \\
e^{\varphi} \widetilde{F}_{\psi_2 \phi r } &= - 4 \vk^3   \ c_{F}^{-1} \ \frac{ \rho ^4 r \sin ^2\zeta}{1+\vk ^2 r^2}, \qquad
&e^{\varphi} \widetilde{F}_{\psi_2 \phi_1 \xi } &= +4 \vk^3   \ c_{F}^{-1} \ \frac{ \rho ^4 r^4 \sin^2\zeta \sin \xi \cos \xi}{1+\vk ^2 r^4 \sin ^2\xi}, \\
\end{aligned}
\ee
\be\label{eq:curved-comp-F5}
\begin{aligned}
e^{\varphi} \widetilde{F}_{t\psi_2\psi_1\zeta\rho } &=  \frac{ 4    \ c_{F}^{-1}  \rho ^3 \sin\zeta \cos\zeta}{\left(1-\vk ^2 \rho ^2\right) \left(1+\vk ^2 \rho ^4 \sin ^2\zeta\right)}, \
&e^{\varphi} \widetilde{F}_{t\psi_1\zeta\rho\phi_2 } &= -\frac{4 \vk ^4   \ c_{F}^{-1}\rho ^5 r^4 \sin \zeta \cos \zeta \sin ^2\xi }{\left(1-\vk ^2 \rho ^2\right) \left(1+\vk ^2 \rho ^4 \sin ^2\zeta\right)}, 
\\
e^{\varphi} \widetilde{F}_{t\psi_2\rho\phi r } &= -\frac{4 \vk ^2 \ c_{F}^{-1} \rho ^3 r \sin ^2\zeta}{\left(1-\vk ^2 \rho ^2\right) \left(1+\vk ^2 r^2\right)},
&e^{\varphi} \widetilde{F}_{t\psi_2\rho\phi_1\xi } &= +\frac{4 \vk ^2 \ c_{F}^{-1} \rho ^3 r^4   \sin ^2\zeta \sin \xi \cos\xi}{\left(1-\vk ^2 \rho ^2\right) \left(1+\vk ^2 r^4 \sin ^2\xi\right)}, 
\\
e^{\varphi} \widetilde{F}_{t\rho\phi\phi_2 r } &= -\frac{4 \vk ^2 \ c_{F}^{-1} \rho  r^3 \sin ^2\xi}{\left(1-\vk ^2 \rho ^2\right) \left(1+\vk ^2 r^2\right)}, 
&e^{\varphi} \widetilde{F}_{t\rho \phi_2 \phi_1\xi } &= -\frac{4 \vk ^2 \ c_{F}^{-1} \rho  r^4 \sin   \xi \cos\xi}{\left(1-\vk ^2 \rho ^2\right) \left(1+\vk ^2 r^4 \sin ^2\xi\right)}.
\end{aligned}
\ee
Let us present another method that can be used to derive the same results for the background RR fields, without the need of computing the Lagrangian at quadratic order in fermions.

\section{Kappa-symmetry}\label{sec:kappa-symm-eta-def-quad-theta}
The Lagrangian of the deformed model is invariant under kappa-symmetry, as proved in~\cite{Delduc:2013qra,Delduc:2014kha}.
Let us first briefly describe what happens when we switch off the deformation.
Local transformations are implemented on the coset by multiplication of the group elements from the \emph{right}.
A kappa-transformation is a local fermionic transformation. We then implement it with $\text{exp}[\varepsilon(\sigma,\tau)]$, where $\varepsilon$ is a local fermionic parameter that takes values in the algebra $\psu(2,2|4)$.
Multiplication of a coset representative $\alg{g}$ gives
\be\label{eq:right-ferm-action}
\alg{g}\cdot \text{exp}(\varepsilon) = \alg{g}'\cdot \alg{h}\,,
\ee
where $\alg{g}'$ is a new element of the coset, and $\alg{h}$ is a compensating element of $\text{SO}(4,1)\times \text{SO}(5)$ needed to remain in the coset.
A generic $\varepsilon$ will not leave the action invariant.
However, taking~\cite{Arutyunov:2009ga}
\be\label{eq:undef-eps-kappa-tr}
\begin{aligned}
\varepsilon &=  \frac{1}{2} (\gamma^{\a\b} \delta^{IJ}- \epsilon^{\a\b}\sigma_3^{IJ}) \left( \gen{Q}^I\kappa_{J\a}  A_\b^{(2)}  + A_\b^{(2)} \gen{Q}^I \kappa_{J\a} \right),
\end{aligned}
\ee
supplemented by the corresponding variation of the worldsheet metric, it is possible to show that indeed the action does not change under this transformation. The parameters $\kappa_{1\a}$ and $\kappa_{2\a}$---whose spinor indices we are omitting---introduced to define $\varepsilon$ are independent local quantities, parameterising odd elements of degree $1$ and $3$ respectively.

In the deformed case one can still prove the existence of a local fermionic symmetry of the form~\eqref{eq:right-ferm-action}, meaning that the parameter $\varepsilon$ is related to the \emph{infinitesimal} variation of the coset representative as
\be
\delta_\kappa \alg{g}=\alg{g}\cdot \varepsilon\,.
\ee
However, the definition~\eqref{eq:undef-eps-kappa-tr} of the parameter $\varepsilon$ has to be deformed in order to get invariance of the action, in particular it will no longer lie just in the odd part of the algebra, but it will have a non-trivial projection also on bosonic generators. 
It is written in terms of an odd element $\varrho$ as~\cite{Delduc:2013qra}
\be\label{eq:eps-op-rho-kappa}
\varepsilon = \op \varrho, \qquad \varrho= \varrho^{(1)} +\varrho^{(3)} .
\ee
where $\op$ is the operator defined in~\eqref{eq:defin-op-def-supercoset} and the two projections $\varrho^{(k)}$ are\footnote{Comparing to~\cite{Delduc:2013qra} we have dropped the factor of $i$ because we use ``anti-hermitian'' generators.}
\be\label{eq:def-varrho-kappa-def}
\begin{aligned}
\varrho^{(1)} &= \frac{1}{2} (\gamma^{\a\b} - \epsilon^{\a\b}) \left( \gen{Q}^1\kappa_{1\a}  \left(\op^{-1} A_\b \right)^{(2)}  + \left(\op^{-1} A_\b \right)^{(2)} \gen{Q}^1\kappa_{1\a} \right),\\
\varrho^{(3)} &=  \frac{1}{2} (\gamma^{\a\b} + \epsilon^{\a\b}) \left( \gen{Q}^2\kappa_{2\a}  \left(\optilde^{-1} A_\b \right)^{(2)}  + \left(\optilde^{-1} A_\b \right)^{(2)} \gen{Q}^2\kappa_{2\a} \right),
\end{aligned}
\ee
where we defined
\be
\optilde=\mathbf{1} + \eta R_{\alg{g}} \circ \widetilde{d}\,.
\ee
In Appendix~\ref{app:standard-kappa-sym} we compute explicitly the form of the variations on bosonic and fermionic fields given the above definitions, and we show that they do not have the standard form. However, after implementing the field redefinitions of  Section~\ref{sec:canonical-form}---needed to put the Lagrangian in the standard Green-Schwarz form---we find that also the kappa-variations become indeed standard
\be\label{eq:kappa-var-32-eta-def}
\begin{aligned}
 \delta_{\kappa}X^M &= - \frac{i}{2} \ \bar{\T}_I \delta^{IJ} \widetilde{e}^{Mm}  \G_m  \delta_{\kappa} \T_J + \mathcal{O}(\T^3),
\\
\delta_{\kappa} \T_I &= -\frac{1}{4} (\delta^{IJ} \gamma^{\a\b} - \sigma_3^{IJ} \epsilon^{\a\b})  \widetilde{e}_{\b}^m  \G_m  \widetilde{K}_{\a J}+ \mathcal{O}(\T^2),
\end{aligned}
\ee
where
\be
\widetilde{K} \equiv \left( \begin{array}{c} 0 \\ 1 \end{array} \right) \otimes \widetilde{\kappa},
\ee
and $\tilde{\kappa}$ is related to $\kappa$ as in~\eqref{eq:def-kappa-tilde-k-symm}.
It is interesting to look also at the kappa-variation for the worldsheet metric, as this provides an independent method to derive the couplings of the fermions to the background fields, already identified from the Lagrangian.
The variation is given by~\cite{Delduc:2013qra}
\be\label{eq:defin-kappa-var-ws-metric}
\delta_{\kappa}\g^{\a\b}=\frac{1-\eta^2}{2} \Str\left( \Upsilon \left[\gen{Q}^1\kappa^{\a}_{1+},P^{(1)}\circ \widetilde{\op}^{-1}( A^{\b}_+ ) \right]
+\Upsilon \left[\gen{Q}^2\kappa^{\a}_{2-},P^{(3)}\circ {\op}^{-1}( A^{\b}_- ) \right] \right)\,,
\ee
where $\Upsilon=\text{diag}(\mathbf{1}_4,-\mathbf{1}_4)$ and the projections of a vector $V_\a$ are defined as
\be
V^\a_{\pm}= \frac{\g^{\a\b}\pm \epsilon^{\a\b}}{2} V_\b\,.
\ee
As we show in Appendix~\ref{app:standard-kappa-sym}, after taking into account the field redefinitions performed to get a standard action, we find a standard kappa-variation also for the worldsheet metric
\be
\begin{aligned}
\delta_\kappa \g^{\a\b}&=2i\Bigg[ 
\bar{\widetilde{K}}^\a_{1+} \widetilde{D}^{\b 1J}_+\Theta_J+\bar{\widetilde{K}}^\a_{2-} \widetilde{D}^{\b 2J}_-\Theta_J
 \Bigg]+ \mathcal{O}(\T^3) \\
&= 2i\ \Pi^{IJ\, \a\a'}\Pi^{JK\, \b\b'}
\ \bar{\widetilde{K}}_{I\a'}\widetilde{D}^{KL}_{\b'}\Theta_{L}+ \mathcal{O}(\T^3),
\end{aligned}
\ee
where we have defined
\be
\Pi^{IJ\, \a\a'}\equiv\frac{\delta^{IJ}\g^{\a\a'}+\sigma_3^{IJ}\epsilon^{\a\a'}}{2}\,.
\ee
The operator $\widetilde{D}$ is the one already identified from the computation of the Lagrangian. It is given in Eq.~~\eqref{eq:deform-D-op}, and in particular we find the same RR fields as in the previous section.

\section{Discussion}\label{sec:discuss-eta-def-background}
From the Lagrangian at quadratic order in fermions and the kappa-symmetry variation of the worldsheet metric, we have read off couplings to tensors that we want to interpret as the field strengths of the RR fields.
In this section we show that the results that we have obtained are \emph{not} compatible with the Bianchi identities and the equations of motion of supergravity.

Let us start by looking at the Bianchi identity for $\widetilde{F}^{(1)}$
\be
\pa_M \widetilde{F}_N- \ _{(M\leftrightarrow N)}=0\,,
\implies
\pa_M\left(e^{\varphi}\widetilde{F}_N\right) -\pa_M\varphi\, e^{\varphi}\widetilde{F}_N-\ _{(M\leftrightarrow N)}=0.
\ee
We prefer to rewrite it in the second form, because we only know the combination $e^{\varphi}\widetilde{F}_M$. In particular we obtain
\be\label{eq:Bianchi-F1-expl}
\pa_M\left(e^{\varphi}\widetilde{F}_{\psi_2}\right) -\pa_M\varphi\, e^{\varphi}\widetilde{F}_{\psi_2}- \ _{(M\leftrightarrow \psi_2)}=0\,,
\qquad
\pa_M\left(e^{\varphi}\widetilde{F}_{\phi_2}\right) -\pa_M\varphi\, e^{\varphi}\widetilde{F}_{\phi_2}- \ _{(M\leftrightarrow \phi_2)}=0\,,
\ee
because from~\eqref{eq:curved-comp-F1} we know that the only non-vanishing components are $\widetilde{F}_{\psi_2},\widetilde{F}_{\phi_2}$.
Moreover, using the fact that the combinations $e^{\varphi}\widetilde{F}_{\psi_2},e^{\varphi}\widetilde{F}_{\phi_2}$ depend just on $\zeta,\rho,\xi,r$ we immediately find that the derivatives of the dilaton $\varphi$ should satisfy the following equations
\be
\begin{aligned}
\pa_{t}\varphi = 0\,, \quad
\pa_{\psi_1}\varphi = 0\,, \quad
\pa_{\phi}\varphi = 0\,, \quad
\pa_{\phi_1}\varphi = 0\,, \\
e^{\varphi}\widetilde{F}_{\phi_2}\, \pa_{\psi_2}\varphi =e^{\varphi}\widetilde{F}_{\psi_2}\, \pa_{\phi_2}\varphi\,, \\
\pa_M\varphi=\frac{1}{e^{\varphi}\widetilde{F}_{\psi_2}}\, \pa_M\left(e^{\varphi}\widetilde{F}_{\psi_2}\right) =
\frac{1}{e^{\varphi}\widetilde{F}_{\phi_2}}\, \pa_M\left(e^{\varphi}\widetilde{F}_{\phi_2}\right) \quad M=\zeta,\rho,\xi,r\,.
\end{aligned}
\ee
The last equation comes from the compatibility of the two equations that we obtain from~\eqref{eq:Bianchi-F1-expl} for $M=\zeta,\rho,\xi,r$.
A consequence of this compatibility is the equation
\be
\pa_M \log \left(  \rho ^4 \sin ^2 \zeta \right) = - \pa_M \log \left(r^4 \sin ^2\xi\right)\,, \qquad M=\zeta,\rho,\xi,r\,,
\ee
which is clearly not satisfied.
We then conclude that the results are not compatible with the Bianchi identity for $\widetilde{F}^{(1)}$.

Failure to satisfy the equations of motion of type IIB supergravity is easily seen by considering the equation
\be\label{eq-combination-sugra-not-satisf}
\pa_{P}\left(\sqrt{-\widetilde{G}}\ e^{-2\varphi}\widetilde{H}^{MNP}\right)-\sqrt{-\widetilde{G}}\  \widetilde{F}^{MNP}\widetilde{F}_P-{1\ov 6}\sqrt{-\widetilde{G}}\ \widetilde{F}^{MNPQR} \widetilde{F}_{PQR}=0\,,
\ee
that is obtained by combining the equation of motion for the NSNS two form~\eqref{eq:eom-B} and the RR two-form~\eqref{eq:eom-C2}.
If we select \eg the indices $(M,N)=(t,\rho)$, the first term---containing $\widetilde{H}$ and the unknown factor with the dilaton---drops out.
We get then just an algebraic equation that we can evaluate using the information at our disposal. 
To avoid writing curved indices, we write it explicitly in terms of tangent indices\footnote{To transform the curved indices $M,N$ into tangent indices it it enough to multiply the equation by the proper vielbein components~\eqref{eq:def-vielb-comp}. Summed indices can be translated from curved to tangent without affecting the result.}
\be
\begin{aligned}
 \widetilde{F}^{041}\widetilde{F}_{1}+ \widetilde{F}^{046}\widetilde{F}_{6} \\
+ \widetilde{F}^{04123}\widetilde{F}_{123}+ \widetilde{F}^{04236}\widetilde{F}_{236}+ \widetilde{F}^{04159}\widetilde{F}_{159}+ \widetilde{F}^{04178}\widetilde{F}_{178}+ \widetilde{F}^{04569}\widetilde{F}_{569}+ \widetilde{F}^{04678}\widetilde{F}_{678}=0
\end{aligned}
\ee
Multiplying by $e^{2\varphi}$ and using~\eqref{eq:flat-comp-F1},~\eqref{eq:flat-comp-F3} and~\eqref{eq:flat-comp-F5} we obtain that the left-hand-side of the above equation is
\be
\frac{16(1+\vk^2) \vk \rho}{1-\vk^2 \rho}\neq 0\,.
\ee
We conclude that the equations of motion of type IIB supergravity are not satisfied.
It is natural to wonder whether there exist field redefinitions at the level of the $\sigma$-model action that can cure this problem. We now proceed by first discussing this possibility, and then by studying two special limits of the $\eta$-deformed model.

\subsection{On field redefinitions}\label{sec:field-red-q-def}
In Section~\ref{sec:canonical-form} we were able to transform the original Lagrangian into the canonical form and, as we have just observed, the RR couplings that we have derived do not satisfy the supergravity equations of motion. However, the NSNS couplings are properly reproduced in the quadratic fermionic action, as they are compatible with the results of the bosonic Lagrangian. We are motivated to ask whether further field redefinitions could be performed 
which exclusively change  the RR content of the theory. It appears to be rather difficult to answer this question in full generality. We will argue however 
that  no such field redefinition exists which is continuous in the deformation parameter. 

We work in the formulation with 32-dimensional fermions $\T_I$ obeying the Majorana and Weyl conditions, see appendix \ref{sec:10-dim-gamma}. We start by considering 
a generic rotation of fermions\footnote{One could imagine more complicated redefinitions like $\Theta_I \to U_{IJ} \Theta_J+V_{IJ}^{\a}\pa_{\a}\T_J$, etc. 
They were not needed to bring the original Lagrangian to the canonical form and we do not consider them here. These redefinitions will generate higher derivative terms in the action,
whose cancellation would imply further stringent constraints on their possible form.} 
\be
\label{eq:FIJ}
\begin{aligned}
\Theta_I &\to U_{IJ} \Theta_J,\quad \bar\Theta_I \to  \bar\Theta_J\bar U_{IJ}\, , \quad  \bar U_{IJ} =- \G_0 U_{IJ}^\dagger\G_0\, , 
\end{aligned}
\ee
where $U_{IJ}$ are rotation matrices which can depend on bosonic fields. We write $U_{IJ}$ as an expansion over a complete basis in the space of $2\times 2$-matrices 
\be
\label{eq:FIJ1}
\begin{aligned}
U_{IJ} &\equiv \delta^{IJ} U_\delta + \sigma_1^{IJ} U_{\sigma_1} + \eps^{IJ} U_{\epsilon} + \sigma_3^{IJ} U_{\sigma_3} = \sum_{a=0}^3\as_a^{IJ}U_a\,,\\
\bar U_{IJ} &= \delta^{IJ} \bar U_\delta + \sigma_1^{IJ} \bar U_{\sigma_1} + \eps^{IJ} \bar U_{\epsilon} + \sigma_3^{IJ} \bar U_{\sigma_3}= \sum_{a=0}^3\as_a^{IJ}\bar U_a\, ,
\end{aligned}
\ee
where we have introduced 
$$
 \as_0^{IJ}=\delta^{IJ}\,,\quad \as_1^{IJ}=\s_1^{IJ}\,,\quad \as_2^{IJ}=\eps^{IJ}\,,\quad \as_3^{IJ}=\s_3^{IJ}\, .
 $$
The objects $U_{a}$ and $\bar{U}_{a}$ are $32\times 32$-matrices and they can be expanded over the complete basis generated by $\Gamma^{(r)}$ and identity, see appendix \ref{sec:10-dim-gamma}
for the definition and properties of $\Gamma^{(r)}$. Further, we  require that the transformation $U_{IJ}$ preserves chirality and the Majorana condition. Conservation of chirality implies that 
the $\Gamma$-matrices appearing in the expansion of $U_{IJ}$ must commute with $\Gamma_{11}$, {\it i.e.} the expansion involves $\Gamma^{(r)}$ of even rank only
\be
\begin{aligned}\label{eq:def-redefinition}
U_a &=  f_a\,{\mathbb I}_{32} + {1\ov2}f^{mn}_a\G_{mn} + {1\ov24}f^{klmn}_a\G_{klmn} \,,\\
\bar U_a &= \bar f_a\,{\mathbb I}_{32} + {1\ov2}\bar f^{mn}_a\G_{mn} + {1\ov24}\bar f^{klmn}_a\G_{klmn} \,.
\end{aligned}
\ee
In this expansion matrices of rank six, eight and ten are missing, because by virtue of duality relations they can be re-expressed via matrices of lower rank. 
The Majorana condition imposes the requirement 
\begin{equation}\label{eq:Maj}
\G_0U^\dagger_{IJ}\G_0=\mathcal{C}U^t_{IJ}\mathcal{C}\, ,
\end{equation}
which implies that the coefficients $f$ are real. Coefficients of $\bar{U}_a$ are then given by 
\begin{equation}
\bar f_{a} = f_a\,,\quad  \bar f_{a}^{mn} =- f_a^{mn}\,,\quad  \bar f_{a}^{klmn} = f_a^{klmn}\,.
\end{equation}
Let us note that combining equations  \eqref{eq:FIJ} and \eqref{eq:Maj}, we get 
\be\la{eq:symFIJ}
\mathcal{C}\bar{U}^t_{IJ}\mathcal{C}= -U_{IJ}\,, \quad \text{ and } \quad \mathcal{C}U^t_{IJ}\mathcal{C}= -\bar{U}_{IJ}\, .
\ee
The total number of degrees of freedom in the rotation matrix is 
$$
4\cdot\Big(1+\frac{10\cdot 9}{2}+\frac{10\cdot 9\cdot 8\cdot 7}{4!}\Big)=2^{10}=(16+16)^2 \, ,
$$
which is precisely the dimension of ${\rm GL}(32,{\mathbb R})$. This correctly reflects the freedom to perform general linear transformations on 
32 real fermions of type IIB. 

\smallskip

Under these rotations the part of the Lagrangian containing derivatives on fermions transforms as
\begin{equation}
\begin{aligned}
&&(\gamma^{\a\b}\de^{IJ}+\eps^{\a\b}\sigma_3^{IJ})\bar{\T}_I \tilde{e}_{\a}^m\G_m\pa_{\beta}\T_{I} \to \\
\nonumber
&&~~~~~~~~~~\to (\gamma^{\a\b}\de^{IJ}+\eps^{\a\b}\sigma_3^{IJ})\Big( \bar{\Theta}_K \, \bar{U}_{IK} \, \widetilde{e}^m_\a  \G_m U_{JL} \, \pa_\b \Theta_L +   \bar{\Theta}_K \, \bar{U}_{IK} \, \widetilde{e}^m_\a \G_m  (\pa_\b U_{JL})\Theta_L  \Big)\, .
\end{aligned}
\end{equation}
The requirement that under rotations the part with $\pa\T$ remains unchanged  can be formulated as the following conditions on $U_{IJ}$:
\be\label{eq:redef_cond}
\begin{aligned}
\T_K\de^{IJ}  \bar{U}_{IK} \,  \G_m U_{JL}\pa\T_L &=\T_K\delta_{KL}\G_m \pa\T_L+{\rm removable~terms}\, , \\
\T_K\sigma_3^{IJ}\bar{U}_{IK} \,  \G_m U_{JL}\pa\T_L&=\T_K\sigma_3^{KL}\G_m\pa\T_L+{\rm removable~terms}\, ,
\end{aligned}
\ee
where ``removable terms" means terms which can be removed by shifting bosons in the bosonic action by fermion bilinears, similarly to what was done in~\eqref{eq:red-bos}. 
In the following it is enough to analyse the first equation in \eqref{eq:redef_cond}. Let us collect all terms on its right hand side that are removable by shifting bosons into a tensor $M_{KL,m}$, where the indices $K,L$ should be multiplied by proper fermions and $m$ is an index in the tangent space. This tensor has the following symmetry property\footnote{Notice that to exhibit  this symmetry property, one has to transpose also the indices $K,L$, on top of transposition in the $32\times32$ space.}
\be\la{eq:sym1}
\mathcal{C}\left(M_{KL,m}\right)^t\mathcal{C}=- M_{LK,m}\,,
\ee
that we need to impose if we want the shift of bosons to be non-vanishing.
Note that the tensor in the canonical kinetic term has exactly the opposite symmetry property
\be\la{eq:sym2}
\mathcal{C}(\delta_{KL} \G_m^t)\mathcal{C}=\delta_{LK} \G_m\, .
\ee
Putting this information together, let us consider the first equation in~\eqref{eq:redef_cond} written as
\be
\bar{U}_{IK}\G_mU_{IL}= \delta_{KL}\G_m+M_{KL,m}\,.
\ee
We take transposition and we multiply by $\mathcal{C}$ from the left and from the right
\be
\begin{aligned}
\mathcal{C}\left(\bar{U}_{IK}\G_mU_{IL}\right)^t \mathcal{C}
&= \delta_{KL}\mathcal{C}\left(\G_m\right)^t\mathcal{C}+\mathcal{C}\left(M_{KL,m}\right)^t\mathcal{C},
\end{aligned}
\ee
and further manipulate as
\be
\begin{aligned}
\mathcal{C}\left(U_{IL}\right)^t \mathcal{C}\cdot  \mathcal{C}\left(\G_m\right)^t \mathcal{C}\cdot \mathcal{C}\left(\bar{U}_{IK}\right)^t \mathcal{C}
&= \delta_{KL}\mathcal{C}\left(\G_m\right)^t\mathcal{C}+\mathcal{C}\left(M_{KL,m}\right)^t\mathcal{C}\, .
\end{aligned}
\ee
With the help of eqs.\eqref{eq:symFIJ} , \eqref{eq:sym1} and \eqref{eq:sym2} and relabelling the indices $K$ and $L$, we get 
\be
\begin{aligned}
\bar{U}_{IK}\G_mU_{IL}= \delta_{KL}\G_m-M_{KL,m}\, ,
\end{aligned}
\ee
which shows that $M_{KL,m}=0$, that is this structure cannot appear because it is incompatible with the symmetry properties of the rotated kinetic term.
It is clear that the same considerations are also applied to the second equation in \eqref{eq:redef_cond}, where $\sigma_3^{IJ}$ replaces $\delta^{IJ}$.
Thus, the rotation matrix $U_{IJ}$ must satisfy a more stringent system of equations.
To write these equations without having to deal with indices of the $2\times 2$ space, we can introduce $64\times 64$ matrices 
\be
U\equiv \sum_{a=0}^3 s_a \otimes U_a\,,
\qquad
\bar{U}\equiv \sum_{a=0}^3 s_a^t \otimes \bar{U}_a=-\sum_{a=0}^3 s_a^t \otimes \mathcal{C}U_a^t\mathcal{C}\, ,
\ee
which allow us to write the constraints as
\be
\begin{aligned}\la{eq:stringent1}
\Pi_-\Big(\bar{U} \,(\mathbf{1}_2 \otimes \G_m)U &- \mathbf{1}_2 \otimes \G_m\Big)\Pi_+=0\,,
\\
\Pi_-\Big(\bar{U} \,( \sigma_3 \otimes \G_m)U &- \sigma_3 \otimes \G_m\Big)\Pi_+=0\, .
\end{aligned}
\ee
Here we are multiplying by the two projectors $\Pi_{\pm}={\mathbf 1}_2\otimes \frac{1}{2}({\mathbf 1}_{32}\pm \Gamma_{11})$ to account for the chirality of the fermions.
We assume that $U$ is a smooth function of $\eta$, and that for small values of the deformation parameter it can be expanded as
\be
U=\mathbf{1}_{64}+\eta^r u+o(\eta^{r})\, .
\ee
Here $\eta^r$ is the first non-trivial order of the contribution, and $o(\eta^{r})$ denotes subleading terms.
At order $\eta^r$ we get a system of linear equations for $u$
\be
\begin{aligned}\la{stringent2}
\Pi_-\Big(\bar{u} \,(\mathbf{1}_2 \otimes \G_m) &+(\mathbf{1}_2 \otimes \G_m)u\Big)\Pi_+=0\,,
\\
\Pi_-\Big(\bar{u} \,( \sigma_3 \otimes \G_m) &+ ( \sigma_3 \otimes \G_m)u\Big)\Pi_+=0\,.
\end{aligned}
\ee
This system appears to have no solution which acts non-trivially on chiral fermions. Thus, non-trivial field redefinitions of the type we considered here 
do not exist.  Whether equation \eqref{eq:stringent1} has solutions which do not depend on $\eta$ is unclear to us.  Finally, let us mention that similar considerations on field redefinitions can be done by considering the kappa-symmetry transformations of the bosonic and fermionic coordinates, and of the worldsheet metric. Doing so we get to the same conclusion found here.

\subsection{Mirror model and Maldacena-Russo background}
In this section we want to take special limits of our results, to make contact with other findings appeared in the literature.
We first study a particular $\vk \to \infty$ limit of {\etaadsfive}, as used in~\cite{Arutyunov:2014cra}. 
There it was shown that this limit implemented on the spacetime metric for the deformed model yields the metric for the mirror model of {\adsfive}.
It was then shown that it is possible to complete this metric to a IIB supergravity background, by supplementing it with a dilaton and a five-form flux.
We have to first rescale the bosonic coordinates as~\cite{Arutyunov:2014cra}
\be
t \to \frac{t}{\vk},
\qquad
\rho \to \frac{\rho}{\vk},
\qquad
\phi \to \frac{\phi}{\vk},
\qquad
r \to \frac{r}{\vk},
\ee
and then send $\vk\to \infty$.
The vielbein components $e^m_\a$ are of order $\mathcal{O}(1/\vk)$ in this limit. We get the following components for $e^m_M$ 
\be
\begin{aligned}
e^0_t &= +\frac{1}{\sqrt{1-\rho^2}},
\quad
e^1_{\psi_2} &=  - \rho \sin \zeta,
\quad
e^2_{\psi_1} &=  - \rho \cos \zeta,
\quad
e^3_{\zeta} &=  - \rho ,
\quad
e^4_{\rho} &= - \frac{1}{\sqrt{1-\rho^2}},
\\
e^5_{\phi} &= + \frac{1}{\sqrt{1+r^2}},
\quad
e^6_{\phi_2} &= - r \sin \xi,
\quad
e^7_{\phi_1} &= - r \cos \xi,
\quad
e^8_{\xi} &= -r,
\quad
e^9_r &= - \frac{1}{\sqrt{1+r^2}},
\end{aligned}
\ee
where we omit powers of $\vk$.
This is compatible with the metric of the mirror background~\cite{Arutyunov:2014cra}.
The $B$-field vanishes in this limit.

For the RR fields we have to keep those components---when we specify tangent indices---that are of order $\mathcal{O}(\vk)$ in this limit. This is to compensate the power $1/\vk$ coming from the vielbein that multiplies them in the definition~\eqref{eq:deform-D-op} of $\widetilde{D}_\a^{IJ}$. The components that survive are
\be
\begin{aligned}
e^\varphi \, F_{123} = - \frac{4 \rho}{\sqrt{1-\rho^2}\sqrt{1+r^2}}\, ,
\qquad
e^\varphi \, F_{678} = - \frac{4 r}{\sqrt{1-\rho^2}\sqrt{1+r^2}}\, ,
\\
e^\varphi \, F_{01234} = +\frac{4 }{\sqrt{1-\rho^2}\sqrt{1+r^2}}\, ,
\qquad
e^\varphi \, F_{04678} = - \frac{4 \rho r}{\sqrt{1-\rho^2}\sqrt{1+r^2}}\, .
\end{aligned}
\ee
Here we are omitting powers of $\vk$. For $F^{(5)}$ one has to take into account also the components that are dual to the ones written above, using~\eqref{eq:sel-duality-F5-curved}.
This result does not match with~\cite{Arutyunov:2014cra}, where the proposed background has vanishing $F^{(3)}$ and an $F^{(5)}$ along different directions.
Checking~\eqref{eq-combination-sugra-not-satisf}, we find that the RR fields obtained in this limit are again not compatible with the equations of motion of supergravity.

\bigskip 

Studying a particular $\vk\to 0$ limit of the results that we have obtained, we can show that we reproduce the Maldacena-Russo (MR) background~\cite{Maldacena:1999mh}. 
This background was constructed with the motivation of studying the large-$N$ limit of non-commutative gauge theories.
We will show agreement with our results both at the level of the NSNS and the RR sector.

We first rescale the coordinates parameterising the deformed AdS space as
\be
t\to \sqrt{\vk} \, t\,,
\quad
\psi_2\to \frac{\sqrt{\vk}}{\sin \zeta_0} \, \psi_2\,,
\quad
\psi_1\to \frac{\sqrt{\vk}}{\cos \zeta_0} \, \psi_1\,,
\quad
\zeta\to\zeta_0+ \sqrt{\vk} \, \zeta\,,
\quad
\rho\to \frac{\rho}{\sqrt{\vk}}\,,
\ee
and then send $\vk\to 0$.
Because we have not rescaled the coordinates on the deformed S$^5$, the corresponding part of the metric just reduces to the usual metric on S$^5$, and the components of the $B$-field in those directions will vanish.
On the other hand, for the part originating from the deformed AdS$_5$ we get a result different from the undeformed case. 
In this limit the complete metric and $B$-field are
\be\label{eq:Malda-Russo-sph-coord}
\begin{aligned}
{\rm d}s^2_{(\text{MR})}&=\rho^2\left(-{\rm d}t^2+ {\rm d}\psi_2^2\right)
  + \frac{\rho^2}{1+\rho^4\sin^2\zeta_0}\left( {\rm d}\psi_1^2+ {\rm d}\zeta^2\right) 
  +\frac{{\rm d}\rho^2}{\rho^2}
+{\rm d}s^2_{\text{S}^5}\,,
\\
{B}_{(\text{MR})} &= +  \frac{\rho^4 \sin \zeta_0}{1+ \rho^4\sin^2 \zeta_0} {\rm d}\psi_1\wedge{\rm d}\zeta .
\end{aligned}
\ee
These equations should be compared with (2.7) of~\cite{Maldacena:1999mh}, using the following identification of the coordinates and the parameters on the two sides
\begin{center}
\begin{tabular}{r|ccccc|c}
Here  & $t$ & $\psi_2$ & $\psi_1$ & $\zeta$ & $\rho$ & $\sin\zeta_0$
\\
\hline
There & $\tilde{x}_0$ & $\tilde{x}_1$ & $\tilde{x}_2$ & $\tilde{x}_3$  & $u$ & $a^2$
\end{tabular}
\end{center}
When we repeat the same limiting procedure on the components of the RR fields found for the $\eta$-deformation of {\adsfive}---see~\eqref{eq:flat-comp-F1},~\eqref{eq:flat-comp-F3} and~\eqref{eq:flat-comp-F5}---we find that the axion vanishes, and only one component of $F^{(3)}$ and one of $F^{(5)}$ (plus its dual) survive. 
These components---when we specify tangent indices---multiplied by the exponential of the dilaton are
\be
e^{\varphi} F_{014} = \frac{4\rho^2\sin \zeta_0}{\sqrt{1+ \rho^4\sin^2 \zeta_0}}\,,
\qquad
e^{\varphi} F_{01234} = \frac{4}{\sqrt{1+ \rho^4\sin^2 \zeta_0}}\,.
\ee
If we take the dilaton to be equal to 
\be
\varphi=\varphi_0-\frac{1}{2}\log ( 1+ \rho^4\sin^2 \zeta_0)\,,
\ee
where $\varphi_0$ is a constant, we then find that the non-vanishing components for the RR fields---in tangent and curved indices---are
\be
\begin{aligned}
& F_{014} = e^{-\varphi_0}\, 4\rho^2\sin \zeta_0\,,
\qquad
&& F_{01234} = e^{-\varphi_0}\,4\,,
\\
& F_{t\psi_2\rho} = e^{-\varphi_0}\, 4\rho^3\sin \zeta_0\,,
\qquad
&& F_{t\psi_2 \psi_1\zeta\rho} = e^{-\varphi_0}\, \frac{4\rho^3}{1+\rho^4\sin^2\zeta_0}\,.
\end{aligned}
\ee
Also the results that we obtain for the dilaton and the RR fields are in perfect agreement with (2.7) of~\cite{Maldacena:1999mh}.
It is very interesting that despite the incompatibility with type IIB supergravity for generic values of the deformation parameter, there exists a certain limit---different from the undeformed {\adsfive}---where this compatibility is restored.

\subsection{Concluding remarks}
In~\cite{Arutyunov:2015qva} the action that we have derived here was used to compute the tree-level scattering elements for excitations on the worldsheet.
In addition to the results collected in Section~\ref{sec:pert-bos-S-mat-eta-ads5s5}---where only interactions among bosons were considered--- it was then possible to derive also the scattering elements that involve two fermions in the asymptotic states of $2\to 2$ processes\footnote{The terms Fermion+Fermion$\to$Fermion+Fermion are missing in the computation, since their derivation requires the Lagrangian quartic in fermions.}.
The derivation of~\cite{Arutyunov:2015qva} shows that the T-matrix obtained using the methods reviewed in Section~\ref{sec:decomp-limit} cannot be factorised into two copies as in~\eqref{eq:factoris-rule-T-mat-AdS5}, see Section~\ref{sec:pert-bos-S-mat-eta-ads5s5} for the discussion in the bosonic sector.
However, there exists a unitary transformation of the basis of two-particle states thanks to which the T-matrix can be factorised into two copies, as desired.
This change of basis is of a particular type, as it is not one-particle factorisable.
Each of the two copies that compose the T-matrix matches with the large tension expansion of the all-loop S-matrix invariant under $\psu_q(2|2)$, and it naturally satisfies the classical Yang-Baxter equation.

The fact that we can prove compatibility with the $q$-deformed S-matrix is a nice further check of our results.
On the other hand, it is not clear why this compatibility is not immediate, given that a change of the two-particle basis is needed.
One possible explanation may lie in the choice of the R-operator that is used to define the deformation.
Our current choice~\eqref{Rop} corresponds to the standard Dynkin diagram of $\psu(2,2|4)$. However, it is believed that only the ``all-loop'' Dynkin diagram can be used to write the Bethe-Yang equations for the undeformed model.
It might be that defining the deformation through an R-operator which is related to the ``all-loop'' Dynkin diagram would give automatically the T-matrix in the factorised form, with no need of changing the basis of two-particle states.
It is natural to wonder whether applying this redefinition---which is in fact $\eta$-independent and it is a symmetry of the undeformed model---necessary to get a factorised T-matrix could also cure the problem of compatibility with type IIB\footnote{This transformation being non-local, it would first be important to check that it does not produce non-local terms in the action.}.

It is of course possible that no cure exists and that the $\eta$-model just does not correspond to a solution of supergravity, as it may be suggested by the findings of~\cite{Hoare:2015wia}. 
There it was shown that the metric and the $B$-field of a T-dual version of the $\eta$-model ---where abelian T-duality is implemented along all six isometric directions---could be completed to a set of background fields\footnote{The corresponding fluxes are purely imaginary, a fact which is attributed to having done a T-duality along the time direction.} satisfying the equations of motion of type IIB.
The peculiarity of this solution is that the dilaton of the T-dual model has a linear dependence on the isometric directions, which forbids to undo the T-dualities and get back to a background for the $\eta$-model. However, it is interesting that this dependence is compensated by the RR fields in such a way that the combination $e^{\varphi}F$ still preserves the isometries. The classical Green-Schwarz action is then still invariant under shifts of all these coordinates, and one can study what happens under the standard T-duality transformations when ignoring the issue with the dilaton.
The resulting background fields are precisely the ones of the $\eta$-model derived in~\cite{Arutyunov:2015qva} and presented here.
Thanks to this formal T-duality relation to a supergravity background, it was then suggested that---while not being Weyl invariant---the $\eta$-model should be UV finite at one loop.
This interpretation was further investigated in~\cite{Arutyunov:2015mqj}, where it was shown that the background fields extracted from the $\eta$-model satisfy a set of second-order equations which should follow from scale invariance of the $\sigma$-model. The $\eta$-model seems to be special since it satisfies also a set of first order equations---a modification of the ones of type IIB supergravity---which should follow from kappa-symmetry.

Let us mention that the methods applied here were used to study also the $\lambda$-deformation of~\cite{Hollowood:2014qma} and deformations based on R-matrices satisfying the classical Yang-Baxter equation~\cite{Matsumoto:2014cja}.
The $\lambda$-deformation of AdS$_2\times$S$^2$ was considered in~\cite{Borsato:2016zcf}, where the RR fields were extracted by looking at the kappa-symmetry variations. In that case the result is found to be a solution of type IIB, suggesting that the $\lambda$-deformation does not break the Weyl invariance of the original $\sigma$-model.
A case-by-case study for some Yang-Baxter deformations of \adsfive \ was carried on in~\cite{Kyono:2016jqy}, providing examples where compatibility with IIB is either realised or not depending on the choice of the R-matrix.

It would be interesting to investigate these deformed models more generally with the goal of identifying the properties that ensure compatibility with supergravity.

\appendix
\chapter{Equations of motion of type IIB supergravity}\label{app:IIBsugra}
In this appendix we collect the action and the equations of motion of type IIB supergravity, as taken from~\cite{IIBbackgroundnotes}.
We use these conventions in particular for Section~\ref{sec:fermions-type-IIB} and~\ref{sec:discuss-eta-def-background}.

In type IIB supergravity we have Neveu-Schwarz--Neveu-Schwarz (NSNS) and \\
Ramond-Ramond (RR) fields. In particular
\begin{itemize}
\item[] \textbf{NSNS}: these are the metric $G_{MN}$, the dilaton $\varphi$, and the anti-symmetric two-form $B_{MN}$ with field strength $H_{MNP}$; 
\item[] \textbf{RR}: these are the axion $\chi$, the anti-symmetric two-form $C_{MN}$, and the anti-symmetric four-form $C_{MNPQ}$. 
\end{itemize}
The RR field strengths are defined as
\bea
&&F_{M}=\pa_{M}\chi\, , \\
&&F_{MNP}=3\pa_{[M}C_{NP]} +\chi H_{MNP}\, , \\
&&F_{MNPQR}=5\pa_{[M}C_{NPQR]}-15(B_{[MN}\pa_{P}C_{QR]}-C_{[MN}\pa_{P}B_{QR]})\, .\eea
Square brackets $[,]$ are used to denote the anti-symmetrizer, \eg
\be
H_{MNP}=3\pa_{[M}B_{NP]}=\frac{3}{3!}\sum_{\pi}(-1)^{\pi}\pa_{\pi(M)}B_{\pi(N) \pi(P)}=\pa_{M}B_{NP}+\pa_{N}B_{PM}+\pa_{P}B_{MN}\, ,
\ee
where we have to sum over all permutations $\pi$ of indices $M$, $N$ and $P$, and the sign $(-1)^{\pi}$ is $+1$ for even and $-1$ for odd permutations.
The equations of motion of type IIB supergravity in the {\it string frame} may be found by first varying the action~\cite{Dall'Agata:1997ju,Dall'Agata:1998va}
\be
\begin{aligned}
S=\frac{1}{2\kappa^2}\int {\rm d}^{10}X \Bigg[\sqrt{-G}\Bigg(& e^{-2\varphi}\Big(R+4\pa_{M}\varphi\pa^{M}\varphi-\frac{1}{12}H_{MNP}H^{MNP}\Big) - \\
&-\frac{1}{2}\pa_{M}\chi \pa^{M}\chi -\frac{1}{12}F_{MNP}F^{MNP}-\frac{1}{4\cdot 5!}F_{MNPQR}F^{MNPQR}
\Bigg)+ \,   \\
&+\frac{1}{8\cdot 4!}\eps^{M_1\ldots M_{10}}C_{M_1M_2M_3M_4}\pa_{M_5}B_{M_6M_7}\pa_{M_8}C_{M_9M_{10}}\Bigg]\, ,
\end{aligned}
\ee
and after that by imposing the self-duality condition for the five-form
\be\label{eq:sel-duality-F5-curved}
F_{M_1M_2M_3M_4M_5}=+\frac{1}{5!}\sqrt{-G}\eps_{M_1\ldots M_{10}}F^{M_6M_7M_8M_9M_{10}}\,.
\ee
Here $G$ is the determinant of the metric, $R$ the Ricci scalar, and for the anti-symmetric tensor $\eps$ we choose the convention $\eps^{0\ldots 9}=1$ and $\eps_{0\ldots 9}=-1$.
Let us write the equations of motion for all the fields.

\medskip

\noindent 
{\bf Equation for the dilaton $\varphi$}  
\be\label{eq:eom-dilaton}
4\pa^{M}\varphi\pa_{M}\varphi-4\pa^{M}\pa_{M}\varphi
-4\pa_{M}G^{MN}\pa_{N}\varphi-2\pa_{M}G_{PQ}G^{PQ}\pa^{M}\varphi=R-\frac{1}{12}H_{MNP}H^{MNP}\, .
\ee
Note that $\pa_{M}G_{PQ}G^{PQ} = 2\pa_{M}\log\sqrt{-G}$.

\noindent 
{\bf Equation for the two-form $B_{MN}$}

\be\label{eq:eom-B}
2\pa_{M}\Big(\sqrt{-G}(e^{-2\varphi}H^{MNP}+\chi F^{MNP})\Big)+\sqrt{-G}F^{NPQ RS} \pa_{Q}C_{RS}=0
\ee
This equation has been rewritten using~\eqref{eq:eom-C4}.

\noindent 
{\bf Equation for the axion $\chi$}  

\be\label{eq:eom-axion}
\pa_{M}\Big(\sqrt{-G}\pa^{M}\chi\Big)=+\frac{1}{6}\sqrt{-G}F_{MNP}H^{MNP}\, .
\ee

\noindent 
{\bf Equation for the two-form $C_{MN}$}

\be\label{eq:eom-C2}
\pa_{M}(\sqrt{-G}F^{MNP})-\frac{1}{6}\sqrt{-G}F^{NPQ RS}H_{QRS}=0
\ee

\noindent 
{\bf Equation for the four-form $C_{MNPQ}$} 
\be\label{eq:eom-C4}
\pa_{N}\left( \sqrt{-G} F^{NM_1M_2M_3M_4}\right)=-\frac{1}{36}\epsilon^{M_1\ldots M_4 M_5\ldots M_{10}}H_{M_5M_6M_7}F_{M_8M_9M_{10}}\, .
\ee

\noindent 
{\bf Einstein equations} 
\be
R_{MN}-\frac{1}{2}G_{MN}R=T_{MN}\, ,
\ee
where the stress tensor is
\be
\begin{aligned}
T_{MN}=&G_{MN}\Bigg[2\pa^{ P}(\pa_{ P}\varphi)-2G^{PQ}\Gamma^{R}_{PQ}\pa_{R}\varphi-2\pa_{P}\varphi\pa^{P}\varphi\\
&-\frac{1}{24}H_{PQR}H^{PQR}-\frac{1}{4}e^{2\varphi}F_{P}F^{P}-\frac{1}{24}e^{2\varphi}F_{PQR}F^{PQR}\Bigg] \\
&-2\pa_{M}\pa_{N}\varphi+2\Gamma^{P}_{MN}\pa_{P}\varphi\\
&+\frac{1}{4}H_{MPQ}H_{N}^{\ PQ}+\frac{1}{2}e^{2\varphi}F_{M}F_{N}
+\frac{1}{4}e^{2\varphi}F_{MPQ}F_{N}^{\ PQ}+\frac{1}{4\cdot 4!}e^{2\varphi}F_{MPQRS}F_{N}^{\ PQRS}\, , \\
\end{aligned}
\ee
and the Christoffel symbol is 
\be
\Gamma^{P}_{MN}=\frac{1}{2}G^{PQ}(\pa_{M}G_{NQ}+\pa_{N}G_{MQ}-\pa_{Q}G_{MN})\, .
\ee

\chapter{\adsthree}\label{app:AdS3}

\section{Gauge-fixed action for {\adsthree} at order $\theta^2$}\label{app:gauge-fixed-action-T4}
In this section we explain how to obtain the action at quadratic order in fermions for the superstring on the pure RR {\adsthree} background, following~\cite{Borsato:2014hja}.
The bosonic action is given by Eq.~\eqref{eq:bos-str-action}, where in our coordinates the spacetime metric for {\adsthree} reads as
\be
\begin{aligned}
{\rm d}s^2 &=  -\left(\frac{1 + \frac{z_1^2 + z_2^2}{4}}{1 - \frac{z_1^2 + z_2^2}{4}}\right)^2 {\rm d}t^2 + \frac{1}{\left(1 - \frac{z_1^2 + z_2^2}{4}\right)^2} ( {\rm d}z_1^2 + {\rm d}z_2^2 ) \\
&+\left(\frac{1 - \frac{y_3^2 + y_4^2}{4}}{1 + \frac{y_3^2 + y_4^2}{4}}\right)^2 {\rm d}\phi^2 + \frac{1}{\left(1 + \frac{y_3^2 + y_4^2}{4}\right)^2} ( {\rm d}y_3^2 + {\rm d}y_4^2 )\\
& + {\rm d}x_i {\rm d}x_i \,.
\end{aligned} 
\ee
We consider the case of vanishing $B$-field.
Coordinates $t,z_1,z_2$ parameterise AdS$_3$, and $t$ is the time coordinate.
Coordinates $\phi,y_3,y_4$ parameterise S$^3$, and $\phi$ is an angle that we will use, together with $t$ to create light-cone coordinates.
Coordintates $x^6,x^7,x^8,x^9$ parameterise the torus.
We prefer to enumerate the coordinates as
\be
X^0=t,\ 
X^1=z_1,\ 
X^2=z_2 ,\ 
X^3=y_3,\ 
X^4=y_4,\ 
X^5=\phi,\ 
X^i=x_i \text{ for } i=6,\ldots,9\,,
\ee
and to use a diagonal vielbein
\be
\begin{aligned}
e^0_t = \frac{1 + \frac{z_1^2 + z_2^2}{4}}{1 - \frac{z_1^2 + z_2^2}{4}}\,,\quad
e^1_{z_1}=e^2_{z_2}=\frac{1}{1 - \frac{z_1^2 + z_2^2}{4}}\,,\\
e^5_\phi = \frac{1 - \frac{y_3^2 + y_4^2}{4}}{1 + \frac{y_3^2 + y_4^2}{4}}\,,\quad
e^3_{y_3}=e^4_{y_4}=\frac{1}{1 + \frac{y_3^2 + y_4^2}{4}}\,,\\
e^i_{x_i}=1\quad i=6,\ldots,9.
\end{aligned} 
\ee
In order to avoid confusion, we use letters to denote explicitly curved indices on vielbein components, etc.
We will never distinguish between upper or lower indices for the coordinates $z_i\equiv z^i,y_i\equiv y^i,x_i\equiv x^i$.
To write down the fermionic action we first define the ten-dimensional Gamma-matrices\footnote{This basis is obtained by permuting the third and fourth spaces in the tensor products defining the Gamma-matrices that we find after implementing the change of basis explained in (2.55) of~\cite{Borsato:2014hja}.} 
\begin{equation}
    \newcommand{\I}{\mathrlap{\,\gen{1}}\hphantom{\sigma_a}}
    \newcommand{\II}{\mathrlap{\,\gen{1}}\hphantom{\gamma^A}}
  \begin{aligned}
&    \Gamma^0 = -i\sigma_1 \otimes  \sigma_3  \otimes \sigma_2 \otimes  \sigma_3  \otimes  \I , \qquad
&&    \Gamma^1 = +\sigma_1 \otimes  \sigma_1  \otimes   \sigma_2  \otimes \I\otimes  \I ,\\
 &   \Gamma^2 = +\sigma_1 \otimes  \sigma_2  \otimes  \sigma_2  \otimes \sigma_3 \otimes  \I , \qquad
&&    \Gamma^3 = +\sigma_1 \otimes  \sigma_1  \otimes \sigma_1   \otimes \sigma_1 \otimes \I ,\\
 &   \Gamma^4 = -\sigma_1 \otimes  \sigma_1  \otimes   \sigma_1  \otimes\sigma_2 \otimes \I , \qquad
&&    \Gamma^5 = -\sigma_1 \otimes \I  \otimes \sigma_1   \otimes \sigma_3 \otimes \I ,\\
 &   \Gamma^6 = +\sigma_1 \otimes   \I  \otimes  \sigma_3    \otimes \I \otimes \sigma_1 , \qquad
&&    \Gamma^7 = +\sigma_1 \otimes \I    \otimes  \sigma_3  \otimes \I  \otimes \sigma_2 ,\\
 &   \Gamma^8 = +\sigma_1 \otimes \I   \otimes \sigma_3  \otimes \I   \otimes \sigma_3 , \qquad
&&    \Gamma^9 = -\sigma_2 \otimes  \I   \otimes  \I  \otimes  \I \otimes \I .
  \end{aligned}
\end{equation}
For all Gamma-matrices we define the antisymmetric product by
\begin{equation}
  \Gamma^{m_1 m_2 \dotsm m_n} = \frac{1}{n!} \sum_{\pi \in S_n} (-1)^{\pi} \Gamma^{m_{\pi(1)}} \Gamma^{m_{\pi(2)}} \dotsm \Gamma^{m_{\pi(n)}} ,
\end{equation}
where the sum runs over all permutations of the indices and $(-1)^{\pi}$ denotes the signature of the permutation. 
For convenience, let us write down explicitly some higher-rank Gamma-matrices that may be obtained by the above definitions
\begin{equation}
    \newcommand{\I}{\mathrlap{\,\gen{1}}\hphantom{\sigma_a}}
  \begin{aligned}
    \Gamma^{1234} &= +\I \otimes  \sigma_3 \otimes    \I\otimes \I  \otimes  \I , \\
    \Gamma^{6789} &= + \sigma_3 \otimes \I \otimes \sigma_3\otimes  \I \otimes \I , \\
    \Gamma = \Gamma^{0123456789} &=  {+} \sigma_3 \otimes \I \otimes \I\otimes \I  \otimes \I .
  \end{aligned}
\end{equation}
The Gamma-matrices satisfy
\begin{equation}
  (\Gamma^m)^t = - \mathcal{C} \Gamma^m \mathcal{C}^{-1} , \qquad
  (\Gamma^m)^\dag = - \Gamma^0 \Gamma^m (\Gamma^0)^{-1} , \qquad
  (\Gamma^m)^* = + \mathcal{B} \Gamma^m \mathcal{B}^{-1} , \qquad
\end{equation}
where
\begin{equation}
  \mathcal{C} = -i\sigma_2 \otimes \sigma_3 \otimes \sigma_2 \otimes \sigma_1 \otimes \sigma_2 , \qquad
  \mathcal{B} = -\Gamma^0 \, \mathcal{C} .
\end{equation}
It is useful to note the relations
\begin{equation}
  \begin{gathered}
    (\Gamma^0)^\dag \Gamma^0 = \mathcal{C}^\dag \mathcal{C} = \mathcal{B}^\dag \mathcal{B} = \gen{1} , \qquad
    \mathcal{B}^t = \mathcal{C} (\Gamma^0)^\dag , \\
    \mathcal{C}^\dag = - \mathcal{C} = + \mathcal{C}^t , \qquad
    (\Gamma^0)^\dag = - \Gamma^0 = + (\Gamma^0)^t , \qquad
    \mathcal{B}^\dag = + \mathcal{B} = + \mathcal{B}^t , \\
    \mathcal{C} = - \Gamma^{01479} , \qquad
    \mathcal{B} = + \sigma_3 \otimes \gen{1} \otimes \sigma_1 \otimes \sigma_2 \otimes \sigma_2 = - \Gamma^{1479} , \\
    \mathcal{B} \Gamma \mathcal{B}^\dag = \Gamma^* .
  \end{gathered}
\end{equation}
The two sets of 32-component Majorana-Weyl spinors labelled by $I=1,2$ satisfy the conditions
\begin{equation}
\Gamma \Theta_I = +\Theta_I,\qquad
  \Theta^*_I = \mathcal{B} \Theta_I , \qquad
  \bar{\Theta}_I =  \Theta^t_I \mathcal{C},
\end{equation}
to give a total of 32 real fermions.
The action at quadratic order in fermions is given by~\eqref{eq:IIB-action-theta2}, where the operator ${D}^{IJ}_\a$ in this case is 
\be
\begin{aligned}
{D}^{IJ}_\a  = 
\delta^{IJ} \left( \pa_\a  -\frac{1}{4} {\omega}^{mn}_\a \G_{mn}  \right)
+\frac{1}{4} \sigma_1^{IJ} (\G^{012}+\G^{345}) \ {e}^m_\a \G_m.
\end{aligned}
\ee

\subsection{Linerarly realised supersymmetries}
The background possesses a total of 16 real supersymmetries.
It is more convenient to redefine the fermions introduced above, such that these supersymmetries are realised as linear shifts of fermionic components.
For a background realised by a supercoset, the original form of the action would correspond to the choice $\alg{g}=\alg{g}_{\text{bos}}\cdot\alg{g}_{\text{fer}}$ for the coset element. 
The redefinition we perform here would allow us to obtain to the choice $\alg{g}=\alg{g}_{\text{fer}}\cdot\alg{g}_{\text{bos}}$ for the coset element. 
For convenience we first redefine the fermions as
\begin{equation}\label{eq:introd-vartheta}
  \Theta_{1} =\vartheta_1 + \vartheta_2,
  \qquad
  \Theta_{2} = \vartheta_1 - \vartheta_2 ,
\end{equation}
and then introduce fermions $\vartheta^\pm_I$ as
\begin{equation}\label{eq:theta-chi-eta-def}
  \begin{aligned}
    \vartheta_1 &= \frac{1}{2} ( 1 + \Gamma^{012345} ) \hat{M} \vartheta^+_1 + \frac{1}{2} ( 1 - \Gamma^{012345} ) \hat{M} \vartheta^-_1 , \\
    \vartheta_2 &= \frac{1}{2} ( 1 + \Gamma^{012345} ) \check{M} \vartheta^+_2 + \frac{1}{2} ( 1 - \Gamma^{012345} ) \check{M} \vartheta^-_2 .
  \end{aligned}
\end{equation}
The projectors $\frac{1}{2} ( 1 \pm \Gamma^{012345} )$ make sure that we are again using a total of 32 real fermions.
Here $\hat{M}$ and $\check{M}$ are $32\times 32$ matrices
\begin{equation}\label{eq:MN-10d-expressions}
  \hat{M} = M_0 M_t , \qquad
  \check{M} = M_0^{-1} M_t^{-1} ,
\end{equation}
where
\begin{equation}\label{eq:def-matr-M0}
  \begin{aligned}
    M_0 &= \frac{1}{\sqrt{\bigl( 1 - \frac{z_1^2 + z_2^2}{4} \bigr) \bigl( 1 + \frac{y_3^2 + y_4^2}{4} \bigr)}}
    \Bigl( {1} - \frac{1}{2} (z_{1} \Gamma^{1}+z_{2} \Gamma^{2}) \Gamma^{012} \Bigr)
    \Bigl( {1} - \frac{1}{2} (y_{3} \Gamma^{3}+y_{4} \Gamma^{4}) \Gamma^{345} \Bigr) ,
    \\
    M_0^{-1} &= \frac{1}{\sqrt{\bigl( 1 - \frac{z_1^2 + z_2^2}{4} \bigr) \bigl( 1 + \frac{y_3^2 + y_4^2}{4} \bigr)}}
    \Bigl( {1} + \frac{1}{2} (z_{1} \Gamma^{1}+z_{2} \Gamma^{2}) \Gamma^{012} \Bigr)
    \Bigl( {1} + \frac{1}{2} (y_{3} \Gamma^{3}+y_{4} \Gamma^{4}) \Gamma^{345} \Bigr) ,
  \end{aligned}
\end{equation}
and
\begin{equation}
  M_t = e^{-\frac{1}{2} ( t \, \Gamma^{12} + \phi \, \Gamma^{34} )} , \qquad
  M_t^{-1} = e^{+\frac{1}{2} ( t \, \Gamma^{12} + \phi \, \Gamma^{34} )} .
\end{equation}
It is useful to see how these fermionic redefinitions are implemented on the Gamma-matrices
\begin{equation}
  \hat{M}^{-1} \, \Gamma_m \, \hat{M} \, e_M^m = \Gamma_m \, \hat{\mathcal{M}}^m{}_n e_M^n ,\qquad
  \check{M}^{-1} \, \Gamma_m \, \check{M} \, e_M^m = \Gamma_m \, \check{\mathcal{M}}^m{}_n e_M^n ,
\end{equation}
where $\hat{\mathcal{M}}_n{}^m$ and $\check{\mathcal{M}}_n{}^m$ are components of orthogonal matrices.
They rotate non-trivially only the indices $m=0,\ldots, 5$ of AdS$_3\times$S$^3$, and they act as the identity for directions tangent to the torus.
In particular, they can be reabsorbed in the definition of the vielbein, to produce\footnote{The conventions for the symbols ``check'' or ``hat'' should not be confused with the ones used in the chapters discussing the $\eta$-deformation of {\adsfive}, where they refer to AdS$_5$ and S$^5$. Here they refer to the matrices $\check{M},\hat{M}$ defined before.}
\begin{equation}
  \hat{e}_M^m = \hat{\mathcal{M}}^m{}_n \, e_M^n \,, \qquad 
  \check{e}_M^m = \check{\mathcal{M}}^m{}_n e_M^n \,.
\end{equation}
Explicitly, the $10\times 10$ matrices $\hat{e},\check{e}$ whose components are $  \hat{e}_M^m,  \check{e}_M^m$ are 
\be
\hat{e}= \hat{e}_{\text{AdS}_3} \oplus \hat{e}_{\text{S}^3} \oplus \gen{1}_4\,,
\qquad
\check{e}= \check{e}_{\text{AdS}_3} \oplus \check{e}_{\text{S}^3} \oplus \gen{1}_4\,,
\ee
where we have defined
\begin{equation}\label{eq:E-components-AdS3}
  \begin{aligned}
    \hat{e}_{\text{AdS}_3} &=
    \begin{pmatrix}
      1 & 0 & 0 \\
      0 & + \cos t & + \sin t \\
      0 & - \sin t & + \cos t
    \end{pmatrix} \cdot
    \begin{pmatrix}
      + 1 & + z_2 & - z_1 \\
      + \frac{z_2}{1 + \frac{z_1^2 + z_2^2}{4}} & 1 - \frac{z_1^2 - z_2^2}{4} & - \frac{z_1 z_2}{2} \\
      - \frac{z_1}{1 + \frac{z_1^2 + z_2^2}{4}} & - \frac{z_1 z_2}{2} & 1 + \frac{z_1^2 - z_2^2}{4}
    \end{pmatrix} ,
    \\
    \check{e}_{\text{AdS}_3} &=
    \begin{pmatrix}
      1 & 0 & 0 \\
      0 & + \cos t & - \sin t \\
      0 & + \sin t & + \cos t
    \end{pmatrix} \cdot
    \begin{pmatrix}
      + 1 & - z_2 & + z_1 \\
      - \frac{z_2}{1 + \frac{z_1^2 + z_2^2}{4}} & 1 - \frac{z_1^2 - z_2^2}{4} & - \frac{z_1 z_2}{2} \\
      + \frac{z_1}{1 + \frac{z_1^2 + z_2^2}{4}} & - \frac{z_1 z_2}{2} & 1 + \frac{z_1^2 - z_2^2}{4}   
    \end{pmatrix} ,
  \end{aligned}
\end{equation}
and
\begin{equation}
  \begin{aligned}
\label{eq:E-components-S3}
  \hat{e}_{\text{S}^3} &=
  \begin{pmatrix}
    + \cos\phi & + \sin\phi & 0 \\
    - \sin\phi & + \cos\phi & 0 \\
    0 & 0 & 1
  \end{pmatrix} \cdot
  \begin{pmatrix}
    1 + \frac{y_3^2 - y_4^2}{4} & + \frac{y_3 y_4}{2} & -\frac{y_4}{1 - \frac{y_3^2 + y_4^2}{4}} \\
    + \frac{y_3 y_4}{2} & 1 - \frac{y_3^2 - y_4^2}{4} & +\frac{y_3}{1 - \frac{y_3^2 + y_4^2}{4}} \\
    + y_4 & - y_3 & 1
  \end{pmatrix} \,,
\\
  \check{e}_{\text{S}^3} &=
  \begin{pmatrix}
    + \cos\phi & - \sin\phi & 0 \\
    + \sin\phi & + \cos\phi & 0 \\
    0 & 0 & 1
  \end{pmatrix} \cdot
  \begin{pmatrix}
    1 + \frac{y_3^2 - y_4^2}{4} & + \frac{y_3 y_4}{2} & +\frac{y_4}{1 - \frac{y_3^2 + y_4^2}{4}} \\
    + \frac{y_3 y_4}{2} & 1 - \frac{y_3^2 - y_4^2}{4} & -\frac{y_3}{1 - \frac{y_3^2 + y_4^2}{4}} \\
    - y_4 & + y_3 & 1
  \end{pmatrix} .
  \end{aligned}
\end{equation}
It is then possible to check that the sum of the bosonic and fermionic Lagrangians are invariant under the following supersymmetry transformations
\begin{equation}\label{eq:susy-transformations}
\begin{aligned}
&  \delta\vartheta^-_I = \epsilon_I \,, 
  \qquad 
 && \delta\vartheta^+_I = 0 \,,
\\
&  \delta\hat{e}_\alpha^m =\delta\check{e}_\alpha^m = -i \bar{\epsilon}_I \Gamma^m \partial_\alpha \vartheta^-_I \quad m=0,\ldots,5\,,
  \qquad
 && \delta\hat{e}_\alpha^m =\delta\check{e}_\alpha^m = 0 \quad m=6,\ldots,9\,,
\end{aligned}
\end{equation}
at first order in $\vartheta^\pm_I$.

\subsection{Gauge-fixed action}
In the previous subsection we showed which is the most convenient choice for the fermions in order to achieve a simple form of the supersymmetry variations.
In this subsection we perform a different fermionc field redefinition.
This is necessary to get fermions that are not charged under the two isometries corresponding to shifts of the coordinates $t,\phi$.
This is needed to fix light-cone kappa-gauge later.
For a background realised as a supercoset, this would correspond to the choice $\alg{g}=\Lambda(t,\phi)\cdot\alg{g}_{\text{fer}}\cdot\alg{g}'_{\text{bos}}$ for the coset representative, where $\Lambda(t,\phi)$ is a group element parameterised by $t,\phi$ only, while $\alg{g}'_{\text{bos}}$ by the transverse bosonic coordinates.

Starting from the fermions appearing in~\eqref{eq:introd-vartheta}, we define fermions $\eta_I,\chi_I$
\begin{equation}\label{eq:theta-chi-eta-def-t}
  \begin{aligned}
    \vartheta_1 &= 
    \frac{1}{2} ( 1 + \Gamma^{012345} ) \mathrlap{M_0 \chi_1}\hphantom{M_0^{-1} \chi_2} 
    + \frac{1}{2} ( 1 - \Gamma^{012345} ) M_0 \eta_1
    \\
    \vartheta_2 &= \frac{1}{2} ( 1 + \Gamma^{012345} ) M_0^{-1} \chi_2 + \frac{1}{2} ( 1 - \Gamma^{012345} ) M_0^{-1} \eta_2 ,
  \end{aligned}
\end{equation}
where the matrices $M_0,M_0^{-1}$ may be read in~\eqref{eq:def-matr-M0}.
As previously, the correct number of fermions is ensured by the presence of the projectors.
After introducing bosonic light-cone coordinates as in Section~\ref{sec:Bos-string-lcg} we impose kappa-gauge like in Section~\ref{sec:fermions-type-IIB}
\begin{equation}\label{eq:bmn-lc-kappa-gauge}
  \Gamma^+ \eta_I = 0 , \qquad
  \Gamma^+ \chi_I = 0 , \qquad
  \Gamma^{\pm} = \frac{1}{2} \bigl( \Gamma^5 \pm \Gamma^0 \bigr) ,
\end{equation}
to keep only a total of 16 real fermions.
The redefinition~\eqref{eq:theta-chi-eta-def-t} and the condition imposed by the kappa-gauge allows us to require that the fermions satisfy
\begin{equation}
    \Gamma^{1234} \chi_I = +\chi_I \,,\qquad 
    \Gamma^{6789} \chi_I = +\chi_I \,,\qquad
    \Gamma^{1234} \eta_I = -\eta_I \,, \qquad
    \Gamma^{6789} \eta_I = -\eta_I \,.
\end{equation}
It is then natural to write them as
\be
\begin{aligned}
\chi_I &= \left(\begin{array}{c}1\\0\end{array}\right) \otimes \left(\begin{array}{c}1\\0\end{array}\right) \otimes \left(\begin{array}{c}1\\0\end{array}\right) \otimes
( \chi_I )^{\underline{a} b} \,,
\\
\eta_I &= \left(\begin{array}{c}1\\0\end{array}\right) \otimes \left(\begin{array}{c}0\\1\end{array}\right) \otimes \left(\begin{array}{c}0\\1\end{array}\right) \otimes
( \eta_I )^{\underline{\dot{a}} \dot{b}}  \,.
\end{aligned}
\ee
As explained in Section~\ref{sec:fermions-type-IIB}, from the light-cone gauge-fixed action we can read the Hamiltonian of the gauge-fixed model.
The Hamiltonian at quadratic order in the fields is written in~\eqref{eq:quadr-Hamilt-fields-T4}.
To obtain that expression we have actually rewritten our bosonic and fermionic coordinates as follows.
We first introduce complex coordinates to parameterise the transverse directions of AdS$_3$ and $S^3$ 
\begin{equation}
Z=-z_2+i\,z_1\,,\qquad
\bar{Z}=-z_2-i\,z_1\,,\qquad\qquad
Y=-y_3-i\,y_4\,,\qquad
\bar{Y}=-y_3+i\,y_4\,,
\end{equation}
together with the corresponding conjugate momenta
\begin{equation}
P_Z=\frac{1}{2}\dot{Z},\qquad
P_{\bar{Z}}=\frac{1}{2}\dot{\bar{Z}}\,,\qquad\qquad
P_Y=\frac{1}{2}\dot{Y},\qquad
P_{\bar{Y}}=\frac{1}{2}\dot{\bar{Y}}\,.
\end{equation}
Similarly, for the four directions in the torus we define the complex combinations
\begin{equation}
X^{12}= x_8 - i \,x_9 \,,
\quad
X^{21}= -x_8 -i \,x_9\,,
\qquad
X^{11}= -x_6 + i\, x_7\,,
\quad
X^{22}= -x_6 - i\, x_7\,,
\end{equation}
 and the conjugate momenta
\begin{equation}
P_{\dot{a}a}= \frac{1}{2}\epsilon_{\dot{a}\dot{b}}\epsilon_{ab}\dot{X}^{\dot{b}b}.
\end{equation}
Upon quantisation, these fields satisfy the canonical commutation relations
\begin{equation}\label{eq:comm-rel-bos}
\begin{aligned}
&[Z(\tau,\sigma_1),P_{\bar{Z}}(\tau,\sigma_2)]=[\bar{Z}(\tau,\sigma_1),P_{Z}(\tau,\sigma_2)]=i\,\delta(\sigma_1-\sigma_2)\,,\\
&[Y(\tau,\sigma_1),P_{\bar{Y}}(\tau,\sigma_2)]=[\bar{Y}(\tau,\sigma_1),P_{Y}(\tau,\sigma_2)]=i\,\delta(\sigma_1-\sigma_2)\,,\\
&[X^{{\dot{a}a}}(\tau,\sigma_1), P_{{\dot{b}b}}(\tau,\sigma_2)]=i\,\delta^{{\dot{a}}}_{\
{\dot{b}}}\,\delta^{{a}}_{\
{b}}\,\delta(\sigma_1-\sigma_2)\,.
\end{aligned}
\end{equation}
For the massive fermions we define the various components as
\begin{equation}
\big(\eta_1\big)^{\underline{\dot{a}}\dot{{a}}}=\left(
\begin{array}{cc}
-e^{+i\pi/4}\,\bar{\eta}_{\smallL 2} & -e^{+i\pi/4}\,
\bar{\eta}_{\smallL 1} \\
\phantom{-}e^{-i\pi/4}\,\eta_{\smallL}^{\ 1} &
-e^{-i\pi/4}\,\eta_{\smallL}^{\ 2}
\end{array}
\right),
\qquad
\big(\eta_2\big)^{\underline{\dot{a}}\dot{{a}}}=\left(
\begin{array}{cc}
\phantom{-}e^{-i\pi/4}\,\eta_{\smallR 2} &
\phantom{-}e^{-i\pi/4}\,\eta_{\smallR 1}\\
-e^{+i\pi/4}\,\bar{\eta}_{\smallR}^{\ 1} &
\phantom{-}e^{+i\pi/4}\,
\bar{\eta}_{\smallR}{}^{ 2}
\end{array}
\right),
\end{equation}
where the signs and the factors of~$e^{\pm i\pi/4}$ are introduced for later
convenience. Similarly, for the massless fermions we write
\begin{equation}
\big(\chi_1\big)^{\underline{a}a}=\left(
\begin{array}{cc}
-e^{+i\pi/4}\bar{\chi}_{+2} & \phantom{-}e^{+i\pi/4}
\bar{\chi}_{+1} \\
-e^{-i\pi/4}\chi_{+}^{\ 1} & -e^{-i\pi/4}\chi_{+}^{\ 2}
\end{array}
\right), \qquad
\big(\chi_2\big)^{\underline{a}a}=\left(
\begin{array}{cc}
\phantom{-}e^{-i\pi/4}\chi_{-}^{\ 1} &
\phantom{-}e^{-i\pi/4} \chi_{-}^{\ 2} \\
-e^{+i\pi/4}\bar{\chi}_{-2} &
\phantom{-}e^{+i\pi/4}\bar{\chi}_{-1}
\end{array}
\right).
\end{equation}
The canonical anti-commutation relations are
\begin{equation}\label{eq:comm-rel-ferm}
\begin{aligned}
&\ \{\bar{\eta}_{\smallL{\dot{a}}}(\sigma_1),{\eta}_{\sL}^{\ {\dot{b}}}(\sigma_2)\}=
\{\bar{\eta}_{\smallR}^{\ {\dot{b}}}(\sigma_1),{\eta}_{\smallR{\dot{a}}}(\sigma_2)\}=\delta_{{\dot{a}}}^{\
{\dot{b}}}\,\delta(\sigma_1-\sigma_2),\\
&\{\bar{\chi}_{+{a}}(\sigma_1),{\chi}_{+}^{{b}}(\sigma_2)\}=
\{\bar{\chi}_{-{a}}(\sigma_1),\chi_{-}^{{b}}(\sigma_2)\}=\delta_{{a}}^{\
{b}}\,\delta(\sigma_1-\sigma_2),
\end{aligned}
\end{equation}
It is possible to derive the (super)currents associated to the isometries of the model.
In general the conserved charges are divided into \emph{kinematical} and \emph{dynamical}. The former do not depend on the light-cone coordinate $x^-$, while the latter do.
Given their definition, kinematical charges commute with the total light-cone momentum $P_-$ of~\eqref{eq:total-light-cone-mom-P+P-}.
Another way to look at the conserved charges is to see if they also do or do not depend on $x^+=\tau$.
Because ${\rm d}\gen{Q}/{\rm d}\tau = \pa \gen{Q}/\pa\tau + \{\gen{H},\gen{Q}\}=0$, it is clear that only the charges without an explicit time-dependence commute with the Hamiltonian.
These are actually the charges we are interested in.
In particular, of the sixteen real conserved supercharges, only eight of them commute with the Hamiltonian.
They turn out to be dynamical.
For {\adsthree} they have been presented at first order in fermions and third order in bosons in~\cite{Borsato:2014hja}.
In~\eqref{eq:supercharges-quadratic-T4} we write them at first order in fermions and first order in bosons.

\section{Oscillator representation}\label{app:oscillators}
Here we introduce creation and annihilation operators.
We will use them to rewrite the conserved charges of Chapter~\ref{sec:SymmetryAlgebraT4} forming the algebra $\mathcal{A}$. We first define the following  wave-function parameters
\begin{equation}
\omega(p,m)=\sqrt{m^2+p^2},\quad
f(p,m)=\sqrt{\frac{\omega(p,m)+|m|}{2}},\quad
g(p,m)=-\frac{p}{2f(p,m)},
\end{equation}
that satisfy
\begin{equation}
\omega(p,m)=f(p,m)^2+g(p,m)^2,\qquad
|m|=f(p,m)^2-g(p,m)^2,
\end{equation}
and the short-hand notation
\begin{equation}\begin{aligned}
\omega_p=\omega(p,\pm1)\qquad
f_p=f(p,\pm1),\qquad
g_p=g(p,\pm1),\\
\tilde{\omega}_p=\omega(p,0),\qquad\ \ 
\tilde{f}_p=f(p,0),\qquad\ \ 
\tilde{g}_p=g(p,0).\ 
\end{aligned}
\end{equation}
For the massive bosons we take
\begin{equation}
\begin{aligned}
a_{\sL z} (p) &= \frac{1}{\sqrt{2\pi}} \int
\frac{{\rm d}\sigma}{2\sqrt{\omega_p}} \left(\omega_p \bar{Z} +2i P_{\bar{Z}}
\right) e^{-i p \sigma},\\
a_{\sR z} (p)&= \frac{1}{\sqrt{2\pi}} \int \frac{{\rm d}\sigma}{2\sqrt{\omega_p}}
\left(\omega_p Z + 2i P_{Z} \right) e^{-i p \sigma}, \\
\\
a_{\sL y}(p) &= \frac{1}{\sqrt{2\pi}} \int
\frac{{\rm d}\sigma}{2\sqrt{\omega_p}} \left(\omega_p \bar{Y} + 2i P_{\bar{Y}}
\right) e^{-i p \sigma},\\
a_{\sR y}(p) &= \frac{1}{\sqrt{2\pi}} \int \frac{{\rm d}\sigma}{2\sqrt{\omega_p}}
\left(\omega_p Y + 2i P_{Y} \right) e^{-i p \sigma}, \\
\end{aligned}
\end{equation}
and for the massless bosons
\begin{equation}
\begin{aligned}
a_{{\dot{a}a}} (p) &= \frac{1}{\sqrt{2\pi}} \int
\frac{{\rm d}\sigma}{2\sqrt{\tilde{\omega}_p}} \left(\tilde{\omega}_p X_{{\dot{a}a}} + 2i
P_{{\dot{a}a}} \right) e^{-i p \sigma}.
\end{aligned}
\end{equation}
The corresponding creation operators are found by taking the complex conjugate of the above expressions and are indicated by a dagger. For massless bosons we have in particular $(a_{{\dot{a}a}})^*=a^{{\dot{a}a}\dagger}$.
\begin{equation}
\begin{gathered}
[a_{\sL\, z}(p_1),a^{\dagger}_{\sL\, z}(p_2)]=
[a_{\sR\, z}(p_1),a^{\dagger}_{\sR\, z}(p_2)]=
\delta(p_1-p_2)\,,\\
[a_{\sL\, y}(p_1),a^{\dagger}_{\sL\, y}(p_2)]=
[a_{\sR\, y}(p_1),a^{\dagger}_{\sR\, y}(p_2)]=
\delta(p_1-p_2)\,,\\
[a^{{\dot{a}a}}(p_1),a^{\dagger}_{{\dot{b}b}}(p_2)]=\delta^{{\dot{a}}}_{\ {\dot{b}}}\,\delta^{{a}}_{\ {b}}\,
\delta(p_1-p_2)\,.
\end{gathered}
\end{equation}
The ladder operators for massive fermions are defined as
\begin{equation}
\label{eq:fermion-par-massive}
\begin{aligned}
d_{\sL {\dot{a}}}(p) &=+ \frac{e^{+i \pi / 4}}{\sqrt{2\pi}} \int
\frac{{\rm d}\sigma}{\sqrt{\omega_p}} \ \epsilon_{{\dot{a}\dot{b}}}\left( f_p \,
{\eta}^{\ {\dot{b}}}_{\smallL}
+i g_p \, \bar{\eta}^{\ {\dot{b}}}_{\smallR} \right) e^{-i p \sigma}, \\
d_{\sR}^{\ {\dot{a}}}(p) &=- \frac{e^{+i \pi / 4}}{\sqrt{2\pi}} \int
\frac{{\rm d}\sigma}{\sqrt{\omega_p}} \ \epsilon^{{\dot{a}\dot{b}}} \left( f_p \, \eta_{\smallR{\dot{b}}}
+i g_p \, \bar{\eta}_{\smallL{\dot{b}}} \right) e^{-i p \sigma}. \\
\end{aligned}
\end{equation}
while for massless fermions we take
\begin{equation}
\label{eq:fermion-par-massless}
\begin{aligned}
\tilde{d}_{{a}}(p)&= \frac{e^{-i \pi / 4}}{\sqrt{2\pi}} \int
\frac{{\rm d}\sigma}{\sqrt{\tilde{\omega}_p}} \left( \tilde{f}_p  \bar{\chi}_{+{a}} -i
\tilde{g}_p \, \epsilon_{{ab}}
{\chi}_{-}^{\ {b}} \right) e^{-i p \sigma},
\\
{d}_{{a}}(p) &= \frac{e^{+i \pi / 4}}{\sqrt{2\pi}} \int
\frac{{\rm d}\sigma}{\sqrt{\tilde{\omega}_p}} \left( \tilde{f}_p\, \epsilon_{{ab}}
{\chi}_{+}^{\ {b}} -i \tilde{g}_p\, \bar{\chi}_{-{a}}\right) e^{-i p \sigma}.
\end{aligned}
\end{equation}
Also in this case the creation operators are found by taking $({d}_{{a}})^*={d}^{{a}\dagger}$.
The anti-commutation relations are
\begin{equation}
\begin{gathered}
\{d_{\sL}^{\ {\dot{a}}\,\dagger}(p_1),d_{\sL {\dot{b}}}(p_2)\}=
\{d_{\sR {\dot{b}}}^{\,\dagger}(p_1),d_{\sR}^{\ {\dot{a}}}(p_2)\}=\delta_{{\dot{b}}}^{\ {\dot{a}}}\,\delta(p_1-p_2)\,,
\\
\{\tilde{d}^{{a}\,\dagger}(p_1),\tilde{d}_{{b}}(p_2)\}=
\{d^{\ {a}\,\dagger}(p_1),d_{{b}}(p_2)\}=\delta_{{b}}^{\ {a}}\,\delta(p_1-p_2)\,.
\end{gathered}
\end{equation}
Using these definitions we can rewrite the conserved charges in terms of creation and annihilation operators, to obtain~\eqref{eq-central-charges-osc-T4} and~\eqref{eq-supercharges-osc-T4}.

\section{Explicit S-matrix elements}\label{app:S-mat-explicit}
Here we write the action of the $\psu(1|1)^4_{\ce}$-invariant S-matrix on two-particle states. 
\subsection{The massive sector}
In the massive sector, when we scatter two Left excitations we get
\begingroup
\addtolength{\jot}{1ex}
\begin{equation}\label{eq:massive-S-matrix-LL}
\begin{aligned}
\mathcal{S}^{\sL\sL}\check{\otimes}\mathcal{S}^{\sL\sL} \ket{Y^{\sL}_p Y^{\sL}_q} =& 
A^{\sL\sL}_{pq} A^{\sL\sL}_{pq} \ket{Y^{\sL}_q Y^{\sL}_p } , \\
\mathcal{S}^{\sL\sL}\check{\otimes}\mathcal{S}^{\sL\sL} \ket{Y^{\sL}_p \eta^{\sL\, \dot{a}}_q} =& 
A^{\sL\sL}_{pq} B^{\sL\sL}_{pq} \ket{\eta^{\sL\, \dot{a}}_q Y^{\sL}_p } 
+A^{\sL\sL}_{pq} C^{\sL\sL}_{pq} \ket{Y^{\sL}_q \eta^{\sL\, \dot{a}}_p } , \\
\mathcal{S}^{\sL\sL}\check{\otimes}\mathcal{S}^{\sL\sL} \ket{Y^{\sL}_p Z^{\sL}_q} =& 
B^{\sL\sL}_{pq} B^{\sL\sL}_{pq} \ket{Z^{\sL}_q Y^{\sL}_p } 
+C^{\sL\sL}_{pq} C^{\sL\sL}_{pq} \ket{Y^{\sL}_q Z^{\sL}_p } 
+\epsilon^{\dot{a}\dot{b}} B^{\sL\sL}_{pq} C^{\sL\sL}_{pq} \ket{\eta^{\sL\, \dot{a}}_q \eta^{\sL\, \dot{b}}_p } 
, \\
\mathcal{S}^{\sL\sL}\check{\otimes}\mathcal{S}^{\sL\sL} \ket{\eta^{\sL\, \dot{a}}_p Y^{\sL}_q} =& 
A^{\sL\sL}_{pq} D^{\sL\sL}_{pq} \ket{Y^{\sL}_q \eta^{\sL\, \dot{a}}_p } 
+A^{\sL\sL}_{pq} E^{\sL\sL}_{pq} \ket{\eta^{\sL\, \dot{a}}_q Y^{\sL}_p } , \\
\mathcal{S}^{\sL\sL}\check{\otimes}\mathcal{S}^{\sL\sL} \ket{\eta^{\sL\, \dot{a}}_p \eta^{\sL\, \dot{b}}_q} =& 
-B^{\sL\sL}_{pq} D^{\sL\sL}_{pq} \ket{\eta^{\sL\, \dot{b}}_q \eta^{\sL\, \dot{a}}_p } 
+C^{\sL\sL}_{pq} E^{\sL\sL}_{pq} \ket{\eta^{\sL\, \dot{a}}_q \eta^{\sL\, \dot{b}}_p } 
\\
&+\epsilon^{\dot{a}\dot{b}}C^{\sL\sL}_{pq} D^{\sL\sL}_{pq} \ket{Y^{\sL}_q Z^{\sL}_p } 
+\epsilon^{\dot{a}\dot{b}}B^{\sL\sL}_{pq} E^{\sL\sL}_{pq} \ket{Z^{\sL}_q Y^{\sL}_p } 
, \\
\mathcal{S}^{\sL\sL}\check{\otimes}\mathcal{S}^{\sL\sL} \ket{\eta^{\sL\, \dot{a}}_p Z^{\sL}_q} =& 
-B^{\sL\sL}_{pq} F^{\sL\sL}_{pq} \ket{Z^{\sL}_q \eta^{\sL\, \dot{a}}_p } 
+C^{\sL\sL}_{pq} F^{\sL\sL}_{pq} \ket{\eta^{\sL\, \dot{a}}_q Z^{\sL}_p } , \\
\mathcal{S}^{\sL\sL}\check{\otimes}\mathcal{S}^{\sL\sL} \ket{Z^{\sL}_p Y^{\sL}_q} =& 
D^{\sL\sL}_{pq} D^{\sL\sL}_{pq} \ket{Y^{\sL}_q Z^{\sL}_p } 
+E^{\sL\sL}_{pq} E^{\sL\sL}_{pq} \ket{Z^{\sL}_q Y^{\sL}_p } 
+\epsilon^{\dot{a}\dot{b}}D^{\sL\sL}_{pq} E^{\sL\sL}_{pq} \ket{\eta^{\sL\, \dot{a}}_q \eta^{\sL\, \dot{b}}_p } 
, \\
\mathcal{S}^{\sL\sL}\check{\otimes}\mathcal{S}^{\sL\sL} \ket{Z^{\sL}_p \eta^{\sL\, \dot{a}}_q} =& 
-D^{\sL\sL}_{pq} F^{\sL\sL}_{pq} \ket{\eta^{\sL\, \dot{a}}_q Z^{\sL}_p } 
+E^{\sL\sL}_{pq} F^{\sL\sL}_{pq} \ket{Z^{\sL}_q \eta^{\sL\, \dot{a}}_p } , \\
\mathcal{S}^{\sL\sL}\check{\otimes}\mathcal{S}^{\sL\sL} \ket{Z^{\sL}_p Z^{\sL}_q} =& 
F^{\sL\sL}_{pq} F^{\sL\sL}_{pq} \ket{Z^{\sL}_q Z^{\sL}_p } , \\
\end{aligned}
\end{equation}
\endgroup
Scattering a Left and a Right excitation yields\footnote{To have a better notation, we prefer to raise the $\alg{su}(2)$ index of Right fermions with $\epsilon^{\dot{a}\dot{b}}$.}
\begingroup
\addtolength{\jot}{1ex}
\begin{equation}\label{eq:massive-S-matrix-LR}
\begin{aligned}
\mathcal{S}^{\sL\sR}\check{\otimes}\mathcal{S}^{\sL\sR} \ket{Y^{\sL}_p Y^{\sR}_q} =& 
A^{\sL\sR}_{pq} A^{\sL\sR}_{pq} \ket{Y^{\sR}_q Y^{\sL}_p } 
-B^{\sL\sR}_{pq} B^{\sL\sR}_{pq} \ket{Z^{\sR}_q Z^{\sL}_p } 
+\epsilon^{\dot{a}\dot{b}}A^{\sL\sR}_{pq} B^{\sL\sR}_{pq} \ket{\eta^{\sR\, \dot{a}}_q \eta^{\sL\, \dot{b}}_p } 
, \\
\mathcal{S}^{\sL\sR}\check{\otimes}\mathcal{S}^{\sL\sR} \ket{Y^{\sL}_p \eta^{\sR\, \dot{a}}_q} =& 
A^{\sL\sR}_{pq} C^{\sL\sR}_{pq} \ket{\eta^{\sR\, \dot{a}}_q Y^{\sL}_p } 
-B^{\sL\sR}_{pq} C^{\sL\sR}_{pq} \ket{Z^{\sR}_q \eta^{\sL\, \dot{a}}_p } , \\
\mathcal{S}^{\sL\sR}\check{\otimes}\mathcal{S}^{\sL\sR} \ket{Y^{\sL}_p Z^{\sR}_q} =& 
C^{\sL\sR}_{pq} C^{\sL\sR}_{pq} \ket{Z^{\sR}_q Y^{\sL}_p } , \\
\mathcal{S}^{\sL\sR}\check{\otimes}\mathcal{S}^{\sL\sR} \ket{\eta^{\sL\, \dot{a}}_p Y^{\sR}_q} =& 
A^{\sL\sR}_{pq} D^{\sL\sR}_{pq} \ket{Y^{\sR}_q \eta^{\sL\, \dot{a}}_p } 
-B^{\sL\sR}_{pq} D^{\sL\sR}_{pq} \ket{\eta^{\sR\, \dot{a}}_q Z^{\sL}_p } , \\
\mathcal{S}^{\sL\sR}\check{\otimes}\mathcal{S}^{\sL\sR} \ket{\eta^{\sL\, \dot{a}}_p \eta^{\sR\, \dot{b}}_q} =& 
+A^{\sL\sR}_{pq} E^{\sL\sR}_{pq} \ket{\eta^{\sR\, \dot{b}}_q \eta^{\sL\, \dot{a}}_p } 
-B^{\sL\sR}_{pq} F^{\sL\sR}_{pq} \ket{\eta^{\sR\, \dot{a}}_q \eta^{\sL\, \dot{b}}_p } \\
&-\epsilon^{\dot{a}\dot{b}}A^{\sL\sR}_{pq} F^{\sL\sR}_{pq} \ket{Y^{\sR}_q Y^{\sL}_p } 
+\epsilon^{\dot{a}\dot{b}}B^{\sL\sR}_{pq} E^{\sL\sR}_{pq} \ket{Z^{\sR}_q Z^{\sL}_p } 
, \\
\mathcal{S}^{\sL\sR}\check{\otimes}\mathcal{S}^{\sL\sR} \ket{\eta^{\sL\, \dot{a}}_p Z^{\sR}_q} =& 
-E^{\sL\sR}_{pq} C^{\sL\sR}_{pq} \ket{Z^{\sR}_q \eta^{\sL\, \dot{a}}_p } 
+C^{\sL\sR}_{pq} F^{\sL\sR}_{pq} \ket{\eta^{\sR\, \dot{a}}_q Y^{\sL}_p } , \\
\mathcal{S}^{\sL\sR}\check{\otimes}\mathcal{S}^{\sL\sR} \ket{Z^{\sL}_p Y^{\sR}_q} =& 
D^{\sL\sR}_{pq} D^{\sL\sR}_{pq} \ket{Y^{\sR}_q Z^{\sL}_p } , \\
\mathcal{S}^{\sL\sR}\check{\otimes}\mathcal{S}^{\sL\sR} \ket{Z^{\sL}_p \eta^{\sR\, \dot{a}}_q} =& 
-D^{\sL\sR}_{pq} E^{\sL\sR}_{pq} \ket{\eta^{\sR\, \dot{a}}_q Z^{\sL}_p } 
+D^{\sL\sR}_{pq} F^{\sL\sR}_{pq} \ket{Y^{\sR}_q \eta^{\sL\, \dot{a}}_p } 
, \\
\mathcal{S}^{\sL\sR}\check{\otimes}\mathcal{S}^{\sL\sR} \ket{Z^{\sL}_p Z^{\sR}_q} =& 
E^{\sL\sR}_{pq} E^{\sL\sR}_{pq} \ket{Z^{\sR}_q Z^{\sL}_p } 
-F^{\sL\sR}_{pq} F^{\sL\sR}_{pq} \ket{Y^{\sR}_q Y^{\sL}_p } 
-\epsilon^{\dot{a}\dot{b}}E^{\sL\sR}_{pq} F^{\sL\sR}_{pq} \ket{\eta^{\sR\, \dot{a}}_q \eta^{\sL\, \dot{b}}_p } , \\
\end{aligned}
\end{equation}
\endgroup

\subsection{The mixed-mass sector}
In the case of left massive excitations that scatter with massless excitations transforming in the $\varrho_{\sL} \otimes \widetilde{\varrho}_{\sL}$ representation of $\alg{psu}(1|1)^4_{\ce}$ we find
\begingroup
\addtolength{\jot}{1ex}
\begin{equation}\label{eq:mixed-S-matrix-L}
\begin{aligned}
\mathcal{S}^{\sL\sL}\check{\otimes}\mathcal{S}^{\sL\tilde{\sL}} \ket{Z^{\sL}_p T^{\dot{a}a}_q} =& - F^{\sL\sL}_{pq}D^{\sL\sL}_{pq} \ket{T^{\dot{a}a}_q Z^{\sL}_p } - F^{\sL\sL}_{pq}E^{\sL\sL}_{pq} \ket{\widetilde{\chi}^{a}_q \eta^{\sL \dot{a}}_p}, \\
\mathcal{S}^{\sL\sL}\check{\otimes}\mathcal{S}^{\sL\tilde{\sL}} \ket{Y^{\sL}_p T^{\dot{a}a}_q} =& + A^{\sL\sL}_{pq}B^{\sL\sL}_{pq} \ket{T^{\dot{a}a}_q Y^{\sL}_p } - A^{\sL\sL}_{pq}C^{\sL\sL}_{pq} \ket{ \chi^{a}_q \eta^{\sL \dot{a}}_p }, \\
\mathcal{S}^{\sL\sL}\check{\otimes}\mathcal{S}^{\sL\tilde{\sL}} \ket{\eta^{\sL \dot{a}}_p \widetilde{\chi}^{a}_q} =& + F^{\sL\sL}_{pq}B^{\sL\sL}_{pq} \ket{\widetilde{\chi}^a_q \eta^{\sL \dot{a}}_p } + F^{\sL\sL}_{pq}C^{\sL\sL}_{pq} \ket{ T^{\dot{a}a}_q Z^{\sL}_p }, \\
\mathcal{S}^{\sL\sL}\check{\otimes}\mathcal{S}^{\sL\tilde{\sL}} \ket{\eta^{\sL \dot{a}}_p \chi^a_q} =& -A^{\sL\sL}_{pq}D^{\sL\sL}_{pq} \ket{\chi^a_q \eta^{\sL \dot{a}}_p } + A^{\sL\sL}_{pq}E^{\sL\sL}_{pq} \ket{ T^{\dot{a}a}_q Y^{\sL}_p }, \\
\mathcal{S}^{\sL\sL}\check{\otimes}\mathcal{S}^{\sL\tilde{\sL}} \ket{Z^{\sL}_p \widetilde{\chi}^a_q} =& + F^{\sL\sL}_{pq}F^{\sL\sL}_{pq} \ket{\widetilde{\chi}^a_q Z^{\sL}_p }, \\
\mathcal{S}^{\sL\sL}\check{\otimes}\mathcal{S}^{\sL\tilde{\sL}} \ket{Y^{\sL}_p \chi^a_q} =& + A^{\sL\sL}_{pq}A^{\sL\sL}_{pq} \ket{\chi^a_q Y^{\sL}_p }, \\
\mathcal{S}^{\sL\sL}\check{\otimes}\mathcal{S}^{\sL\tilde{\sL}} \ket{Z^{\sL}_p \chi^a_q} =& + D^{\sL\sL}_{pq}D^{\sL\sL}_{pq} \ket{\chi^a_q Z^{\sL}_p } + E^{\sL\sL}_{pq}E^{\sL\sL}_{pq} \ket{\widetilde{\chi}^a_q Y^{\sL}_p } + D^{\sL\sL}_{pq}E^{\sL\sL}_{pq} \, \epsilon_{\dot{a}\dot{b}} \ket{ T^{\dot{a}a}_q \eta^{\sL \dot{b}}_p}, \\
\mathcal{S}^{\sL\sL}\check{\otimes}\mathcal{S}^{\sL\tilde{\sL}} \ket{Y^{\sL}_p \widetilde{\chi}^a_q} =& + B^{\sL\sL}_{pq}B^{\sL\sL}_{pq} \ket{ \widetilde{\chi}^a_q Y^{\sL}_p } + C^{\sL\sL}_{pq}C^{\sL\sL}_{pq} \ket{\chi^a_q Z^{\sL}_p } + B^{\sL\sL}_{pq}C^{\sL\sL}_{pq} \, \epsilon_{\dot{a}\dot{b}} \ket{ T^{\dot{a}a}_q \eta^{\sL \dot{b}}_p}, \\
\mathcal{S}^{\sL\sL}\check{\otimes}\mathcal{S}^{\sL\tilde{\sL}} \ket{ \eta^{\sL \dot{a}}_p T^{\dot{b}a}_q} =& + D^{\sL\sL}_{pq}B^{\sL\sL}_{pq} \ket{ T^{\dot{a}a}_q \eta^{\sL \dot{b}}_p } - E^{\sL\sL}_{pq}C^{\sL\sL}_{pq} \ket{ T^{\dot{b}a}_q \eta^{\sL \dot{a}}_p } \\[-1ex]
& + D^{\sL\sL}_{pq}C^{\sL\sL}_{pq} \, \epsilon^{\dot{a}\dot{b}} \ket{ \chi_q^{a} Z^{\sL}_p} + E^{\sL\sL}_{pq}B^{\sL\sL}_{pq} \, \epsilon^{\dot{a}\dot{b}} \ket{ \widetilde{\chi}_q^{a} Y^{\sL}_p}.
\end{aligned}
\end{equation}
\endgroup
When we scatter a right excitation with a massless one we can write the S-matrix elements as
\begingroup
\addtolength{\jot}{1ex}
\begin{equation}\label{eq:mixed-S-matrix-R}
\begin{aligned}
\mathcal{S}^{\sR\sL}\check{\otimes}\mathcal{S}^{\sR\tilde{\sL}} \ket{Z^{\sR}_p T^{\dot{a}a}_q} =& -D^{\sL\sR}_{pq}E^{\sL\sR}_{pq} \ket{T^{\dot{a}a}_q Z^{\sR}_p } + D^{\sL\sR}_{pq}F^{\sL\sR}_{pq} \ket{\chi^a_q \eta^{\sR \dot{a}}_p}, \\
\mathcal{S}^{\sR\sL}\check{\otimes}\mathcal{S}^{\sR\tilde{\sL}} \ket{Y^{\sR}_p T^{\dot{a}a}_q} =& + A^{\sL\sR}_{pq}C^{\sL\sR}_{pq} \ket{T^{\dot{a}a}_q Y^{\sR}_p } -B^{\sL\sR}_{pq}C^{\sL\sR}_{pq}  \ket{ \widetilde{\chi}^a_q \eta^{\sR \dot{a}}_p }, \\
\mathcal{S}^{\sR\sL}\check{\otimes}\mathcal{S}^{\sR\tilde{\sL}} \ket{\eta^{\sR \dot{a}}_p \chi^a_q} =& -D^{\sL\sR}_{pq}A^{\sL\sR}_{pq} \ket{\chi^a_q \eta^{\sR \dot{a}}_p } + D^{\sL\sR}_{pq}B^{\sL\sR}_{pq} \ket{ T^{\dot{a}a}_q Z^{\sR}_p }, \\
\mathcal{S}^{\sR\sL}\check{\otimes}\mathcal{S}^{\sR\tilde{\sL}} \ket{\eta^{\sR \dot{a}}_p \widetilde{\chi}^a_q} =& + E^{\sL\sR}_{pq}C^{\sL\sR}_{pq} \ket{\widetilde{\chi}^a_q \eta^{\sR \dot{a}}_p } - F^{\sL\sR}_{pq}C^{\sL\sR}_{pq} \ket{ T^{\dot{a}a}_q Y^{\sR}_p }, \\
\mathcal{S}^{\sR\sL}\check{\otimes}\mathcal{S}^{\sR\tilde{\sL}} \ket{Z^{\sR}_p \chi^a_q} =& + D^{\sL\sR}_{pq}D^{\sL\sR}_{pq} \ket{\chi^a_q Z^{\sR}_p }, \\
\mathcal{S}^{\sR\sL}\check{\otimes}\mathcal{S}^{\sR\tilde{\sL}} \ket{Y^{\sR}_p \widetilde{\chi}^a_q} =& + C^{\sL\sR}_{pq}C^{\sL\sR}_{pq} \ket{\widetilde{\chi}^a_q Y^{\sR}_p }, \\
\mathcal{S}^{\sR\sL}\check{\otimes}\mathcal{S}^{\sR\tilde{\sL}} \ket{Z^{\sR}_p \widetilde{\chi}^a_q} =& + E^{\sL\sR}_{pq}E^{\sL\sR}_{pq} \ket{\widetilde{\chi}^a_q Z^{\sR}_p }  -F^{\sL\sR}_{pq}F^{\sL\sR}_{pq} \ket{\chi^a_q Y^{\sR}_p } +F^{\sL\sR}_{pq}E^{\sL\sR}_{pq}  \, \epsilon_{\dot{a}\dot{b}} \ket{ T^{\dot{a}a}_q \eta^{\sR \dot{b}}_p}, \\
\mathcal{S}^{\sR\sL}\check{\otimes}\mathcal{S}^{\sR\tilde{\sL}} \ket{Y^{\sR}_p \chi^a_q} =& + A^{\sL\sR}_{pq}A^{\sL\sR}_{pq} \ket{ \chi^a_q Y^{\sR}_p } - B^{\sL\sR}_{pq}B^{\sL\sR}_{pq}\ket{\widetilde{\chi}^a_q Z^{\sR}_p } - B^{\sL\sR}_{pq}A^{\sL\sR}_{pq} \, \epsilon_{\dot{a}\dot{b}} \ket{ T^{\dot{a}a}_q \eta^{\sR \dot{b}}_p}, \\
\mathcal{S}^{\sR\sL}\check{\otimes}\mathcal{S}^{\sR\tilde{\sL}} \ket{ \eta^{\sR \dot{a}}_p T^{\dot{b}a}_q} =& + B^{\sL\sR}_{pq}F^{\sL\sR}_{pq} \ket{ T^{\dot{a}a}_q \eta^{\sR \dot{b}}_p } - A^{\sL\sR}_{pq}E^{\sL\sR}_{pq} \ket{ T^{\dot{b}a}_q \eta^{\sR \dot{a}}_p } \\[-1ex]
& - B^{\sL\sR}_{pq}E^{\sL\sR}_{pq} \, \epsilon^{\dot{a}\dot{b}} \ket{ \widetilde{\chi}^{a}_q Z^{\sR}_p} + A^{\sL\sR}_{pq}F^{\sL\sR}_{pq} \, \epsilon^{\dot{a}\dot{b}} \ket{ \chi^{a}_q Y^{\sR}_p}.
\end{aligned}
\end{equation}
\endgroup
After taking into account a proper normalisation like in Section~\ref{sec:smat-tensor-prod}, the S-matrix elements for left-massless and right-massless scattering can be related by LR symmetry. In order to do so, one needs to implement it on massive and massless excitations as in equations~\eqref{eq:LR-massive} and~\eqref{eq:LR-massless}.

\subsection{The massless sector}
We write the non-vanishing entries of the two-particle S~matrix in the massless sector. First we focus on the structure fixed by the $\alg{psu}(1|1)^4$ invariance. For this reason we omit the indices corresponding to $\alg{su}(2)_{\circ}$.
\begingroup
\addtolength{\jot}{1ex}
\begin{equation}\label{eq:massless-S-matrix}
\begin{aligned}
\mathcal{S}^{\sL\sL}\check{\otimes}\mathcal{S}^{\tilde{\sL}\tilde{\sL}} \ket{T^{\dot{a}\ }_p T^{\dot{b}\ }_q} =& -C^{\sL\sL}_{pq} E^{\sL\sL}_{pq} \ket{T^{\dot{a}\ }_q T^{\dot{b}\ }_p} +B^{\sL\sL}_{pq}D^{\sL\sL}_{pq} \ket{T^{\dot{b}\ }_q T^{\dot{a}\ }_p} \\[-1ex]
& +\epsilon^{\dot{a}\dot{b}} \left(C^{\sL\sL}_{pq} D^{\sL\sL}_{pq} \ket{\chi^{\ }_q \widetilde{\chi}^{\ }_p} + B^{\sL\sL}_{pq}E^{\sL\sL}_{pq}  \ket{\widetilde{\chi}^{\ }_q \chi^{\ }_p}\right), \\
\mathcal{S}^{\sL\sL}\check{\otimes}\mathcal{S}^{\tilde{\sL}\tilde{\sL}} \ket{T^{\dot{a}\ }_p \widetilde{\chi}^{\ }_q} =&-B^{\sL\sL}_{pq}F^{\sL\sL}_{pq} \ket{\widetilde{\chi}^{\ }_q T^{\dot{a}\ }_p} - C^{\sL\sL}_{pq}F^{\sL\sL}_{pq} \ket{T^{\dot{a}\ }_q \widetilde{\chi}^{\ }_p}, \\
\mathcal{S}^{\sL\sL}\check{\otimes}\mathcal{S}^{\tilde{\sL}\tilde{\sL}} \ket{\widetilde{\chi}^{\ }_p T^{\dot{a}\ }_q} =&-F^{\sL\sL}_{pq}D^{\sL\sL}_{pq}\ket{T^{\dot{a}\ }_q \widetilde{\chi}^{\ }_p} - F^{\sL\sL}_{pq}E^{\sL\sL}_{pq}\ket{\widetilde{\chi}^{\ }_q T^{\dot{a}\ }_p}, \\
\mathcal{S}^{\sL\sL}\check{\otimes}\mathcal{S}^{\tilde{\sL}\tilde{\sL}} \ket{T^{\dot{a}\ }_p \chi^{\ }_q} =&-B^{\sL\sL}_{pq}F^{\sL\sL}_{pq} \ket{\chi^{\ }_q T^{\dot{a}\ }_p} - C^{\sL\sL}_{pq}F^{\sL\sL}_{pq} \ket{T^{\dot{a}\ }_q \chi^{\ }_p}, \\
\mathcal{S}^{\sL\sL}\check{\otimes}\mathcal{S}^{\tilde{\sL}\tilde{\sL}} \ket{\chi^{\ }_p T^{\dot{a}\ }_q} =&-F^{\sL\sL}_{pq}D^{\sL\sL}_{pq}\ket{T^{\dot{a}\ }_q \chi^{\ }_p} - F^{\sL\sL}_{pq}E^{\sL\sL}_{pq}\ket{\chi^{\ }_q T^{\dot{a}\ }_p}, \\
\mathcal{S}^{\sL\sL}\check{\otimes}\mathcal{S}^{\tilde{\sL}\tilde{\sL}} \ket{\widetilde{\chi}^{\ }_p \widetilde{\chi}^{\ }_q} =& -A^{\sL\sL}_{pq}A^{\sL\sL}_{pq} \ket{\widetilde{\chi}^{\ }_q \widetilde{\chi}^{\ }_p}, \\
\mathcal{S}^{\sL\sL}\check{\otimes}\mathcal{S}^{\tilde{\sL}\tilde{\sL}} \ket{\chi^{\ }_p \chi^{\ }_q} =& -A^{\sL\sL}_{pq}A^{\sL\sL}_{pq} \ket{\chi^{\ }_q \chi^{\ }_p}, \\
\mathcal{S}^{\sL\sL}\check{\otimes}\mathcal{S}^{\tilde{\sL}\tilde{\sL}} \ket{\widetilde{\chi}^{\ }_p \chi^{\ }_q} =& -D^{\sL\sL}_{pq}D^{\sL\sL}_{pq}\ket{\chi^{\ }_q \widetilde{\chi}^{\ }_p} - E^{\sL\sL}_{pq}E^{\sL\sL}_{pq}\ket{\widetilde{\chi}^{\ }_q \chi^{\ }_p} - E^{\sL\sL}_{pq} D^{\sL\sL}_{pq} \epsilon_{\dot{a}\dot{b}} \ket{T^{\dot{a}\ }_q T^{\dot{b}\ }_p}, \\
\mathcal{S}^{\sL\sL}\check{\otimes}\mathcal{S}^{\tilde{\sL}\tilde{\sL}} \ket{\chi^{\ }_p \widetilde{\chi}^{\ }_q} =& -D^{\sL\sL}_{pq}D^{\sL\sL}_{pq}\ket{\widetilde{\chi}^{\ }_q \chi^{\ }_p} - E^{\sL\sL}_{pq}E^{\sL\sL}_{pq}\ket{\chi^{\ }_q \widetilde{\chi}^{\ }_p} + E^{\sL\sL}_{pq} D^{\sL\sL}_{pq} \epsilon_{\dot{a}\dot{b}} \ket{T^{\dot{a}\ }_q T^{\dot{b}\ }_p}.
\end{aligned}
\end{equation}
\endgroup
The structure fixed by the $\alg{su}(2)_{\circ}$ symmetry is as follows
\begin{equation}
\mathcal{S}_{\alg{su}(2)} \ket{\mathcal{X}^a_p \mathcal{Y}^b_q} = \frac{1}{1+\varsigma_{pq}} \left( \varsigma_{pq} \ket{\mathcal{Y'}^b_q \mathcal{X'}^a_p} + \ket{\mathcal{Y'}^a_q \mathcal{X'}^b_p}\right),
\end{equation}
where we use $\mathcal{X},\mathcal{Y},\mathcal{X'},\mathcal{Y'}$ to denote any of the excitations that appear above. The antisymmetric function $\varsigma_{pq}$ is further constrained in section~\ref{sec:unitarity-YBe}.
The full S-matrix in the massless sector is then found by combining the structures fixed by $\alg{psu}(1|1)^4_{\ce}$ and $\alg{su}(2)_{\circ}$. The preferred normalisation is found by multiplying each element by the scalar factor as in Section~\ref{sec:smat-tensor-prod}.
This S-matrix automatically satisfies the LR-symmetry, where this is implemented on massless excitations as in (\ref{eq:LR-massless}).

\section{On the solutions to the crossing equations}
\label{app:crossing-AdS3}
In this appendix we collect some useful formulae concerning the solutions to the crossing equations of the massive sector of \adsthree.
We start by proving that the expression for the difference of the phases proposed in Section~\ref{sec:dressing-factors} indeed solves the corresponding crossing equation.

\subsection{The solution for $\theta^-$}
\label{app:solving}
We start by defining the integral
\begin{equation}\label{eq:Phi-}
\begin{aligned}
\Phi^-(x,y) &=\ointc \, \frac{dw}{8\pi} \frac{\text{sign}((w-1/w)/i)}{x-w} \log{\ell^-(y,w)}  \ - x \leftrightarrow y \\
&=\left( \inturl - \intdlr \right)\frac{dw}{8\pi} \frac{1}{x-w} \log{ \ell^-(y,w)} \ - x \leftrightarrow y,
\\
\ell^-(y,w)&\equiv(y-w)\left(1-\frac{1}{yw}\right)
\end{aligned}
\end{equation}
The reader may check that the expressions above match with the ones appearing in the solution for $\chi^-$ presented in~\eqref{eq:chi-}. The statement is that $\chi^-(x,y)$ coincides with $\Phi^-(x,y)$ in the region $|x|>1,\ |y|>1$. Outside this region we have to define $\chi^-$ through a proper analytic continuation, and the two functions stop to coincide.
In particular, the first important property that distinguishes them, and that is of crucial important for the proof is that
\begin{equation}
\label{eq:phi-id}
\Phi^-(x,y)-\Phi^-(1/x,y)= 0.
\end{equation}
To prove it we rewrite $\Phi^-(x,y)$ as
\begin{equation}\label{eq:rewrite-Phi-}
\begin{aligned}
\Phi^-(x,y)&=F(x,y)-F(y,x)\,,\\
F(x,y)&=\Fup(x,y)-\Fdw(x,y)=\inturl f(w,x,y)dw - \intdlr f(w,x,y)dw\,,\\
f(w,x,y)&=\frac{1}{8\pi} \frac{1}{x-w} \log\ell^-(y,w)\,.
\end{aligned}
\end{equation}
Because of the anti-symmetrisation of $x$ and $y$, we first focus on the second entry of the function $F(y,x)$. 
Using $f(w,y,x)-f(w,y,1/x)=0$, we can also show
\begin{equation}
\Fup(y,x)-\Fup(y,1/x)=0\,,\qquad \mbox{and}\qquad
\Fdw(y,x)-\Fdw(y,1/x)=0\,,
\end{equation}
that yields $F(y,x)-F(y,1/x)=0$.
Looking now at the first entry of $F(x,y)$ 
\begin{align}
  \Fdw(1/x,y)&=\intdlr \frac{dw}{8\pi} \frac{1}{1/x-w} \log\ell^-(y,w) \nonumber \\
  &=\inturl \frac{du}{8\pi\,u^2} \frac{1}{1/x-1/u} \log\ell^-(y,u) \label{eq:proof1overx} \\
  &=-\inturl \frac{du}{8\pi} \frac{1}{x-u} \log\ell^-(y,u)
  -\inturl \frac{du}{8\pi} \frac{1}{u} \log\ell^-(y,u) \nonumber \\
  &=-\Fup(x,y)-\phi^-(y)\,, \nonumber
\end{align}
where we used the change of variable $u=1/w$ and we assumed that $|x|\neq1$. Sending $x\to1/x$ in the above equation we get also $\Fup(1/x,y)=-\Fdw(x,y)-\phi^-(y)$.
With all this information we can then show that also for the first entry $F(x,y)-F(1/x,y)=0$, and conclude that~\eqref{eq:phi-id} is proved.

The function $\Phi^-(x,y)$ has another important property, it has a jump discontinuity when we cross values of $|x|=1$. To prove it and calculate the amount of the discontinuity, we consider separately the functions $\Fdw(x,y),\Fup(x,y)$ that were introduced in~\eqref{eq:rewrite-Phi-} as a convenient rewriting.
If we start with $\Fdw(x,y)$, on the one hand it is clear that no discontinuity is encountered when we cross the unit cirlce $|x|=1$ from above the real line $\mathbf{Im}(x)>0$. On the other hand, crossing from below the real line $\mathbf{Im}(x)<0$ we get, using the residue theorem
\begin{equation}
\Fdw(e^{i\varphi+\epsilon},y)=\Fdw(e^{i\varphi-\epsilon},y)+\frac{i}{4}\log \ell^-(y,e^{i\varphi})+O(\epsilon),\qquad  \epsilon>0,\quad -\pi<\varphi<0\,.
\end{equation}
Studying the discontinuity of $\Fdw$ in the second entry, we find a jump both when we cross the lower half or the upper half circles\footnote{These results may be found by first studying the discountinuity of $\pa_x\Fdw(y,x)$, and then find the corresponding primitive.}
\begin{equation}
\begin{aligned}
\Fdw(y,e^{i\varphi+\epsilon})&=\Fdw(y,e^{i\varphi-\epsilon})-\frac{i}{4}\,\log\left(y-e^{i\varphi}\right)+\phi_\uparrow(y)\,,+O(\epsilon),\quad  &&-\pi<\varphi<0\,,
\\
\Fdw(y,e^{i\varphi+\epsilon})&=\Fdw(y,e^{i\varphi-\epsilon})+\frac{i}{4}\,\log\left(\frac{1}{ye^{i\varphi}}-1\right)+\phi_\downarrow(y)+O(\epsilon),\quad &&\phantom{-{}} 0<\varphi<\pi\,,
\end{aligned}
\end{equation}
where $\epsilon>0$, and $\phi_\uparrow(y), \phi_\downarrow(y)$ are functions of $y$, that will not be important for our purposes.
The discountinuities of $\Fup$ are found in the same way, and are equivalent to changing the upper and the lower half circles in the above results.

Thanks to these results, we can compute the values of the discontinuities for $\Phi^-(x,y)$ when we cross the unit circle from below or above the real line\footnote{We have omitted functions that depend on $y$ only. They are not important for us, since they do not contribute to the crossing equations.}
\begin{equation}\label{eq:disc-Phi-minus}
\begin{aligned}
\Phi^-(e^{i\varphi+\epsilon},y)&=\Phi^-(e^{i\varphi-\epsilon},y)-\frac{i}{2}\,\log\ell^-(y,e^{i\varphi})+O(\epsilon),\qquad   &&-\pi<\varphi<0\,,
\\
\Phi^-(e^{i\varphi+\epsilon},y)&=\Phi^-(e^{i\varphi-\epsilon},y)+\frac{i}{2}\,\log\ell^-(y,e^{i\varphi})+O(\epsilon),\qquad   &&\phantom{-{}}0<\varphi<\pi\,.
\end{aligned}
\end{equation}

All this information is what we need to construct a solution of the crossing equation for the difference of the phases. 
We define crossing as an analytic continuation from the physical region $|x|>1,|y|>1$ to the crossed region $|x|<1,|y|>1$, where the path crosses the unit circle below the real line $\mathbf{Im}(x)<0$.
Then we construct $\chi^-$ in such a way that it coincides with $\Phi^-$ in the physical region, but is continuos when we perform a crossing transformation and we go to the crossed region
\begin{equation}
\begin{aligned}
\chi^-(x,y)&\equiv\Phi^-(x,y)\,\qquad &&|x|>1,\quad |y|>1\,,
\\
\chi^-(x,y)&\equiv\Phi^-(x,y)-\frac{i}{2}\,\log\ell^-(y,x)\,\qquad &&|x|<1,\quad |y|>1\,.
\end{aligned}
\end{equation}
According to these definitions and using~\eqref{eq:phi-id} we have 
\begin{equation}
\chi^-(x,y)-\chi^-(1/x,y)=\frac{i}{2}\,\log\ell^-(y,1/x)=\frac{i}{2}\,\log\ell^-(y,x)\,,\qquad |x|>1,\quad |y|>1\,.
\end{equation}
Remembering that $\sigma^-(x^\pm,y^\pm)=\text{exp}(i \theta^-(x^\pm,y^\pm))$ and the relation between $\theta^-$ and $\chi^-$ in ~\eqref{eq:thchi} we find
\begin{equation}
\label{eq:crossdifffinal}
\frac{{\ratiosigma(x,y)}^2}{{\ratiosigma(\bar{x},y)}^2}=\text{exp}\left[-\left(\log\ell^-(x^+,y^+)+\log\ell^-(x^-,y^-)-\log\ell^-(x^+,y^-)-\log\ell^-(x^-,y^+)\right)\right]
\end{equation}
 which proves that we have constructed a solution to~\eqref{eq:cr-ratio}.

\subsection{Singularities of the dressing phases}\label{sec:sing-dressing}
We discuss possible singularities of the dressing phases~$\theta^{\bullet\bullet}(x,y)$ and~$\widetilde{\theta}^{\bullet\bullet}(x,y)$, defined in terms of $\chi^{\bullet\bullet}(x,y),\,\widetilde{\chi}^{\bullet\bullet}(x,y)$ as in~\eqref{eq:solution}. We use results concerning the analytic properties of the BES phase, that is known to be regular in the physical region~\cite{Dorey:2007xn,Arutyunov:2009kf}. We then focus on the deviations from it, and we look for logarithmic singularities that might arise for special relative values of~$x$ and~$y$ in the functions
\begin{equation}
\label{eq:PsiPM}
\Psi^\pm(x,y)=\frac{1}{2}\big(-\Phi^{\text{HL}}(x,y)\pm\Phi^-(x,y)\big)\,,
\end{equation}
that contribute to define the two phases as in~\eqref{eq:solution}.
Here $\Phi^{\text{HL}}(x,y)$ is the integral defining the HL phase in the physical region,
\begin{equation}
\Phi^{\text{HL}}(x,y)=\left( \inturl - \intdlr \right)\frac{dw}{4\pi} \frac{1}{x-w} \log\left({\frac{y-w}{y-1/w}}\right),
\end{equation}
and $\Phi^-(x,y)$ is defined in~\eqref{eq:Phi-}. Because of the above expressions, singularities might arise at~$y=x$ or~$y=1/x$, but an explicit evaluation yields
\begin{equation}
\Psi^\pm(x,y)\big|_{y=x}=0\,,\qquad\Psi^\pm(x,y)\big|_{y=1/x}=\frac{1}{4\pi}\big(4\,\text{Li}_2(x)-\text{Li}_2(x^{2})\big)\,,
\end{equation}
with $|y|>1$. This is enough to conclude that the phases have no singularity at $x=y$, where both variables are in the physical region. 

When $y=1/x$ and $|y|>1$, $x$ lies inside the unit circle, and we have to perform a proper analyitic continuation of the above functions to find the contribution to the phases in the crossed region. 
We continue the phases through the lower half-circle as in Appendix~\ref{app:solving}. The result for $\Phi^-$ may be found in~\eqref{eq:disc-Phi-minus}, while for $\Phi^{\text{HL}}(x,y)$ we get
\begin{equation}
\Phi^{\text{HL}}(e^{i\varphi+\epsilon},y)=\Phi^{\text{HL}}(e^{i\varphi-\epsilon},y)-\frac{i}{2}\,\log\left[\frac{y-e^{i\varphi}}{y-e^{-i\varphi}}\right]+O(\epsilon),\qquad  \epsilon>0,\quad -\pi<\varphi<0\,.
\end{equation}
Putting together this information, we find that there is no singularity in $\widetilde{\chi}^{\bullet\bullet}(x,y)$ for $y=1/x$ and $|y|>1$. On the other hand $\chi^{\bullet\bullet}(x,y)$ has a logarithmic singularity such that
\begin{equation}
\label{eq:chizero}
e^{2i \chi^{\bullet\bullet}(x,y)}\sim \left(y-\frac{1}{x}\right)\,,\qquad\text{for}\quad y\sim 1/x\,.
\end{equation}

\chapter{\etaadsfive}\label{app:etaAdS5}

\section{Appendix for Bosonic \etaadsfive}\label{app:bos-eta-def}
In this appendix we collect some useful results needed in Chapter~\ref{ch:qAdS5Bos} and Chapter~\ref{ch:qAdS5Fer}.
\subsection{Coset elements for the bosonic model}
A very convenient parametrisation  for a bosonic coset element is given by
\be \label{basiccoset}
\ag_{\alg{b}}=\small{\left(
\begin{array}{cc}
 \ag_{\alg{a}}
& 0
\\
0 &
 \ag_{\alg{s}}
\end{array}
\right)}\, ,\quad \ag_{\alg{a}}=\Lambda(\psi_k)\, \Xi(\z)\check\ag_{\r}(\r)\, ,\quad \ag_{\alg{s}}=\Lambda(\p_k)\, \Xi(\xi)\check\ag_{\rm r}(r)\, . \ee 
Here the matrix functions $\Lambda$, $\Xi$ and $\check\ag$ are defined as 
\be \label{eq:Lambda}
\Lambda(\vp_k)=\exp(\sum_{k=1}^3
\frac{i}{2}\vp_k h_k )\,,\quad  \Xi(\vp)=\left(
\begin{array}{cccc}
 \cos\frac{\vp}{2} & \sin\frac{\vp}{2} & 0 & 0 \\
 -\sin\frac{\vp}{2} & \cos\frac{\vp}{2} & 0 & 0 \\
 0 & 0 & \cos\frac{\vp}{2} & -\sin\frac{\vp}{2} \\
 0 & 0 & \sin\frac{\vp}{2} & \cos\frac{\vp}{2} \\
\end{array}
\right)\,, 
\ee
\be \label{checkgrho}
 \check\ag_{\rho}(\r) =
\left(
\begin{array}{cccc}
 \r_+& 0 & 0 &\r_- \\
 0 & \r_+& -\r_-& 0 \\
 0 & -\r_- & \r_+& 0 \\
\r_-& 0 & 0 & \r_+\\
\end{array}
\right) \,,\quad \r_\pm= {\sqrt{\sqrt{\rho ^2+1}\pm1} \ov\sqrt 2}\,,\ee
\be \label{checkgr}
 \check\ag_{r}(r) =
\left(
\begin{array}{cccc}
 r_+& 0 & 0 &i\, r_- \\
 0 & r_+& -i \,r_-& 0 \\
 0 & -i\,r_- & r_+& 0 \\
i\,r_-& 0 & 0 & r_+\\
\end{array}
\right) \,,\quad r_\pm= {\sqrt{1\pm\sqrt{1-r ^2}} \ov\sqrt 2}\,,\ee
where the diagonal matrices $h_i$ are given by
\be
h_1={\rm diag}(-1,1,-1,1) \,,\quad 
h_2={\rm diag}(-1,1,1,-1) \,,\quad
h_3={\rm diag}(1,1,-1,-1) \,.
\ee
The coordinates $t\equiv\psi_3\,,\,\psi_1\,,\,\psi_2\,,\, \z\,,\, \r$ and $\phi\equiv\p_3\,,\,\p_1\,,\,\p_2\,,\, \xi\,,\, r$ are the ones introduced in~\eqref{eq:sph-coord-AdS5} and~\eqref{eq:sph-coord-S5} to parameterise AdS$_5$ and S$^5$.
An alternative choice for the bosonic coset element is the one used in~\cite{Arutyunov:2009ga}. The bosonic coset element would be defined as
\be\label{eq:eucl-bos-coset-el}
\gb'=\Lambda(t,\phi)\cdot \alg{g}(\gen{X})\,,
\ee
where
\be
\Lambda(t,\phi)=\small{\left(
\begin{array}{cc}
\Lambda(t)
& 0
\\
0 &
\Lambda(\phi)
\end{array}
\right)}\,,
\ee
defined in~\eqref{eq:Lambda} and
\be
\alg{g}(\gen{X})=\small{\left(
\begin{array}{cc}
\frac{1}{\sqrt{1-\frac{z^2}{4}}}\left( \gen{1}_4 -\frac{1}{2} z_i \g_i \right)
& 0
\\
0 &
\frac{1}{\sqrt{1+\frac{y^2}{4}}}\left( \gen{1}_4 -\frac{i}{2} y_i \g_i \right)
\end{array}
\right)}\,.
\ee
The gamma matrices $\g_i$ are given in~\eqref{eq:choice-5d-gamma}.
The coordinates $t,z_i$ and $\phi,y_i$ are the ones introduced in~\eqref{eq:embed-eucl-coord-AdS5} and~\eqref{eq:embed-eucl-coord-S5} to parameterise AdS$_5$ and S$^5$.

The difference from~\cite{Arutyunov:2009ga} is that we have changed the sign in front of $z_i,y_i$.
In this way the two coset elements are related by a local transformation
\be
\gb'=\gb\cdot \alg{h}\,\qquad \alg{h}\in \so(4,1)\oplus\so(5)\,,
\ee
proving that the Lagrangian is the same and the two descriptions are equivalent.
Alternatively, one could shift the angles $\psi_i\to\psi+\pi,\ \phi_i\to\phi+\pi$ when relating the two set of coordinates.
We remind that with respect to~\cite{Arutyunov:2009ga} we have also exchanged what we call $\g_1,\g_4$.

\subsection{The operator $\op$ at bosonic order and its inverse}\label{app:bosonic-op-and-inverse}
An important property of the coset representative \eqref{basiccoset} is that the $R_\gb$ operator defined in~\eqref{eq:defin-Rg-bos} is independent of the angles $\psi_k$ and $\p_k$:
\be
R_{\ag_{\alg{b}}}(M) = R_{\check\ag}(M) \,,\quad \check\ag=\small{\left(
\begin{array}{cc}
 \check\ag_{\alg{a}}
& 0
\\
0 &
 \check\ag_{\alg{s}}
\end{array}
\right)}\, ,\quad \check\ag_{\alg{a}}=\Xi(\z)\check\ag_{\r}(\r)\, ,\quad \check\ag_{\alg{s}}=\Xi(\xi)\check\ag_{\rm r}(r)\, . \ee 
We collect the formulas for the action of $1/(1-\e R_\gb \circ d)$---where $d$ is given in~\eqref{eq:defin-op-d-dtilde}---on the projections $M^{(2)}$ and $M_{\rm odd}=M^{(1)}+M^{(3)}$ of an elment $M$ of $\su(2,2|4)$. 
The projections induced by the $\mathbb{Z}_4$ grading are defined in~\eqref{eq:def-proj-Z4-grad}.

The action on odd elements appears to be $\check\ag$-independent
\be
{1\ov 1-\e R_{\check\ag}\circ d}(M_{\rm odd}) ={\mI + \eta R\circ d\ov 1 - \eta^2}(M_{\rm odd}) \,.
\ee
This action on $M^{(2)}$ factorizes into a sum of actions on $M_{\alg{a}}$ and $M_{\alg{s}}$ where $M_{\alg{a}}$ is the upper left $4\times4$ block of $M^{(2)}$, and $M_{\alg{s}}$  is the lower right $4\times4$ block of $M^{(2)}$. One can check that the inverse operator is given by
\be
{1\ov 1-\e R_{\check\ag}\circ d}(M_{\alg{a}}) =\Big(\mI+
{\e^3f_{31}^{\alg{a}}+\e^4f_{42}^{\alg{a}}+\e^5h_{53}^{\alg{a}}\ov (1-c_{\alg{a}}\e^2)(1-d_{\alg{a}}\e^2)} +
{\e R_{\check\ag}\circ d + \e^2 R_{\check\ag}\circ d\circ R_{\check\ag}\circ d\ov 1-c_{\alg{a}}\e^2}\Big)\big(M_{\alg{a}}\big)\,,
\ee
\be
{1\ov 1-\e R_{\check\ag}\circ d}(M_{\alg{s}}) =\Big(\mI+
{\e^3f_{31}^{\alg{s}}+\e^4f_{42}^{\alg{s}}+\e^5h_{53}^{\alg{s}}\ov (1-c_{\alg{s}}\e^2)(1-d_{\alg{s}}\e^2)} +
{\e R_{\check\ag}\circ d + \e^2 R_{\check\ag}\circ d\circ R_{\check\ag}\circ d\ov 1-c_{\alg{s}}\e^2}\Big)\big(M_{\alg{s}}\big)\,.
\ee
Here 
\be
c_{\alg{a}}= \frac{4\rho^2}{\left(1-\eta ^2\right)^2}\,,\quad d_{\alg{a}}=-\frac{4\rho^4 \sin^2\z }{\left(1-\eta^2\right)^2} \,,\quad c_{\alg{s}}= -\frac{4r^2}{\left(1-\eta ^2\right)^2}\,,\quad d_{\alg{s}}=-\frac{4r^4 \sin^2\xi }{\left(1-\eta^2\right)^2} \,,
\ee
\be
f_{k,k-2}^{\alg{a}}(M_{\alg{a}}) =\Big(\big(R_{\check\ag}\circ d\big)^k - c_{\alg{a}}\big(R_{\check\ag}\circ d\big)^{k-2}\Big)(M_{\alg{a}}) \,,
\ee
\be
f_{k,k-2}^{\alg{s}}(M_{\alg{s}}) =\Big(\big(R_{\check\ag}\circ d\big)^k - c_{\alg{s}}\big(R_{\check\ag}\circ d\big)^{k-2}\Big)(M_{\alg{s}}) \,,
\ee
$d_{\alg{a}}$ and $d_{\alg{s}}$ appear in the identities
\be
f_{k+2,k}^{\alg{a}} = d_{\alg{a}} f_{k,k-2}^{\alg{a}} \,,\quad f_{k+2,k}^{\alg{s}} = d_{\alg{s}} f_{k,k-2}^{\alg{s}} \,,\quad k = 4,5,\ldots\,,
\ee
and $h_{53}^{\alg{a}}$ and $h_{53}^{\alg{s}}$ appear in 
\be
h_{53}^{\alg{a}}=f_{53}^{\alg{a}} - d_{\alg{a}} f_{31}^{\alg{a}} \,,\quad h_{53}^{\alg{s}}=f_{53}^{\alg{s}} - d_{\alg{s}} f_{31}^{\alg{s}} \,.
\ee

\subsection{On the bosonic Lagrangian}\label{app:bos-lagr-eta-def}
In Section~\ref{sec:def-bos-model} we have computed the bosonic Lagrangian using the bosonic coset element~\eqref{basiccoset}.
It is also possible to compute the deformed Lagrangian by choosing the coset representative~\eqref{eq:eucl-bos-coset-el}. 
Accordingly, for the metric pieces we obtain 
\bea
\label{adsL}
\L_{\alg{a}}^{G}&=&-\frac{g}{2}(1+\varkappa^2)^{1\ov2}\gamma^{\a\b}\Big[-G_{tt}\pa_{\a}t\pa_{\beta}t+G_{zz}\pa_{\a}z_i\pa_{\beta}z_i+G_{\alg{a}}^{(1)}z_i\pa_{\a}z_iz_j\pa_{\b}z_j+\nonumber \\
&&~~~~~~~~~~~~~~~~~~~~~~~~~~ +G_{\alg{a}}^{(2)}(z_3\pa_{\a}z_4-z_4\pa_{\a}z_3)(z_3\pa_{\b}z_4-z_4\pa_{\b}z_3)
\Big]\, ,  \\
\label{spL}
\L_{\alg{s}}^{G}&=&-\frac{g}{2}(1+\varkappa^2)^{1\ov2}\gamma^{\a\b}\Big[G_{\phi\phi}\pa_{\a}\phi\pa_{\beta}\phi+G_{yy}\pa_{\a}y_i\pa_{\beta}y_i+G_{\alg{s}}^{(1)}y_i\pa_{\a}y_iy_j\pa_{\b}y_j+\nonumber \\
&&~~~~~~~~~~~~~~~~~~~~~~~~~~ +G_{\alg{s}}^{(2)}(y_3\pa_{\a}y_4-y_4\pa_{\a}y_3)(y_3\pa_{\b}y_4-y_4\pa_{\b}y_3)
\Big]\, .
\eea
Here the coordinates $z_i$, $i=1,\ldots,4$, and $t$ parametrize the deformed AdS space, while the coordinates $y_i$, $i=1,\ldots,4$, and the angle $\phi$ parametrize the deformed five-sphere.
The components of the deformed AdS metric in (\ref{adsL}) are\footnote{Note that the coordinates $y_i$ and $z_i$ are different from the ones appearing in the quartic Lagrangian \eqref{Lquart} because the nondiagonal components of the deformed metric do not vanish. }
\bea
\begin{aligned}
\label{Gads}
\hspace{-0.5cm}
G_{tt}&=\frac{(1+z^2/4)^2}{(1-z^2/4)^2-\varkappa^2 z^2}\, , ~~~~~~~~~~~~~~~~~~~
G_{zz}=\frac{(1-z^2/4)^2}{(1-z^2/4)^4+\varkappa^2 z^2(z_3^2+z_4^2)} \, ,
 \\
G_{\alg{a}}^{(1)}&=\varkappa^2G_{tt}G_{zz}\frac{z_3^2+z_4^2+(1-z^2/4)^2}{(1-z^2/4)^2(1+z^2/4)^2}\, , ~~~~~~
G_{\alg{a}}^{(2)}=\varkappa^2 G_{zz}\frac{z^2}{(1-z^2/4)^4}
\, .
\end{aligned}
\eea
For the sphere part the corresponding expressions read
\bea
\begin{aligned}
\label{Gsphere}
\hspace{-0.5cm}
G_{\phi\phi}&=\frac{(1-y^2/4)^2}{(1+y^2/4)^2+\varkappa^2 y^2}\, , ~~~~~~~~~~~~~~~~~~~
G_{yy}=\frac{(1+y^2/4)^2}{(1+y^2/4)^4+\varkappa^2 y^2(y_3^2+y_4^2)} \, ,
 \\
G_{\alg{s}}^{(1)}&=\varkappa^2G_{\phi\phi}G_{yy}\frac{y_3^2+y_4^2-(1+y^2/4)^2}{(1-y^2/4)^2(1+y^2/4)^2}\, , ~~~~~~
G_{\alg{s}}^{(2)}=\varkappa^2 G_{yy}\frac{y^2}{(1+y^2/4)^4}
\, .
\end{aligned}
\eea
Obviously, in the limit $\varkappa\to 0$ the components $G_{\alg{a}}^{(i)}$ and $G_{\alg{s}}^{(i)}$ vanish, and one obtains the metric of the {\adsfive}, {\it c.f.} fomulae (1.145) and (1.146)
in  \cite{Arutyunov:2009ga}.
Finally, for the Wess-Zumino terms the results (up to total derivative terms which do not contribute to the action) are
\bea
\begin{aligned}
\label{WZ}
\mathscr{L}_{\alg{a}}^{B}&=2g\varkappa(1+\varkappa^2)^{1\ov2}\, \eps^{\a\b}\frac{(z_3^2+z_4^2)\pa_{\a}z_1\pa_{\b}z_2}{(1-z^2/4)^4+\varkappa^2 z^2(z_3^2+z_4^2)}\, \\
\mathscr{L}_{\alg{s}}^{B}&=-2g\varkappa(1+\varkappa^2)^{1\ov2}\, \eps^{\a\b}\frac{(y_3^2+y_4^2)\pa_{\a}y_1\pa_{\b}y_2}{(1+y^2/4)^4+\varkappa^2 y^2(y_3^2+y_4^2)}\, .
\end{aligned}
\eea 

\medskip

To find the quartic Lagrangian used for 
computing the bosonic part of the four-particle world-sheet scattering matrix, we first expand the Lagrangian \eqref{eq:bos-lagr-eta-def-Pol} up to quartic order in $\r$, $r$ and their derivatives  
\be
  \begin{aligned}
&\L_{\alg{a}} =-{g\ov2}(1+\varkappa^2)^{1\ov2}\Big( \g^{\a\b}\Big[-\pa_\a t\pa_\b t (1+(1+\varkappa ^2) \rho ^2 (1+\varkappa ^2 \rho
   ^2))+
\pa_\a \r\pa_\b \rho  (1+(\varkappa ^2-1) \rho ^2)     \\
&+
\pa_\a \psi_1\pa_\b\psi_1\rho ^2 \cos
   ^2\z+\pa_\a \psi_2\pa_\b\psi_2
  \rho ^2 \sin ^2\z+\pa_\a \z\pa_\b\z \rho ^2\Big]- \varkappa \eps^{\a\b} \rho ^4 \sin 2 \zeta\pa_\a\psi_1\pa_\b\zeta\Big)\,, \\
  \\
&\L_{\alg{s}} =-{g\ov2}(1+\varkappa^2)^{1\ov2}\Big(  \g^{\a\b}\Big[\pa_\a \p\pa_\b \p
  (1-(1+\varkappa ^2) r^2 (1-\varkappa ^2 r^2))    +\pa_\a r\pa_\b r  (1+(1-\varkappa^2)r^2)  \\
   &+
   \pa_\a \p_1\pa_\b \p_1  r^2 \cos ^2\xi +\pa_\a \p_2\pa_\b \p_2  r^2 \sin^2\xi +\pa_\a \xi\pa_\b \xi  r^2 \Big]+ \varkappa \eps^{\a\b} r^4 \sin 2 \xi \pa_\a\p_1\pa_\b\xi\Big)\, .
\end{aligned}
\ee
Further, we make a shift 
\bea
\label{shift}
\rho\to \rho-\frac{\varkappa^2}{4}\rho^3\, , ~~~~~r\to r+\frac{\varkappa^2}{4}r^3\,\eea
so that the quartic action acquires the form 
 \bea
\L_{\alg{a}} &=&-{g\ov2}(1+\varkappa^2)^{1\ov2}\, \g^{\a\b}\times\\
&&\Big[-\pa_\a t\pa_\b t \Big(1+(1+\varkappa ^2)\rho^2 +\tfrac{1}{2}\varkappa^2 (1+\varkappa ^2) \rho^4\Big)+\pa_\a \r\pa_\b \rho  \Big(1- \rho ^2-\tfrac{\varkappa^2}{2}\rho^4\Big)+ \nonumber\\
&& +\Big(\rho ^2-\tfrac{\varkappa^2}{2}\rho^4\Big) \Big(\pa_\a \psi_1\pa_\b\psi_1\cos
   ^2\z+\pa_\a \psi_2\pa_\b\psi_2
  \sin ^2\z+\pa_\a \z\pa_\b\z \Big)\Big] \nonumber  \\
&&+{g\ov2} \varkappa (1+\varkappa^2)^{1\ov2}\eps^{\a\b} \rho ^4 \sin 2 \zeta\pa_\a\psi_1\pa_\b\zeta\,, \nonumber
  \eea
\bea
\L_{\alg{s}} &=&-{g\ov2}(1+\varkappa^2)^{1\ov2}\, \g^{\a\b}\times\\
&&\Big[\pa_\a \p\pa_\b \p
 \Big(1-(1+\varkappa ^2) r^2+\tfrac{1}{2}\varkappa^2(1+\varkappa^2)r^4\Big)  +\pa_\a r\pa_\b r  \Big(1+r^2+\tfrac{\varkappa^2}{2}r^4\Big)   + \nonumber \\
&&+  \Big(r^2+ \tfrac{\varkappa^2}{2}r^4\Big)\Big(\pa_\a \p_1\pa_\b \p_1   \cos ^2\xi +\pa_\a \p_2\pa_\b \p_2  \sin^2\xi +\pa_\a \xi\pa_\b \xi  \Big)\Big] \nonumber \\
&&-{g\ov2} \varkappa(1+\varkappa^2)^{1\ov2} \eps^{\a\b} r^4 \sin 2 \xi \pa_\a\p_1\pa_\b\xi\, .\nonumber\eea
Changing the spherical coordinates to $(z_i,y_i)$ and expanding the resulting action up to the quartic order in $z$ and $y$ fields we get the quartic Lagrangian \eqref{Lquart}. Notice that the shifts of $\r$ and $r$ in \eqref{shift} were chosen so that the deformed metric expanded up to quadratic order in the fields would be diagonal.

\newpage

\section{The $\psu(2|2)_q$-invariant S-matrix}
\label{app:matrixSmatrix}
The S-matrix compatible with $\psu(2|2)_q$ symmetry \cite{Beisert:2008tw} has been studied in detail in~\cite{Hoare:2011wr,Arutyunov:2012zt,Arutyunov:2012ai,Hoare:2013ysa}.
In this Appendix we recall its explicit form following the same notation as in~\cite{Arutyunov:2012zt}.

Let $E_{ij}\equiv E_i^j$ stand for the standard matrix unities, $i,j=1,\ldots, 4$. We introduce the following definition
\begin{equation}
E_{kilj}=(-1)^{\epsilon(l)\epsilon(k)}E_{ki}\otimes E_{lj}\, ,
\end{equation}
where $\epsilon(i)$ denotes the parity of the index, equal to $0$ for $i=1,2$ (bosons) and to $1$ for $i=3,4$ (fermions). The matrices $E_{kilj}$
are convenient to write down invariants with respect to the action of copies of $\su_q(2)\subset \psu_q(2|2)$. If we introduce
\bea
\Lambda_1&=&E_{1111}+\frac{q}{2}E_{1122}+\frac{1}{2}(2-q^2)E_{1221}+\frac{1}{2}E_{2112}+\frac{q}{2}E_{2211}+E_{2222}\, ,\nonumber\\
\Lambda_2&=&\frac{1}{2}E_{1122}-\frac{q}{2}E_{1221}-\frac{1}{2q}E_{2112}+\frac{1}{2}E_{2211}\, , \nonumber \\
\Lambda_3&=&E_{3333}+\frac{q}{2}E_{3344}+\frac{1}{2}(2-q^2)E_{3443}+\frac{1}{2}E_{4334}+\frac{q}{2}E_{4433}+E_{4444} \, , \nonumber\\
\Lambda_4&=&\frac{1}{2}E_{3344}-\frac{q}{2}E_{3443}-\frac{1}{2q}E_{4334}+\frac{1}{2}E_{4433}\, , \nonumber\\
\Lambda_5&=&E_{1133}+E_{1144}+E_{2233}+E_{2244}\, ,\\
\Lambda_6&=&E_{3311}+E_{3322}+E_{4411}+E_{4422}\, , \nonumber\\
\Lambda_7&=&E_{1324}-qE_{1423}-\frac{1}{q}E_{2314}+E_{2413}\, , \nonumber\\
\Lambda_8&=&E_{3142}-qE_{3214}-\frac{1}{q}E_{4132}+E_{4231}\, , \nonumber\\
\Lambda_9&=&E_{1331}+E_{1441}+E_{2332}+E_{2442}\, , \nonumber\\
\Lambda_{10}&=&E_{3113}+E_{3223}+E_{4114}+E_{4224}\, , \nonumber
\eea
the S-matrix of the $q$-deformed model is given by
\be\la{Sqmat}
S_{12}(p_1,p_2)=\sum_{k=1}^{10}a_k(p_1,p_2)\Lambda_k\, ,
\ee
where the coefficients are
\bea
a_1&=&1\, ,  \nonumber \\
a_2&=&-q+\frac{2}{q}\frac{x^-_1(1-x^-_2x^+_1)(x^+_1-x^+_2)}{x^+_1(1-x^-_1x^-_2)(x^-_1-x^+_2)}\nonumber \\
a_3&=&\frac{U_2V_2}{U_1V_1}\frac{x^+_1-x^-_2}{x^-_1-x^+_2}\nonumber \\
a_4&=&-q\frac{U_2V_2}{U_1V_1}\frac{x^+_1-x^-_2}{x^-_1-x^+_2}+\frac{2}{q}\frac{U_2V_2}{U_1V_1}\frac{x^-_2(x^+_1-x^+_2)(1-x^-_1x^+_2)}{x^+_2(x^-_1-x^+_2)(1-x^-_1x^-_2)}\nonumber \\
a_5&=&\frac{x^+_1-x^+_2}{\sqrt{q}\, U_1V_1(x^-_1-x^+_2)} \nonumber \\
a_6&=&\frac{\sqrt{q}\, U_2V_2(x^-_1-x^-_2)}{x^-_1-x^+_2}  \\
a_7&=&\frac{ig}{2}\frac{(x^+_1-x^-_1)(x^+_1-x^+_2)(x^+_2-x^-_2)}{\sqrt{q}\, U_1V_1(x^-_1-x^+_2)(1-x_1^- x_2^-)\gamma_1\gamma_2}
\nonumber \\
a_8&=&\frac{2i}{g}\frac{U_2V_2\,  x^-_1x^-_2(x^+_1-x^+_2)\gamma_1\gamma_2}{q^{\frac{3}{2}} x^+_1x^+_2(x^-_1-x^+_2)(x^-_1x^-_2-1)}\nonumber \\
a_9&=&\frac{(x^-_1-x^+_1)\gamma_2}{(x^-_1-x^+_2)\gamma_1} \nonumber \\
\nonumber
a_{10}&=&\frac{U_2V_2 (x^-_2-x^+_2)\gamma_1}{U_1V_1(x^-_1-x^+_2)\gamma_2}\, .
\eea
Here the basic variables $x^{\pm}$ parametrizing a fundamental representation of the centrally extended superalgebra $\psu_q(2|2)$ satisfy the following constraint \cite{Beisert:2008tw}
\bea
\label{fc}
\frac{1}{q}\left(x^++\frac{1}{x^+}\right)-q\left(x^-+\frac{1}{x^-}\right)=\left(q-\frac{1}{q}\right)\left(\xi+\frac{1}{\xi}\right)\, ,
\eea
where the parameter $\xi$ is related the coupling constant $g$ as
\bea
\xi=-\frac{i}{2}\frac{g(q-q^{-1})}{\sqrt{1-\frac{g^2}{4}(q-q^{-1})^2}}\, .
\eea
The (squares of) central charges are given by
\bea
U_i^2=\frac{1}{q}\frac{x^+_i+\xi}{x^-_i+\xi}=e^{ip_i}\, , ~~~~V^2_i=q\frac{x^+_i}{x^-_i}\frac{x^-_i+\xi}{x^+_i+\xi} \, ,
\eea
and the parameters $\gamma_i$ are
\bea
\gamma_i=q^{\frac{1}{4}}\sqrt{\frac{ig}{2}(x^-_i-x^+_i)U_iV_i}\, .
\eea
The $q$-deformed dispersion relation ${\cal E}$ takes the form
\be\la{qdisp}
\Bigg(1-\frac{g^2}{4}(q-q^{-1})^2 \Bigg)\Bigg(\frac{q^{{\cal E}/2}-q^{-{\cal E}/2}}{q-1/q}\Bigg)^2-g^2\sin^2\frac{p}{2}=\Bigg(\frac{q^{1/2}-q^{-1/2}}{q-1/q}\Bigg)^2\, .
\ee
Finally, we point out that in the $q$-deformed dressing phase the variable $u$ appears which is given by 
\bea
u(x)=\frac{1}{\qdp}\log\Bigg[-\frac{x+\tfrac{1}{x}+\xi+\tfrac{1}{\xi}}{\xi-\tfrac{1}{\xi}}\Bigg]\, .
\eea

The log of the $q$-deformed Gamma function admits an integral representation valid in the strip $-1< \mathbf{Re}( x)<k$ (with $k>1$)~\cite{Hoare:2011wr}
\be\label{lnGovG}
\begin{aligned}
\log\Gamma_{q^2}(1+x)&=\frac{i\pi x(x-1)}{2k}\\ &\qquad+\int_0^\infty\frac{dt}t\,\frac{e^{-tx}-e^{(x-k+1)t}
-x(e^{-t}-1)(1+e^{(2-k)t})+e^{(1-k)t}-1}{(e^{t}-1)(1-e^{-kt})}\,,
\end{aligned}
\ee
where $q=e^{i\pi/k}$. Seding $k\to \infty$ one recovers the integral representation for the conventional Gamma function.

Writing $k=-i\pi g/\nu$, keeping $\nu$ and $x$ fixed and sending $g\to\infty$, we find that  at leading order
\bea\la{qGamma1}
\log\frac{\Gamma_{q^2}(1+g x)}{\Gamma_{q^2}(1-gx)} &\approx& g\Big(-2 x+2 x \log (g)+x \big(\log (-x)+  \log
   (x)\big)\Big)\\\nonumber
&+&   g{2\pi\ov i \nu} \big(\psi ^{(-2)}(1-{i\nu x\ov\pi} )-\psi^{(-2)}(1+ {i\nu x\ov\pi} )\big)\,,
\eea
where $\psi ^{(-2)}\left(z\right)$ is given by
\be
\psi ^{(-2)}\left(z\right) = \int_0^z\, dt\,  \text{log$\Gamma
   $}\left(t\right)\,.
\ee
A derivation of this formula may be found in Appendix C of~\cite{Arutyunov:2013ega}

\section{Clifford algebra and $\psu(2,2|4)$}\label{app:su224-algebra}
Our preferred basis of $4\times 4$ gamma-matrices is\footnote{Here it was useful to exchange the definition of $\gamma_1, \gamma_4$ from the one of~\cite{Arutyunov:2009ga}.}
\be\label{eq:choice-5d-gamma}
\newcommand{\0}{\color{black!40}0}
\begin{aligned}
\gamma_0  &=
\left(
\begin{array}{cccc}
 \phantom{-}1\phantom{-} & \0 & \0 & \0 \\
 \0 &  \phantom{-}1\phantom{-} & \0 & \0 \\
 \0 & \0 & -1\phantom{-} & \0 \\
 \0 & \0 & \0 & -1\phantom{-} \\
\end{array}
\right) , \\
\gamma_1 &=
\left(
\begin{array}{cccc}
 \0 & \0 & -i\phantom{-} & \0 \\
 \0 & \0 & \0 &  \phantom{-}i\phantom{-} \\
  \phantom{-}i\phantom{-} & \0 & \0 & \0 \\
 \0 & -i\phantom{-} & \0 & \0 \\
\end{array}
\right), \qquad
&\gamma_2 =
\left(
\begin{array}{cccc}
 \0 & \0 & \0 &  \phantom{-}i\phantom{-} \\
 \0 & \0 &  \phantom{-}i\phantom{-} & \0 \\
 \0 & -i\phantom{-} & \0 & \0 \\
 -i\phantom{-} & \0 & \0 & \0 \\
\end{array}
\right), \\
\gamma_3 &=
\left(
\begin{array}{cccc}
 \0 & \0 &  \phantom{-}1\phantom{-} & \0 \\
 \0 & \0 & \0 & \phantom{-}1\phantom{-} \\
  \phantom{-}1\phantom{-} & \0 & \0 & \0 \\
 \0 & \phantom{-}1\phantom{-} & \0 & \0 \\
\end{array}
\right), \qquad
&\gamma_4 =
\left(
\begin{array}{cccc}
 \0 & \0 & \0 & -1\phantom{-} \\
 \0 & \0 &  \phantom{-}1\phantom{-} & \0 \\
 \0 &  \phantom{-}1\phantom{-} & \0 & \0 \\
 -1\phantom{-} & \0 & \0 & \0 \\
\end{array}
\right).
\end{aligned}
\ee
Matrices for AdS$_5$ and S$^5$ in terms of the above gamma-matrices have been defined in~\eqref{eq:gamma-AdS5-S5}.
When we need to write explicitly the matrix indices we use underlined Greek letters for AdS$_5$ ${(\check{\gamma}_m)_{\ul{\a}} }^{\ul{\nu}}$, and underlined Latin letters for S$^5$ ${(\hat{\gamma}_m)_{\ul{a}}}^{\ul{b}}$.
It is useful to consider the matrices 
\be\label{eq:SKC-gm}
\begin{aligned}
\Sigma = 
\left(
\begin{array}{cccc}
 1 & 0 & 0 & 0 \\
 0 &  1 & 0 & 0 \\
 0 & 0 & -1 & 0 \\
 0 & 0 & 0 & -1 \\
\end{array}
\right), \quad
K = 
\left(
\begin{array}{cccc}
 0 & -1 & 0 & 0 \\
 1 & 0 & 0 &  0 \\
 0 & 0 & 0 & -1 \\
 0 & 0 & 1 & 0 \\
\end{array}
\right), \quad
C = 
\left(
\begin{array}{cccc}
  0 & -1 & 0 & 0 \\
 1 & 0 & 0 &  0 \\
 0 & 0 & 0 & 1 \\
 0 & 0 & -1 & 0 \\
\end{array}
\right),
\end{aligned}
\ee
that are defined with upper indices $\Sigma^{\ul{a}\ul{b}},K^{\ul{a}\ul{b}},C^{\ul{a}\ul{b}}$.
Their inverse matrices are then defined with lower indices. They transform the gamma matrices in the following way
\be
\gamma_m^t = K \gamma_m K^{-1},
\ee
\be
\begin{aligned}
\gamma_m^t & = -C \gamma_m C^{-1}, \quad m=1,...,4, \qquad
\gamma_0^t & = C \gamma_0 C^{-1}, \\
\gamma_m^\dagger & = -\Sigma \gamma_m \Sigma^{-1}, \quad m=1,...,4, \qquad
\gamma_0^\dagger & = \Sigma \gamma_0 \Sigma^{-1}.
\end{aligned}
\ee
The matrix $K$---and not $C$---is the charge conjugation matrix for our Clifford algebra. We choose to follow the same notation of~\cite{Arutyunov:2009ga}.
From the last equation one then has $\check{\gamma}_m^\dagger = -\Sigma \check{\gamma}_m \Sigma^{-1}, \quad m=0,...,4$.
For raising and lowering spinor indices we follow the conventions of~\cite{Freedman:2012zz}
\be
\lambda^\a = K^{\a\b} \lambda_\b,
\qquad
\lambda_\a = \lambda^\b K_{\b\a},
\ee
where $K^{\a\b}$ are the components of the matrix $K$, that plays the role of charge conjugation matrix for the Clifford algebra.
We also have
\be
K^{\a\b} K_{\g\b} = \delta^\a_\g,
\qquad
K_{\b\a} K^{\b\g} = \delta_\a^\g,
\qquad
\chi^\a \lambda_\a = - \chi_\a \lambda^\a .
\ee
The five-dimensional gamma matrices satisfy the symmetry properties 
\be\label{eq:symm-prop-5dim-gamma}
\begin{aligned}
(K\gamma^{(r)})^t &= - t_r^\g \ K\gamma^{(r)}\,,
\\
K(\gamma^{(r)})^t K&= - t_r^\g \ \gamma^{(r)}\,,
\qquad
t_0^\g=t_1^\g=+1,\quad  t_2^\g=t_3^\g=-1\,.
\end{aligned}
\ee
Here $\gamma^{(r)}$ denotes the antisymmetrised product of $r$ gamma matrices and the coefficients $t_r^\g$ are the same for AdS and the sphere---we label them with ${}^\g$ to distinguish them from the coefficients of ten-dimensional Gamma matrices.
For the rules concerning Hermitian conjugation we find
\be\label{eq:herm-conj-prop-5dim-gamma}
\begin{aligned}
&\check{\g}_m^\dagger \phantom{{}_n}=+\check{\g}^0\check{\g}_m\check{\g}^0\,,
\qquad
&&\hat{\g}_m^\dagger \phantom{{}_n}=+\hat{\g}_m\,,
\\
&\check{\g}_{mn}^\dagger=+\check{\g}^0\check{\g}_{mn}\check{\g}^0\,,
\qquad
&&\hat{\g}_{mn}^\dagger=-\hat{\g}_{mn}\,,
\end{aligned}
\ee
With these ruels we find useful formulas to take the bar of some expressions
\be
\begin{aligned}
& ((\check{\gamma}_m \otimes \mathbf{1}_4) \theta_I)^\dagger (\check{\gamma}^0 \otimes \mathbf{1}_4) = - \bar{\theta}_I (\check{\gamma}_m \otimes \mathbf{1}_4)  ,
\\
& ((\mathbf{1}_4 \otimes \hat{\gamma}_m) \theta_I)^\dagger (\check{\gamma}^0\otimes \mathbf{1}_4) = + \bar{\theta}_I (\mathbf{1}_4 \otimes \hat{\gamma}_m ) ,
\end{aligned}
\ee
\be
\begin{aligned}
& ((\check{\gamma}_{mn}\otimes \mathbf{1}_4) \theta_I)^\dagger (\check{\gamma}^0\otimes \mathbf{1}_4) = - \bar{\theta}_I (\check{\gamma}_{mn} \otimes \mathbf{1}_4) ,
\\
& ((\mathbf{1}_4 \otimes \hat{\gamma}_{mn}) \theta_I)^\dagger (\check{\gamma}^0\otimes \mathbf{1}_4) = - \bar{\theta}_I (\mathbf{1}_4 \otimes \hat{\gamma}_{mn})  ,
\end{aligned}
\ee
Thanks to~\eqref{eq:symm-prop-5dim-gamma} one can also show that given two Grassmann bi-spinors $\psi_{\ul{\a}\ul{a}},\chi_{\ul{\a}\ul{a}}$ the ``Majorana-flip'' relations are
\be\label{eq:symm-gamma-otimes-gamma}
\bar{\chi} \left(\check{\gamma}^{(r)}\otimes \hat{\gamma}^{(s)} \right) \psi = - t_r^\g t_s^\g \ \bar{\psi} \left(\check{\gamma}^{(r)}\otimes \hat{\gamma}^{(s)} \right) \chi.
\ee
%
Knowing this, it is easy to prove 
\be\label{eq:Maj-flip}
s^{IJ }\bar{\theta}_I \left(\check{\gamma}^{(r)}\otimes \hat{\gamma}^{(s)} \right) \theta_J=0
\qquad \text{ if }
\left\{\begin{array}{ccc} 
s^{IJ} = + s^{JI}& \text{ and } & t_r^\g t_s^\g=+1 \\
s^{IJ} = - s^{JI}& \text{ and } & t_r^\g t_s^\g=-1 \\
 \end{array} \right.\,.
\ee
To conclude we also have
\be
\bar{\psi} \mathcal{D} \lambda =  \bar{\lambda} \mathcal{D} \psi,
\qquad
\bar{\psi}_I D^{IJ} \lambda_J =  \bar{\lambda}_J D^{JI} \psi_I.
\ee
up to a total derivative.

\medskip

Before multiplying the generators by the fermions $\theta$, the commutators between odd and even elements with explicit spinor indices read as
\be
\begin{aligned}
& [\genQind{I}{\a a}{}, \check{\gen{P}}_m] = - \frac{i}{2} \epsilon^{IJ}  \ \genQind{J}{\nu a}{}\ {(\check{\gamma}_m)_{\ul{\nu}}}^{\ul{\a}}, & \qquad
& [\genQind{I}{\a a}{}, \hat{\gen{P}}_m] =  \frac{1}{2} \epsilon^{IJ}  \ \genQind{J}{\a b}{}\ {(\hat{\gamma}_m)_{\ul{b}}}^{\ul{a}}, & \\
& [\genQind{I}{\a a}{}, \check{\gen{J}}_{mn}] = - \frac{1}{2} \delta^{IJ} \ \genQind{J}{\nu a}{}\  {(\check{\gamma}_{mn})_{\ul{\nu}}}^{\ul{\a}}, & \qquad
& [\genQind{I}{\a a}{}, \hat{\gen{J}}_{mn}] =  -\frac{1}{2} \delta^{IJ}  \ \genQind{J}{\a b}{}\ {(\hat{\gamma}_{mn})_{\ul{b}}}^{\ul{a}}, & 
\end{aligned}
\ee
The anti-commutator of two supercharges gives
\be
\begin{aligned}
\{\genQind{I}{\a a}{}, \genQind{J}{\nu b}{}\} =& \delta^{IJ} \left( i\, K^{\ul{\a}\ul{\lambda}} K^{\ul{a}\ul{b}} \ {(\check{\gamma}^m)_{\ul{\lambda}}}^{\ul{\nu}} \, \check{\gen{P}}_m - \, K^{\ul{\a}\ul{\nu}} \, K^{\ul{a}\ul{c}} {(\hat{\gamma}^m)_{\ul{c}}}^{\ul{b}}  \, \hat{\gen{P}}_m -\frac{i}{2} K^{\ul{\a}\ul{\nu}} K^{\ul{a}\ul{b}} \mathbf{1}_8 \right) \\
- &  \frac{1}{2} \epsilon^{IJ} \left( K^{\ul{\a}\ul{\lambda}} K^{\ul{a}\ul{b}} \ {(\check{\gamma}^{mn})_{\ul{\lambda}}}^{\ul{\nu}} \, \check{\gen{J}}_{mn}  - \, K^{\ul{\a}\ul{\nu}} \, K^{\ul{a}\ul{c}} {(\hat{\gamma}^{mn})_{\ul{c}}}^{\ul{b}}  \, \hat{\gen{J}}_{mn} \right),
\end{aligned}
\ee
where the indices $m,n$ are raised with the metric $\eta_{mn}$. For completeness we have written also the term proportional to the identity, since the supermatrices are a realisation of $\alg{su}(2,2|4)$. To obtain $\alg{psu}(2,2|4)$ one just needs to drop the term proportional to $i\mathbf{1}_8$ in the r.h.s. of the anti-commutator.
Similarly, the supertrace of the product of two odd elements read as
\be
\Str[\genQind{I}{\a a}{}\genQind{J}{\nu b}{}] = -2 \epsilon^{IJ} K^{\ul{\a}\ul{\nu}} K^{\ul{a}\ul{b}}
\ee
Remembering that the spinor indices are raised and lowered with the matrix $K$, the last equation can be written also as $\Str[\genQind{I}{\a a}{}\genQind{J}{}{\nu b}] = -2\epsilon^{IJ} \delta^{\ul{\a}}_{\ul{\nu}} \delta^{\ul{a}}_{\ul{b}}$ .

\newpage

\section{Action of $R_\gb$ on bosonic elements}\label{sec:useful-results-eta-def}
The coefficients $\lambda$ introduced in Eq.~\eqref{eq:Rgb-action-lambda}---corresponding to the action of the operator $R_{\gb}$ on bosonic generators---are explicitly
\be\label{eq:lambda11}
\lambda_0^{\ 4} = \lambda_4^{\ 0} = \rho, \qquad \lambda_2^{\ 3} = - \lambda_3^{\ 2} = - \rho^2 \sin \zeta, \qquad 
\lambda_5^{\ 9} =- \lambda_9^{\ 5} = r, \qquad \lambda_7^{\ 8} = - \lambda_8^{\ 7} =  r^2 \sin \xi,
\ee
\be\label{eq:lambda12}
\begin{aligned}
& \lambda_1^{01} =\lambda_2^{02} =\lambda_3^{03} =\lambda_4^{04} = \sqrt{1+\rho^2}, \qquad 
&&  \lambda_6^{56} =\lambda_7^{57} =\lambda_8^{58} =\lambda_9^{59} = -\sqrt{1-r^2}, \\
& \lambda_1^{12} =- \lambda_3^{23}= -\rho \cos \zeta,
&& \lambda_6^{67} =- \lambda_8^{78}= r \cos \xi,\\
& \lambda_2^{34} = - \lambda_3^{24}= -\rho \sqrt{1+\rho^2} \sin \zeta, 
&&  \lambda_7^{89} = - \lambda_8^{79}= r \sqrt{1-r^2} \sin \xi, 
\end{aligned}
\ee
\be\label{eq:lambda21}
\begin{aligned}
& \lambda_{01}^1 = \lambda_{02}^2 = \lambda_{03}^3 = \lambda_{04}^4 = -\sqrt{1+\rho^2}, \qquad
&& \lambda_{56}^6 = \lambda_{57}^7 = \lambda_{58}^8 = \lambda_{59}^9 = \sqrt{1-r^2},\\
& \lambda_{12}^1 = - \lambda_{23}^3 = -\rho \cos \zeta,
&& \lambda_{67}^6 = - \lambda_{78}^8 = -r \cos \xi,\\
& \lambda_{24}^3 = - \lambda_{34}^2 = \rho \sqrt{1+\rho^2} \sin \zeta, 
&& \lambda_{79}^8 = - \lambda_{89}^7 = r \sqrt{1-r^2} \sin \xi,
\end{aligned}
\ee
\be\label{eq:lambda22a}
\begin{aligned}
& \lambda_{01}^{14} = \lambda_{02}^{24} = \lambda_{03}^{34} = \lambda^{01}_{14} = \lambda^{02}_{24} = \lambda^{03}_{34} = -\rho, \qquad
& \lambda_{12}^{13} = - \lambda_{13}^{12} = \sin \zeta, \qquad \\
& \lambda_{12}^{14} = -\lambda_{14}^{12}= -\lambda_{23}^{34}= \lambda_{34}^{23}= - \sqrt{1+ \rho^2} \cos \zeta, \qquad 
& \lambda_{24}^{34} = - \lambda_{34}^{24} = (1+\rho^2) \sin \zeta
\end{aligned}
\ee
\be\label{eq:lambda22s}
\begin{aligned}
& \lambda_{56}^{69} = \lambda_{57}^{79} = \lambda_{58}^{89} = -\lambda^{56}_{69} = -\lambda^{57}_{79} = -\lambda^{58}_{89} = -r, \qquad
& \lambda_{67}^{68} = - \lambda_{68}^{67} = \sin \xi, \qquad \\
& \lambda_{67}^{69} = -\lambda_{69}^{67}= -\lambda_{78}^{89}= \lambda_{89}^{78}= - \sqrt{1-r^2} \cos \xi, \qquad 
& \lambda_{79}^{89} = - \lambda_{89}^{79} = (1-r^2) \sin \xi
\end{aligned}
\ee
They satisfy the properties
\be\label{eq:app-swap-lambda}
{\lambda_m}^n = - \, \eta_{mm'} \eta^{nn'} {\lambda_{n'}}^{m'},
\qquad
\check{\lambda}_m^{np} = \eta_{mm'} \eta^{nn'} \eta^{pp'} \check{\lambda}^{m'}_{n'p'},
\qquad
\hat{\lambda}_m^{np} = -\, \eta_{mm'} \eta^{nn'} \eta^{pp'} \hat{\lambda}^{m'}_{n'p'},
\ee
that are used to simplify some terms in the Lagrangian.

\section{The contribution $\{101\}$ to the fermionic Lagrangian}\label{app:der-Lagr-101}
In this Appendix we show how to write $\lagr_{\{101\}}$ in the form presented in~\eqref{eq:Lagr-101}.
It is easy to see that the insertion of $\opinv_{(0)}$ between two odd currents does not change the fact that the expression is anti-symmetric in $\a,\b$ and we have
\be\label{eq:orig-lagr-101}
\lagr_{\{101\}} = - \frac{\tilde{g}}{2} \epsilon^{\a\b} \left( -\sigma_1^{IK} + \frac{\vk}{1+\sqrt{1+\vk^2}} \delta^{IK} \right) (D^{IJ}_\a \theta_J)^\dagger \bg^0  D^{KL}_\b \theta_L .
\ee
The above contribution contains terms quadratic in $\pa \theta$, a feature that does no match with the generic type IIB action~\eqref{eq:IIB-action-theta2}. These terms remain even when sending the deformation parameter to $0$.
This is not a problem since these terms are of the form $\epsilon^{\a\b} s^{IK} \pa_\a \bar{\theta}_I \pa_\b \theta_K$, where $s^{IK}$ is a generic tensor symmetric in the two indices. Although not vanishing, they can be rewritten and traded for a total derivative $\epsilon^{\a\b} s^{IK} \pa_\a \bar{\theta}_I \pa_\b \theta_K = \pa_\a(\epsilon^{\a\b} s^{IK}  \bar{\theta}_I \pa_\b \theta_K)$, using $\epsilon^{\a\b}\pa_\a\pa_\b=0$. The unwanted terms then do not contribute to the action.
First we note that
\be
(D^{IJ}_\a \theta_J)^\dagger \bg^0 = \delta^{IJ} \left(\pa_\a \bar{\theta}_J + \frac{1}{4} \bar{\theta}_J \omega^{mn}_\a \bg_{mn}  \right) + \frac{i}{2} \epsilon^{IJ} \bar{\theta}_J e^m_\a \bg_m ,
\ee
and using this we show that the contribution to the Lagrangian is
\be
\begin{aligned}
\lagr_{\{101\}} &= - \frac{\tilde{g}}{2}  \epsilon^{\a\b} \left( \sigma_1^{IK}- \frac{\vk}{1+\sqrt{1+\vk^2}} \delta^{IK} \right) \bar{\theta}_J D^{JI}_\a D^{KL}_\b \theta_L \\
&+\pa_\a \left(\frac{\tilde{g}}{2}  \epsilon^{\a\b} \left( \sigma_1^{IK}- \frac{\vk}{1+\sqrt{1+\vk^2}} \delta^{IK} \right) \bar{\theta}_J  D^{KL}_\b \theta_L \right).
\end{aligned}
\ee
The last term is the total derivative that we discard.
The result is only na\"ively quadratic in $D^{IJ}$.
To show it we divide the computation into three terms
\be
\epsilon^{\a\b} s^{IK} \bar{\theta}_L D^{LK}_\a D^{IJ}_\b \theta_J= \text{WZ}_1 + \text{WZ}_2 + \text{WZ}_3
\ee
where the object $s^{IK}$ is introduced to keep the computation as general as possible. We will only assume that it is symmetric in the indices $IK$.
For each of the terms we then get
\be
\begin{aligned}
\text{WZ}_1 & \equiv \epsilon^{\a\b} s^{IK} \bar{\theta}_L \mathcal{D}^{LK}_\a \mathcal{D}^{IJ}_\b \theta_J  \\
& = -\frac{1}{4} \epsilon^{\a\b} s^{JL}  \bar{\theta}_L e^m_\a e^n_\b \bg_{m} \bg_{n}  \theta_J , \\
\text{WZ}_2 & \equiv \frac{i}{2} \epsilon^{\a\b} s^{IK} \bar{\theta}_L  \left( \epsilon^{IJ} \mathcal{D}^{LK}_\a (e^n_\b \bg_n  \theta_J) + \epsilon^{LK} e^m_\a \bg_m  \mathcal{D}^{IJ}_\b \theta_J \right) \\
& = + i \epsilon^{\a\b} s^{IK}  \epsilon^{JI} \bar{\theta}_J e^m_\a \bg_m  \mathcal{D}^{KL}_\b \theta_L, \\
\text{WZ}_3 & \equiv -\frac{1}{4} \epsilon^{\a\b} s^{IK}  \epsilon^{LK}  \epsilon^{IJ} e^m_\a e^n_\b \bar{\theta}_L \bg_m  \bg_n  \theta_J ,
\end{aligned}
\ee
where we used~\eqref{eq:d-veilbein},\eqref{eq:d-spin-conn} and the fact that the covariant derivative $\mathcal{D}$ on the vielbein is zero
\be
\epsilon^{\a\b} \mathcal{D}^{IJ}_\a (e^m_\b \bg_m  \theta ) = \epsilon^{\a\b} e^m_\b \bg_m  \mathcal{D}^{IJ}_\a  \theta .
\ee
The final result for the deformed case is
\be
\begin{aligned}
\lagr_{\{101\}} &= - \frac{\tilde{g}}{2}  \epsilon^{\a\b} \bar{\theta}_L \, i \, e^m_\a \bg_m \left( \sigma_3^{LK}  D^{KJ}_\b \theta_J
-\frac{\vk}{1+\sqrt{1+\vk^2}} \ \epsilon^{LK}  \mathcal{D}^{KJ}_\b \theta_J \right) \\
&= - \frac{\tilde{g}}{2}  \epsilon^{\a\b} \bar{\theta}_I \left( \sigma_3^{IJ}  
 -\frac{\vk}{1+\sqrt{1+\vk^2}} \ \epsilon^{IJ}  \right) \, i \, e^m_\a \bg_m \mathcal{D}_\b \theta_J  + \frac{\tilde{g}}{4}  \epsilon^{\a\b} \bar{\theta}_I  \sigma_1^{IJ}  e^m_\a \bg_m e^n_\b \bg_n  \theta_J .
\end{aligned}
\ee

\newpage

\section{Total Lagrangian and field redefinitions}\label{app:total-lagr-field-red}
For convenience, in this appendix we write down explicitly the Lagrangian that is obtained after the field redefinitions~\eqref{eq:red-fer-2x2-sp} and~\eqref{eq:red-bos} have been done. The bosonic-dependent rotation of the fermions~\eqref{eq:red-ferm-Lor-as} has not been implemented yet. The total Lagrangian can be written as the sum of the contribution with the worldsheet metric $\g^{\a\b}$ and the contribution with $\eps^{\a\b}$: $\mathcal{L}^{\g}+\mathcal{L}^{\eps}$.
The first of these results is
\be\label{eq:lagr-gamma-no-F-red}
\begin{aligned}
\mathcal{L}^{\g} = 
& \frac{\tilde{g}}{2}  \ \g^{\a\b} \bar{\theta}_I 
\Bigg[ 
- \frac{i}{2}  \delta^{IJ}  \bg_n 
 -\frac{i}{2} \vk \sigma_3^{IJ}  {\lambda_{n}}^{p} \bg_{p} 
\Bigg] 
 ({k^n}_{m}+{k_{m}}^n) e^m_\a \pa_\b \theta_J  \\
& - \tilde{g} \g^{\a\b} \left(  -  \pa_\a X^M \ \bar{\theta}_I \ \widetilde{G}_{MN} \left( \pa_\b f^N_{IJ} \right) \  \theta_J 
 - \frac{1}{2}  \partial_\alpha X^M \partial_\beta X^N \pa_P \widetilde{G}_{MN} \ \bar{\theta}_I \, f^P_{IJ} \theta_J \right) \\
&+ \frac{\tilde{g}}{4}  \gamma^{\a\b} (k^p_{\ q}  +{k_{q}}^{p} )e^q_{\a} \ \bar{\theta}_I 
\Bigg[ \frac{i}{4} \delta^{IJ} \bg_p \omega^{rs}_\b\bg_{rs}  \\
&  +\frac{1}{8} \left( -\vk \sigma_1^{IJ} -(-1+\sqrt{1+\vk^2} ) \delta^{IJ} \right)  \lambda_{p}^{mn}  \bg_{mn} \  \omega^{rs}_\b \bg_{rs} \\
& - \frac{1}{2}\left( (-1-2\vk^2+\sqrt{1+\vk^2})\delta^{IJ}-\vk(-1+2\sqrt{1+\vk^2}) \sigma_1^{IJ} \right) \ {\lambda_p}^n\bg_n   e^r_\b \bg_r \ \\
& +\frac{i}{4} (\vk \sigma_3^{IJ} - (-1+\sqrt{1+\vk^2})\epsilon^{IJ}) \ {\lambda_p}^n \bg_n \left( \omega^{rs}_\b\bg_{rs}\right)  \\
& +\frac{1}{2} (\vk \sigma_3^{IJ}+\sqrt{1+\vk^2}\epsilon^{IJ}) \bg_p   e^r_\b \bg_r   \\
& -\frac{i}{4} \left( \vk \sigma_3^{IJ} + (-1+ \sqrt{1+\vk^2})\epsilon^{IJ} \right)  \lambda_{p}^{mn}\bg_{mn} e^r_\b \bg_r
 \Bigg]  \theta_J \\
&  + \frac{\tilde{g}}{8}  \gamma^{\a\b} \vk e^v_\a e^m_\b \, {k^{u}}_v {k_m}^n \, \bar{\theta}_I  \\
\Bigg[  
& 2 (\sqrt{1+\vk^2}\delta^{IJ}+\vk \sigma_1^{IJ}) \left( \bg_u\left(\bg_n +\frac{i}{4} \lambda_n^{pq} \bg_{pq} \right)
- \frac{i}{4}  \lambda^{pq}_{u}\bg_{pq} \bg_n\right) \\
&  + \epsilon^{IJ} \left(  \bg_u {\lambda_n}^p \bg_p 
-   {\lambda_u}^p\bg_p \bg_n \right)  \\
&+\left(-(-1+\sqrt{1+\vk^2}) \delta^{IJ} -\vk \sigma_1^{KI}\right) \left( \bg_u -\frac{i}{2} \bg_{pq}  \lambda_u^{pq} \right) \left(\bg_n +\frac{i}{2} \lambda_n^{rs} \bg_{rs}\right) \\
& + \left((1+2\vk^2-\sqrt{1+\vk^2}) \delta^{IJ} -\vk(1-2\sqrt{1+\vk^2}) \sigma_1^{IJ}\right) {\lambda_u}^p\bg_p  {\lambda_n}^r \bg_r \bigg) \\
& +\left( \vk \sigma_3^{IJ}-(-1+\sqrt{1+\vk^2}) \epsilon^{IJ} \right)  {\lambda_u}^p \bg_p\left(\bg_n +\frac{i}{2} \lambda_n^{rs} \bg_{rs}\right) \\
& + \left( \vk \sigma_3^{IJ}+ (-1+\sqrt{1+\vk^2}) \epsilon^{IJ} \right)\left( \bg_u -\frac{i}{2} \bg_{pq}  \lambda_u^{pq} \right) {\lambda_n}^r \bg_r  
\Bigg] \theta_J.
\end{aligned}
\ee
The WZ contribution reads as
\be\label{eq:lagr-epsilon-no-F-red}
\begin{aligned}
\mathcal{L}^{\epsilon} = 
& - \frac{\tilde{g}}{2}  \epsilon^{\a\b} \bar{\theta}_I  \sigma_3^{IJ}  
   \, i \, e^m_\a \bg_m \pa_\b \theta_J  \\
& - \frac{\tilde{g}}{2}  \ \epsilon^{\a\b} \bar{\theta}_I 
\Bigg[ 
- \frac{i}{2}  \delta^{IJ}  \bg_n 
 -\frac{i}{2} \vk \sigma_3^{IJ}  {\lambda_{n}}^{p} \bg_{p} 
\Bigg] 
 ({k^n}_{m}-{k_{m}}^n) e^m_\a \pa_\b \theta_J  \\
& - \tilde{g} \epsilon^{\a\b}   \left( +  \pa_\a X^M \ \bar{\theta}_I \ \widetilde{B}_{MN} \left( \pa_\b f^N_{IJ} \right) \  \theta_J 
 + \frac{1}{2}   \partial_\alpha X^M \partial_\beta X^N \pa_P \widetilde{B}_{MN} \ \bar{\theta}_I \, f^P_{IJ} \theta_J  \right) \\
&-\frac{\tilde{g}}{4}  \epsilon^{\a\b} \bar{\theta}_I \frac{-1+\sqrt{1+\vk^2}}{\vk} \epsilon^{IJ} 
  \pa_\a X^M  \left(\pa_\b  e^m_M   \right)  i\bg_m \theta_J  \\
& - \frac{\tilde{g}}{8}  \epsilon^{\a\b} \bar{\theta}_I \left(- \sigma_3^{IJ}  
 +\frac{\vk}{1+\sqrt{1+\vk^2}} \ \epsilon^{IJ}  \right) \, i \, e^m_\a \bg_m \omega^{np}_\b \bg_{np} \theta_J  \\
&+ \frac{\tilde{g}}{4}  \epsilon^{\a\b} \bar{\theta}_I  \left( \vk \delta^{IJ} +\sqrt{1+\vk^2} \sigma_1^{IJ} \right)  e^m_\a \bg_m e^n_\b \bg_n  \theta_J \\
&- \frac{\tilde{g}}{4}  \epsilon^{\a\b} (k^p_{\ q} - {k_{q}}^{p} )e^q_{\a} \ \bar{\theta}_I 
\Bigg[ \frac{i}{4} \delta^{IJ} \bg_p \omega^{rs}_\b\bg_{rs}  \\
&  +\frac{1}{8} \left( -\vk \sigma_1^{IJ} -(-1+\sqrt{1+\vk^2} ) \delta^{IJ} \right)  \lambda_{p}^{mn}  \bg_{mn} \  \omega^{rs}_\b \bg_{rs} \\
& - \frac{1}{2}\left( (-1-2\vk^2+\sqrt{1+\vk^2})\delta^{IJ}-\vk(-1+2\sqrt{1+\vk^2}) \sigma_1^{IJ} \right) \ {\lambda_p}^n\bg_n   e^r_\b \bg_r \ \\
& +\frac{i}{4} (\vk \sigma_3^{IJ} - (-1+\sqrt{1+\vk^2})\epsilon^{IJ}) \ {\lambda_p}^n \bg_n \left( \omega^{rs}_\b\bg_{rs}\right)  \\
& +\frac{1}{2} (\vk \sigma_3^{IJ}+\sqrt{1+\vk^2}\epsilon^{IJ}) \bg_p   e^r_\b \bg_r   \\
& -\frac{i}{4} \left( \vk \sigma_3^{IJ} + (-1+ \sqrt{1+\vk^2})\epsilon^{IJ} \right)  \lambda_{p}^{mn}\bg_{mn} e^r_\b \bg_r
 \Bigg]  \theta_J \\
& - \frac{\tilde{g}}{8}  \epsilon^{\a\b} \vk e^v_\a e^m_\b \, {k^{u}}_v {k_m}^n \, \bar{\theta}_I  \\
\Bigg[  
& 2 (\sqrt{1+\vk^2}\delta^{IJ}+\vk \sigma_1^{IJ}) \left( \bg_u\left(\bg_n +\frac{i}{4} \lambda_n^{pq} \bg_{pq} \right)
- \frac{i}{4}  \lambda^{pq}_{u}\bg_{pq} \bg_n\right) \\
&  + \epsilon^{IJ} \left(  \bg_u {\lambda_n}^p \bg_p 
-   {\lambda_u}^p\bg_p \bg_n \right)  \\
&+\left(-(-1+\sqrt{1+\vk^2}) \delta^{IJ} -\vk \sigma_1^{KI}\right) \left( \bg_u -\frac{i}{2} \bg_{pq}  \lambda_u^{pq} \right) \left(\bg_n +\frac{i}{2} \lambda_n^{rs} \bg_{rs}\right) \\
& + \left((1+2\vk^2-\sqrt{1+\vk^2}) \delta^{IJ} -\vk(1-2\sqrt{1+\vk^2}) \sigma_1^{IJ}\right) {\lambda_u}^p\bg_p  {\lambda_n}^r \bg_r \bigg) \\
& +\left( \vk \sigma_3^{IJ}-(-1+\sqrt{1+\vk^2}) \epsilon^{IJ} \right)  {\lambda_u}^p \bg_p\left(\bg_n +\frac{i}{2} \lambda_n^{rs} \bg_{rs}\right) \\
& + \left( \vk \sigma_3^{IJ}+ (-1+\sqrt{1+\vk^2}) \epsilon^{IJ} \right)\left( \bg_u -\frac{i}{2} \bg_{pq}  \lambda_u^{pq} \right) {\lambda_n}^r \bg_r  
\Bigg] \theta_J
\end{aligned}
\ee
The function $f^M_{IJ}(X)$ is defined in~\eqref{eq:def-shift-bos-f}.

To implement the bosonic-dependent redefinition on the fermions~\eqref{eq:red-ferm-Lor-as}, we find more efficient to use~\eqref{eq:transf-rule-gamma-ferm-rot} and write its action on the gamma matrices $\bg$. We have for example the rule
\be
\bar{\theta}_K b^m \g_m \theta_I 
\to
\bar{\theta}_K \bar{U}_{(K)} b^m \g_m U_{(I)} \theta_I 
=
\bar{\theta}_K  b^m (\Lambda_{(K)})_m^{\ n} \g_n \bar{U}_{(K)} U_{(I)} \theta_I \,,
\ee
where we have inserted for convenience the identity $U_{(K)} \bar{U}_{(K)} = \gen{1}$.
To give a couple of examples, it means 
\be
\begin{aligned}
\bar{\theta}_1 b^m \g_m \theta_1  &\to \bar{\theta}_1  b^m {(\Lambda_1)_m}^n \g_n  \theta_1
\\
\bar{\theta}_2 b^m \g_m \theta_1  &\to \bar{\theta}_2  b^m {(\Lambda_2)_m}^n \g_n \bar{U}_{(2)} U_{(1)} \theta_1
\end{aligned}
\ee
The terms with derivatives on fermions become (here $I$ is kept fixed)
\be
\bar{\theta}_I b^m \g_m \pa_\b \theta_I 
\to
\bar{\theta}_I  b^m {(\Lambda_{(I)})_m}^n \g_n \pa_\b \theta_I
+\bar{\theta}_I  b^m {(\Lambda_{(I)})_m}^n \g_n (\bar{U}_{(I)} \pa_\b U_{(I)}) \theta_I .
\ee
The second of these terms will contribute to the coupling to the spin connection and the B-field.

To compute these quantities it is useful to know the action of the derivative on the matrix $U_{(I)}$
\be
\begin{aligned}
\bar{U}_{(I)}^{\alg{a}} {\rm d} U_{(I)}^{\alg{a}}  &=\sigma_{3II}\, \frac{\vk}{2} \left(\frac{  \rho  (2 \sin \zeta {\rm d}\rho +\rho  {\rm d}\zeta  \cos \zeta)}{1+\vk ^2 \rho ^4 \sin ^2\zeta} \check{\g}_{23}+
\frac{  {\rm d}\rho }{1- \vk ^2 \rho ^2} \check{\g}_{04} \right), \\
\bar{U}_{(I)}^{\alg{s}} {\rm d} U_{(I)}^{\alg{s}} &= \sigma_{3II}\, \frac{\vk}{2}\left( -\frac{r (2 \sin \xi  {\rm d} r+r {\rm d} \xi  \cos \xi )}{1+\kappa ^2 r^4 \sin ^2\xi } \hat{\g}_{78}
-\frac{{\rm d} r}{1+\kappa ^2 r^2} \hat{\g}_{59} \right),
\end{aligned}
\ee
and also the results for the multiplication of matrices $U_{(I)}$ 
\be
\begin{aligned}
\bar{U}^{\alg{a}}_{(I)} U^{\alg{a}}_{(J)}&=
\delta_{IJ}\mathbf{1}_4+
\frac{\sigma_{1IJ}(\mathbf{1}_4
-i \vk ^2 \rho ^3 \sin \zeta \, \check{\g}_1)
-\epsilon_{IJ} \vk (\rho^2 \sin \zeta \, \check{\g}_{23}+ \rho \, \check{\g}_{04})
}{\sqrt{1-\vk ^2 \rho ^2} \sqrt{1+\vk ^2 \rho ^4\sin ^2\zeta }} , 
\\
\bar{U}^{\alg{s}}_{(I)} U^{\alg{s}}_{(J)}&=
\delta_{IJ}\mathbf{1}_4+
\frac{\sigma_{1IJ}(\mathbf{1}_4
- \vk ^2 r ^3 \sin \xi \, \hat{\g}_6)
+\epsilon_{IJ}\vk ( r^2 \sin \xi \, \hat{\g}_{78}+ r \, \hat{\g}_{59})
}{\sqrt{1+\vk ^2 r ^2} \sqrt{1+\vk ^2 r ^4\sin ^2\xi }} .
\end{aligned}
\ee
As a comment, sometimes it is useful to use redefined coordinates $\rho',\zeta',r',\xi'$ given by
\be
\rho = \vk^{-1} \sin \rho',
\qquad
\sin \zeta = \vk \frac{\sinh \zeta'}{\sin^2 \rho'},
\qquad
r = \vk^{-1} \sinh r',
\qquad
\sin \xi = \vk \frac{\sinh \xi'}{\sinh^2 r'},
\ee
that help to simplify some expressions.

\section{Standard kappa-symmetry}\label{app:standard-kappa-sym}
In this Appendix we compute explicitly the variation for bosonic and fermionic fields in the deformed model.
We show that using~\eqref{eq:eps-op-rho-kappa} their variation is not the standard one~\eqref{eq:kappa-var-32}. However, after implementing the field redefinitions of Section~\ref{sec:canonical-form}---needed to set the terms with derivatives on fermions in the canonical form---they do become standard.
We actually prefer to impose the equation 
\be\label{eq:kappa-var}
\op^{-1}(\alg{g}^{-1} \delta_\kappa \alg{g}) = \varrho \,,
\ee
coming from~\eqref{eq:eps-op-rho-kappa}, where we also used $\varepsilon\equiv \alg{g}^{-1} \delta_\kappa \alg{g}$. The reason is that the computation is then formally the same as the one done in Section~\ref{sec:inverse-op} to derive the results needed to compute the deformed Lagrangian. We just need to do the substitution $\pa_\a \to - \delta_{\kappa}$.
Let us express the result as a linear combination of generators $\gen{P}_m$ and $\gen{Q}^I$
\be
\op^{-1}(\alg{g}^{-1} \delta_\kappa \alg{g}) = j^m_{\delta_{\kappa}} \gen{P}_m + \gen{Q}^I j_{\delta_{\kappa},I}+j^{mn}_{\delta_{\kappa}} \gen{J}_{mn}\,.
\ee
The contributions of the generators $\gen{J}_{mn}$ will not be important for the discussion.
The coefficients $j^m_{\delta_{\kappa}} , j_{\delta_{\kappa},I}$ are the quantities that we need to compute explicitly to discover the form of the action of the kappa-symmetry variation on the fields. Because $\varrho$---standing in the right hand side of~\eqref{eq:kappa-var}---belongs to the odd part of the algebra $\varrho= \gen{Q}^I \psi_I$, it means that we get the equations
\be
j^m_{\delta_{\kappa}}=0, 
\qquad
j_{\delta_{\kappa},I}=\psi_I.
\ee
We may expand the above equations in powers of $\theta$.
We actually stop at leading order in the expansion, meaning that we will compute
\be\label{eq:order-kappa-var}
\begin{aligned}
& j^m_{\delta_{\kappa}}\sim \left[\# +\mathcal{O}(\theta^2)\right]\delta_{\kappa}X+ \left[\# \theta+\mathcal{O}(\theta^3)\right]\delta_{\kappa}\theta,
\\
& j_{\delta_{\kappa},I}\sim \left[\# +\mathcal{O}(\theta^2)\right]\delta_{\kappa}\theta,
\qquad
\psi \sim \left[\# +\mathcal{O}(\theta^2)\right] \kappa,
\end{aligned}
\ee
where $\#$ stands for functions of the bosons,
in such a way that upon solving the equations we get $\delta_{\kappa}X \sim \# \theta \kappa$ and $\delta_{\kappa}\theta \sim \# \kappa$.

Let us start computing $j^m_{\delta_{\kappa}}$. Because of the deformation, the term inside parenthesis proportional to $\gen{Q}^I$ contributes
\be
\begin{aligned}
j^m_{\delta_{\kappa}} \gen{P}_m 
&= -P^{(2)}\circ \frac{1}{\mathbf{1} - \eta R_{\alg{g}} \circ d} \left[ \left( \delta_{\kappa}X^M e^m_M + \frac{i}{2} \bar{\theta}_I \bg^m \delta_{\kappa} \theta_I + \cdots \right) \gen{P}_m -\gen{Q}^I \delta_{\kappa} \theta_I + \cdots\right]
\\
&=  -\delta_{\kappa}X^M e^m_M  {k_m}^q \ \gen{P}_q  \\
&-\frac{1}{2} \ \bar{\theta}_I  \Bigg[ \delta^{IJ} i \bg_p +
 (-\vk \sigma_1^{IJ} +(-1+\sqrt{1+\vk^2})\delta^{IJ}) \left( i\bg_p +\frac{1}{2}  \lambda^{mn}_{ p} \bg_{mn}  \right) \\
& + i \, (\vk \sigma_3^{IJ} -(-1+\sqrt{1+\vk^2})\epsilon^{IJ})  {\lambda_p}^n \bg_n
\Bigg] \delta_{\kappa} \theta_J \ k^{pq} \ \gen{P}_q + \cdots
\end{aligned}
\ee
Imposing the equation $j^m_{\delta_{\kappa}}=0$ and solving for $ \delta_{\kappa}X^M$ at leading order we get
\be
\begin{aligned}
 \delta_{\kappa}X^M =
- \frac{1}{2} \ \bar{\theta}_I e^{Mp} \Bigg[ &\delta^{IJ} i \bg_p +
 (-\vk \sigma_1^{IJ} +(-1+\sqrt{1+\vk^2})\delta^{IJ}) \left( i\bg_p +\frac{1}{2}  \lambda^{mn}_{ p} \bg_{mn}  \right) \\
& + i \, (\vk \sigma_3^{IJ} -(-1+\sqrt{1+\vk^2})\epsilon^{IJ})  {\lambda_p}^n \bg_n
\Bigg] \delta_{\kappa} \theta_J + \cdots.
\end{aligned}
\ee
The computation for $j_{\delta_{\kappa},I} $ gives simply
\be
\begin{aligned}
\gen{Q}^I j_{\delta_{\kappa},I} 
&= (P^{(1)}+P^{(3)}) \circ \frac{1}{\mathbf{1} - \eta R_{\alg{g}} \circ d} \left[  \gen{Q}^I \delta_{\kappa} \theta_I + \cdots\right]
\\
&=   \frac{1}{2}\left( (1+\sqrt{1+\vk^2})\ \delta^{IJ} -\vk \sigma_1^{IJ} \right) \gen{Q}^J \delta_{\kappa} \theta_I + \cdots.
\end{aligned}
\ee
When we compute the two projections of $\varrho$ as defined in~\eqref{eq:def-varrho-kappa-def} at leading order we can set $\theta=0$.
Then we just have
\be
\begin{aligned}
P^{(2)} \circ \op^{-1} A_\b & = P^{(2)} \circ\op^{-1} \left( e^m_\b \gen{P}_m +\cdots \right) = e_{\b m} k^{mn} \gen{P}_n,
\\
P^{(2)} \circ  \optilde^{-1} A_\b & = P^{(2)} \circ  \optilde^{-1} \left( e^m_\b \gen{P}_m +\cdots \right) = e_{\b m} k^{nm} \gen{P}_n,
\end{aligned}
\ee
where the second result can be obtained from the first one sending $\vk \to - \vk$.
Explicitly
\be
\begin{aligned}
\varrho^{(1)} &=\frac{1}{2} (\gamma^{\a\b} - \epsilon^{\a\b})   e_{\b m} k^{mn} \left( \gen{Q}^1 \gen{P}_n + \gen{P}_n \gen{Q}^1\right)\kappa_{\a 1},\\
\varrho^{(3)} &=  \frac{1}{2} (\gamma^{\a\b} + \epsilon^{\a\b})  e_{\b m} k^{nm} \left( \gen{Q}^2 \gen{P}_n + \gen{P}_n \gen{Q}^2\right)\kappa_{\a 2} ,
\end{aligned}
\ee
A direct computation shows that
\be
\gen{Q}^I \check{\gen{P}}_m + \check{\gen{P}}_m \gen{Q}^I = -\frac{1}{2} \gen{Q}^I \check{\bg}_m,
\qquad
\gen{Q}^I \hat{\gen{P}}_m + \hat{\gen{P}}_m \gen{Q}^I = +\frac{1}{2} \gen{Q}^I \hat{\bg}_m.
\ee
We get
\be
\begin{aligned}
\varrho^{(1)} &=\gen{Q}^1 \psi_1, 
\qquad
 \psi_1=\frac{1}{4} (\gamma^{\a\b} - \epsilon^{\a\b})   \left( -e_{\b m} k^{mn} \check{\bg}_n + e_{\b m} k^{mn} \hat{\bg}_n \right)\kappa_{\a 1},\\
\varrho^{(3)} &=\gen{Q}^2 \psi_2,
\qquad
\psi_2=  \frac{1}{4} (\gamma^{\a\b} + \epsilon^{\a\b})    \left( -e_{\b m} k^{nm} \check{\bg}_n + e_{\b m} k^{nm} \hat{\bg}_n \right)\kappa_{\a 2},
\end{aligned}
\ee
and to conclude we can solve the equation $j_{\delta_{\kappa},I}=\psi_I$ setting
\be
\delta_{\kappa} \theta_I =  \frac{1}{1+\sqrt{1+\vk^2}}\left( (1+\sqrt{1+\vk^2})\delta^{IJ} + \vk \sigma_1^{IJ} \right) \psi_J.
\ee
Setting $\vk=0$ the formulas are simplified to 
\be
\begin{aligned}
 \delta_{\kappa}X^M &= -\frac{i}{2} \ \bar{\theta}_I \delta^{IJ} e^{Mp}    \bg_p  \delta_{\kappa} \theta_J + \cdots,
\\
\delta_{\kappa} \theta_I &=  \psi_I,
\\
\psi_1&=\frac{1}{4} (\gamma^{\a\b} - \epsilon^{\a\b})  \left( -e_{\b}^m  \check{\bg}_m + e_{\b}^m \hat{\bg}_m \right)\kappa_{\a 1},
\\
\psi_2&=  \frac{1}{4} (\gamma^{\a\b} + \epsilon^{\a\b})   \left( -e_{\b}^m  \check{\bg}_m + e_{\b}^m \hat{\bg}_m \right)\kappa_{\a 2},
\end{aligned}
\ee
showing that the kappa-symmetry variation is then the standard one that we expect.
The results for the kappa-variations have to be modified according to the field redefinitions needed to put the lagrangian in canonical form. 
When we rotate the fermions we get that their variation is modified as
\be
\theta_I \to U_{IJ} \theta_J
\implies
\delta_{\kappa} \theta_I \to U_{IJ} \delta_{\kappa} \theta_J + \delta_{\kappa} U_{IJ}  \theta_J,
\ee
and since we are considering $\delta_{\kappa} \theta$ at leading order, in the following we will drop the term containing $\delta_{\kappa} U_{IJ}$.
We first redefine our fermions as
\be
\theta_I \to \frac{\sqrt{1+\sqrt{1+\vk^2}}}{\sqrt{2}} \left(\delta^{IJ} + \frac{\vk}{1+\sqrt{1+\vk^2}} \sigma_1^{IJ} \right) \theta_J.
\ee
and we get
\be
\begin{aligned}
 \delta_{\kappa}X^M &=
- \frac{1}{2} \ \bar{\theta}_I e^{Mp} \Bigg[ \delta^{IJ} i \bg_p 
- (\vk \sigma_1^{IJ} +(-1+\sqrt{1+\vk^2})\delta^{IJ}) \frac{1}{2}  \lambda^{mn}_{ p} \bg_{mn}   \\
& + i \, (\vk \sigma_3^{IJ} -(-1+\sqrt{1+\vk^2})\epsilon^{IJ})  {\lambda_p}^n \bg_n
\Bigg] \delta_{\kappa} \theta_J + \cdots,
\\
\delta_{\kappa} \theta_I &=  \sqrt{\frac{2}{1+\sqrt{1+\vk^2}}} \ \psi_I
\end{aligned}
\ee
When we shift the bosons as $X^M \to X^M +\bar{\theta}_I f^M_{IJ} \theta_J$ their variation is modified to $\delta_{\kappa}X^M \to \delta_{\kappa}X^M +2\bar{\theta}_I f^M_{IJ} \delta_{\kappa}\theta_J +\bar{\theta}_I \delta_{\kappa}f^M_{IJ} \theta_J$. Once again, since we are considering the variation at leading order we drop the term with $\delta_{\kappa}f^M_{IJ}$. We use the definition of the function $f^M_{IJ}$ given in~\eqref{eq:def-shift-bos-f}
 and we conclude that, after the shift of the bosons, their variation is
\be
\begin{aligned}
 \delta_{\kappa}X^M &= -2\bar{\theta}_I f^M_{IJ} \delta_{\kappa}\theta_J -
 \frac{1}{2} \ \bar{\theta}_I e^{Mp} \Bigg[\delta^{IJ} i \bg_p 
- (\vk \sigma_1^{IJ} +(-1+\sqrt{1+\vk^2})\delta^{IJ}) \frac{1}{2}  \lambda^{mn}_{ p} \bg_{mn}   \\
& + i \, (\vk \sigma_3^{IJ} -(-1+\sqrt{1+\vk^2})\epsilon^{IJ})  {\lambda_p}^n \bg_n
\Bigg] \delta_{\kappa} \theta_J + \cdots
\\
&= - \frac{i}{2} \ \bar{\theta}_I e^{Mm} \left( \delta^{IJ} \bg_m 
 +  \vk \sigma_3^{IJ}  {\lambda_m}^n \bg_n
\right) \delta_{\kappa} \theta_J + \cdots.
\end{aligned}
\ee
The shift does not affect $\delta_{\kappa} \theta_I$ at leading order.
The final result is obtained by implementing the bosonic-dependent rotation of the fermions~\eqref{eq:red-ferm-Lor-as}
\be
\begin{aligned}
 \delta_{\kappa}X^M &= - \frac{i}{2} \ \bar{\theta}_I \bar{U}_{(I)} \ e^{Mm} \left( \delta^{IJ} \bg_m 
 +  \vk \sigma_3^{IJ}  {\lambda_m}^n \bg_n
\right) \ U_{(I)} \delta_{\kappa} \theta_J + \cdots \\
&= 
- \frac{i}{2} \ \bar{\theta}_I \delta^{IJ} \widetilde{e}^{Mm}  \bg_m  \delta_{\kappa} \theta_J + \cdots,
\\
\delta_{\kappa} \theta_1 &=  \sqrt{\frac{2}{1+\sqrt{1+\vk^2}}} \left(  \frac{1}{4} (\gamma^{\a\b} - \epsilon^{\a\b}) \bar{U}_{(1)}  \left( -\check{e}_{\b m} k^{mn} \check{\bg}_n + \hat{e}_{\b m} k^{mn} \hat{\bg}_n \right)\kappa_{\a 1} \right)
\\
\delta_{\kappa} \theta_2 &=  \sqrt{\frac{2}{1+\sqrt{1+\vk^2}}} \left(\frac{1}{4} (\gamma^{\a\b} + \epsilon^{\a\b})  \bar{U}_{(2)}  \left( -\check{e}_{\b m} k^{nm} \check{\bg}_n + \hat{e}_{\b m} k^{nm} \hat{\bg}_n \right)\kappa_{\a 2} \right)
\end{aligned}
\ee
The variation of the bosons already appears to be related to the one of the fermions in the standard way.
It has actually the same form as in the undeformed case, where one just puts a tilde to get the deformed quantities.
We can achieve the same also for the variation of the fermions if we use the fact that for both expressions
\be
\begin{aligned}
\sqrt{\frac{2}{1+\sqrt{1+\vk^2}}} \  \bar{U}_{(1)}  \left( -\check{e}_{\b m} k^{mn} \check{\bg}_n + \hat{e}_{\b m} k^{mn} \hat{\bg}_n \right)\kappa_{\a 1} = 
\left( -\widetilde{e}_{\b}^m \check{\bg}_m + \widetilde{e}_{\b}^m  \hat{\bg}_m \right) \widetilde{\kappa}_{\a 1}
\\
\sqrt{\frac{2}{1+\sqrt{1+\vk^2}}} \ \bar{U}_{(2)}  \left( -\check{e}_{\b m} k^{nm} \check{\bg}_n + \hat{e}_{\b m} k^{nm} \hat{\bg}_n \right)\kappa_{\a 2} =
\left( -\widetilde{e}_{\b}^m \check{\bg}_m + \widetilde{e}_{\b}^m  \hat{\bg}_m \right) \widetilde{\kappa}_{\a 2}
\end{aligned}
\ee
where we have inserted the identity $\mathbf{1}=U_{(I)}\bar{U}_{(I)}$ and defined 
\be\label{eq:def-kappa-tilde-k-symm}
 \widetilde{\kappa}_{\a I} \equiv \sqrt{\frac{2}{1+\sqrt{1+\vk^2}}} \ \bar{U}_{(I)} \kappa_{\a I}.
\ee
To summarise we have
\be\label{eq:kappa-var-16}
\begin{aligned}
 \delta_{\kappa}X^M &= - \frac{i}{2} \ \bar{\theta}_I \delta^{IJ} \widetilde{e}^{Mm}  \bg_m  \delta_{\kappa} \theta_J + \cdots,
\\
\delta_{\kappa} \theta_I &=    \widetilde{\psi}_I,
\\
\widetilde{\psi}_1&=\frac{1}{4} (\gamma^{\a\b} - \epsilon^{\a\b})  \left( -\widetilde{e}_{\b}^m  \check{\bg}_m + \widetilde{e}_{\b}^m \hat{\bg}_m \right) \widetilde{\kappa}_{\a 1},
\\
\widetilde{\psi}_2&=  \frac{1}{4} (\gamma^{\a\b} + \epsilon^{\a\b})   \left( -\widetilde{e}_{\b}^m  \check{\bg}_m + \widetilde{e}_{\b}^m \hat{\bg}_m \right)\widetilde{\kappa}_{\a 2}.
\end{aligned}
\ee
Also in the deformed case the kappa-symmetry variations can be written in the standard way.
We can rewrite the kappa-symmetry variations in terms of $32$-dimensional fermions $\T$.
To do it we need to introduce $32$-dimensional spinors $\widetilde{K}$ that have the opposite chirality of $\T$
\be
\widetilde{K} \equiv \left( \begin{array}{c} 0 \\ 1 \end{array} \right) \otimes \widetilde{\kappa}.
\ee
The variations written above are then written as
\be
\begin{aligned}
 \delta_{\kappa}X^M &= - \frac{i}{2} \ \bar{\T}_I \delta^{IJ} \widetilde{e}^{Mm}  \G_m  \delta_{\kappa} \T_J + \cdots,
\\
\delta_{\kappa} \T_I &= -\frac{1}{4} (\delta^{IJ} \gamma^{\a\b} - \sigma_3^{IJ} \epsilon^{\a\b})  \widetilde{e}_{\b}^m  \G_m  \widetilde{K}_{\a J}.
\end{aligned}
\ee
Ten-dimensional Gamma-matrices are defined in Appendix~\ref{sec:10-dim-gamma}.
Let us now look at the kappa-variation of the worldsheet metric, whose expression is given in~\eqref{eq:defin-kappa-var-ws-metric}.
The kappa-variation starts at first order in power of fermions. Then we have to compute
\be
\begin{aligned}
P^{(1)}\circ \widetilde{\op}^{-1}( A^{\b}_+ ) &= 
P^{(1)}\circ \widetilde{\op}^{\text{inv}}_{(0)} (- \gen{Q}^{I} \, D^{\b IJ}_+ \theta_J) + P^{(1)}\circ \widetilde{\op}^{\text{inv}}_{(1)} ( e^{m\b}_+\gen{P}_{m}  )+\mathcal{O}(\theta^3)\,,
\\
P^{(3)}\circ {\op}^{-1}( A^{\b}_- ) &= 
P^{(3)}\circ \opinv_{(0)} (- \gen{Q}^{I} \, D^{\b IJ}_- \theta_J) + P^{(3)}\circ \opinv_{(1)} ( e^{m\b}_-\gen{P}_{m}  )+\mathcal{O}(\theta^3)\,.
\end{aligned}
\ee
Let us start from the last line. We have
\be
\begin{aligned}
& P^{(3)}\circ \opinv_{(0)} (- \gen{Q}^{I} \, D^{\b IJ}_- \theta_J) = -\left(\frac{1}{2} (1+\sqrt{1+\vk^2}) \, \delta^{I2}- \frac{\vk}{2} {\sigma_1}^{I2} \, \right) \gen{Q}^{2} D^{\b IJ}_- \theta_J \\
& P^{(3)}\circ \opinv_{(1)} ( e^{m\b}_-\gen{P}_{m}  ) = -\frac{\vk}{4} \gen{Q}^2 \  e^{m\b}_- {k_m}^n \ \Bigg[ 
 \left((1+\sqrt{1+\vk^2})\delta^{2J} -\vk \sigma_1^{2J}\right) \left(i \bg_n - \frac{1}{2} \lambda_n^{pq} \bg_{pq} \right) \\
 &\qquad\qquad + i \left((1+\sqrt{1+\vk^2}) \epsilon^{2J} + \vk \sigma_3^{2J}\right) {\lambda_n}^p \bg_p 
\Bigg] \theta_J\,.
\end{aligned}
\ee
For the first line we can use that $\widetilde{\op}^{\text{inv}}_{(0)}$ and $\opinv_{(0)}$ coincide on odd elements, while on even elements their action is equivalent to sending $\vk\to-\vk$, and we can write
\be
\widetilde{\op}^{\text{inv}}_{(0)}(\gen{Q}^I)=\opinv_{(0)}(\gen{Q}^I)\,,
\qquad
\widetilde{\op}^{\text{inv}}_{(0)}(\gen{P}_m)=k^n_{\ m} \gen{P}_n +\# \gen{J}\,,
\ee
where $k^n_{\ m}=\eta^{nn'}\eta_{mm'} k_{n'}^{\ m'}$.
On the other hand, the action of $\optilde_{(1)}$ on even elements is minus the one of $\op_{(1)}$
\be
\optilde_{(1)}(\gen{P}_m)=-\op_{(1)}(\gen{P}_m)\,.
\ee
These considerations need to be taken into account when computing the action of $\widetilde{\op}^{\text{inv}}_{(1)}$ on $\gen{P}_{m}$.
Then we find
\be
\begin{aligned}
& P^{(1)}\circ \widetilde{\op}^{\text{inv}}_{(0)} (- \gen{Q}^{I} \, D^{\b IJ}_+ \theta_J) = -\left(\frac{1}{2} (1+\sqrt{1+\vk^2}) \, \delta^{I1}- \frac{\vk}{2} {\sigma_1}^{I1} \, \right) \gen{Q}^{1} D^{\b IJ}_+ \theta_J \\
& P^{(1)}\circ \widetilde{\op}^{\text{inv}}_{(1)} ( e^{m\b}_+\gen{P}_{m}  ) = +\frac{\vk}{4} \gen{Q}^1 \  e^{m\b}_+ {k^n}_m \ \Bigg[ 
 \left((1+\sqrt{1+\vk^2})\delta^{1J} -\vk \sigma_1^{1J}\right) \left(i \bg_n - \frac{1}{2} \lambda_n^{pq} \bg_{pq} \right) \\
 &\qquad\qquad + i \left((1+\sqrt{1+\vk^2}) \epsilon^{1J} + \vk \sigma_3^{1J}\right) {\lambda_n}^p \bg_p 
\Bigg] \theta_J\,.
\end{aligned}
\ee
When computing the commutators in~\eqref{eq:defin-kappa-var-ws-metric}, we should care only about the contribution proportional to the identity operator, as the others yield a vanishing contribution after we multiply by $\Upsilon$ and take the supertrace.

We write the result for the variation of the worldsheet metric, after the redefinition~\eqref{eq:red-fer-2x2-sp} has been done
\be
\begin{aligned}
\delta_\kappa \g^{\a\b} &=  \frac{2i\, \sqrt{2}}{\sqrt{1+\sqrt{1+\vk^2}}} \Bigg[
\bar{\kappa}^\a_{1+} \Bigg( \delta^{1J}\pa^{\b}_+  - \frac{1}{4} \delta^{1J} \omega^{\b mn}_+ \bg_{mn} +\frac{i}{2} (\sqrt{1+\vk^2}\eps^{1J}+\vk \sigma_3^{1J} ) e^{m\b}_+ \bg_m
\\
&-\frac{\vk}{2}  e^{m\b}_+ {k^n}_m \  \Bigg(
\delta^{1J} \left(i \bg_n - \frac{1}{2} \lambda_n^{pq} \bg_{pq} \right) 
+ i \left(\sqrt{1+\vk^2} \epsilon^{1J} + \vk \sigma_3^{1J}\right) {\lambda_n}^p \bg_p 
\Bigg)\Bigg)
\\
& \qquad \qquad+\bar{\kappa}^\a_{2-} \Bigg( \delta^{2J}\pa^{\b}_-  - \frac{1}{4} \delta^{2J} \omega^{\b mn}_- \bg_{mn} +\frac{i}{2} (\sqrt{1+\vk^2}\eps^{2J}+\vk \sigma_3^{2J} ) e^{m\b}_- \bg_m
\\
&+\frac{\vk}{2}  e^{m\b}_- {k_m}^n \  \Bigg(
\delta^{2J} \left(i \bg_n - \frac{1}{2} \lambda_n^{pq} \bg_{pq} \right) 
+ i \left(\sqrt{1+\vk^2} \epsilon^{2J} + \vk \sigma_3^{2J}\right) {\lambda_n}^p \bg_p 
\Bigg)\Bigg)\Bigg] \theta_J.
\end{aligned}
\ee
Here we have written the result in terms of $\bar{\kappa}=\kappa^\dagger\bg^0$.
We do not need to take into account the shift of the bosonic fields~\eqref{eq:red-bos}, since it matters at higher orders in fermions.
To take into account the last fermionic field redefinition and write the final form of the variation of the worldsheet metric, we divide the result into ``diagonal''and ``off-diagonal'', where this is meant in the labels $I,J$ for the fermions
\be
\begin{aligned}
\delta_\kappa \g^{\a\b}|_{\text{diag}} &=  2i \Bigg[
\bar{\tilde{\kappa}}^\a_{1+} \Bigg(\pa^{\b}_+ +\bar{U}_{(1)}\pa^{\b}_+U_{(1)}\\
&\qquad  - \frac{1}{4} \left(\omega^{\b mn}_+ (\Lambda_{(1)})_{m}^{\ m'}(\Lambda_{(1)})_{n}^{\ n'}\bg_{m'n'} -\vk e^{m\b}_+ {k^n}_m \lambda_n^{pq}  (\Lambda_{(1)})_{p}^{\ p'}(\Lambda_{(1)})_{q}^{\ q'}\bg_{p'q'}\right) \\
&\qquad
+\frac{i \vk }{2}  e^{m\b}_+ \left((\Lambda_{(1)})_{m}^{\ m'}\bg_{m'} -{k^n}_m \  \left(
 (\Lambda_{(1)})_{n}^{\ n'}\bg_{n'} +   \vk  {\lambda_n}^p (\Lambda_{(1)})_{p}^{\ p'}\bg_{p'} 
\right)\right)
 \Bigg)\theta_1
\\
& \qquad \bar{\tilde{\kappa}}^\a_{2-} \Bigg(\pa^{\b}_- +\bar{U}_{(2)}\pa^{\b}_-U_{(2)}\\
&\qquad  - \frac{1}{4} \left(\omega^{\b mn}_- (\Lambda_{(2)})_{m}^{\ m'}(\Lambda_{(2)})_{n}^{\ n'}\bg_{m'n'} +\vk e^{m\b}_- {k_m}^n \lambda_n^{pq}  (\Lambda_{(2)})_{p}^{\ p'}(\Lambda_{(2)})_{q}^{\ q'}\bg_{p'q'}\right) \\
&\qquad
-\frac{i \vk }{2}  e^{m\b}_- \left((\Lambda_{(2)})_{m}^{\ m'}\bg_{m'} -{k_m}^n \  \left(
 (\Lambda_{(2)})_{n}^{\ n'}\bg_{n'} -   \vk  {\lambda_n}^p (\Lambda_{(2)})_{p}^{\ p'}\bg_{p'} 
\right)\right)
 \Bigg)\theta_2
\Bigg] ,
\end{aligned}
\ee
\be
\begin{aligned}
\delta_\kappa \g^{\a\b}|_{\text{off-diag}} &=   - \sqrt{1+\vk^2}\Bigg[
\bar{\tilde{\kappa}}^\a_{1+}  \bar{U}_{(1)}U_{(2)} e^{m\b}_+ \left(  (\Lambda_{(2)})_{m}^{\ m'}  \bg_{m'}
-\vk   {k^n}_m \      {\lambda_n}^p(\Lambda_{(2)})_{p}^{\ p'} \bg_{p'} \right)\theta_2
\\
& \qquad \qquad-\bar{\tilde{\kappa}}^\a_{2-}  \bar{U}_{(2)}U_{(1)} e^{m\b}_-\left(  (\Lambda_{(1)})_{m}^{\ m'} \bg_{m'}
+\vk   {k_m}^n \      {\lambda_n}^p(\Lambda_{(1)})_{p}^{\ p'} \bg_{p'} \right) \theta_1
\Bigg].
\end{aligned}
\ee
Looking at the diagonal contribution, we find that the expressions containing rank-1 gamma matrices actually vanish, as they should.
The rest yields exactly the couplings that we expect to spin-connection and $H^{(3)}$
\be
\begin{aligned}
\delta_\kappa \g^{\a\b}|_{\text{diag}}&=2i\Bigg[ 
\bar{\tilde{\kappa}}^\a_{1+} \left( \pa^{\b}_+ -\frac{1}{4} \widetilde{\omega}^{\b mn}_+ \bg_{mn} +\frac{1}{8} e^{m\b}_+ \widetilde{H}_{mnp} \bg^{np}\right)\theta_1\\
&\qquad\qquad
+\bar{\tilde{\kappa}}^\a_{2-} \left( \pa^{\b}_- -\frac{1}{4} \widetilde{\omega}^{\b mn}_- \bg_{mn} -\frac{1}{8} e^{m\b}_- \widetilde{H}_{mnp} \bg^{np}\right)\theta_2
 \Bigg]\,.
\end{aligned}
\ee
When we consider the off-diagonal contribution we find that it gives the correct couplings to the RR fields
\be
\begin{aligned}
\delta_\kappa \g^{\a\b}|_{\text{off-diag}}&=2i\left( -\frac{1}{8} e^{\varphi} \right)\Bigg[ 
\bar{\tilde{\kappa}}^\a_{1+} \left( \bg^n \widetilde{F}^{(1)}_n + \frac{1}{3!}\bg^{npq} \widetilde{F}^{(3)}_{npq}+\frac{1}{2\cdot 5!} \bg^{npqrs} \widetilde{F}^{(5)}_{npqrs}\right) e^{m\b}_+ \bg_m\, \theta_2\\
&\qquad\qquad
+\bar{\tilde{\kappa}}^\a_{2-} \left( -\bg^n \widetilde{F}^{(1)}_n + \frac{1}{3!}\bg^{npq} \widetilde{F}^{(3)}_{npq}-\frac{1}{2\cdot 5!} \bg^{npqrs} \widetilde{F}^{(5)}_{npqrs}\right) e^{m\b}_- \bg_m\, \theta_1
 \Bigg]\,,
\end{aligned}
\ee
where the components of the RR fields are given in~\eqref{eq:flat-comp-F1}-\eqref{eq:flat-comp-F3}-\eqref{eq:flat-comp-F5}.
Putting together these results we find a standard kappa-variation also for the worldsheet metric
\be
\begin{aligned}
\delta_\kappa \g^{\a\b}&=2i\Bigg[ 
\bar{\tilde{\kappa}}^\a_{1+} \widetilde{D}^{\b 1J}_+\theta_J+\bar{\tilde{\kappa}}^\a_{2-} \widetilde{D}^{\b 2J}_-\theta_J
 \Bigg] \\
&= 2i\ \Pi^{IJ\, \a\a'}\Pi^{JK\, \b\b'}
\ \bar{\tilde{\kappa}}_{I\a'}\widetilde{D}^{KL}_{\b'}\theta_{L}.
\end{aligned}
\ee
where we defined 
\be
\Pi^{IJ\, \a\a'}\equiv\frac{\delta^{IJ}\g^{\a\a'}+\sigma_3^{IJ}\epsilon^{\a\a'}}{2}\,.
\ee
The rewriting in terms of 32-dimensional spinors is straightforward.

\section{Ten-dimensional $\Gamma$-matrices}\label{sec:10-dim-gamma}
We use the $4\times 4$ gamma matrices $\check{\g}, \hat{\g}$ to define the $32 \times 32$ gamma matrices
\be\label{eq:def-10-dim-Gamma}
\G_m = \sigma_1 \otimes \check{\g}_m \otimes {\bf 1}_4  , \ \ m=0, \cdots, 4,
\qquad
\G_m =  \sigma_2 \otimes {\bf 1}_4 \otimes \hat{\g}_m , \ \ m=5, \cdots, 9,
\ee
that satisfy $\{\G_m,\G_n\}= 2\eta_{mn}$ and also gives $\G_{11} \equiv \G_0 \cdots \G_9 = \sigma_3 \otimes  {\bf 1}_4 \otimes {\bf 1}_4 $. 
Anti-symmetrised products of gamma-matrices are defined as
$\G_{m_1\cdots m_r} = \frac{1}{r!} \G_{[m_1} \cdots \G_{m_r]}$.
The charge conjugation matrix is defined as $\mathcal{C} \equiv  i\, \sigma_2 \otimes K \otimes K$, and $\mathcal{C}^2=-\mathbf{1}_{32}$.
In the chosen representation, the Gamma matrices satisfy the symmetry properties
\be
\begin{aligned}
(\mathcal{C} \G^{(r)})^t &= - t_r^\G \ \mathcal{C} \G^{(r)},
\\
\mathcal{C}( \G^{(r)})^t\mathcal{C} &= - t_r^\G \ \G^{(r)},
\qquad
t_0^\G =t_3^\G =+1,
\qquad
 t_1^\G =t_2^\G=-1.
\end{aligned}
\ee
Under Hermitian conjugation we find
\be
\G^0 (\G^{(r)})^\dagger \G^0 =
\left\{\begin{array}{c}
+ \G^{(r)}, \quad r=1,2 \text{ mod } 4,
\\
- \G^{(r)}, \quad r=0,3 \text{ mod } 4.
\end{array}
\right.
\ee
The $2 \times 2$ space that sits at the beginning is the space of positive/negative chirality. Given two 4-component spinors $\check{\psi}, \hat{\psi}$ with AdS and sphere spinor indices respectively, a 32-component spinor is constructed as
\be
\Psi_+ = \left( \begin{array}{c} 1 \\ 0 \end{array} \right) \otimes \check{\psi} \otimes \hat{\psi} , 
\qquad
\Psi_- = \left( \begin{array}{c} 0 \\ 1 \end{array} \right) \otimes \check{\psi} \otimes \hat{\psi} , 
\ee
for the case of positive and negative chirality respectively.
In the main text we use 16-components fermions with two spinor indices $\theta_{\ul{\a}\ul{a}}$, and we construct a 32-component Majorana fermion with positive chirality as 
\be\label{eq:def-32-dim-Theta}
\Theta = \left( \begin{array}{c} 1 \\ 0 \end{array} \right) \otimes \theta ,
\qquad
\bar{\Theta} = \Theta^t \mathcal{C} =  \left( \  0 \ , \ 1 \  \right) \otimes\bar{\theta} .
\ee
It is also useful to define $16\times 16$-matrices $\bg_m$ (that we continue to call gamma-matrices, even if they don't satisfy a Clifford algebra) as in~\eqref{eq:def16x16-gamma}  that satisfy
\be
\bar{\Theta}_1 \G_m \Theta_2 \equiv \bar{\theta}_1 \bg_m \theta_2 
\implies
\left\{
\begin{array}{rll}
\bg_m &= \check{\g}_m \otimes {\bf 1}_4, & \quad m=0,\cdots 4,
\\
\bg_m &= {\bf 1}_4 \otimes i\hat{\g}_m, & \quad m=5,\cdots 9,
\end{array}
\right.
\ee
The above formulae explain the reason for the factor of $i$ in the definition of $\bg_m$ for the sphere. In the same way we can explain why there is a $+$ sign and not $-$ in the definition of $\bg_{mn}$ for the sphere, computing\footnote{When we consider even rank $\G$-matrices, we need to insert also an odd rank $\G$-matrix in order not to get 0 when $\Theta_{1,2}$ have the same chirality.}
\be
\bar{\Theta}_1 \G_{p} \G_{mn} \Theta_2 \equiv \bar{\theta}_1 \bg_{p} \bg_{mn} \theta_2 
\implies
\left\{
\begin{array}{rll}
\bg_{mn} &= \check{\g}_{mn} \otimes {\bf 1}_4, & \quad m,n=0,\cdots 4,
\\
\bg_{mn} &= {\bf 1}_4 \otimes \hat{\g}_{mn}, & \quad m,n=5,\cdots 9,
\\
\bg_{mn} &= -\check{\g}_{m} \otimes i\hat{\g}_{n}, & \quad m=0,\cdots 4, \ n=5,\cdots 9.
\end{array}
\right.
\ee
Similarly, for rank-3 Gamma matrices we would obtain
\be
\bar{\Theta}_1 \G_{mnp} \Theta_2 \equiv \bar{\theta}_1 \bg_{mnp} \theta_2 
\implies
\left\{
\begin{array}{rll}
\bg_{mnp} &= \check{\g}_{mnp} \otimes {\bf 1}_4, & \quad m,n,p=0,\cdots 4,
\\
\bg_{mnp} &= {\bf 1}_4 \otimes i\hat{\g}_{mnp}, & \quad m,n,p=5,\cdots 9,
\\
\bg_{mnp} &= \frac{1}{3} \check{\g}_{mn} \otimes i\hat{\g}_{p}, & \quad m,n=0,\cdots 4, \ p=5,\cdots 9,
\\
\bg_{mnp} &= \frac{1}{3} \check{\g}_{p} \otimes \hat{\g}_{mn}, & \quad p=0,\cdots 4, \ m,n=5,\cdots 9.
\end{array}
\right.
\ee

\newpage
\pagestyle{plain}

\chapter*{Acknowledgements}
\addcontentsline{toc}{chapter}{Acknowledgements}

Many people deserve my gratitude for helping me to carry out my PhD in Utrecht.
Here I want to thank in particular the collaborators I have had during those years: Gleb Arutyunov, Sergey Frolov, Olof Ohlsson Sax, Alessandro Sfondrini, Bogdan Stefa\'nski and Alessandro Torrielli.
A special thank goes to my supervisor Gleb Arutyunov, who gave me the chance to be one of his students and offered me his constant guidance and support. 
I wish to thank also Arkady Tseytlin and Kostya Zarembo for granting me the opportunity of continuing to do research.

\vspace{12pt}

The author acknowledges funding from the ERC Advanced grant No.290456.
This review is based on the author's PhD thesis, written at the Institute for Theoretical Physics of Utrecht University and defended on 7th September 2015.
The work was supported by the Netherlands Organization for Scientific Research (NWO) under the VICI grant 680-47-602.  The work was also part of the ERC Advanced grant research programme No.  246974, ``Supersymmetry:  a window to non-perturbative physics'', and of the D-ITP consortium, a program of the NWO that is funded by the Dutch Ministry ofEducation, Culture and Science (OCW). 

\vspace{12pt}

A version of this PhD thesis can be found also in the repository of Utrecht University at
\url{http://dspace.library.uu.nl/handle/1874/323083}
\begin{flushright} 
$\Box$ 
\end{flushright}

\bibliographystyle{nb}
\bibliography{PhDThesis}{}
\addcontentsline{toc}{chapter}{Bibliography}

\end{document}